%% file: RNC_review.tex
\RequirePackage{fix-cm}
\documentclass[12pt]{article}
\pdfoutput=1
\usepackage{authblk} 
\usepackage[margin=3.05cm]{geometry}
\usepackage[utf8]{inputenc}
\usepackage{graphicx}
\usepackage{mathptmx}      
\usepackage{amssymb,amsmath}
\usepackage{enumitem}
\usepackage{graphicx}
\usepackage{pgf}
\usepackage{xcolor}
\usepackage{physics}
\usepackage{verbatim}
\usepackage{cancel}
\usepackage{wrapfig}
\usepackage{comment}
\usepackage{latexsym}
\usepackage{mathtools}
\usepackage{dsfont}
\usepackage{mathrsfs}
\usepackage{float}
\usepackage{framed}
\usepackage{amsfonts}
\usepackage{ytableau}

\usepackage{pgf} 

\definecolor{tealblue}{rgb}{0.21, 0.46, 0.53}
\definecolor{mred}{rgb}{0.5, 0.0, 0.0}
\usepackage[colorlinks=true,linkcolor=tealblue,urlcolor=tealblue,citecolor=tealblue,allcolors=tealblue]{hyperref}

\usepackage[framemethod=default]{mdframed}
\newmdenv[skipabove=10pt,
skipbelow=7pt,
rightline=false,
leftline=true,
topline=false,
bottomline=false,
linecolor=tealblue,
backgroundcolor=tealblue!5,
innerleftmargin=4pt,
innerrightmargin=10pt,
innertopmargin=0pt,
leftmargin=2pt,
rightmargin=0pt,
linewidth=2pt,
innerbottommargin=-10pt]{lbBox}
\newenvironment{importantbox}{\begin{lbBox}\vspace{1 mm}
	} {\vspace{1.5 mm}\end{lbBox}}

\usepackage[
backend=bibtex,
sorting=none,
style=phys,
eprint=true,
doi=false,
biblabel=brackets
]{biblatex}

\bibliography{RNC_review.bib}

\newcommand*{\bfrac}[2]{\genfrac{}{}{0pt}{}{#1}{#2}}
\newcommand{\ds}{\ensuremath{\operatorname{dS}}}

\newcommand{\commie}[1]{}
\newcommand{\ads}{\mathrm{AdS}}
\newcommand{\ess}{\mathbb{S}}
\newcommand{\Vol}{\mathrm{Vol}}

\newcommand{\adsts}{{\ads \times \ess}}

\newcommand{\beq}{\begin{equation}}
\newcommand{\eeq}{\end{equation}}
\newcommand{\bea}{\begin{eqnarray}}
\newcommand{\eea}{\end{eqnarray}}

\numberwithin{equation}{section}

\newenvironment{eqaed}
{\begin{equation}
	\begin{aligned}
}
{ 
	\end{aligned}
	\end{equation}
}

\begin{document}

\title{{\bf Supersymmetry~Breaking~and  Stability~in~String~Vacua}\\ {\small \sc Brane dynamics, bubbles and the swampland}
}






\author{Ivano Basile\thanks{ivano.basile@umons.ac.be}}
\affil{\emph{Service de Physique de l'Univers, Champs et Gravitation\\ \emph{Universit\'{e} de Mons}}\\\emph{Place du Parc 20, 7000 Mons, Belgium}}

\maketitle

\begin{abstract}
We review some aspects of the dramatic consequences of supersymmetry breaking on string vacua. In particular, we focus on the issue of vacuum stability in ten-dimensional string models with broken, or without, supersymmetry, whose perturbative spectra are free of tachyons. After formulating the models at stake, we introduce their unified low-energy effective description and present a number of vacuum solutions to the classical equations of motion. In addition, we present a generalization of previous no-go results for de Sitter vacua in warped flux compactifications. Then we analyze the classical and quantum stability of these vacua, studying linearized field fluctuations and bubble nucleation. Then, we describe how the resulting instabilities can be framed in terms of brane dynamics, examining in particular brane interactions, back-reacted geometries and commenting on a brane-world string construction along the lines of a recent proposal. After providing a summary, we conclude with some perspectives on possible future developments.
\end{abstract}

{\hypersetup{linkcolor=black}
\tableofcontents
}

\commie{
\section{Introduction}
\label{intro}
Your text comes here. Separate text sections with
\section{Section title}
\label{sec:1}
Text with citations \cite{Antonelli:2018qwz}.
\subsection{Subsection title}
\label{sec:2}
as required. Don't forget to give each section
and subsection a unique label (see Sect.~\ref{sec:1}).

\begin{equation}
a^2+b^2=c^2
\end{equation}

\begin{figure}
  \includegraphics{example.eps}
\caption{Please write your figure caption here}
\label{fig:1}       
\end{figure}
%
\begin{figure*}
  \includegraphics[width=0.75\textwidth]{example.eps}
\caption{Please write your figure caption here}
\label{fig:2}       
\end{figure*}
%
\begin{table}
\caption{Please write your table caption here}
\label{tab:1}       
\begin{tabular}{lll}
\hline\noalign{\smallskip}
first & second & third  \\
\noalign{\smallskip}\hline\noalign{\smallskip}
number & number & number \\
number & number & number \\
\noalign{\smallskip}\hline
\end{tabular}
\end{table}
}


\section{Introduction}\label{Section0}

The issue of supersymmetry breaking in string theory is of vital importance, both technically and conceptually. On a foundational level, many of the richest and most illuminating lessons appear obscured by a lack of solid, comprehensive formulations and of befitting means to explore these issues in depth. As a result, unifying guiding principles to oversee our efforts have been elusive, although a variety of successful complementary frameworks~\cite{Polyakov:1981rd, Polyakov:1981re, Callan:1985ia, Maldacena:1997re, Banks:1996vh} hint at a unique, if tantalizing, consistent structure~\cite{Witten:1995ex}. Despite these shortcomings, string theory has surely provided a remarkable breadth of new ideas and perspectives to theoretical physics, and one can argue that its relevance as a framework has thus been established to a large extent, notwithstanding its eventual vindication as a realistic description of our universe. On a more phenomenological level, the absence of low-energy supersymmetry and the extensive variety of mechanisms to break it, and consequently the wide range of relevant energy scales, point to a deeper conundrum, whose resolution would conceivably involve qualitatively novel insights. However, the paradigm of spontaneous symmetry breaking in gauge theories has proven pivotal in model building, both in particle physics and condensed matter physics, and thus it is natural to envision spontaneous supersymmetry breaking as an elegant resolution of these bewildering issues. Yet, in the context of string theory this phenomenon could in principle occur around the string scale, perhaps even naturally so, and while the resulting dramatic consequences have been investigated for a long time, the ultimate fate of these settings appears still largely not under control. 

All in all, a deeper understanding of the subtle issues of supersymmetry breaking in string theory is paramount to progress toward a more complete picture of its underlying foundational principles and more realistic phenomenological models. While approaches based on string world-sheets would appear to offer a more fundamental perspective, the resulting analyses are typically met by gravitational tadpoles, which signal an incongruous starting point of the perturbative expansion and whose resummation entails a number of technical and conceptual subtleties~\cite{Fischler:1986ci, Fischler:1986tb, Dudas:2004nd, Kitazawa:2008hv, Pius:2014gza}. On the other hand, low-energy effective theories appear more tractable in this respect, but connecting the resulting lessons to the underlying microscopic physics tends to be more intricate. A tempting analogy for the present state of affairs would compare current knowledge to the coastline of an unexplored island, whose internal regions remain unscathed by any attempt to further explore them.

Nevertheless, the material presented in this review is motivated by an attempt to shed some light on these remarkably subtle issues. Indeed, as we shall discuss, low-energy effective theories, accompanied by some intuition drawn from well-understood supersymmetric settings, appear to provide the tools necessary to elucidate matters, at least to some extent. A detailed analysis of the resulting models, and in particular of their classical solutions and the corresponding instabilities, suggests that fundamental branes play a crucial rôle in unveiling the microscopic physics at stake. Both the relevant space-time field configurations and their (classical and quantum) instabilities dovetail with a brane-based interpretation, whereby controlled flux compactifications arise as near-horizon limits within back-reacted geometries, strongly warped regions arise as confines of the space-time ``carved out'' by the branes in the presence of runaway tendencies, and instabilities arise from brane interactions. In addition to provide a vantage point to build intuition from, the rich dynamics of fundamental branes offers potentially fruitful avenues of quantitative investigation via world-volume gauge theories and holographic approaches. Furthermore, settings of this type naturally accommodate cosmological brane-world scenarios alongside the simpler bulk cosmologies that have been analyzed, and the resulting models offer a novel and intriguing perspective on the long-standing problem of dark energy in string theory. Indeed, many of the controversies regarding the ideas that have been put forth in this respect~\cite{Kachru:2002gs, Kachru:2003aw, Gautason:2018gln, Hamada:2019ack, Gautason:2019jwq} point to a common origin, namely an attempt to impose static configurations on systems naturally driven toward dynamics. As a result, uncontrolled back-reactions and instabilities can arise, and elucidating the aftermath of their manifestation has proven challenging.

While in supersymmetric settings the lack of a selection principle generates seemingly unfathomable ``landscapes'' of available models, in the absence of supersymmetry their very consistency has been questioned, leading to the formulation of a number of criteria and proposals collectively dubbed ``swampland conjectures''~\cite{Ooguri:2016pdq, Brennan:2017rbf, Obied:2018sgi, Palti:2019pca}. Among the most ubiquitous stands the weak gravity conjecture~\cite{ArkaniHamed:2006dz}, which appears to entail far-reaching implications concerning the nature of quantum-gravitational theories in general. In this review we shall approach matters from a complementary viewpoint, but, as we shall discuss, the emerging lessons resonate with the results of ``bottom-up'' programs of this type. Altogether, the indications that we have garnered appear to portray an enticing, if still embryonic, picture of dynamics as a fruitful selection mechanism for more realistic models and as a rich area to investigate on a more foundational level, and to this end a deeper understanding of high-energy supersymmetry breaking would constitute an invaluable asset to string theory insofar as we grasp it at present.

\paragraph{Synopsis.---}\label{sec:synopsis}

The material presented in this review mainly covers the results of~\cite{Basile:2018irz, Antonelli:2019nar, Basile:2020mpt, Basile:2020xwi} within the larger context of supersymmetry breaking in string theory. Its contents are organized as follows.

We shall begin in Section~\ref{Section1} with an overview of the formalism of vacuum amplitudes in string theory, and the construction of three ten-dimensional string models with broken supersymmetry. These comprise two orientifold models, the $USp(32)$ model of~\cite{Sugimoto:1999tx} and the $U(32)$ model of~\cite{Sagnotti:1995ga, Sagnotti:1996qj}, and the $SO(16) \times SO(16)$ heterotic model of~\cite{AlvarezGaume:1986jb, Dixon:1986iz}, and their perturbative spectra feature no tachyons. Despite this remarkable property, these models also exhibit gravitational tadpoles, whose low-energy imprint includes an exponential potential which entails runaway tendencies. The remainder of this review is focused on investigating the consequences of this feature, and whether interesting phenomenological scenarios can arise as a result.

In Section~\ref{Section2} we shall describe a family of effective theories which encodes the low-energy physics of the string models that we have introduced in Section~\ref{Section1}, and we present a number of solutions to the corresponding equations of motion. In order to balance the runaway effects of the dilaton potential, the resulting field profiles can be warped~\cite{Dudas:2000ff, Antonelli:2019nar} or involve large fluxes~\cite{Mourad:2016xbk}. Then we address the issue of $\ds$ cosmology, considering warped flux compactifications and extending the no-go results of~\cite{Gibbons:1984kp, Maldacena:2000mw}. We conclude discussing how our findings connect with recent swampland conjectures~\cite{Ooguri:2006in, Obied:2018sgi, Ooguri:2018wrx, Lust:2019zwm}.

In Section~\ref{Section3} we shall present a detailed analysis of the classical stability of the Dudas-Mourad solutions of~\cite{Dudas:2000ff} and of the $\adsts$ solutions of~\cite{Mourad:2016xbk}. To this end, we shall derive the linearized equations of motion for field perturbations, and obtain criteria for the stability of modes. In the case of the cosmological Dudas-Mourad solutions, an intriguing instability of the homogeneous tensor mode emerges~\cite{Basile:2018irz}, and we offer as an enticing, if speculative, explanation a potential tendency of space-time toward spontaneous compactification.

In Section~\ref{Section4} we shall turn to the non-perturbative instabilities of the $\ads$ compactifications discussed in Section~\ref{Section2}, in which charged membranes nucleate~\cite{Antonelli:2019nar} reducing the flux in the space-time inside of them. We shall compute the decay rate associated to this process, and frame it in terms of fundamental branes via consistency conditions that we shall derive and discuss.

In Section~\ref{Section5} we shall further develop the brane picture presented in Section~\ref{Section4}, starting from the Lorentzian expansion that bubbles undergo after nucleation. The potential that drives the expansion encodes a renormalized charge-to-tension ratio that is consistent with the weak gravity conjecture. In addition, we shall comment on a string-theoretic embedding of the brane-world scenarios recently revisited in~\cite{Banerjee:2018qey, Banerjee:2019fzz, Banerjee:2020wix, Basile:2020mpt}. Then we shall turn to the gravitational back-reaction of the branes, studying the resulting near-horizon and asymptotic geometries. In the near-horizon limit we shall recover $\adsts$ throats, while the asymptotic region features a ``pinch-off'' singularity at a finite distance, mirroring the considerations of~\cite{Dudas:2000ff}.

In Section~\ref{Section8} we provide a summary and collect some concluding remarks.

\section{String models with broken supersymmetry}\label{Section1}

In this section we introduce the string models with broken supersymmetry that we shall investigate in the remainder of this review. To this end, we begin in Section~\ref{sec:vacuum_amplitudes} with a review of one-loop vacuum amplitudes in string theory, starting from the supersymmetric ten-dimensional models. Then, in Section~\ref{sec:orientifold_models} we introduce orientifold models, or ``open descendants'', within the formalism of vacuum amplitudes, focusing on the $USp(32)$ model~\cite{Sugimoto:1999tx} and the $U(32)$ model~\cite{Sagnotti:1995ga, Sagnotti:1996qj}. While the latter features a non-supersymmetric perturbative spectrum without tachyons, the former is particularly intriguing, since it realizes supersymmetry non-linearly in the open sector~\cite{Antoniadis:1999xk, Angelantonj:1999jh, Aldazabal:1999jr, Angelantonj:1999ms}. Finally, in Section~\ref{sec:heterotic_model} we move on to heterotic models, constructing the non-supersymmetric $SO(16) \times SO(16)$ projection~\cite{AlvarezGaume:1986jb, Dixon:1986iz}. The material presented in this section is largely based on~\cite{Angelantonj:2002ct}. For a more recent review, see~\cite{Mourad:2017rrl}.

\subsection{Vacuum amplitudes}\label{sec:vacuum_amplitudes}

Vacuum amplitudes probe some of the most basic aspects of quantum systems. In the functional formulation, they can be computed evaluating the effective action $\Gamma$ on vacuum configurations. While in the absence of supersymmetry or integrability exact results are generally out of reach, their one-loop approximation only depends on the perturbative excitations around a classical vacuum. In terms of the corresponding mass operator $M^2$, one can write integrals over Schwinger parameters of the form
\begin{eqaed}\label{eq:gamma_str_mod}
	\Gamma = - \, \frac{\text{Vol}}{2 \left( 4\pi \right)^{\frac{D}{2}}} \int_{\Lambda^{-2}}^\infty \frac{dt}{t^{\frac{D}{2} + 1}} \, \text{STr} \left( e^{- t M^2} \right) \, ,
\end{eqaed}
where $\text{Vol}$ is the volume of (Euclidean) $D$-dimensional space-time, and the supertrace $\mathrm{Str}$ sums over \textit{signed} polarizations, \textit{i.e.} with a minus sign for fermions. The UV divergence associated to small values of the world-line proper time $t$ is regularized by the cut-off scale $\Lambda$.

\begin{figure}[!th]
	\centering
	\includegraphics[scale=0.15]{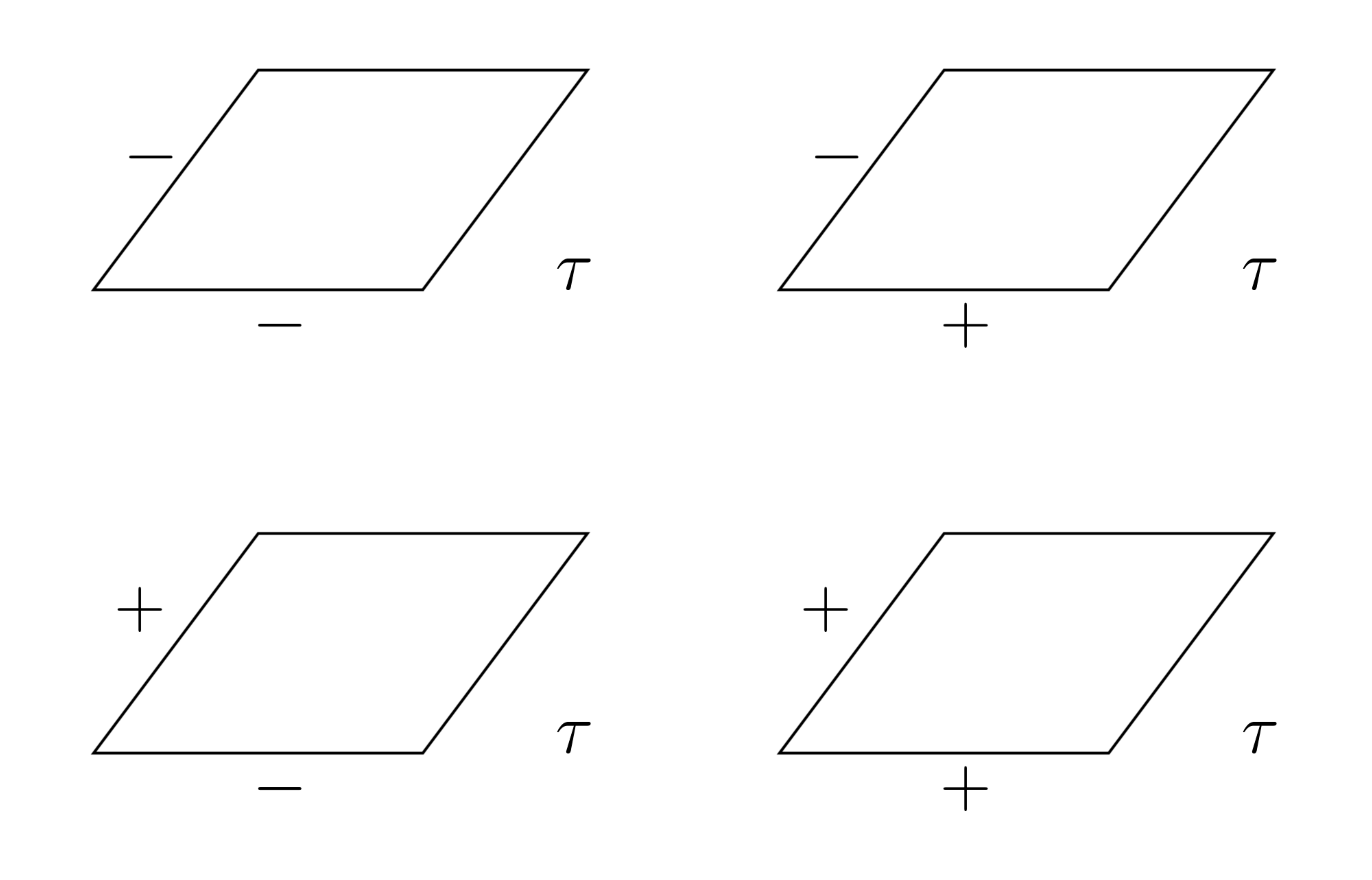}
	\caption[Spin_structures]{inequivalent spin structures on the torus, specified by a choice of periodic ($-$) or anti-periodic ($+$) conditions along each independent cycle.}
	\label{fig:spin_structures}
\end{figure}

Due to modular invariance\footnote{We remark that, in this context, modular invariance arises as the residual gauge invariance left after fixing world-sheet diffeomorphisms and Weyl rescalings. Hence, violations of modular invariance would result in gauge anomalies.}, one-loop vacuum amplitudes in string theory can be recast as integrals over the moduli space of Riemann surfaces with vanishing Euler characteristic, and the corresponding integrands can be interpreted as partition functions of the world-sheet conformal field theory. Specifically, in the case of a torus with modular parameter $q \equiv e^{2\pi i \tau}$, in the RNS light-cone formalism one ought to consider\footnote{We work in ten space-time dimensions, since non-critical string perturbation theory entails a number of challenges.} (combinations of) the four basic traces
\begin{eqaed}\label{eq:z--_z+-_mod}
	Z_{(- -)}(\tau) & \equiv \text{Tr}_{\text{NS}} \, q^{L_0} = \frac{\prod_{m=1}^\infty \left(1 + q^{m - \frac{1}{2} }\right)^8}{q^{\frac{1}{2}} \prod_{n=1}^\infty \left(1 - q^n\right)^8 } \, , \\
	Z_{(+ -)}(\tau) & \equiv \text{Tr}_{\text{R}} \, q^{L_0} = 16 \, \frac{\prod_{m=1}^\infty \left(1 + q^m \right)^8}{\prod_{n=1}^\infty \left(1 - q^n\right)^8 } \, , \\
	Z_{(- +)}(\tau) & \equiv \text{Tr}_{\text{NS}} \left( (-1)^F \, q^{L_0} \right) = \frac{\prod_{m=1}^\infty \left(1 - q^{m - \frac{1}{2}} \right)^8}{q^{\frac{1}{2}} \prod_{n=1}^\infty \left( 1 - q^n \right)^8} \, , \\
	Z_{(+ +)}(\tau) & \equiv \text{Tr}_{\text{R}} \left( (-1)^F \, q^{L_0} \right) = 0 \, ,
\end{eqaed}
which arise from the four spin structures depicted in fig.~\ref{fig:spin_structures}. The latter two correspond to ``twisted'' boundary conditions for the world-sheet fermions, and are implemented inserting the fermion parity operator $(-1)^F$. While $Z_{(+ +)}$ vanishes, its structure contains non-trivial information about perturbative states, and its modular properties are needed in order to build consistent models.

The modular properties of the traces in eq.~\eqref{eq:z--_z+-_mod} can be highlighted recasting them in terms of the \textit{Dedekind $\eta$ function}
\begin{eqaed}\label{eq:dedekind_eta_mod}
	\eta(\tau) \equiv q^{\frac{1}{24}} \prod_{n=1}^\infty \left(1 - q^n \right) \, ,
\end{eqaed}
which transforms according to
\begin{eqaed}\label{eq:eta_t_s_mod}
	\eta(\tau + 1) = e^{\frac{i \pi}{12}} \, \eta(\tau) \, , \qquad \eta\left( - \, \frac{1}{\tau} \right) = \left(-i \tau \right)^{\frac{1}{2}} \eta(\tau)
\end{eqaed}
under the action of the generators 
\begin{eqaed}\label{eq:s_t_generators}
	T \; : \; \tau \; \to \; \tau + 1 \, , \qquad S \; : \; \tau \; \to \; - \, \frac{1}{\tau}
\end{eqaed}
of the modular group on the torus, and the Jacobi $\vartheta$ functions. The latter afford both the series representation~\cite{whittaker1996course}
\begin{eqaed}\label{eq:jacobi_theta_sum_mod}
	\vartheta \left[\bfrac{\alpha}{\beta}\right] \!\left( z | \tau \right) \equiv \sum_{n \in \mathbb{Z}} q^{\frac{1}{2} \left( n + \alpha \right)^2} e^{2\pi i \left(n + \alpha \right) \left(z - \beta \right)}
\end{eqaed}
and the infinite product representation
\begin{eqaed}\label{eq:jacobi_theta_product_mod}
	\vartheta \left[\bfrac{\alpha}{\beta}\right] \!\left( z | \tau \right) = \; & e^{2\pi i \alpha \left(z - \beta \right)} \, q^{\frac{\alpha^2}{2}} \prod_{n=1}^\infty \left( 1 - q^n \right) \\
	& \times \left(1 + q^{n + \alpha - \frac{1}{2}} \, e^{2\pi i \left(z - \beta \right)} \right) \left(1 + q^{n - \alpha - \frac{1}{2}} \, e^{-2\pi i \left(z - \beta \right)} \right) \, ,
\end{eqaed}
and they transform under the action of $T$ and $S$ according to
\begin{eqaed}\label{eq:modulart_modulars_mod}
	\vartheta \left[\bfrac{\alpha}{\beta}\right] \!\left( z | \tau + 1 \right) & = e^{-i\pi \alpha \left( \alpha + 1 \right)} \, \vartheta \left[\bfrac{\alpha}{\beta - \alpha - \frac{1}{2}}\right] \!\left( z | \tau \right) \, , \\
	\vartheta \left[\bfrac{\alpha}{\beta}\right] \!\left( \frac{z}{\tau} \bigg| - \, \frac{1}{\tau} \right) & = \left(-i \tau \right)^{\frac{1}{2}} \, e^{-2\pi i \alpha \beta + \frac{i \pi z^2}{\tau}} \, \vartheta \left[\bfrac{- \beta}{\alpha} \right] \!\left( z | \tau \right) \, .
\end{eqaed}
Therefore, both the Dedekind $\eta$ function and the Jacobi $\vartheta$ functions are modular forms of weight $\frac{1}{2}$. In particular, we shall make use of $\vartheta$ functions evaluated at $z = 0$ and $\alpha \, ,\beta \in \{0 \, , \, \frac{1}{2} \}$, which are commonly termed Jacobi constants\footnote{Non-vanishing values of the argument $z$ of Jacobi $\vartheta$ functions are nonetheless useful in string theory. They are involved, for instance, in the study of string perturbation theory on more general backgrounds and D-brane scattering.}. Using these ingredients, one can recast the traces in eq.~\eqref{eq:z--_z+-_mod} in the form
\begin{eqaed}\label{eq:z_theta1_theta2_mod}
	Z_{(- -)}(\tau) & = \frac{\vartheta^4 \left[\bfrac{0}{0}\right] \!\left( 0 | \tau \right)}{\eta^{12}(\tau)} \, , \qquad Z_{(+ -)}(\tau) = \frac{\vartheta^4 \left[\bfrac{0}{\frac{1}{2} }\right] \!\left( 0 | \tau \right)}{\eta^{12}(\tau)} \, , \\
	Z_{(- +)}(\tau) & = \frac{\vartheta^4 \left[\bfrac{\frac{1}{2}}{0}\right] \!\left( 0 | \tau \right)}{\eta^{12}(\tau)} \, , \qquad Z_{(+ +)}(\tau) = \frac{\vartheta^4 \left[\bfrac{\frac{1}{2}}{\frac{1}{2}}\right] \!\left( 0 | \tau \right)}{\eta^{12}(\tau)} \, ,
\end{eqaed}
and, in order to obtain the corresponding (level-matched) torus amplitudes, one is to integrate products of left-moving holomorphic and right-moving anti-holomorphic contributions over the fundamental domain $\mathcal{F}$ with respect to the modular invariant measure $\frac{d^2\tau}{\Im(\tau)^2}$. The absence of the UV region from the fundamental domain betrays a striking departure from standard field-theoretic results, and arises from the gauge-fixing procedure in the Polyakov functional integral.

All in all, modular invariance is required by consistency, and the resulting amplitudes are constrained to the extent that the perturbative spectra of consistent models are fully determined. In order to elucidate their properties, it is quite convenient to introduce the characters of the level-one affine $\mathfrak{so}(2n)$ algebra
\begin{eqaed}\label{eq:o2n_v2n_s2n_c2n}
	O_{2n} & \equiv \frac{\vartheta^n \left[\bfrac{0}{0}\right] \!\left( 0 | \tau \right) + \vartheta^n \left[\bfrac{0}{\frac{1}{2}}\right] \!\left( 0 | \tau \right)}{2\eta^n(\tau)} \, , \\
	V_{2n} & \equiv \frac{\vartheta^n \left[\bfrac{0}{0}\right] \!\left( 0 | \tau \right) - \vartheta^n \left[\bfrac{0}{\frac{1}{2}}\right] \!\left( 0 | \tau \right)}{2\eta^n(\tau)} \, , \\
	S_{2n} & \equiv \frac{\vartheta^n \left[\bfrac{\frac{1}{2}}{0}\right] \!\left( 0 | \tau \right) + i^{- n} \, \vartheta^n \left[\bfrac{\frac{1}{2}}{\frac{1}{2}}\right] \!\left( 0 | \tau \right)}{2\eta^n(\tau)} \, , \\
	C_{2n} & \equiv \frac{\vartheta^n \left[\bfrac{\frac{1}{2}}{0}\right] \!\left( 0 | \tau \right) - i^{- n} \, \vartheta^n \left[\bfrac{\frac{1}{2}}{\frac{1}{2}}\right] \!\left( 0 | \tau \right)}{2\eta^n(\tau)} \, ,
\end{eqaed}
which comprise contributions from states pertaining to the four conjugacy classes of $SO(2n)$. Furthermore, they also inherit the modular properties from $\vartheta$ and $\eta$ functions, reducing the problem of building consistent models to matters of linear algebra\footnote{We remark that different combinations of characters reflect different projections at the level of the Hilbert space.}. While $n = 4$ in the present case, the general expressions can also encompass heterotic models, whose right-moving sector is built from $26$-dimensional bosonic strings. As we have anticipated, these expressions ought to be taken in a formal sense: if one were to consider their actual value, one would find for instance the \textit{numerical} equivalence $S_8 = C_8$, while the two corresponding sectors of the Hilbert space are distinguished by the chirality of space-time fermionic excitations. Moreover, a remarkable identity proved by Jacobi~\cite{whittaker1996course} implies that
\begin{eqaed}\label{eq:aequatio_mod}
	V_8 = S_8 = C_8 \, .
\end{eqaed}
This peculiar identity was referred to by Jacobi as \textit{aequatio identica satis abstrusa}, but in the context of superstrings its meaning becomes apparent: it states that string models built using an $SO(8)$ vector and a $SO(8)$ Majorana-Weyl spinor, which constitute the degrees of freedom of a ten-dimensional supersymmetric Yang-Mills multiplet, contain equal numbers of bosonic and fermionic excited states at all levels. In other words, it is a manifestation of space-time supersymmetry in these models.

\subsubsection{Modular invariant closed-string models}\label{sec:type_ii_0}

Altogether, only four torus amplitudes built out of the $\mathfrak{so}(8)$ characters of eq.~\eqref{eq:o2n_v2n_s2n_c2n} satisfy the constraints of modular invariance and spin-statistics\footnote{In the present context, spin-statistics amounts to positive (resp. negative) contributions from space-time bosons (resp. fermions).}. They correspond to type IIA and type IIB superstrings,
\begin{eqaed}\label{eq:ii_mod}
	\mathcal{T}_{\text{IIA}} \; & : \; \left(V_8 - C_8 \right) \overline{\left(V_8 - S_8 \right)} \, , \\
	\mathcal{T}_{\text{IIB}} \; & : \; \left(V_8 - S_8 \right) \overline{\left(V_8 - S_8 \right)} \, ,
\end{eqaed}
which are supersymmetric, and to two non-supersymmetric models, termed type 0A and type 0B,
\begin{eqaed}\label{eq:0_mod}
	\mathcal{T}_{\text{0A}} \; & : \; O_8 \, \overline{O_8} + V_8 \, \overline{V_8} + S_8 \, \overline{C_8} + C_8 \, \overline{S_8}  \, , \\
	\mathcal{T}_{\text{0B}} \; & : \; O_8 \, \overline{O_8} + V_8 \, \overline{V_8} + S_8 \, \overline{S_8} + C_8 \, \overline{C_8}  \, ,
\end{eqaed}
where we have refrained from writing the volume prefactor and the integration measure
\begin{eqaed}\label{eq:torus_measure}
	\int_{\mathcal{F}} \frac{d^2\tau}{\tau_2^6} \, \frac{1}{\abs{\eta(\tau)}^{16}} \, , \qquad \tau_2 \equiv \Im(\tau)
\end{eqaed}
for clarity. We shall henceforth use this convenient notation. Let us remark that the form of~\eqref{eq:ii_mod} translates the chiral nature of the type IIB superstring into its world-sheet symmetry between the left-moving and the right-moving sectors\footnote{Despite this fact the type IIB superstring is actually anomaly-free, as well as all five supersymmetric models owing to the Green-Schwarz mechanism~\cite{Green:1984sg}. This remarkable result was a considerable step forward in the development of string theory.}.

\subsection{Orientifold models}\label{sec:orientifold_models}

The approach that we have outlined in the preceding section can be extended to open strings, albeit with one proviso. Namely, one ought to include all Riemann surfaces with vanishing Euler characteristic, including the Klein bottle, the annulus and the Möbius strip. 

\begin{figure}[ht!]
	\centering
	\includegraphics[scale=0.35]{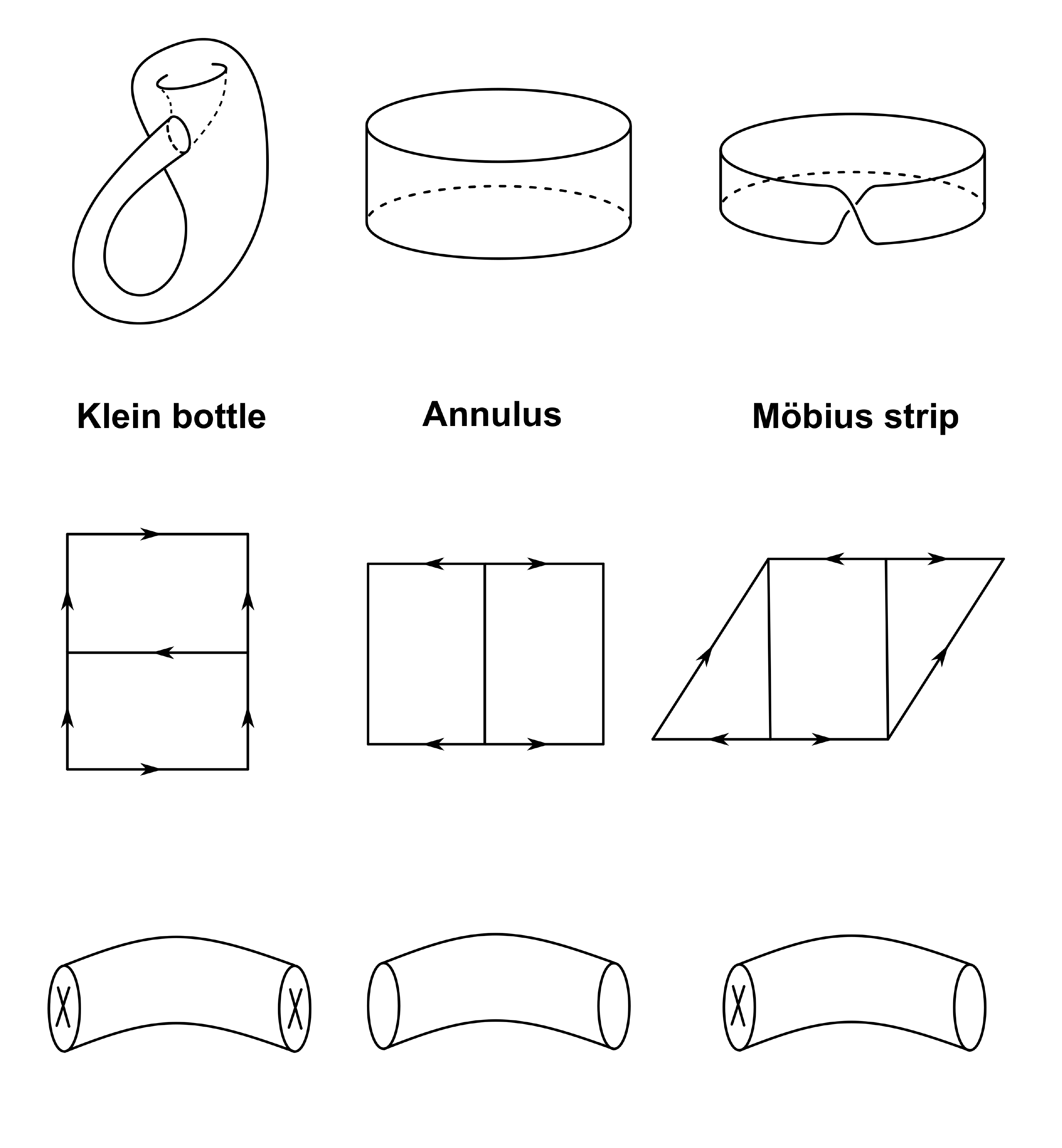}
	\caption[Surfaces]{the string world-sheet topologies (excluding the torus) which contribute to the one-loop vacuum amplitude, and the corresponding fundamental polygons. From the point of view of open strings, they can be associated to boundary conditions with boundaries or cross-caps. The corresponding space-time picture involves D-branes or orientifold planes.}
	\label{fig:surfaces}
\end{figure}

To begin with, the orientifold projection dictates that the contribution of the torus amplitude be halved and added to (half of) the Klein bottle amplitude $\mathcal{K}$. Since the resulting amplitude would entail gauge anomalies due to the Ramond-Ramond (R-R) tadpole, one ought to include the annulus amplitude $\mathcal{A}$ and Möbius strip amplitude $\mathcal{M}$, which comprise the contributions of the open sector and signal the presence of D-branes. The corresponding modular parameters are built from the covering tori of the fundamental polygons, depicted in fig.~\ref{fig:surfaces}, while the Möbius strip amplitude involves ``hatted'' characters that differ from the ordinary one by a phase\footnote{The ``hatted'' characters appear since the modular paramater of the covering torus of the Möbius strip is not real, and they ensure that states contribute with integer degeneracies.}. so that in the case of the type I superstring
\begin{eqaed}\label{eq:klein_annulus_mobius_mod}
	\mathcal{K} \; & : \; \frac{1}{2} \, \frac{\left( V_8 - S_8 \right) \!\left(2i \tau_2\right)}{\eta^8\!\left(2i \tau_2 \right)} \, , \\
	\mathcal{A} \; & : \; \frac{N^2}{2} \, \frac{\left( V_8 - S_8 \right) \!\left(\frac{i\tau_2}{2}\right)}{\eta^8\!\left(\frac{i\tau_2}{2} \right)} \, , \\
	\mathcal{M} \; & : \; \frac{\varepsilon \, N}{2} \, \frac{\left( \widehat{V}_8 - \widehat{S}_8 \right) \!\left(\frac{i\tau_2}{2} + \frac{1}{2}\right)}{\widehat{\eta}^8\!\left(\frac{i\tau_2}{2} + \frac{1}{2} \right)} \, ,
\end{eqaed}
where the sign $\epsilon$ is a reflection coefficient and $N$ is the number of Chan-Paton factors. Here, analogously as in the preceding section, we have refrained from writing the volume prefactor and the integration measure
\begin{eqaed}\label{eq:open_measure}
	\int_0^\infty \frac{d\tau_2}{\tau_2^6} \, ,
\end{eqaed}
for clarity. At the level of the closed spectrum, the projection symmetrizes the NS-NS sector, so that the massless closed spectrum rearranges into the minimal ten-dimensional $\mathcal{N} = (1,0)$ supergravity multiplet, but anti-symmetrizes the R-R sector, while the massless open spectrum comprises a super Yang-Mills multiplet. It is instructive to recast the ``loop channel'' amplitudes of eq.~\eqref{eq:klein_annulus_mobius_mod} in the ``tree-channel'' using a modular transformation. The resulting amplitudes describe tree-level exchange of closed-string states, and read
\begin{eqaed}\label{eq:klein_annulus_mobius_tree_mod}
	\widetilde{\mathcal{K}} & = \frac{2^5}{2} \int_0^\infty d\ell \, \frac{\left( V_8 - S_8 \right) \!\left(i\ell\right)}{\eta^8\!\left(i\ell \right)} \, , \\
	\widetilde{\mathcal{A}} & = \frac{2^{-5} \, N^2}{2} \int_0^\infty d\ell \, \frac{\left( V_8 - S_8 \right) \!\left(i\ell\right)}{\eta^8\!\left(i\ell \right)} \, , \\
	\widetilde{\mathcal{M}} & = \frac{2 \, \varepsilon \, N}{2} \int_0^\infty d\ell \, \frac{\left( \widehat{V}_8 - \widehat{S}_8 \right) \!\left(i\ell + \frac{1}{2}\right)}{\widehat{\eta}^8\!\left(i\ell + \frac{1}{2} \right)} \, .
\end{eqaed}
The UV divergences of the loop-channel amplitudes are translated into IR divergences, which are associated to the $\ell \to \infty$ regime of the integration region. Physically they describe the exchange of zero-momentum massless modes, either in the NS-NS sector or in the R-R sector, and the corresponding coefficients can vanish on account of the tadpole cancellation condition
\begin{eqaed}\label{eq:tadpole_super_mod}
	\frac{2^5}{2} + \frac{2^{-5} \, N^2}{2} + \frac{2 \, \varepsilon \, N}{2} = \frac{2^{-5}}{2} \left( N + 32 \, \varepsilon \right)^2 = 0 \, .
\end{eqaed}
Let us stress that these conditions apply both to the NS-NS sector, where they grant the absence of a gravitational tadpole, and to the R-R sector, where they grant R-charge neutrality and thus anomaly cancellation via the Green-Schwarz mechanism. The unique solution to eq.~\eqref{eq:tadpole_super_mod} is $N = 32$ and $\varepsilon = -1$, \textit{i.e.} the $SO(32)$ type I superstring. The corresponding space-time interpretation involves $32$ $\text{D}9$-branes\footnote{Since the $\text{D}9$-branes are on top of the $\text{O}9_-$-plane, counting conventions can differ based on whether one includes ``image'' branes.} and an $\text{O}9_-$-plane, which has \textit{negative} tension and charge.

\subsubsection{The Sugimoto model: brane supersymmetry breaking}\label{sec:sugimoto_model}

On the other hand, introducing an $\text{O}9_+$-plane with positive tension and charge one can preserve the R-R tadpole cancellation while generating a non-vanishing NS-NS tadpole, thus breaking supersymmetry at the string scale. At the level of vacuum amplitudes, this is reflected in a sign change in the Möbius strip amplitude, so that now
\begin{importantbox}
	\begin{eqaed}\label{eq:mobius_bsb_mod}
		\mathcal{M}_{\text{BSB}} \; : \; \frac{\varepsilon \, N}{2} \, \frac{\left( \widehat{V}_8 + \widehat{S}_8 \right) \!\left(\frac{i\tau_2}{2} + \frac{1}{2}\right)}{\widehat{\eta}^8\!\left(\frac{i\tau_2}{2} + \frac{1}{2} \right)} \, .
	\end{eqaed}
\end{importantbox}
The resulting tree-channel amplitudes are given by
\begin{eqaed}\label{eq:mobius_bsb_tree_mod}
	\widetilde{\mathcal{M}}_{\text{BSB}} & = \frac{2 \, \varepsilon \, N}{2} \int_0^\infty d\ell \, \frac{\left( \widehat{V}_8 + \widehat{S}_8 \right) \!\left(i\ell + \frac{1}{2}\right)}{\widehat{\eta}^8\!\left(i\ell + \frac{1}{2} \right)} \, ,
\end{eqaed}
from which the R-R tadpole condition now requires that $\varepsilon = 1$ and $N = 32$, \textit{i.e.} a $USp(32)$ gauge group. However, one is now left with a NS-NS tadpole, and thus at low energies runaway exponential potential of the type
\begin{eqaed}\label{eq:runaway_potential_string_frame}
	T \int d^{10} x \, \sqrt{- g_S} \, e^{- \phi}
\end{eqaed}
emerges in the string frame, while its Einstein-frame counterpart is
\begin{eqaed}\label{eq:runaway_potential_einstein_frame}
	T \int d^{10} x \, \sqrt{- g} \, e^{\gamma \phi} \, , \qquad \gamma = \frac{3}{2} \, .
\end{eqaed}
Exponential potentials of the type of eq.~\eqref{eq:runaway_potential_einstein_frame} are smoking guns of string-scale supersymmetry breaking, and we shall address their effect on the resulting low-energy physics in following sections. Notice also that the fermions are in the anti-symmetric representation of $USp(32)$, which is reducible. The corresponding singlet is a very important ingredient: it is the Goldstino that is to accompany the breaking of supersymmetry, while the closed spectrum is supersymmetric to lowest order and contains a ten-dimensional gravitino. The relevant low-energy interactions manifest an expected structure \textit{\`{a} la} Volkov-Akulov~\cite{Dudas:2000nv}, but a complete understanding of the super-Higgs mechanism in this ten-dimensional context remains elusive~\cite{Dudas:2000ff, Dudas:2001wd}.

All in all, a supersymmetric closed sector is coupled to a non-supersymmmetric open sector, which lives on $32$ $\overline{\text{D}9}$-branes where supersymmetry is non-linearly realized\footnote{The original works can be found in~\cite{Sagnotti:1987tw,Pradisi:1988xd,Horava:1989vt,Horava:1989ga,Bianchi:1990yu,Bianchi:1990tb,Bianchi:1991eu,Sagnotti:1992qw}. For reviews, see~\cite{Dudas:2000bn,Angelantonj:2002ct,Mourad:2017rrl}.}~\cite{Dudas:2000nv,Pradisi:2001yv,Kitazawa:2018zys} in a manner reminiscent of the Volkov-Akulov model, and due to the runaway potential of eq.~\eqref{eq:runaway_potential_string_frame} the effective space-time equations of motion do not admit Minkowski solutions. The resulting model is a special case of more general $\text{D}9$-$\overline{\text{D}9}$ branes systems, which were studied in~\cite{Sugimoto:1999tx}, and the aforementioned phenomenon of ``brane supersymmetry breaking'' (BSB) was investigated in detail in~\cite{Antoniadis:1999xk,Angelantonj:1999jh,Aldazabal:1999jr,Angelantonj:1999ms}. On the phenomenological side, the peculiar behavior of BSB also appears to provide a rationale for the low-$\ell$ lack of power in the Cosmic Microwave Background~\cite{Sagnotti:2015asa,Gruppuso:2015xqa,Gruppuso:2017nap,Mourad:2017rrl}.

While the presence of a gravitational tadpole is instrumental in breaking supersymmetry in a natural fashion, in its presence string theory back-reacts dramatically\footnote{In principle, one could address these phenomena by systematic vacuum redefinitions~\cite{Fischler:1986ci, Fischler:1986tb, Dudas:2004nd, Kitazawa:2008hv, Pius:2014gza}, but carrying out the program at high orders appears prohibitive.} on the original Minkowski vacuum, whose detailed fate appears, at present, largely out of computational control. Let us remark that these difficulties are not restricted to this type of scenarios. Indeed, while a variety of supersymmetry-breaking mechanisms have been investigated, they are all fraught with conceptual and technical obstacles, and primarily with the generic presence of instabilities, which we shall address in detail in Section~\ref{Section3} and Section~\ref{Section4}. Although these issues are ubiquitous in settings of this type, it is worth mentioning that string-scale supersymmetry breaking in particular appears favored by anthropic arguments~\cite{Susskind:2004uv, Douglas:2004qg}.

\subsubsection{The type \texorpdfstring{$0'\text{B}$}{0'B} string}\label{sec:0'b}

Let us now describe another instance of orientifold projection which leads to non-tachyonic perturbative spectra, starting from the type 0B model\footnote{The corresponding orientifold projections of the type 0A model were also investigated. See~\cite{Angelantonj:2002ct}, and references therein.} described by eq.~\eqref{eq:0_mod}. There are a number of available projections, encoded in different choices of the Klein bottle amplitude. Here we focus on
\begin{eqaed}\label{eq:0'b_klein_mod}
	\mathcal{K}_{0'\text{B}} \; : \; \frac{1}{2} \left( - \, O_8 + V_8 + S_8 - C_8 \right) \, ,
\end{eqaed}
which, in contrast to the more standard projection defined by the combination $O_8+V_8-S_8-C_8$, implements anti-symmetrization in the $O_8$ and $C_8$ sectors. This purges tachyons from the spectrum, and thus the resulting model, termed type ``$0'\text{B}$'', is particularly intriguing. The corresponding tree-channel amplitude is given by
\begin{eqaed}\label{eq:0'b_klein_tree_mod}
	\widetilde{K}_{0'\text{B}} = - \, \frac{2^6}{2} \int_0^\infty d\ell \, C_8 \, .
\end{eqaed}
In order to complete the projection one is to specify the contributions of the open sector, consistently with anomaly cancellation. Let us consider a family of solution that involves two Chan-Paton charges, and is described by~\cite{Sagnotti:1996qj}
\begin{eqaed}\label{eq:0'b_annulus_mobius_mod}
	\mathcal{A}_{0'\text{B}} \; & : \; n \, \overline{n} \, V_8 - \, \frac{n^2 + \overline{n}^2}{2} \, C_8 \, , \\
	\mathcal{M}_{0'\text{B}} \; & : \; \frac{n + \overline{n}}{2} \, \widehat{C}_8 \, .
\end{eqaed}
This construction is a special case of a more general four-charge solution~\cite{Sagnotti:1996qj}, and involves complex ``eigencharges'' $n \, ,\overline{n}$ with corresponding unitary gauge groups. Moreover, while we kept the two charges formally distinct, consistency demands $n = \overline{n}$, while the tadpole conditions fix $n = 32$, and the resulting model has a $U(32)$ gauge group\footnote{Strictly speaking, the anomalous $U(1)$ factor carried by the corresponding gauge vector disappears from the low-lying spectrum, thus effectively reducing the group to $SU(32)$.}. As in the case of the $USp(32)$ model, this model admits a space-time description in terms of orientifold planes, now with vanishing tension, and the low-energy physics of both non-supersymmetric orientifold models can be captured by effective actions that we shall discuss in Section~\ref{Section2}. In addition to orientifold models, the low-energy description can also encompass the non-supersymmetric heterotic model, which we shall now discuss in detail, with a simple replacement of numerical coefficients in the action.

\subsection{Heterotic strings}\label{sec:heterotic_model}

Heterotic strings are remarkable hybrids of the bosonic string and superstrings, whose existence rests on the fact that the right-moving sector and the left-moving sector are decoupled. Indeed, their right-moving sector can be built using the $26$-dimensional bosonic string\footnote{One can alternatively build heterotic right-moving sectors using ten-dimensional strings with auxiliary fermions.}, while their left-moving sector is built using the ten-dimensional superstring. In order for these costructions to admit a sensible space-time interpretation, $16$ of the $26$ dimensions pertaining to the right-moving sector are compactified on a torus defined by a lattice $\Lambda$, of which there are only two consistent choices, namely the weight lattices of $SO(32)$ and $E_8 \times E_8$. These groups play the rôle of gauge groups of the two corresponding supersymmetric heterotic models, aptly dubbed ``HO'' and ``HE'' respectively. Their perturbative spectra are concisely captured by the torus amplitudes
\begin{eqaed}\label{eq:ho_he_mod}
	\mathcal{T}_{\text{HO}} \; & : \; \left(V_8 - S_8 \right) \overline{\left( O_{32} + S_{32} \right)} \, , \\
	\mathcal{T}_{\text{HE}} \; & : \; \left(V_8 - S_8 \right) \overline{\left( O_{16} + S_{16} \right)}^2 \, ,
\end{eqaed}
which feature $\mathfrak{so}(16)$ and $\mathfrak{so}(32)$ characters in the right-moving sector. As in the case of type II superstrings, these two models can be related by T-duality, which in this context acts as a projection onto states with even fermion number in the right-moving (``internal'') sector. However, a slightly different projection yields the non-supersymmetric heterotic string of~\cite{AlvarezGaume:1986jb, Dixon:1986iz}, which we shall now describe.

\subsubsection{The non-supersymmetric heterotic model}\label{eq:non-susy_het_model}

Let us consider a projection of the HE theory onto the states with even total fermion number. At the level of one-loop amplitudes, one is to halve the original torus amplitude and add terms obtained changing the signs in front of the $S$ characters, yielding the two ``untwisted'' contributions
\begin{eqaed}\label{eq:h++_h+-_mod}
	\mathcal{T}_{(++)} \; & : \; \frac{1}{2} \left(V_8 - S_8 \right) \overline{\left( O_{16} + S_{16} \right)}^2 \, , \\
	\mathcal{T}_{(+-)} \; & : \; \frac{1}{2} \left(V_8 + S_8 \right) \overline{\left( O_{16} - S_{16} \right)}^2 \, .
\end{eqaed}
The constraint of modular invariance under $S$, which is lacking at this stage, further leads to the addition of the image of $\mathcal{T}_{+-}$ under $S$, namely
\begin{eqaed}\label{eq:h-+_mod}
	\mathcal{T}_{(-+)} \; & : \; \frac{1}{2} \left(O_8 - C_8 \right) \overline{\left( V_{16} + C_{16} \right)}^2 \, .
\end{eqaed}
The addition of $\mathcal{T}_{-+}$ now spoils invariance under $T$ transformations, which is restored adding
\begin{eqaed}\label{eq:h--_mod}
	\mathcal{T}_{(--)} \; & : \; - \, \frac{1}{2} \left(O_8 + C_8 \right) \overline{\left( V_{16} - C_{16} \right)}^2 \, .
\end{eqaed}
All in all, the torus amplitude arising from this projection of the HE theory yields a theory with a manifest $SO(16) \times SO(16)$ gauge group, and whose torus amplitude finally reads
\begin{importantbox}
	\begin{eqaed}\label{eq:h_soxso_mod}
		\mathcal{T}_{SO(16) \times SO(16)} \; & : \; O_8 \, \overline{\left( V_{16} \, C_{16} + C_{16} \, V_{16} \right)} \\
		& \quad + V_8 \, \overline{\left( O_{16} \, O_{16} + S_{16} \, S_{16} \right)} \\
		& \quad - S_8 \, \overline{\left( O_{16} \, S_{16} + S_{16} \, O_{16} \right)} \\
		& \quad - C_8 \, \overline{\left( V_{16} \, V_{16} + C_{16} \, C_{16} \right)} \, .
	\end{eqaed}
\end{importantbox}
The massless states originating from the $V_8$ terms comprise the gravitational sector, constructed out of the bosonic oscillators, as well as a $(\mathbf{120},\mathbf{1}) \oplus (\mathbf{1},\mathbf{120})$ multiplet of $SO(16) \times SO(16)$, \textit{i.e.} in the adjoint representation of its Lie algebra, while the $S_8$ terms provide spinors in the $(\mathbf{1},\mathbf{128}) \oplus (\mathbf{128},\mathbf{1})$ representation. Furthermore, the $C_8$ terms correspond to right-handed $(\mathbf{16},\mathbf{16})$ spinors. The terms in the first line of eq.~\eqref{eq:h_soxso_mod} do not contribute at the massless level, due to level matching and the absence of massless states in the corresponding right-moving sector. In particular, this entails the absence of tachyons from this string model, but the vacuum energy does not vanish\footnote{In some orbifold models, it is possible to obtain suppressed or vanishing leading contributions to the cosmological constant~\cite{Dienes:1990ij, Dienes:1990qh, Kachru:1998hd, Angelantonj:2004cm, Abel:2015oxa, Abel:2017rch}.}, since it is not protected by supersymmetry. Indeed, up to a volume prefactor its value can be computed integrating eq.~\eqref{eq:h_soxso_mod} against the measure of eq.~\eqref{eq:torus_measure}, and, since the resulting string-scale vacuum energy couples with the gravitational sector in a universal fashion\footnote{At the level of the space-time effective action, the vacuum energy contributes to the string-frame cosmological constant. In the Einstein frame, it corresponds to a runaway exponential potential for the dilaton.}, its presence also entails a dilaton tadpole, and thus a runaway exponential potential for the dilaton. In the Einstein frame, it takes the form
\begin{eqaed}\label{eq:runaway_potential_het}
	T \int d^{10}x \, \sqrt{- g} \, e^{\gamma \phi} \, , \qquad \gamma = \frac{5}{2} \, , 
\end{eqaed}
and thus the effect of the gravitational tadpoles on the low-energy physics of both the orientifold models of Section~\ref{sec:orientifold_models} and the $SO(16) \times SO(16)$ heterotic model can be accounted for with the same type of exponential dilaton potential. On the
phenomenological side, this model has recently sparked some interest in non-supersymmetric
model building~\cite{Abel:2015oxa, Abel:2017vos}\footnote{In the same spirit, three-generation non-tachyonic heterotic models were constructed in~\cite{Ashfaque:2015vta}. Recently, lower-dimensional non-tachyonic models have been realized compactifying ten-dimensional tachyonic superstrings~\cite{Faraggi:2019fap, Faraggi:2019drl}.} in Calabi-Yau compactifications~\cite{Blaszczyk:2015zta}, and in Section~\ref{Section2} we shall investigate in detail the consequences of dilaton tadpoles on space-time.

\section{Non-supersymmetric vacuum solutions}\label{Section2}

In this section we investigate the low-energy physics of the string models that we have described in Section~\ref{Section1}, namely the non-supersymmetric $SO(16) \times SO(16)$ heterotic model~\cite{AlvarezGaume:1986jb,Dixon:1986iz}, whose first quantum correction generates a dilaton potential, and two orientifold models, the non-supersymmetric $U(32)$ type $0'\text{B}$ model~\cite{Sagnotti:1995ga,Sagnotti:1996qj} and the $USp(32)$ model~\cite{Sugimoto:1999tx} with ``Brane Supersymmetry Breaking'' (BSB)~\cite{Antoniadis:1999xk,Angelantonj:1999jh,Aldazabal:1999jr,Angelantonj:1999ms}, where a similar potential reflects the tension unbalance present in the vacuum. To begin with, in Section~\ref{sec:low-energy_eft} we discuss the low-energy effective action that we shall consider. Then we proceed to discuss some classes of solutions of the equations of motion. Specifically, in Section~\ref{sec:no-flux_solutions} we present the Dudas-Mourad solutions of~\cite{Dudas:2000ff}, which comprise nine-dimensional static compactifications on warped intervals and ten-dimensional cosmological solutions. In Section~\ref{sec:flux_compactifications} we introduce fluxes, which lead to parametrically controlled Freund-Rubin~\cite{Freund:1980xh} compactifications~\cite{Mourad:2016xbk, Antonelli:2019nar}, and we show that, while the string models at stake admit only $\ads$ solutions of this type, in a more general class of effective theories $\ds$ solutions always feature an instability of the radion mode. In Section~\ref{Section7} we complete our discussion on $\ds$ solutions, examining general warped flux compactifications and extending previous no-go results, connecting them to recent swampland conjectures. 

\subsection{The low-energy description}\label{sec:low-energy_eft}

Let us now present the effective (super)gravity theories related to the string models at stake. For the sake of generality, we shall often work with a family of $D$-dimensional effective gravitational theories, where the bosonic fields include a dilaton $\phi$ and a $(p+2)$-form field strength $H_{p+2} = dB_{p+1}$. Using the ``mostly plus'' metric signature, the (Einstein-frame) effective actions
\begin{importantbox}
	\begin{eqaed}\label{eq:action}
		S = \frac{1}{2\kappa_D^2}\int d^D x \, \sqrt{-g} \, \left( R - \frac{4}{D-2} \left(\partial \phi \right)^2 - V(\phi) - \frac{f(\phi)}{2(p+2)!}\, H_{p+2}^2 \right)
	\end{eqaed}
\end{importantbox}
subsume all relevant cases\footnote{This effective field theory can also describe non-critical strings~\cite{Silverstein:2001xn, Maloney:2002rr}, since the Weyl anomaly can be saturated by the contribution of an exponential dilaton potential.}, and whenever needed we specialize them according to
\begin{eqaed}\label{eq:potential_form-coupling}
	V(\phi) = T \, e^{\gamma \phi} \, , \qquad f(\phi) = e^{\alpha \phi} \, ,
\end{eqaed}
which capture the lowest-order contributions in the string coupling for positive\footnote{The case $\gamma = 0$, which at any rate does not arise in string perturbation theory, would not complicate matters further.} $\gamma$ and $T$. In the orientifold models, the dilaton potential arises from the non-vanishing NS-NS tadpole at (projective-)disk level, while in the heterotic model it arises from the torus amplitude. The massless spectrum of the corresponding string models also includes Yang-Mills fields, whose contribution to the action takes the form
\begin{eqaed}\label{eq:gauge_action}
	S_{\text{gauge}} = - \, \frac{1}{2\kappa_D^2} \int d^D x \, \sqrt{-g} \left( \frac{w(\phi)}{4} \, \text{Tr} \, \mathcal{F}_{MN} \, \mathcal{F}^{MN} \right)
\end{eqaed}
with $w(\phi)$ an exponential, but we shall not consider them. Although $\ads$ compactifications supported by non-Abelian gauge fields, akin to those discussed in Section~\ref{sec:flux_compactifications}, were studied in~\cite{Mourad:2016xbk}, their perturbative corners appear to forego the dependence on the non-Abelian gauge flux. On the other hand, an $\ads_3 \times \ess^7$ solution of the heterotic model with no counterpart without non-Abelian gauge flux was also found~\cite{Mourad:2016xbk}, but it is also available in the supersymmetric case.

The (bosonic) low-energy dynamics of both the $USp(32)$ BSB model and the $U(32)$ type $0'\text{B}$ model is encoded in the Einstein-frame parameters
\begin{eqaed}\label{eq:bsb_electric_params}
	D = 10 \, , \quad p = 1 \, , \quad \gamma = \frac{3}{2} \, , \quad \alpha = 1 \, ,
\end{eqaed}
whose string-frame counterpart stems from the effective action\footnote{In eq.~\eqref{eq:string_frame_action_bsb} we have used the notation $F_3 = dC_2$ in order to stress the Ramond-Ramond (RR) origin of the field strength.}~\cite{Dudas:2000nv}
\begin{eqaed}\label{eq:string_frame_action_bsb}
	S_{\text{orientifold}} = \frac{1}{2\kappa_{10}^2}\int d^{10} x \, \sqrt{-g_S} \, \left( e^{-2\phi} \left[ R + 4 \left(\partial \phi \right)^2 \right] - T \, e^{-\phi} - \frac{1}{12}\, F_{3}^2 \right) \, .
\end{eqaed}
The $e^{-\phi}$ factor echoes the (projective-)disk origin of the exponential potential for the dilaton, and the coefficient $T$ is given by
\begin{eqaed}\label{eq:T_bsb}
	T = 2\kappa_{10}^2 \times 64 \, T_{\text{D}9} = \frac{16}{\pi^2 \, \alpha'}
\end{eqaed}
in the BSB model, reflecting the cumulative contribution of $16$ $\overline{\text{D}9}$-branes and the orientifold plane~\cite{Sugimoto:1999tx}, while in the type $0'\text{B}$ model $T$ is half of this value.

On the other hand, the $SO(16) \times SO(16)$ heterotic model of~\cite{AlvarezGaume:1986jb} is described by
\begin{eqaed}\label{eq:het_electric_params}
	D = 10 \, , \quad p = 1 \, , \quad \gamma = \frac{5}{2} \, , \quad \alpha = -1 \, ,
\end{eqaed}
corresponding to the string-frame effective action
\begin{equation}\label{eq:string_frame_action_het}
S_{\text{heterotic}} = \frac{1}{2\kappa_{10}^2}\int d^{10} x \, \sqrt{-g_S} \, \left( e^{-2\phi} \left[ R + 4 \left(\partial \phi \right)^2 - \frac{1}{12}\, H_{3}^2\right] - T  \right) \, ,
\end{equation}
which contains the Kalb-Ramond field strength $H_3$ and the one-loop cosmological constant $T$, which was estimated in~\cite{AlvarezGaume:1986jb}. One can equivalently dualize the Kalb-Ramond form and work with the Einstein-frame parameters
\begin{eqaed}\label{eq:het_magnetic_params}
	D = 10 \, , \quad p = 5 \, , \quad \gamma = \frac{5}{2} \, , \quad \alpha = 1 \, .
\end{eqaed}
One may wonder whether the effective actions of eq.~\eqref{eq:action} can be reliable, since the dilaton potential contains one less power of $\alpha'$ with respect to the other terms. The $\ads$ landscapes that we shall present in Section~\ref{sec:flux_compactifications} contain weakly coupled regimes, where curvature corrections and string loop corrections are expected to be under control, but their existence rests on large fluxes. While in the orientifold models the vacua are supported by R-R fluxes, and thus a world-sheet formulation appears subtle, the simpler nature of the NS-NS fluxes in the heterotic model is balanced by the quantum origin of the dilaton tadpole\footnote{At any rate, it is worth noting that world-sheet conformal field theories on $\ads_3$ backgrounds have been related to WZW models, which can afford $\alpha'$-exact algebraic descriptions~\cite{Maldacena:2000hw}.}. On the other hand, the solutions discussed in Section~\ref{sec:no-flux_solutions} do not involve fluxes, but their perturbative corners do not extend to the whole space-time.

The equations of motion stemming from the action in eq.~\eqref{eq:action} are
\begin{eqaed}\label{eq:eoms_eft}
	R_{MN} & = \widetilde{T}_{MN} \, , \\
	\frac{8}{D-2} \, \Box \, \phi \, - V'(\phi) - \, \frac{f'(\phi)}{2(p+2)!} \, H_{p+2}^2 & = 0 \, , \\
	d \star (f(\phi) \, H_{p+2}) & = 0 \, ,
\end{eqaed}
where the trace-reversed stress-energy tensor
\begin{eqaed}\label{eq:trace-reversed_stress}
	\widetilde{T}_{MN} \equiv T_{MN} - \frac{1}{D-2} \, {T^A}_A \, g_{MN}
\end{eqaed}
is defined in terms of the standard stress-energy tensor $T_{MN}$, and with our conventions
\begin{eqaed}\label{eq:stress_definition}
	T_{MN} \equiv - \, \frac{\delta S_{\text{matter}}}{\delta g^{MN}} \, .
\end{eqaed} 
From the effective action of eq.~\eqref{eq:action}, one obtains
\begin{eqaed}\label{eq:stress_tensor}
	\widetilde{T}_{MN} = \; & \frac{4}{D-2} \, \partial_M \phi \, \partial_N \phi + \frac{f(\phi)}{2(p+1)!} \left(H_{p+2}^2\right)_{MN} \\
	& + \frac{g_{MN}}{D-2} \left( V - \frac{p+1}{2(p+2)!} \, f(\phi) \, H_{p+2}^2 \right) \, ,
\end{eqaed}
where $\left(H_{p+2}^2\right)_{MN} \equiv H_{M A_1 \dots A_{p+1}} \, {H_N}^{A_1 \dots A_{p+1}}$. In the following sections, we shall make extensive use of eqs.~\eqref{eq:eoms_eft} and~\eqref{eq:stress_tensor} to obtain a number of solutions, both with and without fluxes.

\subsection{Solutions without flux}\label{sec:no-flux_solutions}

Let us now describe in detail the Dudas-Mourad solutions of~\cite{Dudas:2000ff}. They comprise static solutions with nine-dimensional Poincar\'e symmetry\footnote{For a similar analysis of a T-dual version of the $USp(32)$ model, see~\cite{Blumenhagen:2000dc}.}, where one dimension is compactified on an interval, and ten-dimensional cosmological solutions.

\subsubsection{Static Dudas-Mourad solutions}\label{sec:static_solutions}

Due to the presence of the dilaton potential, the maximal possible symmetry available to static solutions is nine-dimensional Poincar\'e symmetry, and therefore the most general solution of this type is a warped product of nine-dimensional Minkowski space-time, parametrized by coordinates $x^\mu$, and a one-dimensional internal space, parametrized by a coordinate $y$. As we shall discuss in Section~\ref{Section5}, in the absence of fluxes the resulting equations of motion can be recast in terms of an integrable Toda-like dynamical system, and the resulting Einstein-frame solution reads
\begin{eqaed}\label{eq:dm_orientifold_einstein_mod}
	ds_\text{orientifold}^2 & = \abs{\alpha_\text{O} \, y^2}^{\frac{1}{18}} \, e^{- \frac{\alpha_\text{O} y^2}{8}} \, dx_{1,8}^2 + e^{- \frac{3}{2} \Phi_0} \, \abs{\alpha_\text{O} \, y^2}^{- \frac{1}{2}} \, e^{- \frac{9 \alpha_\text{O} y^2}{8}} dy^2 \, , \\
	\phi & = \frac{3}{4} \, \alpha_\text{O} \, y^2 + \frac{1}{3} \, \log \abs{\alpha_\text{O} \, y^2} + \Phi_0
\end{eqaed}
for the orientifold models, where here and in the remainder of this review
\begin{eqaed}\label{eq:minkowski_p_mod}
	dx_{1,p}^2 \equiv \eta_{\mu \nu} \, dx^\mu \, dx^\nu
\end{eqaed}
is the $(p+1)$-dimensional Minkowski metric. The absolute values in eq.~\eqref{eq:dm_orientifold_einstein_mod} imply that the geometry is described by the coordinate patch in which $y \in (0,\infty)$. The corresponding Einstein-frame solution of the heterotic model reads
\begin{eqaed}\label{eq:dm_heterotic_einstein_mod}
	ds_\text{heterotic}^2 = \, & \left(\sinh \abs{\sqrt{\alpha_\text{H}} \, y}\right)^{\frac{1}{12}} \left(\cosh \abs{\sqrt{\alpha_\text{H}} \, y}\right)^{- \frac{1}{3}} dx_{1,8}^2 \\
	& + e^{- \frac{5}{2} \Phi_0} \left(\sinh \abs{\sqrt{\alpha_\text{H}} \, y}\right)^{- \frac{5}{4}} \left(\cosh \abs{\sqrt{\alpha_\text{H}} \, y}\right)^{-5} dy^2 \, , \\
	\phi & = \frac{1}{2} \, \log \, \sinh \abs{\sqrt{\alpha_{\text{H}}} \, y} + 2 \, \log \, \cosh \abs{\sqrt{\alpha_{\text{H}}} \, y} + \Phi_0 \, .          
\end{eqaed}
In eqs.~\eqref{eq:dm_orientifold_einstein_mod} and~\eqref{eq:dm_heterotic_einstein_mod} the scales $\alpha_\text{O,H} \equiv \frac{T}{2}$, while $\Phi_0$ is an arbitrary integration constant. As we shall explain in Section~\ref{Section5}, the internal spaces parametrized by $y$ are actually intervals of finite length, and the geometry contains a weakly coupled region in the middle of the parametrically wide interval for $g_s \equiv e^{\Phi_0} \ll 1$. Moreover, the isometry group appears to be connected to the presence of uncharged $8$-branes~\cite{Antonelli:2019nar}. 

It is convenient to recast the two solutions in terms of conformally flat metrics, so that one is led to consider expressions of the type
\begin{eqaed}\label{eq:dm_conformal_mod}
	ds^2 & = e^{2 \Omega(z)} \left( dx_{1,8}^2 + dz^2 \right) \, , \qquad
	\phi = \phi(z) \, ,
\end{eqaed}
In detail,  for the orientifold models the coordinate $z$ is obtained integrating the relation
\begin{eqaed}\label{eq:dz_orientifold_mod}
	dz = \abs{\alpha_\text{O} \, y^2}^{- \frac{5}{18}} \, e^{- \frac{3}{4} \Phi_0} \, e^{- \frac{\alpha_{\text{O}} y^2}{2}} \, dy \, ,
\end{eqaed}
while
\begin{eqaed}\label{eq:omega_orientifold_mod}
	e^{2 \Omega(z)} = \abs{\alpha_\text{O} \, y^2}^{\frac{1}{18}} \, e^{- \frac{\alpha_{\text{O}} y^2}{8}} \, .
\end{eqaed}
On the other hand, for the heterotic model
\begin{eqaed}\label{eq:dz_heterotic_mod}
	dz = e^{- \frac{5}{4} \Phi_0} \left(\sinh \abs{\sqrt{\alpha_\text{H}} \, y}\right)^{- \frac{2}{3}} \left(\cosh \abs{\sqrt{\alpha_\text{H}} \, y}\right)^{- \frac{7}{3}} \, dy \, ,
\end{eqaed}
and the corresponding conformal factor reads
\begin{eqaed}\label{eq:omega_heterotic_mod}
	e^{2 \Omega(z)} = \left(\sinh \abs{\sqrt{\alpha_\text{H}} \, y}\right)^{\frac{1}{12}} \left(\cosh \abs{\sqrt{\alpha_\text{H}} \, y}\right)^{- \frac{1}{3}} \, .
\end{eqaed}
Notice that one is confronted with an interval whose (string-frame) finite length is proportional to $\frac{1}{\sqrt{g_s \, \alpha_{\text{O}}}}$ and  $\frac{1}{\sqrt{g_s^2 \, \alpha_{\text{H}}}}$ in the two cases, but which hosts a pair of curvature singularities at its two ends, with a local string coupling $e^{\phi}$ that is weak at the former and strong at the latter. Moreover, the parameters $\alpha_\text{O,H}$ are proportional to the dilaton tadpoles, and therefore as one approaches the supersymmetric case the internal length diverges\footnote{The supersymmetry-breaking tadpoles cannot be sent to zero in a smooth fashion. However, it is instructive to treat them as parameters, in order to highlight their rôle.}. Despite these shortcomings, one can still attempt to assess the qualitative importance of string loop corrections studying integrable potentials~\cite{Pelliconi:2021eak, Fre:2013vza}.

\subsubsection{Cosmological Dudas-Mourad solutions}\label{sec:cosmological_solutions}

The cosmological counterparts of the static solutions of eqs.~\eqref{eq:dm_orientifold_einstein_mod} and~\eqref{eq:dm_heterotic_einstein_mod} can be obtained via the analytic continuation $y \; \to \; i t$, and consequently under $z \; \to \; i \eta$ in conformally flat coordinates. For the orientifold models, one thus finds
\begin{eqaed}\label{eq:dm_orientifold_cosm_mod}
	ds_{\text{orientifold}}^2 & = \abs{\alpha_{\text{O}} \, t^2}^{\frac{1}{18}} \, e^{\frac{\alpha_\text{O} t^2}{8}} \, d\mathbf{x}^2 - e^{- \frac{3}{2} \Phi_0} \, \abs{\alpha_{\text{O}} \, t^2}^{- \frac{1}{2}} \, e^{\frac{9 \alpha_\text{O} t^2}{8}} \, dt^2 \, , \\
	\phi & = - \, \frac{3}{4} \, \alpha_\text{O} \, t^2 + \frac{1}{3} \, \log \abs{\alpha_{\text{O}} \, t^2} + \Phi_0 \, ,
\end{eqaed}
where the parametric time $t$ takes values in $(0,\infty)$, as usual for a decelerating cosmology with an initial singularity. The corresponding solution of the heterotic model reads
\begin{eqaed}\label{eq:dm_heterotic_cosm_mod}
	ds_{\text{heterotic}}^2 = \, & \left(\sin \abs{\sqrt{\alpha_{\text{H}}} \, t}\right)^{\frac{1}{12}} \left(\cos \abs{\sqrt{\alpha_{\text{H}}} \, t}\right)^{- \frac{1}{3}} d\mathbf{x}^2 \\
	& - e^{- \frac{5}{2} \Phi_0} \left(\sin \abs{\sqrt{\alpha_{\text{H}}} \, t}\right)^{- \frac{5}{4}} \left(\cos \abs{\sqrt{\alpha_{\text{H}}} \, t}\right)^{-5} dt^2 \, , \\
	\phi & = \frac{1}{2} \, \log \sin \abs{\sqrt{\alpha_{\text{H}}} \, t} + 2 \, \log \cos \abs{\sqrt{\alpha_{\text{H}}} \, t} + \Phi_0 \, ,
\end{eqaed}
where now $0 < \sqrt{\alpha_\text{H}} \, t < \frac{\pi}{2}$. Both cosmologies have a nine-dimensional Euclidean symmetry, and in both cases, as shown in~\cite{Dudas:2010gi}, the dilaton is forced to emerge from the initial singularity climbing up the potential. In this fashion it reaches an upper bound before it begins its descent, and thus the local string coupling is bounded and parametrically suppressed for $g_s \ll 1$. 

As in the preceding section, it is convenient to recast these expressions in conformal time according to
\begin{eqaed}\label{eq:dm_conformal_cosm_mod}
	ds^2 & = e^{2 \Omega(\eta)} \left(d\mathbf{x}^2 - d\eta^2 \right) \, , \\
	\phi & = \phi(\eta) \, ,
\end{eqaed}
and for the orientifold models the conformal time $\eta$ is obtained integrating the relation
\begin{eqaed}\label{eq:eta_orientifold_mod}
	d\eta = \abs{\alpha_{\text{O}} \, t^2}^{- \frac{5}{18}} \, e^{- \frac{3}{4} \Phi_0} \, e^{\frac{\alpha_\text{O} t^2}{2}} \, dt \, ,
\end{eqaed}
while the conformal factor reads
\begin{eqaed}\label{eq:omega_orientifold_cosm_mod}
	e^{2 \Omega(\eta)} = \abs{\alpha_{\text{O}} \, t^2}^{\frac{1}{18}} \, e^{\frac{\alpha_{\text{O}} t^2}{8}} \, .
\end{eqaed}
On the other hand, for the heterotic model
\begin{eqaed}\label{eq:eta_heterotic_mod}
	d\eta = \left(\sin \abs{\sqrt{\alpha_{\text{H}}} \, t}\right)^{- \frac{2}{3}} \left(\cos \abs{\sqrt{\alpha_{\text{H}}} \, t}\right)^{- \frac{7}{3}} e^{- \frac{5}{4} \Phi_0} \, dt \, ,
\end{eqaed}
and
\begin{eqaed}\label{eq:omega_heterotic_cosm_mod}
	e^{2 \Omega(\eta)} = \left(\sin \abs{\sqrt{\alpha_{\text{H}}} \, t}\right)^{\frac{1}{12}} \left(\cos \abs{\sqrt{\alpha_{\text{H}}} \, t}\right)^{- \frac{1}{3}} \, .
\end{eqaed}
In both models one can choose the range of $\eta$ to be $(0,\infty)$, with the initial singularity at the origin, but in this case the future singularity is not reached in a finite proper time. Moreover, while string loops are in principle under control for $g_s \ll 1$, curvature corrections are expected to be relevant at the initial singularity~\cite{Condeescu:2013gaa}.

\subsection{Flux compactifications}\label{sec:flux_compactifications}

While the Dudas-Mourad solutions that we have discussed in the preceding section feature the maximal amount of symmetry available in the string models at stake, they are fraught with regions where the low-energy effective theory of eq.~\eqref{eq:action} is expected to be unreliable. In order to address this issue, in this section we turn on form fluxes, and study Freund-Rubin compactifications. While the parameters of eq.~\eqref{eq:bsb_electric_params} and~\eqref{eq:het_magnetic_params} allow only for $\ads$ solutions, it is instructive to investigate the general case in detail. To this effect, we remark that the results presented in the following sections apply to general $V(\phi)$ and $f(\phi)$, up to the replacement
\begin{eqaed}\label{eq:v_f_non-exp}
	\gamma \; \to \; \frac{V'(\phi_0)}{V(\phi_0)} \, , \qquad \alpha \; \to \; \frac{f'(\phi_0)}{f(\phi_0)} \, ,
\end{eqaed}
since the dilaton is stabilized to a constant value $\phi_0$.

\subsubsection{Freund-Rubin solutions}\label{sec:freund-rubin_solutions}

Since \textit{a priori} both electric and magnetic fluxes may be turned on, let us fix the convention that $\alpha > 0$ in the frame where the field strength $H_{p+2}$ is a $(p+2)$-form. With this convention, the dilaton equation of motion implies that a Freund-Rubin solution\footnote{The Laplacian spectrum of the internal space $\mathcal{M}_q$ can have some bearing on perturbative stability.} can only exist with an electric flux, and is thus of the form $X_{p+2} \times \mathcal{M}_q$. Here $X_{p+2}$ is Lorentzian and maximally symmetric with curvature radius $L$, while $\mathcal{M}_q$ is a compact Einstein space with curvature radius $R$. The corresponding ansatz takes the form
\begin{eqaed}\label{eq:adsxs_ansatz}
	ds^2 & = L^2 \, ds_{X_{p+2}}^2 + R^2 \, ds_{\mathcal{M}_q}^2 \, , \\
	H_{p+2} & = c \, \Vol_{X_{p+2}} \, , \\
	\phi & = \phi_0 \, ,
\end{eqaed}
where $ds_{X_{p+2}}^2$ is the unit-radius space-time metric and $\Vol_{X_{p+2}}$ denotes the canonical volume form on $X_{p+2}$ with radius $L$. The dilaton is stabilized to a \textit{constant} value by the electric form flux on internal space\footnote{The flux $n$ in eq.~\eqref{eq:electric_flux} is normalized for later convenience, although it is not dimensionless nor an integer.},
\begin{eqaed}\label{eq:electric_flux}
	n = \frac{1}{\Omega_q} \int_{\mathcal{M}_q} f \star H_{p+2} = c \, f \, R^q \, ,
\end{eqaed}
whose presence balances the runaway tendency of the dilaton potential. Here $\Omega_q$ denotes the volume of the unit-radius internal manifold. Writing the Ricci tensor
\begin{eqaed}\label{eq:ricci_tensors_freund_rubin}
	R_{\mu \nu} & = \sigma_X \, \frac{p+1}{L^2} \, g_{\mu \nu} \, , \\
	R_{ij} & = \sigma_{\mathcal{M}} \, \frac{q-1}{R^2} \, g_{ij}
\end{eqaed}
in terms of $\sigma_X \, , \, \sigma_{\mathcal{M}} \in \{-1 \, , \, 0 \, , \, 1 \}$, the geometry exists if and only if 
\begin{eqaed}\label{eq:adsxs_existence_conditions}
	\sigma_{\mathcal{M}} = 1 \, , \qquad \alpha > 0 \, , \qquad q > 1 \, , \qquad \sigma_X \left( \left(q-1\right) \frac{\gamma}{\alpha} - 1 \right) < 0 \, ,
\end{eqaed}
and using eq.~\eqref{eq:potential_form-coupling} the values of the string coupling $g_s = e^{\phi_0}$ and the curvature radii $L \, , \, R$ are given by
\begin{importantbox}
	\begin{eqaed}\label{eq:ads_s_solution}
		c & = \frac{n}{g_s^\alpha R^q} \, , \\
		g_s^{(q-1)\gamma-\alpha} & = \left(\frac{(q-1)(D-2)}{\left(1+\frac{\gamma}{\alpha} \left(p+1 \right) \right)T}\right)^q \,\frac{2\gamma T}{\alpha n^2} \, , \\
		R^{2\frac{\left(q-1\right)\gamma-\alpha}{\gamma}} & = \left( \frac{\alpha + \left(p+1\right) \gamma}{(q-1)(D-2)}\right)^{\frac{\alpha+\gamma}{\gamma}} \left(\frac{T}{\alpha}\right)^{\frac{\alpha}{\gamma}}\frac{n^2}{2\gamma} \, , \\
		L^2 & = - \, \sigma_X \, R^2 \left(\frac{p+1}{q-1} \cdot \frac{\left(p+1\right)\gamma+ \alpha}{\left(q-1\right)\gamma- \alpha}\right) \equiv \frac{R^2}{A} \, .
	\end{eqaed}
\end{importantbox}
From eq.~\eqref{eq:ads_s_solution} one can observe that the ratio of the curvature radii is a constant independent on $n$ but is not necessarily unity, in contrast with the case of the supersymmetric $\ads_5 \times \ess^5$ solution of type IIB supergravity. Furthermore, in the actual string models the existence conditions imply $\sigma_X = -1$, \text{i.e.} an $\ads_{p+2} \times \mathcal{M}_q$ solution.

These solutions exhibit a number of interesting features. To begin with, they only exist in the presence of the dilaton potential, and indeed they have no counterpart in the supersymmetric case for $p \neq 3$. Moreover, the dilaton is constant, but in contrast to the supersymmetric $\ads_5 \times \ess^5$ solution its value is not a free parameter. Instead, the solution is entirely fixed by the flux parameter $n$. Finally, in the case of $\ads$ the large-$n$ limit always corresponds to a perturbative regime where both the string coupling and the curvatures are parametrically small, thus suggesting that the solution reliably captures the dynamics of string theory for its special values of $p$ and $q$. As a final remark, let us stress that only one sign of $\alpha$ can support a vacuum with \textit{electric} flux threading the internal manifold. However, models with the opposite sign admit vacua with \textit{magnetic} flux, which can be included in our general solution dualizing the form field, and thus also inverting the sign of $\alpha$. No solutions of this type exist if $\alpha = 0$, which is the case relevant to the back-reaction of $\text{D}3$-branes in the type $0'\text{B}$ model. Indeed, earlier attempts in this respect~\cite{Angelantonj:1999qg, Angelantonj:2000kh,Dudas:2000sn} were met by non-homogeneous deviations from $\ads_5$, which are suppressed, but not uniformly so, in large-$n$ limit\footnote{Analogous results in tachyonic type $0$ strings were obtained in~\cite{Klebanov:1998yya}.}.

\subsubsection{In the orientifold models: \texorpdfstring{$\ads_3 \times \mathcal{M}_7$}{AdS3 x M7} solutions}\label{sec:ads3xm7}

For later convenience, let us present the explicit solution in the case of the two orientifold models. Since $\alpha = 1$ in this case, they admit $\ads_3 \times \mathcal{M}_7$ solutions with electric flux, and in particular $\mathcal{M}_7 = \ess^7$ ought to correspond to near-horizon geometries of $\text{D}1$-brane stacks, according to the microscopic picture that we shall discuss in Section~\ref{Section4} and Section~\ref{Section5}. On the other hand, while $\text{D}5$-branes are also present in the perturbative spectra of these models~\cite{Dudas:2001wd}, they appear to behave differently in this respect, since no corresponding $\ads_7 \times \ess_3$ vacuum exists\footnote{This is easily seen dualizing the three-form in the orientifold action \eqref{eq:bsb_electric_params}, which inverts the sign of $\alpha$, in turn violating the condition of eq.~\eqref{eq:adsxs_existence_conditions}.}. Using the values in eq.~\eqref{eq:bsb_electric_params}, one finds
\begin{eqaed}\label{eq:bsb_solution}
	g_s & = 3 \times 2^{\frac{7}{4}} \, T^{-\frac{3}{4}} n^{-\frac{1}{4}} \, , \\
	R & = 3^{-\frac{1}{4}} \times 2^{-\frac{5}{16}} \, T^{\frac{1}{16}} \, n^{\frac{3}{16}} \, , \\
	L^2 & = \frac{R^2}{6} \, .
\end{eqaed}
Since every parameter in this $\ads_3 \times \mathcal{M}_7$ solution is proportional to a power of $n$, one can use the scalings
\begin{equation}\label{eq:bsb_scalings}
g_s \propto n^{-\frac{1}{4}} \, , \qquad R \propto n^{\frac{3}{16}}
\end{equation}
to quickly derive some of the results that we shall present in Section~\ref{Section4}.

\subsubsection{In the heterotic model: \texorpdfstring{$\ads_7 \times \mathcal{M}_3$}{AdS7 x M3} solutions}\label{sec:ads7xm3}

The case of the heterotic model is somewhat subtler, since the physical parameters of eq.~\eqref{eq:het_electric_params} only allow for solutions with \textit{magnetic flux},
\begin{eqaed}\label{eq:het_magnetic_flux}
	n = \frac{1}{\Omega_3} \int_{\mathcal{M}_3} H_3 \, .
\end{eqaed}
The corresponding microscopic picture, which we shall discuss in Section~\ref{Section4} and Section~\ref{Section5}, would involve $\text{NS}5$-branes, while the dual electric solution, which would be associated to fundamental heterotic strings, is absent. Dualities of the strong/weak type could possibly shed light on the fate of these fundamental strings, but their current understanding in the non-supersymmetric context is limited\footnote{Despite conceptual and technical issues, non-supersymmetric dualities connecting the heterotic model to open strings have been explored in~\cite{Blum:1997cs,Blum:1997gw}. Similar interpolation techniques have been employed in~\cite{Faraggi:2009xy}. A non-perturbative interpretation of non-supersymmetric heterotic models has been proposed in~\cite{Faraggi:2007tj}.}.

In the present case the Kalb-Ramond form lives on the internal space, so that dualizing it one can recast the solution in the form of eq.~\eqref{eq:ads_s_solution}, using the values in eq.~\eqref{eq:het_magnetic_params} for the parameters. The resulting $\ads_7 \times \mathcal{M}_3$ solution is described by
\begin{eqaed}\label{eq:het_solution}
	g_s & = 5^{\frac{1}{4}} \, T^{-\frac{1}{2}} n^{-\frac{1}{2}} \, , \\
	R & = 5^{-\frac{5}{16}} \, T^{\frac{1}{8}} \, n^{\frac{5}{8}} \, , \\
	L^2 & = 12 \, R^2 \, ,
\end{eqaed}
so that the relevant scalings are
\begin{equation}\label{eq:het_scalings}
g_s \propto n^{-\frac{1}{2}} \, , \qquad R \propto n^{\frac{5}{8}} \, .
\end{equation}

As a natural generalization of the Freund-Rubin solutions that we have described in the preceding section, one can consider flux compactifications on products of Einstein spaces~\cite{Basile:2020xwi}. The resulting multi-flux landscapes appear considerably more complicated to approach analytically, but can feature regimes where some of the internal curvatures are parametrically smaller than the other factors, including space-time~\cite{Lust:2020npd}. However, this type of scale separation does not reduce the effective space-time dimension at low energies, which appears to resonate with the results of~\cite{Lust:2020npd} and with recent conjectures regarding scale separation in the absence of supersymmetry~\cite{Gautason:2015tig, Lust:2019zwm}\footnote{For recent results on the issue of scale separation in supersymmetric $\ads$ compactifications, see~\cite{Marchesano:2020qvg}. For a more general approach, see~\cite{DeLuca:2021mcj}.}.

As a final remark, it is worth noting that the stability properties of multi-flux landscapes appear qualitatively different from the those of single-flux landscapes. This issue has been addressed in~\cite{Brown:2010bc} in the context of models with no exponential dilaton potentials.

\subsection{de Sitter cosmology: no-gos and brane-worlds}\label{Section7}

In this section we address the possibility of $\ds$ flux compactifications, starting from the Freund-Rubin case that we have described in the preceding section. In order to assess whether $\ds$ are allowed in the string models discussed in Section~\ref{Section2}, in Section~\ref{sec:warped_flux_compactifications} we examine general warped flux compactifications, along the lines of~\cite{Dasgupta:1999ss, Giddings:2001yu}, and we obtain conditions that fix the (sign of the) resulting cosmological constant in terms of the parameters of the model, generalizing the results of~\cite{Gibbons:1984kp, Maldacena:2000mw, Giddings:2001yu, Hertzberg:2007wc} to models with exponential potentials. In Section~\ref{sec:tcc_dsc} we discuss how our results connect to recent swampland conjectures~\cite{Ooguri:2006in, Obied:2018sgi, Ooguri:2018wrx, Lust:2019zwm, Lee:2019xtm, Lee:2019wij, Baume:2020dqd, Perlmutter:2020buo}.

The issue of $\ds$ configurations in string theory has proven to be remarkably challenging, to the extent that the most well-studied constructions~\cite{Kachru:2003aw} have been subject to thorough scrutiny and discussion. We shall not attempt to provide a comprehensive account of this extensive subject and its state of affairs, since our focus in the present case lies on higher-dimensional approaches~\cite{Koerber:2007xk, Moritz:2017xto, Kallosh:2018nrk, Bena:2018fqc, Gautason:2018gln, Hamada:2019ack, Gautason:2019jwq} and, in particular, in the search for new solutions~\cite{Danielsson:2009ff, Cordova:2018dbb, Blaback:2019zig, Cribiori:2019clo, Andriot:2019wrs, Shukla:2019dqd, Shukla:2019akv, Cordova:2019cvf, Andriot:2020wpp, Andriot:2020vlg, Bena:2020qpa, Andriot:2021rdy,DeLuca:2021pej}. Specifically, the issue at stake is whether the ingredients provided by string-scale supersymmetry breaking can allow for $\ds$ compactifications. While a number of parallels between lower-dimensional anti-brane uplifts and the ten-dimensional BSB scenario discussed in Section~\ref{Section1} appear encouraging to this effect, as we shall see shortly the presence of exponential potentials does not ameliorate the situation, insofar as (warped) flux compactifications are concerned. On the other hand, as explained in~\cite{Basile:2020mpt, Basile:2020xwi}, the very presence of exponential potentials allows for intriguing brane-world scenarios within the $\ads$ landscapes discussed in Section~\ref{Section2}, whose non-perturbative instabilities, addressed in Section~\ref{Section4}, play a crucial rôle in this respect.

\subsubsection{No-go for de Sitter compactifications: first hints}\label{sec:hessian_freund-rubin}

From the general Freund-Rubin solution one can observe that $\ds$ Freund-Rubin compactifications exist only whenever\footnote{The same result was derived in~\cite{Montero:2020rpl}.}
\begin{eqaed}\label{eq:de-sitter_freund_rubin_condition}
	(q-1) \, \frac{\gamma}{\alpha} - 1 < 0 \, .
\end{eqaed}
However, this requirement also implies the existence of perturbative instabilities. This can be verified studying fluctuations of the $(p+2)$-dimensional metric, denoted by $\widetilde{ds}^2_{p+2}(x)$, and of the radion $\psi(x)$, writing
\begin{eqaed}\label{eq:radion_ansatz}
	ds^2 = e^{- \frac{2q}{p} \psi(x)} \, \widetilde{ds}_{p+2}^2(x) + R_0^2 \, e^{2\psi(x)} \, ds^2_{\mathcal{M}_q}
\end{eqaed}
with $R_0$ an arbitrary reference radius, thus selecting the $(p+2)$-dimensional Einstein frame. The corresponding effective potential for the dilaton and radion fields
\begin{eqaed}\label{eq:dilaton-radion_potential}
	\mathcal{V}(\phi, \psi) & = V(\phi) \, e^{- \frac{2q}{p} \psi} - \frac{q(q-1)}{R_0^2} \, e^{- \frac{2(D-2)}{p} \psi} + \frac{n^2}{2R_0^{2q}} \, \frac{e^{- \frac{q(p+1)}{p} \psi}}{f(\phi)} \\
	& \equiv \mathcal{V}_T + \mathcal{V}_{\mathcal{M}} + \mathcal{V}_n
\end{eqaed}
reproduces the Freund-Rubin solution when extremized\footnote{Notice that, in order to derive eq.~\eqref{eq:dilaton-radion_potential} substituting the ansatz of eq.~\eqref{eq:radion_ansatz} in the action, the flux contribution is to be expressed in the magnetic frame, since the correct equations of motion arise varying $\phi$ and $B_{p+1}$ independently, while the electric-frame ansatz relates them.}, and identifies three contributions: the first arises from the dilaton tadpole, the second arises from the curvature of the internal space, and the third arises from the flux. Since each contribution is exponential in both $\phi$ and $\psi$, extremizing $\mathcal{V}$ one can express $\mathcal{V}_{\mathcal{M}}$ and $\mathcal{V}_n$ in terms of $\mathcal{V}_T$, so that
\begin{eqaed}\label{eq:on-shell_potential}
	\mathcal{V} = \frac{p}{D-2} \left(1 - \left( q - 1 \right) \frac{\gamma}{\alpha} \right) \mathcal{V}_T \, ,
\end{eqaed}
which is indeed positive whenever eq.~\eqref{eq:de-sitter_freund_rubin_condition} holds. Moreover, the same procedure also shows that the determinant of the corresponding Hessian matrix is proportional to $(q-1) \, \frac{\gamma}{\alpha} - 1$, so that de Sitter solutions always entail an instability.

While the results of our analysis, presented in~\cite{Basile:2020mpt}, resonate with the ones of~\cite{Montero:2020rpl}, one could wonder whether similar conditions hold for more general $\ds$ settings, \text{e.g.} for fluxes threading cycles of complicated internal manifolds.

\subsubsection{Warped flux compactifications: no-go results}\label{sec:warped_flux_compactifications}

In order to address the problem of $\ds$ solutions to low-energy effective theories with exponential potentials in more generality, let us consider a compactification of the $D$-dimensional theory discussed in Section~\ref{Section2} on a $d_Y$-dimensional closed manifold $Y$ parametrized by coordinates $y^i$, while the $d_X$-dimensional space-time is parametrized by coordinates $x^\mu$. Excluding the Freund-Rubin compactifications, which we have already described in~\ref{Section2}, in the models of interest the space-time dimension does not match the rank of the form field strength, and thus there cannot be any electric flux. Since at any rate one can dualize the relevant forms, we shall henceforth work in the magnetic frame, which in our convention involves a $q$-form field strength with the coupling $f(\phi) = e^{-\alpha \phi}$ to the dilaton, and we shall seek configurations where $H_q$ is supported on $Y$, and where each field only depends on the $y^i$. Writing the metric
\begin{eqaed}\label{eq:warped_metric}
	ds^2 = e^{2b u(y)} \, \widehat{ds}^2(x) + e^{2 u(y)} \, \widetilde{ds}^2(y) \, ,
\end{eqaed}
with $b = - \, \frac{d_Y}{d_X-2}$ in order to select the $d_X$-dimensional Einstein frame, one finds that sufficiently well-behaved functions $h(y)$ satisfy
\begin{eqaed}\label{eq:laplacian_zeros}
	\int_Y d^{d_Y}y \sqrt{\widetilde{g}} \, e^{2b u(y)} \, \Box_D h(y) = 0 \, , \qquad \int_Y d^{d_Y}y \sqrt{\widetilde{g}} \, \Delta_Y h(y) = 0 \, ,
\end{eqaed}
where $\Box_D$ and $\Delta_Y$ denote the $D$-dimensional d'Alembert operator and the Laplacian operator on $Y$ respectively. Furthermore, let us define
\begin{eqaed}\label{eq:warped_integrals}
	\mathcal{I}_V \equiv \int_Y d^{d_Y}y \sqrt{\widetilde{g}} \, e^{2b u(y)} \, V > 0 \, , \qquad
	\mathcal{I}_H \equiv \int_Y d^{d_Y}y \sqrt{\widetilde{g}} \, e^{2b u(y)} \, \frac{f}{q!} \, H_q^2 > 0
\end{eqaed}
for convenience. Using these relations, integrating the equation of motion for the dilaton yields
\begin{eqaed}\label{eq:integrated_dilaton_eq}
	\mathcal{I}_H = \frac{2\gamma}{\alpha} \, \mathcal{I}_V \, .
\end{eqaed}
In order to proceed, let us collect general results concerning warped products. Let us consider a multiple warped product described by a metric of the type
\begin{eqaed}\label{eq:N-fold_metric}
	ds^2 = \widetilde{ds}^2(x) + \sum_I \, e^{2 a_I(x)} \, \widehat{ds}^2_{(I)} \, ,
\end{eqaed}
where the dimensions of the $I$-th internal space is denoted by $q_I$. The Ricci tensor is then block-diagonal, and its space-time components read
\begin{eqaed}\label{eq:N-fold_ricci_ext}
	R_{\mu \nu} = \widetilde{R}_{\mu\nu} - \sum_I \, q_I \left(\nabla_\mu \nabla_\nu a_I + (\nabla_\mu a_I) (\nabla_\nu a_I) \right) \, ,
\end{eqaed}
while its internal components in the $I$-th internal space read
\begin{eqaed}\label{eq:N-fold_ricci_int}
	R^{(I)}_{ij} = \widehat{R}^{(I)}_{ij} - e^{2 a_I(x)} \left( \Delta \, a_I + \sum_J \, q_J \left(\partial^\mu a_J\right) \left(\partial_\mu a_I\right) \right) \widehat{g}^{(I)}_{ij} \, ,
\end{eqaed}
where $\Delta$ denotes the Laplacian operator associated to space-time and we have kept the notation signature-independent for the sake of generality. Employing eqs.~\eqref{eq:N-fold_ricci_ext} and~\eqref{eq:N-fold_ricci_int}, the space-time Ricci tensor takes the form
\begin{eqaed}\label{eq:warped_efe}
	R_{\mu \nu} & = \widehat{R}_{\mu \nu} - b \, e^{-2u} \left( \Delta_Y u - \frac{2(D-2)}{d_X-2} \, \abs{\nabla u}^2 \right) \, g_{\mu \nu} \\
	& = \widehat{R}_{\mu \nu} - \frac{d_Y}{2(D-2)} \, e^{-2b u} \, \Delta_Y \left( e^{- \frac{2(D-2)}{d_X-2} u} \right) \, .
\end{eqaed}
Hence, assuming a maximally symmetric space-time with $\widehat{R}_{\mu \nu} = \frac{2\Lambda}{d_X-2} \, \widehat{g}_{\mu \nu}$, integrating the space-time Einstein equations finally yields~\cite{Basile:2020mpt}
\begin{importantbox}
	\begin{eqaed}\label{eq:integrated_efe}
		\text{vol}(Y) \, \Lambda & = \frac{d_X-2}{2(D-2)} \left( \mathcal{I}_V - \frac{q-1}{2} \, \mathcal{I}_H \right) \\& = \frac{d_X-2}{2(D-2)} \left( 1 - \left(q-1\right) \frac{\gamma}{\alpha} \right) \mathcal{I}_V \, , 
	\end{eqaed}
\end{importantbox}
where $\text{vol}(Y) \equiv \int_Y \sqrt{\widetilde{g}}$ is the (unwarped) volume of $Y$. This result\footnote{As we have anticipated, eq.~\eqref{eq:integrated_efe} can be thought of as a generalization of the no-go results of~\cite{Gibbons:1984kp, Maldacena:2000mw} to models with exponential potentials.} shows that the existence condition for de Sitter Freund-Rubin compactifications actually extends to general warped flux compactifications as well, thus excluding this class of solutions for the string models that we have studied in the preceding sections. All in all, the no-go result of eq.~\eqref{eq:integrated_efe} shows that \textit{the effective action of eqs.~\eqref{eq:action} and~\eqref{eq:potential_form-coupling}, including an exponential dilaton potential, does not admit $\ds$ warped flux compactifications of the form of eq.~\eqref{eq:warped_metric} whenever $(q-1) \, \frac{\gamma}{\alpha} > 1$}. In particular, this inequality holds for the string models described by eqs.~\eqref{eq:bsb_electric_params} and~\eqref{eq:het_electric_params}, for which the contribution of the gravitational tadpole does not suffice to obtain $\ds$ vacua. This result can be further extended including the presence of localized sources~\cite{Basile:2020mpt, Basile:2020xwi}, in the spirit of~\cite{Giddings:2001yu}.

\subsubsection{Relations to swampland conjectures}\label{sec:tcc_dsc}

\begin{figure}[!ht]
	\centering
	\includegraphics[width=0.42\textwidth]{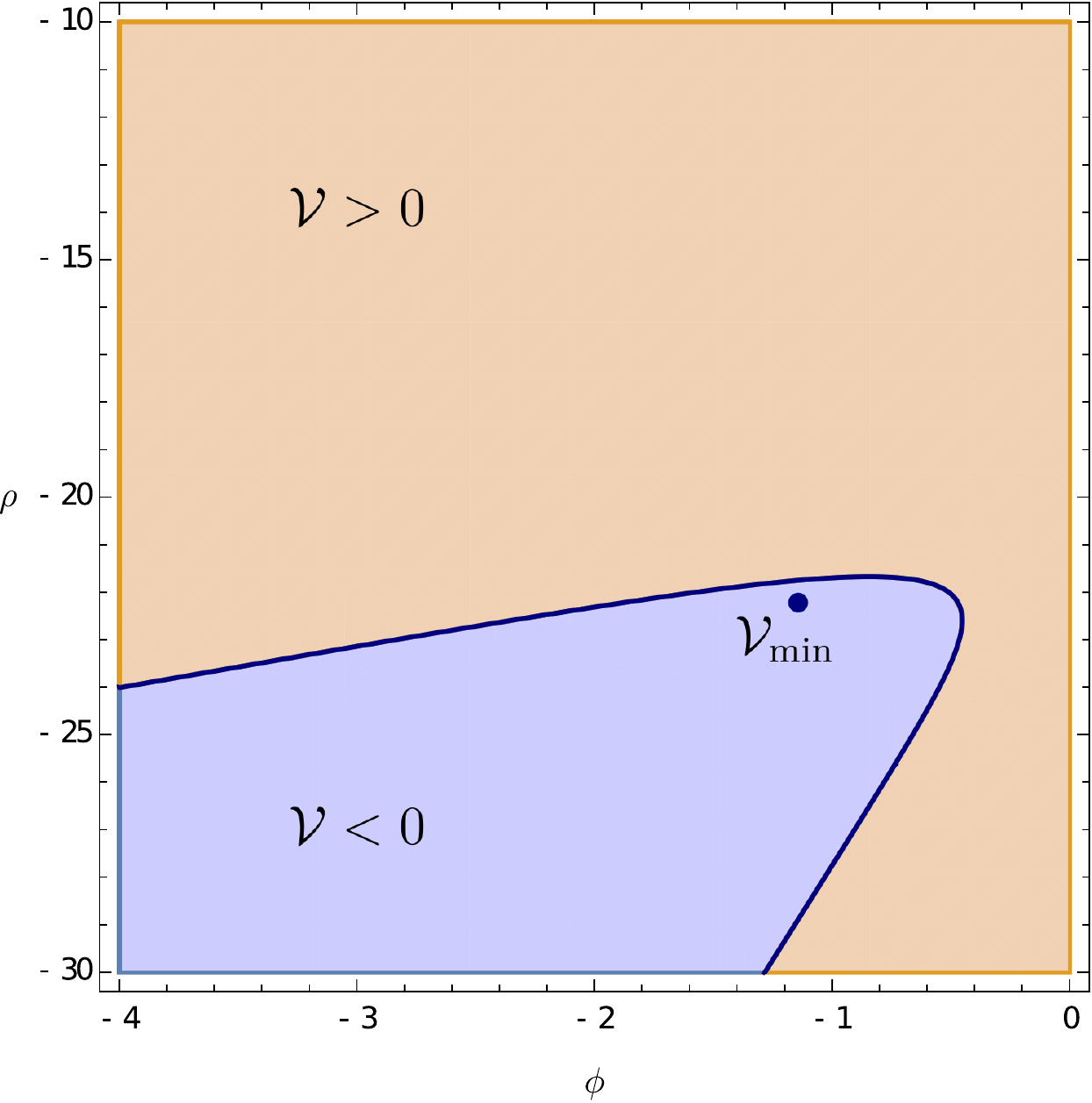}
	\hspace{25pt}
	\includegraphics[width=0.42\textwidth]{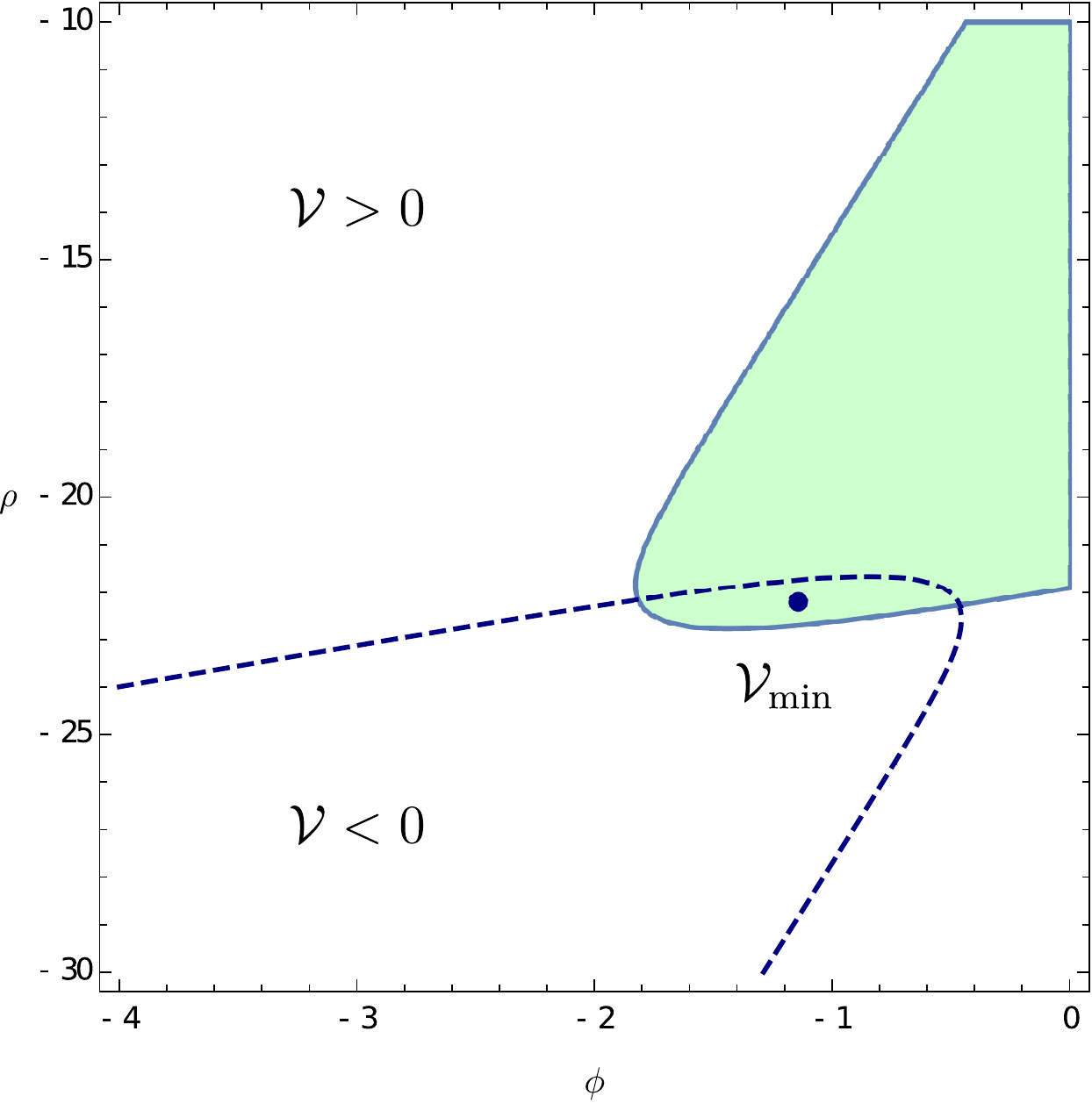}
	\caption{plots of the sign of the potential of eq.~\eqref{eq:dilaton-radion_ds} in units of $T$, with its minimum marked, and of the signature of its Hessian matrix in the orientifold models. Left: regions where the potential is positive (orange) and negative (blue), for $n = 10^6$. Right: region where its Hessian matrix is positive-definite (green).}
	\label{fig:v_bsb}
\end{figure}

\begin{figure}[!ht]
	\centering
	\includegraphics[width=0.42\textwidth]{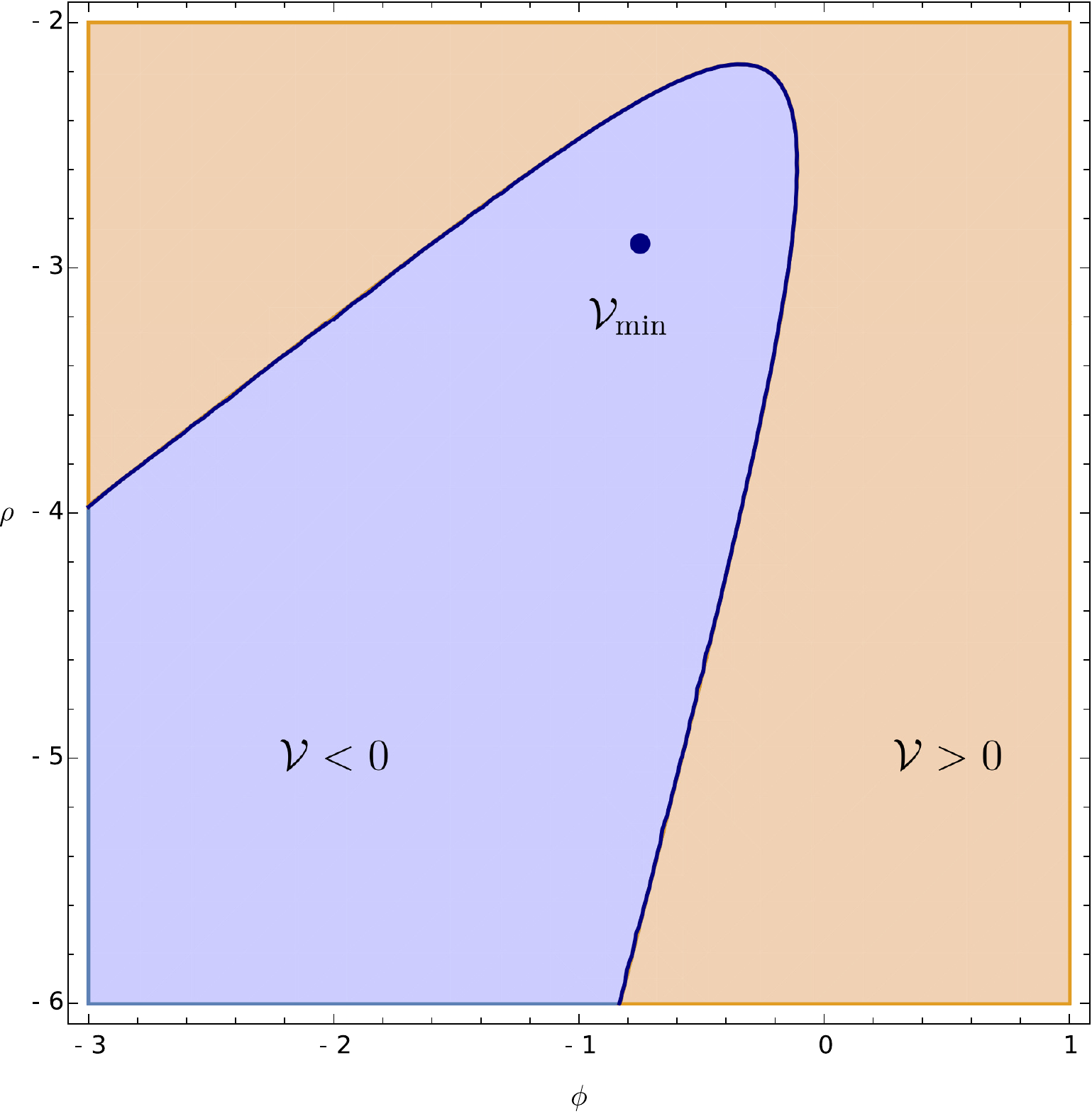}
	\hspace{25pt}
	\includegraphics[width=0.42\textwidth]{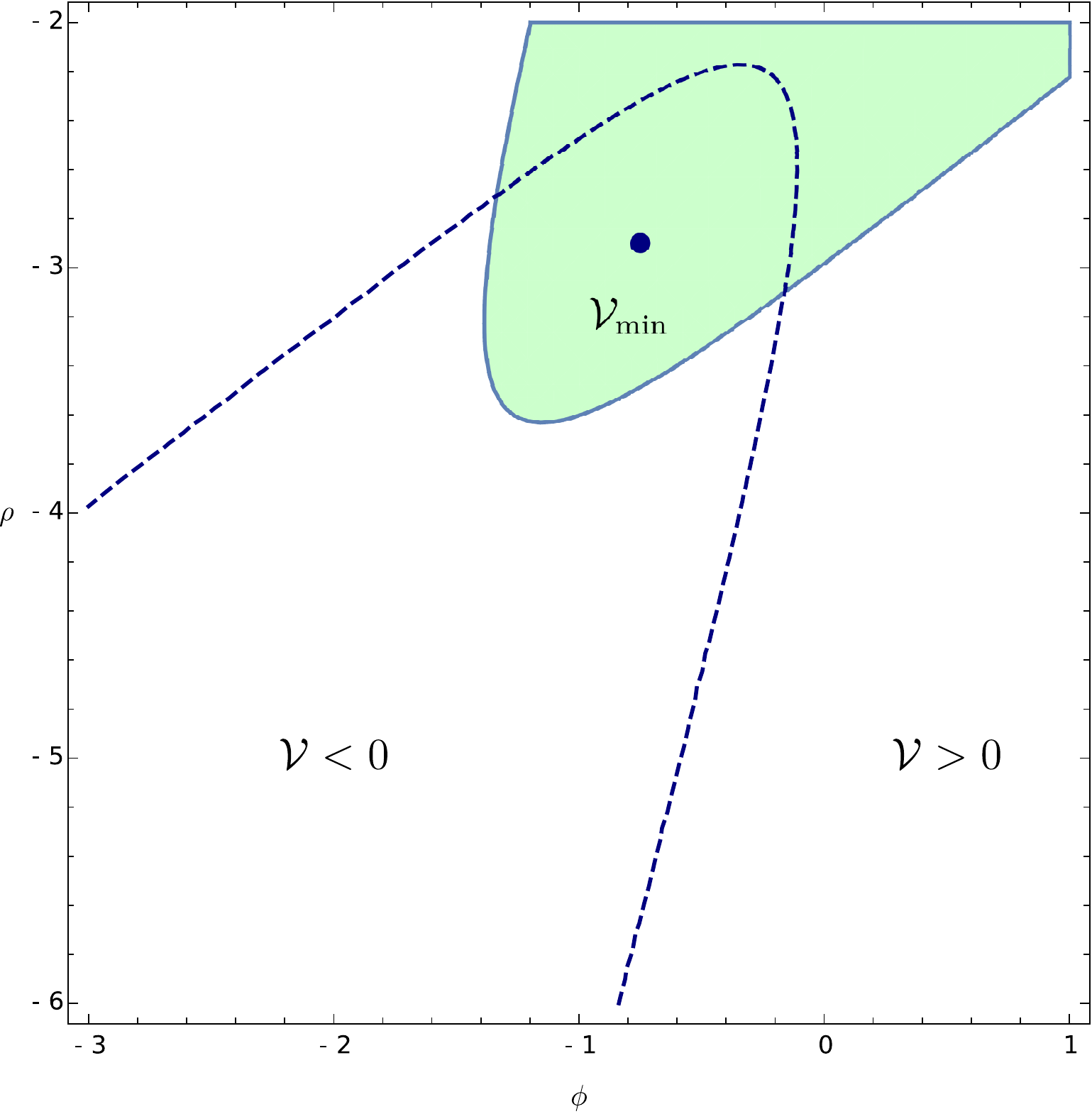}
	\caption{plots of the sign of the potential of eq.~\eqref{eq:dilaton-radion_ds} in units of $T$, with its minimum marked, and of the signature of its Hessian matrix in the heterotic model. Left: regions where the potential is positive (orange) and negative (blue), for $n = 10$. Right: region where its Hessian matrix is positive-definite (green).}
	\label{fig:v_het}
\end{figure}

Let us now comment on whether our results support the recent conjectures concerning the existence of $\ds$ solutions in string theory~\cite{Obied:2018sgi, Ooguri:2018wrx}, showing that the ratio $\frac{\abs{\nabla \mathcal{V}}}{\mathcal{V}}$ is bounded from below by an $\mathcal{O}\!\left(1\right)$ constant $c$ whenever the effective potential $\mathcal{V} > 0$. Extending our preceding arguments to the effect that $\ds$ Freund-Rubin compactifications are unstable in the dilaton-radion sector, let us recall the corresponding (magnetic-frame) effective potential, whose relevant features are highlighted in fig.~\ref{fig:v_bsb} (resp. fig.~\ref{fig:v_het}) for the orientifold models (resp. for the heterotic model), reads
\begin{eqaed}\label{eq:dilaton-radion_ds}
	\mathcal{V}(\phi, \psi) & = V(\phi) \, e^{- \frac{2q}{p} \psi} - \frac{q(q-1)}{R^2} \, e^{- \frac{2(D-2)}{p} \psi} + \frac{n^2}{2R^{2q}} \, f(\phi) \, e^{- \frac{q(p+1)}{p} \psi} \, ,
\end{eqaed}
where we have shifted the radion in order to place its on-shell value to zero, and we have replaced $R_0 \; \to \; R$ accordingly. Then, introducing the canonically normalized radion $\rho$, defined by
\begin{eqaed}\label{eq:canonically_norm_radion}
	- \, \frac{q}{p} \, \psi \equiv \sqrt{\frac{q}{2 \, p \left(D-2\right)}} \, \rho \, ,
\end{eqaed}
the ratio of interest takes the form
\begin{eqaed}\label{eq:grad_v_v}
	\frac{\abs{\grad \mathcal{V}}}{\mathcal{V}} = \frac{\sqrt{\left(\partial_\phi \mathcal{V} \right)^2 + \left(\partial_\rho \mathcal{V} \right)^2}}{\mathcal{V}} \, ,
\end{eqaed}
while shifting $\phi$ one can also do away with the remaining parametric dependence on the dimensionless combination $\nu \equiv n \, T^{\frac{q - 1}{2}}$. Altogether, the resulting ratios depend only on $\phi$ and $\rho$, and we have minimized them numerically imposing the constraint\footnote{The constraint $\mathcal{V} > 0$ can also be recast in terms of $\phi$ and $\rho$ only, with no parametric dependence left.} $\mathcal{V} > 0$, finding approximately $2$ (resp. $2.5$) for the orientifold models (resp. the heterotic model). This result resonates with the $\ds$ swampland conjecture of~\cite{Obied:2018sgi, Ooguri:2018wrx}, showing that in this case $\ds$ solutions are behind an $\mathcal{O}\!\left(1\right)$ ``barrier'' in the sense of eq.~\eqref{eq:grad_v_v}. 

The above considerations can be extended to the more general warped flux compactifications that we have discussed in Section~\ref{sec:warped_flux_compactifications}. In this case, in terms of the canonically normalized dilaton and radion fields\footnote{Notice that, in order to canonically normalize the radion, one needs to rescale the field $\psi(x)$ that we have introduced in Section~\ref{Section2}.} $\phi \, , \rho$, the effective potential is given by
\begin{eqaed}\label{eq:warped_eff_pot}
	\mathcal{V}(\phi, \rho) = \mathcal{I}_V \, e^{2k \rho} - \int_Y d^{d_Y}y \, \sqrt{- \widetilde{g}} \, e^{2bu(y)} R_{\mathcal{M}_q} \, e^{\frac{2k(D-2)}{d_Y} \rho}  + \frac{1}{2} \, \mathcal{I}_H \, e^{2k(d_X - 1) \rho} \, ,
\end{eqaed}
where we have introduced
\begin{eqaed}\label{eq:canon_rescaling}
	k \equiv \sqrt{\frac{d_Y}{2\left(d_X-2\right)\left(D-2\right)}}
\end{eqaed}
in order to canonically normalize $\rho$. Using the integrals defined in eq.~\eqref{eq:warped_integrals}, one can recast the potential of eq.~\eqref{eq:warped_eff_pot} in terms of its derivatives according to
\begin{eqaed}\label{eq:pot_as_derivatives}
	\mathcal{V} & = \frac{d_Y (d_X - 1 )}{\alpha ( D - 2 )} \, \partial_\phi \mathcal{V} + \frac{d_Y}{2 k ( D - 2 )} \, \partial_\rho \mathcal{V} \\
	& + \frac{d_X - 2}{D - 2} \left( 1 - \left(d_Y - 1 \right) \frac{\gamma}{\alpha} \right) \mathcal{I}_V \, e^{2 k \rho} \, ,
\end{eqaed}
and, since $d_Y \geq q$ in order to allow for magnetic fluxes, one finds that
\begin{eqaed}\label{eq:eff_pot_ineq}
	\frac{d_Y (d_X - 1 )}{\alpha ( D - 2 )} \, \partial_\phi \mathcal{V} + \frac{d_Y}{2 k ( D - 2 )} \, \partial_\rho \mathcal{V} \geq \mathcal{V}
\end{eqaed}
holds off-shell whenever the no-go result discussed in Section~\ref{sec:warped_flux_compactifications} applies. Then, applying the Cauchy-Schwartz inequality one arrives at
\begin{eqaed}\label{eq:cauchy_schwartz}
	\sqrt{\left(\partial_\phi \mathcal{V} \right)^2 + \left(\partial_\rho \mathcal{V} \right)^2} \geq \frac{\sqrt{2} \, \alpha \left(D - 2\right)}{\sqrt{d_Y \left(2 \, d_Y \left(d_X - 1 \right)^2 + \alpha^2 \left(D - 2\right) \left(d_X - 2\right) \right)}} \, \mathcal{V} \, ,
\end{eqaed}
which whenever $\mathcal{V} > 0$ provides an $\mathcal{O}\!\left(1\right)$ lower bound $c$ for the ratio of eq.~\eqref{eq:grad_v_v}.

This result, along with the further developments of~\cite{Basile:2020mpt}, constitutes non-trivial evidence for a number of swampland conjectures in top-down non-supersymmetric settings. It would be interesting to investigate additional swampland conjectures in the absence of supersymmetry and the resulting constraints on phenomenology~\cite{Abel:2015oxa, Abel:2017vos, Garg:2018reu, March-Russell:2020lkq}. In~\cite{Basile:2020mpt} we have also investigated the `Transplanckian Censorship conjecture'~\cite{Bedroya:2019snp, Bedroya:2019tba, Andriot:2020lea} and pointed out possible realizations of the `distance conjecture'~\cite{Ooguri:2006in, Ooguri:2018wrx, Lust:2019zwm, Lee:2019xtm, Lee:2019wij, Baume:2020dqd, Perlmutter:2020buo}, identifying Kaluza-Klein states as the relevant tower of states that become massless at infinite distance in field space. A more detailed analysis would presumably require a deeper knowledge of the geometry of the moduli spaces which can arise in non-supersymmetric compactifications, albeit our arguments rest solely on the existence of the ubiquitous dilaton-radion sector. It would be also interesting to address whether the `Distant Axionic String conjecture'~\cite{Lanza:2020qmt, Lanza:2021qsu}, which predicts the presence of axionic strings within any infinite-distance limit in field space, holds also in non-supersymmetric settings.

\section{Classical stability: perturbative analysis}\label{Section3}

In this section we investigate in detail the classical stability of the solutions that we have described in the preceding section, presenting the results of~\cite{Basile:2018irz}. To this end, we derive the linearized equations of motion for field fluctuations around each background, and we study the resulting conditions for stability. In Section~\ref{sec:dudas-mourad_stability} we study fluctuations around the Dudas-Mourad solutions, starting from the static case, and subsequently applying our results to the cosmological case in Section~\ref{sec:cosmological_case}. Intriguingly, in this case a logarithmic instability of the homogeneous tensor mode suggests a tendency toward dynamical compactification\footnote{An analogous idea in the context of higher-dimensional $\ds$ space-times was put forth in~\cite{Carroll:2009dn}.}. Then, in Section~\ref{sec:ads_stability} we proceed to the $\adsts$ solutions\footnote{A family of non-supersymmetric $\ads_7$ solutions of the type IIA superstring was recently studied in~\cite{Apruzzi:2013yva}, and its stability properties were investigated in~\cite{Apruzzi:2019ecr}.}, deriving the linearized equations of motion and comparing the resulting masses to the Breitenlohner-Freedman bounds. While the $\ads$ compactifications that we have obtained in the preceding section allow for general Einstein internal spaces, choosing the sphere simplifies the analysis of tensor and vector perturbations. Moreover, as we shall argue in Section~\ref{Section5}, the case of $\adsts$ appears to relate to near-horizon geometries sourced by brane stacks.

\subsection{Stability of static Dudas-Mourad solutions}\label{sec:dudas-mourad_stability}

Let us begin deriving the linearized equations of motion for the static Dudas-Mourad solutions that we have presented in the preceding section. The equations of interest are now
\begin{eqaed}\label{eq:dm_eom_mod}
	\Box \, \phi \, - V'(\phi) & = 0 \, , \\
	R_{MN} + \frac{1}{2} \, \partial_M \phi \, \partial_N \phi + \frac{1}{8} \, g_{MN} \, V & = 0 \, ,
\end{eqaed}
and the corresponding perturbed fields take the form
\begin{eqaed}\label{eq:dm_perturbed_fields}
	ds^2 & = e^{2 \Omega(z)} \left( \eta_{MN} + h_{MN}(x,z) \right) dx^M \, dx^N \, , \\
	\phi & = \phi(z) + \varphi(x,z) \, .
\end{eqaed}
As a result, the perturbed Ricci curvature can be extracted from
\begin{eqaed}\label{eq:dm_perturbed_ricci}
	R^{(1)}_{MN} = \; & 8 \, \nabla_M \nabla_N \Omega + \left(\eta_{MN} + h_{MN}\right) \nabla^A \nabla_A \Omega \\
	& - 8 \left(\nabla_M \Omega \, \nabla_N \Omega - \left(\eta_{MN} + h_{MN}\right) \nabla^A \Omega \, \nabla_A \Omega \right) \\
	& + \frac{1}{2} \left( \left( \Box_9 + \partial_z^2\right) h_{MN} - \nabla_M\left(\nabla \cdot h\right)_N - \nabla_N\left(\nabla \cdot h \right)_M + \nabla_M \nabla_N {h^A}_A \right) \, ,
\end{eqaed}
an expression valid up to first order in the perturbations. Here and henceforth $\Box_9$ denotes the d'Alembert operator pertaining to Minkowski slices, while in the following we shall denote derivatives $\partial_z$ with respect to $z$ by $f' \equiv \partial_z f$ (except for the dilaton potential $V$). In addition, covariant derivatives do not involve $\Omega$, and thus refer to $\eta_{MN}\,+\,h_{MN}$, which is also used to raise and lower indices. Up to first order the metric equations of motion thus read
\begin{eqaed}\label{eq:dm_efe_perturbed}
	R^{(1)}_{MN} & + \frac{1}{2} \, \partial_M \phi \, \partial_N \phi + \frac{1}{2} \, \partial_M \phi \, \partial_N \varphi + \frac{1}{2} \, \partial_M \varphi \, \partial_N \phi \\
	& + \frac{1}{8} \, e^{2 \Omega} \left( \left(\eta_{MN} + h_{MN} \right) V + \eta_{MN} \, V' \, \varphi \right) = 0 \, ,
\end{eqaed}
and combining this result with the dilaton equation of motion in eq.~\eqref{eq:dm_eom_mod} yields the unperturbed equations of motion
\begin{eqaed}\label{eq:dm_unperturbed_eom}
	\Omega'' + 8 \left(\Omega'\right)^2 + \frac{1}{8} \, e^{2 \Omega} \, V & = 0 \, , \\
	9 \, \Omega'' + \frac{1}{8} \, e^{2 \Omega} \, V + \frac{1}{2} \left(\phi'\right)^2 & = 0 \, , \\
	\phi'' + 8 \, \Omega' \, \phi' - e^{2 \Omega} \, V' & = 0 \, ,
\end{eqaed}
where $V$ and $V'$ shall henceforth denote the potential and its derivative computed on the classical vacuum.
Notice that the first two equations can be equivalently recast in the form
\begin{eqaed}\label{eq:dm_simplified_eom}
	72 \left(\Omega'\right)^2 - \, \frac{1}{2} \left(\phi'\right)^2 + e^{2 \Omega} \, V & = 0 \, , \\
	8 \left( \Omega'' - \left(\Omega'\right)^2 \right) + \frac{1}{2} \left(\phi'\right)^2 & = 0 \, ,
\end{eqaed}
and that the equation of motion for $\phi$ is a consequence of these.

All in all, eq.~\eqref{eq:dm_perturbed_ricci} finally leads to
\begin{eqaed}\label{eq:dm_perturbed_eom}
	- \, \frac{1}{8} \, e^{2 \Omega} \, \eta_{\mu \nu} \, V' \, \varphi = \, & - \, 4 \, \Omega' \left(\partial_\mu h_{\nu 9} + \partial_\nu h_{\mu 9} - h'_{\mu \nu} \right)  \\
	& - \eta_{\mu \nu} \bigg[\left( \Omega'' + 8 \left( \Omega'\right)^2 \right) h_{99} \\	
	& + \Omega' \left( \partial_\alpha h ^{\alpha 9} - \, \frac{1}{2} \left({h'^\alpha}_\alpha - h'_{99} \right) \right) \bigg] \\
	& + \frac{1}{2} \bigg[ \Box_9 \, h_{\mu \nu} + h''_{\mu \nu} - \partial_\mu \left(\partial_\alpha {h^\alpha}_\nu + h'_{\nu 9} \right) \\
	& - \partial_\nu \left(\partial_\alpha {h^\alpha}_\mu + h'_{\mu 9} \right) \bigg] - \frac{1}{2} \, \partial_\mu \partial_\nu \left({h^\alpha}_\alpha + h_{99} \right) \, , \\	
	- \, \frac{1}{2} \, \phi' \, \partial_\mu \varphi = \, & - \, 4 \, \Omega' \, \partial_\mu h_{99} \\
	& + \frac{1}{2} \left( \Box_9 \, h_{\mu 9} - \partial_\mu \partial_\alpha {h^\alpha}_9 - \partial_\alpha {h'^\alpha}_\mu + \partial_\mu {h'^\alpha}_\alpha \right) \, , \\
	- \, \phi' \, \varphi' - \, \frac{1}{8} \, e^{2 \Omega} \left( V \, h_{99} + V' \, \varphi\right) = \, & - \, 4 \, \Omega' \, h'_{99} - \Omega' \left( \partial_\alpha {h^\alpha}_9 - \, \frac{1}{2} \left({h'^\alpha}_\alpha - h'_{99} \right) \right) \\
	& + \frac{1}{2} \left(\Box_9 \, h_{99} - 2 \, \partial_\alpha {h'^\alpha}_9 + {h''^\alpha}_\alpha \right) \, ,
\end{eqaed}
while the perturbed dilaton equation of motion reads
\begin{eqaed}\label{eq:dm_dilaton_perturbed}
	\Box_9 \, \varphi & + \varphi'' + 8 \, \Omega' \, \varphi' + \phi' \left( \frac{1}{2} \, {h'^\alpha}_\alpha - \, \frac{1}{2} \, h'_{99} - \partial_\alpha {h^\alpha}_9 - 8 \, \Omega' \, h_{99} \right) \\
	& - \phi'' \, h_{99} - e^{2 \Omega} \, V'' \, \varphi = 0 \, .
\end{eqaed}
Starting from eqs.~\eqref{eq:dm_perturbed_eom} and~\eqref{eq:dm_dilaton_perturbed} we shall now proceed separating perturbations into tensor, vector and scalar modes.

\subsubsection{Tensor and vector perturbations}\label{sec:static_tensor_vector}

Tensor perturbations are simpler to study, and to this end one only allows a transverse trace-less $h_{\mu\nu}$. After a Fourier transform with respect to $x$ one is thus led to
\begin{eqaed}\label{eq:dm_tensor_pert}
	h''_{\mu \nu} + 8 \, \Omega' \, h'_{\mu \nu} + m^2 \, h_{\mu \nu} = 0 \, ,
\end{eqaed}
where $m^2 \equiv - \, p^\mu \, p^\nu \, \eta_{\mu \nu}$, which defines a Schr\"odinger-like problem along the lines of eq.~\eqref{eq:dm_second_order_eq}, with $b = 0$ and $a = 8 \, \Omega'$. Hence, with Dirichlet or Neumann boundary conditions the argument of Section~\ref{sec:static_scalar} applies, and one obtains a discrete spectrum of masses. Moreover, one can verify that there is a normalizable mode with $h_{\mu\nu}$ independent of $z$, which signals that at low energies gravity is effectively nine-dimensional\footnote{The same conclusion can be reached computing the effective nine-dimensional Newton constant~\cite{Dudas:2000ff}.}.

Vector perturbations entail some mixings, since in this case they originate from transverse $h_{\mu 9}$ and from the trace-less combination
\begin{eqaed}\label{eq:dm_traceless_combinations}
	h_{\mu \nu} = \partial_\mu \Lambda_\nu + \partial_\nu \Lambda_\mu \, ,
\end{eqaed}
so that
\begin{eqaed}\label{eq:dm_divergenceless}
	\partial^\mu \Lambda_\mu = 0 \, .
\end{eqaed}
The relevant vector combination
\begin{eqaed}\label{eq:dm_c-vector}
	C_\mu = h_{\mu 9} - \Lambda'_\mu
\end{eqaed}
satisfies the two equations
\begin{eqaed}\label{eq:dm_c-eqs}
	\left(p_\mu \, C_\nu + p_\nu \, C_\mu \right)' + 8 \, \Omega' \left( p_\mu \, C_\nu + p_\nu \, C_\mu \right) & = 0 \, , \\
	m^2 \, C_\mu & = 0 \, ,
\end{eqaed}
the first of which is clearly solved by
\begin{eqaed}\label{eq:dm_c-vector_sol}
	C_\mu = C_\mu^{(0)} \, e^{- 8 \Omega} \, ,
\end{eqaed}
with a constant $C_\mu^{(0)}$. In analogy with the preceding discussion, one might be tempted to identify a massless vector. However, one can verify that, contrary to the case of tensors, this is not associated to a normalizable zero mode. The result is consistent with standard expectations from Kaluza-Klein theory, since the internal manifold has no translational isometry.

\subsubsection{Scalar perturbations}\label{sec:static_scalar}

The scalar perturbations are defined by\footnote{We reserve the symbol $B$ for scalar perturbations of the form field, which we shall introduce in Section~\ref{sec:ads_stability}.}
\begin{eqaed}\label{eq:scalar_pert}
	h_{\mu \nu} = \eta_{\mu \nu} \, e^{i p \cdot x} \, A(z) \, , \qquad h_{\mu 9} = i p_\mu \, D(z) \, e^{i p \cdot x} \, , \qquad h_{99} = e^{i p \cdot x} \, C(z) \, ,
\end{eqaed}
with $p \cdot x \equiv p^\mu \, x^\nu \, \eta_{\mu\nu}$, so that altogether the four scalars $A$, $C$, $D$ and $\phi$ obey the linearized equations
\begin{eqaed}\label{eq:dm_linearized_eom_mod}
	- \, \frac{1}{8} \, e^{2 \Omega} \, V' \, \varphi = \, & - \Omega' \left(m^2 \, D - \frac{1}{2} \left(17 \, A' - C' \right) \right) \\
	& + \frac{1}{2} \left( m^2 \, A + A'' \right) - C \left( \Omega'' + 8 \left( \Omega' \right)^2 \right) \, , \\
	- \phi' \, \varphi' - \, \frac{1}{8} \, e^{2 \Omega} \left( V \, C + V' \, \varphi \right) \! = \, & - \Omega' \left(m^2 \, D - \, \frac{9}{2} \left(A' - C'\right) \right) \\
	& + \frac{1}{2} \left( m^2 \left(C - 2 \, D' \right) + 9 \, A'' \right) \, , \\
	7 \, A + C - 2 \, D' - 16 \, \Omega' \, D = \, & \, 0 \, , \\
	4 \, \Omega' \, C - 4 \, A' - \frac{1}{2} \, \phi' \, \varphi = \, & \, 0 \, .
\end{eqaed}
Notice that some of the metric equations, the third one and the fourth one above, are constraints, and that there is actually another constraint that obtains combining the first and the last so as to remove $A''$. Moreover, the dilaton equation of motion is a consequence of these.

The system, however, has a residual local gauge invariance, a diffeomorphism of the type
\begin{eqaed}\label{eq:dm_residual_diff}
	z' = z + \epsilon(x,z) \, ,
\end{eqaed}
which is available in the presence of a single internal dimension and implies
\begin{eqaed}\label{eq:dm_diff_der}
	dz = dz' \left( 1 - \, \frac{d\epsilon}{dz'} \right) - dx^\mu \, \partial_\mu \epsilon \, .
\end{eqaed}
Taking into account the original form of the metric, which in terms of the scalar perturbations of eq.~\eqref{eq:scalar_pert} reads
\begin{eqaed}\label{eq:dm_metric_scalars}
	ds^2 = e^{2 \Omega} \left( \left(1 + A\right) dx_{1,8}^2 + 2 \, dz \, dx^\mu \, \partial_\mu D + \left(1 + C\right) dz^2 \right) \, ,
\end{eqaed}
one can thus identify the transformations
\begin{eqaed}\label{eq:dm_scalar_diff}
	A \; & \to \; A - 2 \, \Omega' \, \epsilon \, , \\
	C \; & \to \; C - 2 \, \Omega' \, \epsilon - 2 \, \epsilon' \, , \\
	D \; & \to \; D - \epsilon \, , \\
	\varphi \; & \to \; - \phi' \, \epsilon \, .
\end{eqaed}
Notice that $D$ behaves as a St\"uckelberg field, and can be gauged away, leaving only one scalar degree of freedom after taking into account the constraints, as expected from Kaluza-Klein theory. After gauging away $D$ the third equation of eq.~\eqref{eq:dm_linearized_eom_mod} implies that
\begin{eqaed}\label{eq:dm_c-a_constraint}
	C = - 7 \, A \, ,
\end{eqaed}
while the third equation of eq.~\eqref{eq:dm_linearized_eom_mod} implies that
\begin{eqaed}\label{eq:dm_dilaton_constraint}
	\varphi = - \, \frac{8}{\phi'} \left(A' + 7 \, \Omega' \, A \right) \, .
\end{eqaed}
Substituting these expressions in the first equation of eq.~\eqref{eq:dm_linearized_eom_mod} finally leads to a second-order eigenvalue equation for $m^2$:
\begin{importantbox}
	\begin{eqaed}\label{eq:dm_eigenvalue_eq}
		A'' + \left( 24 \, \Omega' - \, \frac{2}{\phi'} \, e^{2 \Omega} \, V' \right) A' + \left( m^2 - \, \frac{7}{4} \, e^{2 \Omega} \, V - 14 \, e^{2 \Omega} \, \Omega' \, \frac{V'}{\phi'} \right) A = 0 \, .
	\end{eqaed}
\end{importantbox}
There is nothing else, since differentiating the fourth equation of eq.~\eqref{eq:dm_linearized_eom_mod} and using eq.~\eqref{eq:dm_simplified_eom} gives
\begin{eqaed}\label{eq:dm_redundant}
	- \phi' \, \varphi' = - 8 \, A'' - 120 \, \Omega' \, A' + 8 \, e^{2 \Omega} \, \frac{V'}{\phi'} \, A' + 7 \, e^{2 \Omega} \left(V + 8 \, \Omega' \, \frac{V'}{\phi'} \right) A \, .
\end{eqaed}
Taking this result into account, one can verify that the second equation of eq.~\eqref{eq:dm_linearized_eom_mod} also leads to eq.~\eqref{eq:dm_eigenvalue_eq}, whose properties we now turn to discuss.

The issue at stake is the stability of the solution, which in this case reflects itself in the sign of $m^2$: a negative value would signal a tachyonic instability in the nine-dimensional Minkowski space, and one can show that the solution corresponding the lowest-order level potentials is stable, in both the orientifold and heterotic models. To this end, let us recall that a generic second-order equation of the type
\begin{eqaed}\label{eq:dm_second_order_eq}
	f''(z) + a(z) \, f'(z) + \left(m^2 - b(z) \right) f(z) = 0
\end{eqaed}
can be turned into a Schr\"odinger-like form via the transformation
\begin{eqaed}\label{eq:dm_schrodinger_trans}
	f(z) = \Psi(z) \, e^{- \frac{1}{2} \int a dz} \, .
\end{eqaed}
One is thus led to
\begin{eqaed}\label{eq:dm_schrodinger_eq}
	\Psi'' + \left(m^2 - b - \, \frac{a'}{2} - \, \frac{a^2}{4} \right) \Psi = 0 \, ,
\end{eqaed}
and tracing the preceding steps one can see that $\Psi \in L^2$. Eq.~\eqref{eq:dm_schrodinger_eq} can be conveniently discussed connecting it to a more familiar problem of the type
\begin{eqaed}\label{eq:dm_schroedinger_eigs}
	\widehat{H} \, \Psi = m^2 \, \Psi \, , \qquad \widehat{H} \equiv b + \mathcal{A}^\dagger \mathcal{A} \, ,
\end{eqaed}
with
\begin{eqaed}\label{eq:dm_creation_annihilation}
	\mathcal{A} \equiv - \, \frac{d}{dz} + \frac{a}{2} \, , \qquad \mathcal{A}^\dagger \equiv \frac{d}{dz} + \frac{a}{2} \, .
\end{eqaed}
Once these relations are supplemented with Dirichlet or Neumann conditions at each end in $z$, one can conclude that in all these cases the operator
\begin{eqaed}\label{eq:a_adagger_positive}
	\mathcal{A}^\dagger \mathcal{A} \geq 0 \, .
\end{eqaed}
All in all, positive $b$ then implies positive values of $m^2$, and this condition is indeed realized for the static Dudas-Mourad solutions, since
\begin{eqaed}\label{eq:dm_b_function}
	b = \frac{7}{4} \, e^{2 \Omega} \left(V + 8 \, \Omega' \, \frac{V'}{\phi'} \right) \, ,
\end{eqaed}
and the corresponding $V \propto e^{\frac{3}{2} \phi}$, so that
\begin{eqaed}\label{eq:dm_b_orientifold}
	b = \frac{7}{4} \, e^{2 \Omega} \, V \left(1 + 12 \, \frac{\Omega'}{\phi'} \right) \, .
\end{eqaed}
The ratio of derivatives can be computed in terms of the $y$ coordinate using the expressions that we have presented in the preceding section, yielding
\begin{eqaed}\label{eq:dm_b_orientifold_spec}
	b = \frac{7 \, e^{2 \Omega} \, V}{1 + \frac{9}{4} \, \alpha_{\text{O}} \, y^2} \geq 0 \, .
\end{eqaed}
For the heterotic model $V \propto e^{\frac{5}{2} \phi}$, so that
\begin{eqaed}\label{eq:dm_b_heterotic}
	b = \frac{7}{4} \, e^{2 \Omega} \, V \left(1 + 20 \, \frac{\Omega'}{\phi'} \right) \, .
\end{eqaed}
Making use of the explicit solutions that we have presented in the preceding section, one thus finds
\begin{eqaed}\label{eq:dm_b_heterotic_spec}
	b = \frac{8}{3} \, e^{2 \Omega} \, V \, \frac{1 - \, \frac{1}{2} \, \tanh^2 \left(\sqrt{\alpha_\text{H}} \, y \right)}{1 + 4 \, \tanh^2 \left(\sqrt{\alpha_\text{H}} \, y \right)} \geq 0 \, ,
\end{eqaed}
which is again non negative, so that both nine-dimensional Dudas-Mourad solutions are perturbatively stable solutions of the respective Einstein-dilaton systems for all allowed choices of boundary conditions at the ends of the interval. The presence of regions where curvature or string loop corrections are expected to be relevant, however, makes the lessons of these results less evident for string theory.

As a final comment, let us mention that one can repeat the calculations that we have presented in $D$ dimensions without further difficulties, and one finds
\begin{eqaed}\label{eq:dm_general_dim_stability}
	b = 2 \left(D-3\right) e^{2 \Omega} \left( \frac{V}{D-2} + \frac{D-2}{8} \, \Omega' \, \frac{V'}{\phi'} \right) \geq 0 \, ,
\end{eqaed}
so that the resulting solutions are perturbatively stable in any dimension.

\subsection{Stability of cosmological Dudas-Mourad solutions}\label{sec:cosmological_case}

Let us now turn to the issue of perturbative stability of the Dudas-Mourad cosmological solutions that we have presented in the preceding section. The following analysis is largely analogous to the one of the preceding section, and we shall begin discussing tensor perturbations, which reveal an interesting feature in the homogeneous case.

\subsubsection{Tensor perturbations: an intriguing instability}\label{sec:tensor_perturbations_cosmological}

The issue at stake, here and in the following sections, is whether solutions determined by arbitrary initial conditions provided some time after the initial singularity can grow in the future evolution of the universe.
This can be ascertained rather simply at large times, which translate into large values of the conformal time $\eta$, where many expressions simplify. Moreover, for finite values of $\eta$ the geometry is regular, and the coefficients in eq.~\eqref{eq:dm_cosm_tensor_pert} are bounded, so that the solutions are also not singular. However, a growth of order $\mathcal{O}\!\left(1\right)$ is relevant for perturbations, and therefore we shall begin with the late-time asymptotics and then, at the end of the section, we shall also approach the problem globally.

In the ten-dimensional orientifold and heterotic models of interest, performing spatial Fourier transforms and proceeding as in the preceding section, one can show that tensor perturbations evolve according to
\begin{eqaed}\label{eq:dm_cosm_tensor_pert}
	h''_{ij} + 8 \, \Omega' \, h'_{ij} + \mathbf{k}^2 \, h_{ij} = 0 \, ,
\end{eqaed}
where ``primes'' denote derivatives with respect to the conformal time $\eta$.
Let us begin observing that, for all exponential potentials
\begin{eqaed}\label{eq:exp_potential_cosm}
	V = T \, e^{\gamma \phi}
\end{eqaed}
with $\gamma \geq \frac{3}{2}$, and therefore for the potentials pertaining to the orientifold models, which have $\gamma=\frac{3}{2}$ and are ``critical'' in the sense of~\cite{Dudas:2010gi}, but also for the heterotic model, which has $\gamma_E=\frac{5}{2}$ and is ``super-critical'' in the sense of~\cite{Dudas:2010gi}, the solutions of the background equations
\begin{eqaed}\label{eq:dm_cosm_background_eom}
	\Omega'' + 8 \left( \Omega' \right)^2 - \, \frac{1}{8} \, e^{2 \Omega} \, V & = 0 \, , \\
	9 \, \Omega'' - \, \frac{1}{8} \, e^{2 \Omega} \, V + \frac{1}{2} \left(\phi'\right)^2 & = 0 \, , \\
	\phi'' + 8 \, \Omega' \, \phi' + \gamma \, e^{2 \Omega} \, V & = 0
\end{eqaed}
are dominated, for large values of $\eta$, by
\begin{eqaed}\label{eq:dm_cosm_large-eta}
	\phi \sim - \, \frac{3}{2} \, \log \left(\sqrt{\alpha_\text{H}} \, \eta \right) \, , \qquad \Omega \sim \frac{1}{8} \, \log \left(\sqrt{\alpha_\text{H}} \, \eta \right) \, .
\end{eqaed}
In the picture of~\cite{Dudas:2010gi}, in this region the scalar field has overcome the turning point and is descending the potential, so that the (super)gravity approximation is expected to be reliable, but the potential contribution is manifestly negligible only in the ``super-critical'' case, where $e^{2\Omega}\,V$ decays faster than $\frac{1}{\eta^2}$ for large $\eta$. However, the result also applies for $\gamma=\frac{3}{2}$, which marks the onset of the ``climbing behavior''. This can be appreciated retaining subleading terms, which results in
\begin{eqaed}\label{eq:dm_cosm_large-eta_sub}
	\phi & \sim - \, \frac{3}{2} \, \log \left(\sqrt{\alpha_\text{O}} \, \eta \right) - \, \frac{5}{6} \, \log \log \left(\sqrt{\alpha_\text{O}} \, \eta \right) \, , \\
	\Omega & \sim \frac{1}{8} \, \log \left(\sqrt{\alpha_\text{O}} \, \eta \right) + \frac{1}{8} \, \log \log \left(\sqrt{\alpha_\text{O}} \, \eta \right) \, ,
\end{eqaed}
so that the potential decays as
\begin{eqaed}\label{eq:dm_cosm_potential_decay}
	e^{2 \Omega} \, V \sim \frac{T}{2 \, \alpha_\text{O} \, \eta^2 \, \log \left(\sqrt{\alpha_\text{O}} \, \eta \right)} \, ,
\end{eqaed}
which is faster than $\frac{1}{\eta^2}$. Notice that a similar behavior, but with the scalar climbing up the potential, also emerges for small values of $\eta$, for which
\begin{eqaed}\label{eq:dm_cosm_small-eta}
	\phi \sim \frac{3}{2} \, \log \left(\sqrt{\alpha_\text{O,H}} \, \eta \right) \, , \qquad \Omega \sim \frac{1}{8} \, \log \left(\sqrt{\alpha_\text{O,H}} \, \eta \right)
\end{eqaed}
for all $\gamma \geq \frac{3}{2}$, and thus in all orientifold and heterotic models of interest. However, these expressions are less compelling, since they concern the onset of the climbing phase. The potential is manifestly subleading for small values of $\eta$, but curvature corrections, which are expected to be relevant in this region, are not taken into account. In conclusion, for $\gamma \geq \frac{3}{2}$ and for large values of $\eta$ eq.~\eqref{eq:dm_cosm_large-eta} holds and eq.~\eqref{eq:dm_cosm_tensor_pert}, which describes tensor perturbations, therefore approaches
\begin{eqaed}\label{eq:dm_cosm_large-eta_tensor_pert}
	h''_{ij} + \frac{1}{\eta} \, h'_{ij} \sim - \, \mathbf{k}^2 \, h_{ij} \, .
\end{eqaed}
Consequently, for $\mathbf{k} \neq 0$
\begin{eqaed}\label{eq:dm_cosm_tensor_pert_k}
	h_{ij} \sim A_{ij} \, J_0\left(k \eta\right) + B_{ij} \, Y_0 \left(k \eta\right) \, ,
\end{eqaed}
and the oscillations are damped for large times, so that no instabilities arise.

\begin{figure}[!ht]
	\begin{center}
		\includegraphics[width=90mm]{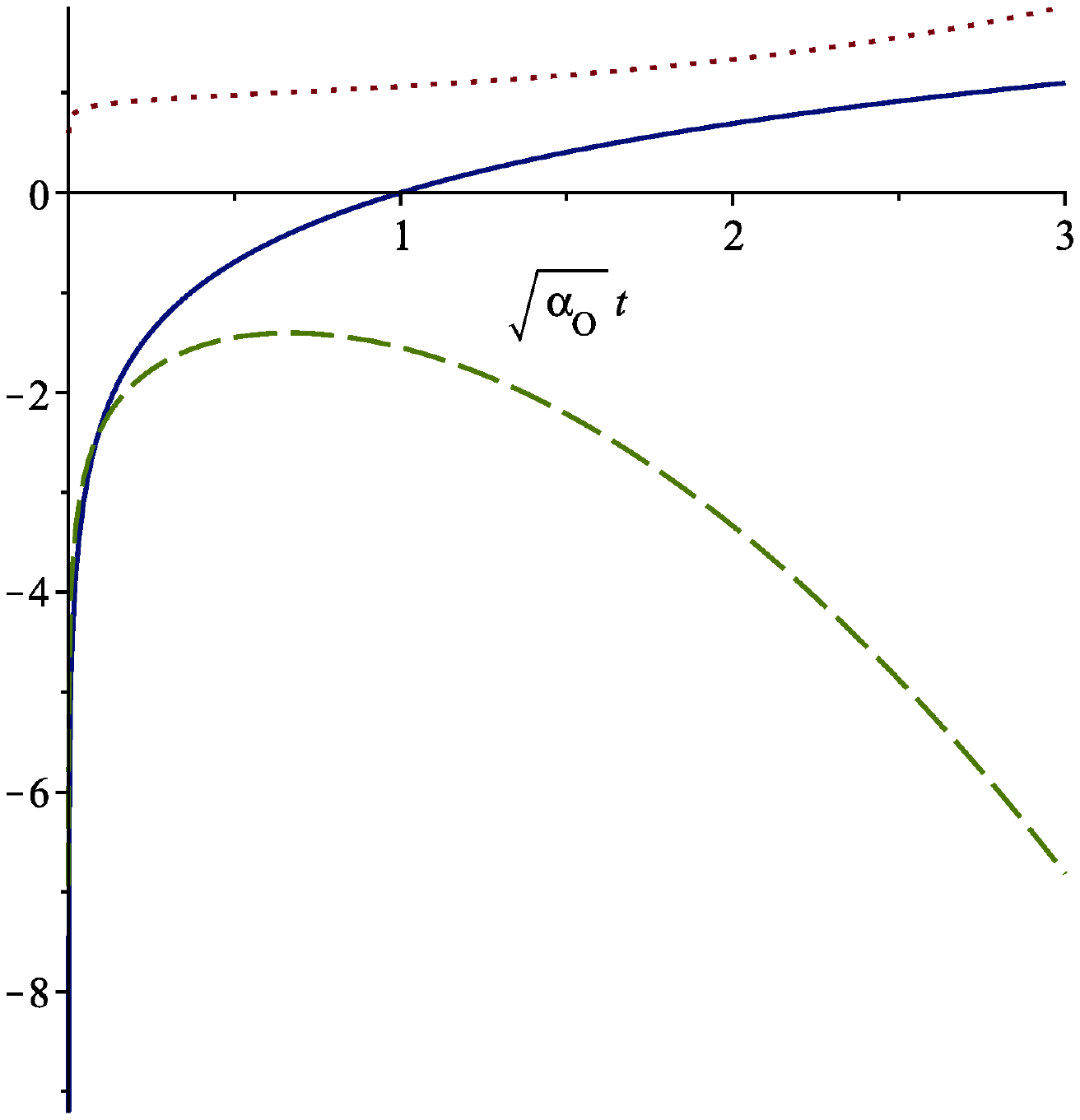}
		\vspace*{-5truecm}
	\end{center}
	\caption{the scale factor $e^\Omega$ (red, dotted), the unstable homogeneous tensor mode (blue) and the dilaton $\phi$ (green, dashed) as functions of the parametric time $\sqrt{\alpha_\text{O}} \, t$.}
	\label{fig:instability_tensor}
\end{figure}

On the other hand, an intriguing behavior emerges for ${\mathbf{k}}  = 0$. In this case the solution of eq.~\eqref{eq:dm_cosm_large-eta_tensor_pert} implies that
\begin{eqaed}\label{eq:k0_instability}
	h_{ij} \sim A_{ij} + B_{ij} \, \log \left( \frac{\eta}{\eta_0} \right) \, ,
\end{eqaed}
and therefore spatially homogeneous tensor perturbations experience in general a logarithmic growth. This result indicates that homogeneity is preserved while isotropy is generally violated in the ten-dimensional ``climbing-scalar'' cosmologies~\cite{Dudas:2010gi} that emerge in string theory with broken supersymmetry. One can actually get a global picture of the phenomenon: the linearized equation of motion for $\mathbf{k} = 0$ can be solved in terms of the parametric time $t$, and one finds
\begin{eqaed}\label{eq:k0_instability_t_orientifold}
	h_{ij} = A_{ij} + B_{ij} \, \log \left(\sqrt{\alpha_\text{O}} \, t \right)
\end{eqaed}
for the orientifold models, while
\begin{eqaed}\label{eq:k0_instability_t_heterotic}
	h_{ij} = A_{ij} + B_{ij} \, \log \tan \left(\sqrt{\alpha_\text{H}} \, t \right)
\end{eqaed}
for the heterotic model. These results are qualitatively similar, if one takes into account the limited range of $t$ in the heterotic model, and typical behaviors are displayed in fig.~\ref{fig:instability_tensor}.

The general lesson is that perturbations acquire $\mathcal{O}\!\left(1\right)$ variations toward the end of the climbing phase, where curvature corrections do not dominate the scene anymore, thus providing support to the present analysis. This result points naturally to an awaited tendency toward lower-dimensional space-times, albeit without a selection criterion for the resulting dimension\footnote{This result resonates at least with some previous investigations~\cite{Kim:2011cr, Anagnostopoulos:2017gos} of matrix models related to the type IIB superstring~\cite{Ishibashi:1996xs}.}. While perturbation theory is at most a clue to this effect, the resulting picture appears enticing, and moreover the dynamics becomes potentially richer and more stable in lower dimensions, where other branes that become space-filling can inject an inflationary phase devoid of this type of instability~\cite{Basile:2018irz}.

\subsubsection{Scalar perturbations}\label{sec:scalar_perturbations_cosmological}

Scalar perturbations exhibit a very different behavior in the presence of the exponential potentials of eq.~\eqref{eq:exp_potential_cosm} with $\gamma \geq \frac{3}{2}$. Our starting point is now the analytic continuation of eq.~\eqref{eq:dm_eigenvalue_eq} with respect to $z \to i \,\eta$, which reads
\begin{importantbox}
	\begin{eqaed}\label{eq:dm_cosm_scalar_pert}
		A'' + \left(24 \, \Omega' + 2 \, e^{2 \Omega} \, \frac{V'}{\phi'} \right) A' + \left(\mathbf{k}^2 + \frac{7}{4} \, e^{2 \Omega} \, V + 14 \, e^{2 \Omega} \, \Omega' \, \frac{V'}{\phi'} \right) A = 0 \, .
	\end{eqaed}
\end{importantbox}
As in eq.~\eqref{eq:dm_cosm_tensor_pert}, we have also replaced $m^2$ with $- \, \mathbf{k}^2$, which originates from a spatial Fourier transform, and ``primes'' denote again derivatives with respect to the conformal time $\eta$. As we have stressed in the preceding section, the potential is subdominant in eq.~\eqref{eq:dm_cosm_scalar_pert} for $\gamma \geq \frac{3}{2}$, which leads to the asymptotic behaviors of eq.~\eqref{eq:dm_cosm_small-eta} during the climbing phase, and of eq.~\eqref{eq:dm_cosm_large-eta} during the descending phase. As a result, during the latter eq.~\eqref{eq:dm_cosm_scalar_pert} reduces to
\begin{eqaed}\label{eq:dm_cosm_large-eta_scalar}
	A '' + \frac{3}{\eta} \, A + \mathbf{k}^2 \, A = 0 \, ,
\end{eqaed}
whose general solution takes the form
\begin{eqaed}\label{eq:large-eta_scalar_sol}
	A = A_1 \, \frac{J_1 \left( k \eta \right)}{\eta} + A_2 \, \frac{Y_1 \left( k \eta \right)}{\eta} \, ,
\end{eqaed}
with $A_1$, $A_2$ constants. For $\mathbf{k} \neq 0$ the amplitude always decays proportionally to $\eta^{- \frac{3}{2}}$, while for $\mathbf{k} = 0$ the two independent solutions of eq.~\eqref{eq:dm_cosm_scalar_pert} are dominated by
\begin{eqaed}\label{eq:k0_scalar}
	A = A_3 + \frac{A_4}{\eta^2} \, ,
\end{eqaed}
with $A_3$, $A_4$ constants. Therefore, scalar perturbations do not grow in time, even for the homogeneous mode with $\mathbf{k} = 0$, for $\gamma \geq \frac{3}{2}$, and thus, in particular, for the orientifold models and for the heterotic model. Similar results can be obtained studying the perturbative stability of linear dilaton backgrounds, both in the static case and in the cosmological case~\cite{Basile:2018irz}.

\subsection{Stability of \texorpdfstring{$\ads$}{AdS} flux compactifications}\label{sec:ads_stability}

In this section we discuss the perturbative stability of the $\ads$ flux compatifications that we have presented in the preceding section. In order to simplify the analysis of tensor and vector perturbations, we shall work with internal spheres, but the resulting equations for scalar perturbations are independent of this choice\footnote{The stability analysis of scalar perturbations can also be carried out in general dimensions and for general parameters without additional difficulties, but we have not found such generalizations particularly instructive in the context of this review.}, insofar as the internal space is Einstein. In the following we shall work in the duality frames where $p = 1$, which is the electric frame in the orientifold models, for which $\alpha = 1$, and the magnetic frame in the heterotic model, for which $\alpha = -1$. Let us begin from the orientifold models, writing the perturbations
\begin{eqaed}\label{eq:ads_pert}
	g_{MN} = g^{(0)}_{MN} + h_{MN} \, , \qquad \phi & = \phi_0 + \varphi \, , \qquad B_{MN} = B^{(0)}_{MN} + \frac{e^{- \alpha \phi_0}}{c} \, b_{MN} \, ,
\end{eqaed}
where the background metric is split as
\begin{eqaed}\label{eq:metric_split}
	ds^2_{(0)} & = L^2 \, \lambda_{\mu \nu} \, dx^\mu \, dx^\nu + R^2 \, \gamma_{ij} \, dy^i \, dy^j \, ,
\end{eqaed}
and linearizing the resulting equations of motion. We shall also make use of the convenient relations
\begin{eqaed}\label{eq:mss_commutators}
	[\nabla_\mu \, , \nabla_\nu ] \, V_\rho & = \frac{1}{L^2} \left( \lambda_{\nu \rho} V_\mu - \lambda_{\mu \rho} V_\nu \right) \, , \\
	[\nabla_i \, , \nabla_j ] \, V_k & = - \, \frac{1}{R^2} \left( \gamma_{jk} V_i - \gamma_{ik} V_j \right) \, , \\
\end{eqaed}
valid for maximally symmetric spaces. The linearized equations of motion for the form field are\footnote{Here and in the following $\epsilon$ denotes the Levi-Civita tensor, which includes the metric determinant prefactor.}
\begin{eqaed}\label{eq:ads_pert_form_eqs}
	\Box_{10} \, b_{\mu \nu} & - \nabla_\mu \nabla^M b_{M \nu} - \nabla_\nu \nabla^M b_{\mu M} + \frac{2}{L^2} \, b_{\mu \nu} \\
	& + 4 \, \mathcal{R}^+_\text{O} \, \epsilon_{\mu \nu \rho} \left(\alpha \, \nabla^\rho \phi - \nabla^i {h_i}^\rho - \frac{1}{2} \, \nabla^\rho \lambda \cdot h + \frac{1}{2} \, \nabla^\rho \gamma \cdot h \right) = 0 \, , \\
	\Box_{10} \, b_{\mu i} & - \nabla_\mu \nabla^M b_{M i} - \nabla_i \nabla^M b_{\mu M} + 2 \, \mathcal{R}^-_\text{O} \, b_{\mu i} + 4 \, \mathcal{R}^+_\text{O} \, \epsilon_{\alpha \beta \mu} \, \nabla^\alpha {h^\beta}_i = 0 \, , \\
	\Box_{10} \, b_{ij} & - \nabla_i \nabla^M b_{Mj} - \nabla_j \nabla^M b_{iM} - \frac{10}{R^2} \, b_{ij} = 0 \, ,
\end{eqaed}
where, here and in the following, the ten-dimensional d'Alembert operator
\begin{eqaed}\label{eq:laplacian_split}
	\Box_{10} = \Box + \nabla^2
\end{eqaed}
is split in terms of the $\ads$ and sphere contributions, and we have defined
\begin{eqaed}\label{eq:radii_combination}
	\mathcal{R}^\pm_\text{O} \equiv \frac{1}{L^2} \pm \frac{3}{R^2}
\end{eqaed}
for convenience. Similarly, the linearized equation of motion for the dilaton is
\begin{eqaed}\label{eq:ads_pert_dilaton_eq}
	\Box_{10} \, \varphi - V''_0 \, \varphi + 2 \, \mathcal{R}^+_\text{O} \left(\alpha^2 \, \varphi - \alpha \, \lambda \cdot h \right) - \, \frac{\alpha}{2} \, \epsilon^{\mu \nu \rho} \, \nabla_\mu b_{\nu \rho} = 0 \, .
\end{eqaed}
Finally, the linearized Einstein equations rest on the linearized Ricci tensor
\begin{eqaed}\label{eq:lin_ricci}
	R_{MN}^{(1)} & = R^{(0)}_{MN} + \frac{1}{2} \left(\Box \, h_{MN} - \nabla_M \left(\nabla \cdot h\right)_N - \nabla_N \left(\nabla \cdot h\right)_M + \nabla_M \nabla_N {h^A}_A \right) \\
	& + \frac{1}{2} \, {R^{(0) \, A}}_M \, h_{AN} + \frac{1}{2} \, {R^{(0) \, A}}_N \, h_{AM} - {R^{(0) \, A}}_M{^B}_N \, h_{AB} \, ,
\end{eqaed}
and read
\begin{eqaed}\label{eq:ads_pert_metric_eqs}
	\Box_{10} \, h_{\mu \nu} & + \frac{2}{L^2} \, h_{\mu \nu} - \nabla_\mu \left( \nabla \cdot h \right)_\nu - \nabla_\nu \left( \nabla \cdot h \right)_\mu + \nabla_\mu \nabla_\nu \left(\lambda \cdot h + \gamma \cdot h\right) \\
	& + \lambda_{\mu \nu} \left(- \, \frac{5 \alpha}{2} \, \mathcal{R}^+_\text{O} \, \varphi - 3 \, \mathcal{R}^-_\text{O} \, \lambda \cdot h - \, \frac{3}{4} \, \epsilon^{\alpha \beta \gamma} \, \nabla_{\alpha} b_{\beta \gamma} \right) = 0 \, , \\
	\Box_{10} \, h_{\mu i} & + 2 \, \mathcal{R}^+_\text{O} \, h_{\mu i} - \nabla_\mu \left(\nabla \cdot h\right)_i - \nabla_i \left(\nabla \cdot h\right)_\mu + \nabla_\mu \nabla_i \left(\lambda \cdot h + \gamma \cdot h \right) \\
	& + \frac{1}{2} \, {\epsilon^{\alpha \beta}}_\mu \left(\nabla_i b_{\alpha \beta} + \nabla_\alpha b_{\beta i} + \nabla_\beta b_{i \alpha} \right) = 0 \, , \\
	\Box_{10} \, h_{ij} & - \frac{2}{R^2} \, h_{ij} - \nabla_i \left( \nabla \cdot h\right)_j - \nabla_j \left(\nabla \cdot h \right)_i + \nabla_i \nabla_j \left( \lambda \cdot h + \gamma \cdot h\right) \\
	& + \gamma_{ij} \left(\frac{2}{R^2} \, \gamma \cdot h + \mathcal{R}^+_\text{O} \left(\frac{3\alpha}{2} \, \varphi - \lambda \cdot h \right) - \, \frac{1}{4} \, \epsilon^{\alpha \beta \gamma} \, \nabla_\alpha b_{\beta \gamma} \right) = 0 \, ,
\end{eqaed}
where $\lambda \cdot h$ and $\gamma \cdot h$ denote the partial traces of the metric perturbation with respect to $\ads$ and the internal sphere. In all cases and models, the perturbations depend on the $\ads$ coordinates $x^\mu$ and on the sphere coordinates $y^i$, and they will be expanded in terms of the corresponding spherical harmonics\footnote{Choosing a different internal space would require knowledge of its (tensor) Laplacian spectrum.}, whose structure is briefly reviewed in Appendix~\ref{sec:tensor_harmonics}. For instance, expanding internal scalars with respect to $\ess^n$ spherical harmonics will always result in expressions of the type
\begin{eqaed}\label{eq:tensor_harmonics_decomp}
	h_{\mu \nu}(x,y) = \sum_\ell h_{\mu \nu \, , \, I_1 \dots I_\ell}(x) \, \mathcal{Y}_{(n)}^{I_1 \dots I_\ell}(y) \, ,
\end{eqaed}
where $I_i=1,\ldots ,n$ and $h_{\mu\nu,\,I_1 \ldots I_\ell}(x)$ is totally symmetric and trace-less in the Euclidean $I_i$ labels. However, the eigenvalues of the internal Laplace operator $\nabla^2$ will only depend on $\ell$. Hence, for the sake of brevity, we shall leave the internal labels implicit, although in some cases we shall refer to their ranges when counting multiplicities. For tensors in internal space there are some additional complications. For example, expanding mixed metric components one obtains expressions of the type
\begin{eqaed}\label{eq:tensor_harmonics_vectors}
	h_{\mu i}(x,y) = \sum_\ell h_{\mu J \, , \, I_1 \dots I_\ell}(x) \, \mathcal{Y}_{(n) \, i}^{I_1 \dots I_\ell \, , \, J}(y) \, ,
\end{eqaed}
where $h_{\mu J,\,I_1 \ldots I_\ell}(x)$ corresponds to a ``hooked'' Young tableau of mixed symmetry and $\ell \geq 1$, as explained in Appendix~\ref{sec:tensor_harmonics}. Here the ${\cal Y}_{(n)\,i}$ are vector spherical harmonics, and we shall drop all internal labels, for brevity, also for the internal tensors that we shall consider.

In the heterotic model the linearized equations of motion for the form field read
\begin{eqaed}\label{eq:ads_pert_form_eqs_heterotic}
	\Box_{10} \, b_{ij} & - \nabla_i \nabla^M b_{M j} - \nabla_j \nabla^M b_{i M} - \frac{2}{R^2} \, b_{ij} \\
	& + 4 \, \mathcal{R}^+_\text{H} \, \epsilon_{ijk} \left(\alpha \, \nabla^k \phi - \nabla^\alpha {h_\alpha}^k - \frac{1}{2} \, \nabla^k \gamma \cdot h + \frac{1}{2} \, \nabla^k \lambda \cdot h \right) = 0 \, , \\
	\Box_{10} \, b_{i \mu} & - \nabla_i \nabla^M b_{M \mu} - \nabla_\mu \nabla^M b_{i M} + 2 \, \mathcal{R}^-_\text{H} \, b_{i \mu} + 4 \, \mathcal{R}^+_\text{H} \, \epsilon_{kli} \, \nabla^k {h^l}_\mu = 0 \, , \\
	\Box_{10} \, b_{\mu \nu} & - \nabla_\mu \nabla^M b_{M\nu} - \nabla_\nu \nabla^M b_{\mu M} + \frac{10}{L^2} \, b_{\mu \nu} = 0 \, ,
\end{eqaed}
where now
\begin{eqaed}\label{eq:radii_combination_het}
	\mathcal{R}^\pm_\text{H} \equiv \frac{3}{L^2} \pm \frac{1}{R^2} \, ,
\end{eqaed}
while the linearized equation of motion for the dilaton is
\begin{eqaed}\label{eq:ads_pert_dilaton_eq_het}
	\Box_{10} \, \varphi - V''_0 \, \varphi - 2 \, \mathcal{R}^+_\text{H} \left(\alpha^2 \, \varphi - \alpha \, \gamma \cdot h \right) - \, \frac{\alpha}{2} \, \epsilon^{ijk} \, \nabla_i b_{jk} = 0 \, .
\end{eqaed}
Finally, the linearized Einstein equations rest on eq.~\eqref{eq:lin_ricci} and read
\begin{eqaed}\label{eq:ads_pert_metric_eqs_het}
	\Box_{10} \, h_{ij} & - \frac{2}{R^2} \, h_{ij} - \nabla_i \left( \nabla \cdot h \right)_j - \nabla_j \left( \nabla \cdot h \right)_i + \nabla_i \nabla_j \left(\lambda \cdot h + \gamma \cdot h\right) \\
	& + \gamma_{ij} \left(\frac{5 \alpha}{2} \, \mathcal{R}^+_\text{H} \, \varphi - 3 \, \mathcal{R}^-_\text{H} \, \gamma \cdot h - \, \frac{1}{4} \, \epsilon^{klm} \, \nabla_{k} b_{lm} \right) \\
	& + \frac{1}{2} \, {\epsilon^{kl}}_i \left(\nabla_j b_{kl} + \nabla_k b_{lj} + \nabla_l b_{jk} \right) + \left(i \leftrightarrow j \right) = 0 \, , \\
	\Box_{10} \, h_{i \mu} & - 2 \, \mathcal{R}^+_\text{H} \, h_{i \mu} - \nabla_i \left(\nabla \cdot h\right)_\mu - \nabla_\mu \left(\nabla \cdot h\right)_i + \nabla_i \nabla_\mu \left(\lambda \cdot h + \gamma \cdot h \right) \\
	& + \frac{1}{2} \, {\epsilon^{kl}}_i \left(\nabla_\mu b_{kl} + \nabla_k b_{l \mu} + \nabla_l b_{\mu k} \right) = 0 \, , \\
	\Box_{10} \, h_{\mu \nu} & + \frac{2}{L^2} \, h_{\mu \nu} - \nabla_\mu \left( \nabla \cdot h\right)_\nu - \nabla_\nu \left(\nabla \cdot h \right)_\mu + \nabla_\mu \nabla_\nu \left( \lambda \cdot h + \gamma \cdot h\right) \\
	& + \lambda_{\mu \nu} \left(- \, \frac{2}{L^2} \, \lambda \cdot h - \mathcal{R}^+_\text{H} \left(\frac{3\alpha}{2} \, \varphi - \gamma \cdot h \right) - \, \frac{1}{4} \, \epsilon^{ijk} \, \nabla_i b_{jk} \right) = 0 \, .
\end{eqaed}
In order to simplify the linearized equations of motion for tensor, vector and scalar perturbations it is convenient to introduce (minus) the eigenvalues of the scalar Laplacian on the unit $\ess^n$,
\begin{eqaed}\label{eq:scalar_laplacian_eigvals}
	\Lambda_n \equiv \ell \left(\ell + n - 1\right) \, , \qquad \ell \in \{0 \, , 1 \, , 2 \, , \dots \} \, ,
\end{eqaed}
as well as the two parameters
\begin{eqaed}\label{eq:sigma_tau_orientifold}
	\sigma_3 \equiv 1 + 3 \, \frac{L^2}{R^2} = \frac{3}{2} \, , \qquad \tau_3 \equiv L^2 \, V''_0 = \frac{9}{2}
\end{eqaed}
for the orientifold models, and
\begin{eqaed}\label{eq:sigma_tau_heterotic}
	\sigma_7 \equiv 3 + \frac{L^2}{R^2} = 15 \, , \qquad \tau_7 \equiv L^2 \, V''_0 = 75
\end{eqaed}
for the heterotic model. These parameters are related to the first and second derivatives of the dilaton tadpole potential evaluated on the background solutions, and thus we shall explore the stability of these solutions varying their values. While in principle including curvature corrections or string loop corrections would modify the values in eqs.~\eqref{eq:sigma_tau_orientifold} and~\eqref{eq:sigma_tau_heterotic}, one could expect that the differences would be subleading in the regime of validity of the present analysis, which corresponds to large fluxes.

\subsubsection{Tensor and vector perturbations in \texorpdfstring{$\ads$}{AdS}}\label{sec:tensor_vector_ads}

Let us now move on to study tensor and vector perturbations, starting from the orientifold models. Following standard practice, we classify them referring to their behavior under the isometry group $SO(2,2) \times SO(8)$ of the $\ads_3 \times \ess^7$ background. In this fashion, the possible unstable modes violate the Breitenlohner-Freedman (BF) bounds, which depend on the nature of the fields involved and correspond, in general, to finite negative values of (properly defined) squared $\ads$ masses. Indeed, as reviewed in~\cite{Basile:2018irz}, care must be exercised in order to identify the proper masses to which the bounds apply~\cite{Breitenlohner:1982jf, Lu:2011qx}, since in general they differ from the eigenvalues of the corresponding $\ads$ d'Alembert operator. In particular, aside from the case of scalars, massless field equations always exhibit gauge invariance.

\subsubsection{Tensor perturbations}\label{sec:tensor_pert_ads}

Let us begin considering tensor perturbations, which result from transverse trace-less $h_{\mu\nu}$, with all other perturbations vanishing. The corresponding equations of motion
\begin{eqaed}\label{eq:ads_tensor_pert_orientifold}
	\left( \Box - \frac{\Lambda_7 \left(\sigma_3 - 1\right)}{3 L^2} \right) h_{\mu \nu}  + \frac{2}{L^2} \, h_{\mu \nu} = 0 \, ,
\end{eqaed}
where we have replaced the internal radius $R$ with the $\ads$ radius $L$ using eq.~\eqref{eq:sigma_tau_orientifold}, is obtained expanding the perturbations in spherical harmonics using the results summarized in Appendix~\ref{sec:tensor_harmonics}. These harmonics are eigenfunctions of the internal Laplacian in eq.~\eqref{eq:laplacian_split}. In order to properly interpret this result, however, it is crucial to observe that the massless tensor equation in $\ads$ is the one determined by gauge invariance. In fact, the linearized Ricci tensor determined by eq.~\eqref{eq:lin_ricci} is not invariant under linearized diffeomorphisms of the $AdS$ background, since
\begin{eqaed}\label{eq:diff_massless}
	\delta_\xi R_{\mu \nu} = \frac{2}{L^2} \left( \nabla_\mu \xi_\nu + \nabla_\nu \xi_\mu \right) \, .
\end{eqaed}
However, the fluxes that are present endow, consistently, the stress-energy tensor with a similar behavior, and $\ell = 0$ in eq.~\eqref{eq:ads_tensor_pert_orientifold} corresponds precisely to massless modes. Thus, as expected from Kaluza-Klein theory, eq.~\eqref{eq:ads_tensor_pert_orientifold} describes a massless field for $\ell = 0$, and an infinite tower of massive ones for $\ell > 0$. These perturbations are all consistent with the BF bound, and therefore no instabilities are present in this sector.

There are also (space-time) scalar excitations resulting from the trace-less part of $h_{ij}$ that is also divergence-less, which is a tensor with respect to the internal rotation group and thus $\ell \geq 2$. According to the results in Appendix~\ref{sec:tensor_harmonics}, they satisfy
\begin{eqaed}\label{eq:ads_tensor_s_pert_orientifold}
	\left( L^2 \, \Box - \frac{\Lambda_7 \left(\sigma_3 - 1\right)}{3} \right) h_{ij} = 0 \, ,
\end{eqaed}
so that their squared masses are all positive. Finally, there are massive $b_{ij}$ perturbations, which are divergence-less and satisfy
\begin{eqaed}\label{eq:ads_b_s_pert_orientifold}
	\left( L^2 \, \Box - \frac{\left(\Lambda_7 + 8\right) \left(\sigma_3 - 1\right)}{3} \right) b_{ij} = 0 \, ,
\end{eqaed}
where again $\ell \geq 2$.

The corresponding tensor perturbations in the heterotic model satisfy
\begin{eqaed}\label{eq:ads_tensor_pert_heterotic}
	\left( L^2 \, \Box - \Lambda_3 \left(\sigma_7 - 3\right) \right) h_{\mu \nu}  + \frac{2}{L^2} \, h_{\mu \nu} = 0 \, ,
\end{eqaed}
which, for $\ell = 0$, describes a massless field, accompanied by a tower of Kaluza-Klein fields for higher $\ell$. Hence, once again there are no instabilities in this sector.

Analogously to the case of the orientifold models, there are massive (space-time) scalar excitations resulting from the trace-less part of $h_{ij}$ that is also divergence-less, which satisfy
\begin{eqaed}\label{eq:ads_tensor_s_pert_heterotic}
	\left( L^2 \, \Box - \Lambda_3 \left(\sigma_7 - 3\right) \right) h_{ij} = 0 \, ,
\end{eqaed}
so that the results in Appendix~\ref{sec:tensor_harmonics} imply that again no instabilities are present. There are also no instabilities arising from transverse $b_{\mu\nu}$ excitations, which satisfy
\begin{eqaed}\label{eq:ads_b_s_pert_heterotic}
	\left( L^2 \, \Box - \Lambda_3 \left(\sigma_7 - 3\right) + 10\right) b_{\mu \nu} = 0 \, ,
\end{eqaed}
so that the lowest ones, corresponding to $\ell = 0$, are massless.

\subsubsection{Vector perturbations}\label{sec:vector_perturbations}

The analysis of vector perturbations is slightly more involved, due to mixings between $h_{\mu i}$ and $b_{\mu i}$ induced by fluxes. The relevant equations are
\begin{eqaed}\label{eq:ads_vector_pert_orientifold}
	\Box_{10} \, b_{\mu i} + 2 \, \mathcal{R}^-_\text{O} \, b_{\mu i} + 4 \, \mathcal{R}^+_\text{O} \, \epsilon_{\alpha \beta \mu} \, \nabla^\alpha {h^\beta}_i & = 0 \, , \\
	\Box_{10} \, h_{\mu i} + 2 \, \mathcal{R}^+_\text{O} \, h_{\mu i} + \frac{1}{2} \, {\epsilon^{\alpha \beta}}_\mu \left(\nabla_\alpha b_{\beta i} + \nabla_\beta b_{i \alpha} \right) & = 0 \, ,
\end{eqaed}	
where $h_{\mu i}$ and $b_{\mu i}$ are divergence-less in both indices. It is now possible to write
\begin{eqaed}\label{eq:b_to_f}
	b_{\mu i} = \epsilon_{\alpha \beta \mu} \, \nabla^\alpha {F^\beta}_i \, ,
\end{eqaed}
but this does not determine $F_i^\beta$ uniquely, since the redefinitions
\begin{eqaed}\label{eq:f_redundancy}
	{F^\beta}_i \; \to \; {F^\beta}_i + \nabla^\beta w_i
\end{eqaed}	
do not affect $b_{\mu i}$. The divergence-less $b_{\mu i}$ of interest, in particular, corresponds to a $F_i^\beta$ that is divergence-less in its internal index $i$, and divergence-less $w_i$ do not affect this condition. One is thus led to the system\footnote{In all these expressions that refer to vector perturbations $\ell \geq 1$, as described in Appendix~\ref{sec:tensor_harmonics}.}
\begin{importantbox}
	\begin{eqaed}\label{eq:ads_pert_vector_system_orientifold}
		\left( L^2 \, \Box - \frac{\Lambda_7 + 5}{3} \left(\sigma_3 - 1\right) + 2 \right) {F_i}^\mu + 4 \, \sigma_3 \, {h_i}^\mu & = 0 \, , \\
		\left( L^2 \, \Box - \frac{\Lambda_7 + 5}{3} \left(\sigma_3 - 1\right) - 2 \right) {h_i}^\mu + \frac{\Lambda_7 + 5}{3} \left( \sigma_3 - 1 \right) {F_i}^\mu & = 0 \, .
	\end{eqaed}
\end{importantbox}
Due to the redundancy expressed by eq.~\eqref{eq:f_redundancy}, the system in eq.~\eqref{eq:ads_pert_vector_system_orientifold} could in principle accommodate a source term of the type $\nabla^\mu\, {\widetilde w}_i$. However, its contribution can be absorbed by a redefinition according to eq.~\eqref{eq:f_redundancy}, and thus we shall henceforth neglect it. Similar arguments apply to the ensuing analysis of scalar perturbations. The eigenvalues of the resulting mass matrix\footnote{We use the convention in which the mass matrix $\mathcal{M}^2$ appears alongside the d'Alembert operator in the combination $\Box - \mathcal{M}^2$.}, here and henceforth expressed in units of $\frac{1}{L^2}$, are thus
\begin{eqaed}\label{eq:ads_vector_masses_orientifold}
	\frac{\Lambda_7 + 5}{3} \left(\sigma_3 - 1 \right) \pm 2 \sqrt{\frac{\Lambda_7 + 5}{3} \left(\sigma_3 - 1\right) \sigma_3 + 1} \, .
\end{eqaed}
In order to refer to the BF bound~\cite{Breitenlohner:1982jf, Lu:2011qx}, one should add $2$ to these expressions and compare the result with zero~\cite{Basile:2018irz}. All in all, there are no modes below the BF bound in this sector, and thus no instabilities. The vector modes lie above it for $\ell > 1$ for $\sigma_3 > 1$, while they are massless for $\ell = 1$ and all allowed values of $\sigma_3 > 1$, and also, for all $\ell$, in the singular case where $\sigma_3 = 1$, which would translate into a seven-sphere of infinite radius. For $\ell = 1$ there are $28$ massless vectors corresponding to one of the eigenvalues above. Indeed, according to the results in Appendix~\ref{sec:tensor_harmonics} they build up a second-rank anti-symmetric tensor in the internal vector indices, and therefore an adjoint multiplet of $SO(8)$ vectors. This counting is consistent with Kaluza-Klein theory and reflects the internal symmetry of $\ess^7$, although the massless vectors originate from mixed contributions of the metric and the two-form field in the present case.

The above considerations extend to the heterotic model, for which we let
\begin{eqaed}\label{eq:b_to_f_het}
	b_{i \mu} = \epsilon_{ijk} \, \nabla^j {F_\mu}^k \, ,
\end{eqaed}
which is transverse in internal space. The resulting system reads
\begin{importantbox}
	\begin{eqaed}\label{eq:ads_vector_pert_system_het}
		\left( L^2 \, \Box - \left(\ell + 1\right)^2 \left(\sigma_7 - 3\right) + 6\right) {F_{\mu}}^i + 4 \, \sigma_7 \, {h_\mu}^i & = 0 \, , \\
		\left( L^2 \, \Box - \left(\ell + 1\right)^2 \left(\sigma_7 - 3\right) - 6\right) {h_{\mu}}^i + \left(\ell + 1\right)^2 \left( \sigma_7 - 3\right) {F_\mu}^i & = 0 \, ,
	\end{eqaed}
\end{importantbox}
and the eigenvalues of the corresponding mass matrix are given by
\begin{eqaed}\label{eq:ads_vector_masses_het}
	\left(\ell + 1\right)^2 \left(\sigma_7 - 3\right) \pm 2 \sqrt{\left(\ell + 1\right)^2 \left(\sigma_7 - 3\right) \sigma_7 + 9} \, .
\end{eqaed}
In order to refer to the BF bound~\cite{Breitenlohner:1982jf, Lu:2011qx} one should add $6$ to these expressions and compare the result with $-4$~\cite{Basile:2018irz}. Hence, there are no modes below the BF bound in this sector. The vector modes are massive for $\ell > 1$ in the region $\sigma_7 > 3$, while they become massless for $\ell = 1$ and all allowed values of $\sigma_7 > 3$, and for all values of $\ell$ in the singular limit $\sigma_7 = 3$, which would correspond to a three-sphere of infinite radius. All in all, for $\ell = 1$ there are $6$ massless vectors arising from one of the two eigenvalues above, and according to the results in Appendix~\ref{sec:tensor_harmonics} they build up an second-rank anti-symmetric tensor in the internal vector indices, and therefore an adjoint multiplet of $SO(4)$ vectors. The counting is consistent with Kaluza-Klein theory and with the internal symmetry of $S^3$, although the massless vectors originate once again from mixed contributions of the metric and the two-form field. In light of these results, one could expect that choosing a different internal space with non-trivial isometries would not result in instabilities of tensor or vector modes, since tensors are decoupled and the gauge invariance arising from Kaluza-Klein arguments underpins massless modes.

\subsubsection{Scalar perturbations in \texorpdfstring{$\ads$}{AdS}}\label{sec:scalar_ads}

Let us now discuss scalar perturbations. Since there are seven independent such perturbations in the present cases, the analysis of the resulting systems is more involved with respect to the case of tensor and vector perturbations. While the results in this section can be obtained using a suitable gauge fixing of the metric, we shall proceed along the lines of~\cite{Basile:2018irz}, where algebraic constraints arise from the Einstein equations.

\subsubsection{Scalar perturbations in the orientifold models}\label{sec:scalar_ads_orientifold}

Let us now focus on scalar perturbations in the orientifold models. To begin with, $b_{\mu\nu}$ contributes to scalar perturbations, as one can verify letting
\begin{eqaed}\label{eq:b_scalar_contribution}
	b_{\mu \nu} = \epsilon_{\mu \nu \rho} \, \nabla^\rho B \, ,
\end{eqaed}
an expression that satisfies identically
\begin{eqaed}\label{eq:divless_b}
	\nabla^\mu b_{\mu \nu} = 0 \, .
\end{eqaed}
On the other hand, they do not arise from $b_{\mu i}$ and $b_{ij}$, since the corresponding contributions would be pure gauge. On the other hand, scalar metric perturbations can be parametrized as
\begin{eqaed}\label{eq:scalar_metric}
	h_{\mu \nu} & = \lambda_{\mu \nu} \, A \, , \\
	h_{\mu i} & = R^2 \, \nabla_\mu \nabla_i D \, , \\
	h_{ij} & = \gamma_{ij} \, C \, ,
\end{eqaed}
up to a diffemorphism with independent parameters along $\ads_3$ and $\ess^7$ directions. The linearized equations of motion for $b_{\mu\nu}$ yield
\begin{eqaed}\label{eq:b_equation_scalar}
	\Box_{10} \, B + 4 \, \mathcal{R}^+_\text{O} \left(\alpha \, \varphi - R^2 \, \nabla^2 D - \frac{3}{2} \, A + \frac{7}{2} \, C\right) = 0 \, ,
\end{eqaed}
where $\nabla^2$ denotes the internal background Laplacian, according to the decomposition of eq.~\eqref{eq:laplacian_split}. Expanding with respect to spherical harmonics, so that $\nabla^2 \; \to \; - \, \frac{\Lambda_7}{R^2}$, eq.~\eqref{eq:b_equation_scalar} becomes (an $\ads$ derivative of\footnote{The overall derivative can be removed on account of suitable boundary conditions.})
\begin{eqaed}\label{eq:b_equation_scalar_eigs}
	\left( \Box - \frac{\Lambda_7}{R^2} \right) B + 4 \, \mathcal{R}^+_\text{O} \left(\alpha \, \varphi + \Lambda_7 \, D - \frac{3}{2} \, A + \frac{7}{2} \, C\right) = 0 \, .
\end{eqaed}
Notice that a redefinition $B \; \to \; B + \delta B(y)$, where $\delta B(y)$ depends only on internal coordinates, would not affect $b_{\mu\nu}$ in eq.~\eqref{eq:b_scalar_contribution}. As a result, while eqs.~\eqref{eq:b_equation_scalar} and~\eqref{eq:b_equation_scalar_eigs} could in principle contain a source term, this can be eliminated taking this redundancy into account. Similar considerations apply for the heterotic model.

In a similar fashion, the linearized equation of motion for the dilaton becomes
\begin{eqaed}\label{eq:dilaton_equation_eigs}
	\left( \Box - \frac{\Lambda_7}{R^2} - V''_0 \right) \varphi + 2 \, \mathcal{R}^+_\text{O} \left(\alpha^2 \, \varphi - 3 \, \alpha \, A\right) + \alpha \, \Box \, B = 0 \, ,
\end{eqaed}
where the last term can be eliminated using eq.~\eqref{eq:b_equation_scalar_eigs}. Analogously, the linearized Einstein equations take the form
\begin{eqaed}\label{eq:scalar_metric_eqs}
	\lambda_{\mu \nu} \left[ \left(\Box - \frac{\Lambda_7}{R^2} - \frac{4}{L^2} \right) A + \mathcal{R}^+_\text{O} \left(\frac{7\alpha}{2} \, \varphi + 21 \, C + 6 \, \Lambda_7 \, D \right) - \, \frac{3 \, \Lambda_7}{2 \, R^2} \, B \right] \\
	\qquad + \nabla_\mu \nabla_\nu \left(A + 7 \, C + 2 \, \Lambda_7 \, D \right) = 0 \, , \\
	\qquad \nabla_\mu \nabla_i \left(12 \, D - B + 2 \, A + 6 \, C \right) = 0 \, , \\
	\gamma_{ij} \left[\left(\Box - \frac{\Lambda_7 + 9}{R^2} - \frac{7}{L^2} \right) C - \frac{\Lambda_7}{2} \left(4 \, \mathcal{R}^+_\text{O} \, D - \frac{1}{R^2} \, B \right) - \, \frac{\alpha}{2} \, \mathcal{R}^+_\text{O} \, \varphi \right] \\
	\qquad + \nabla_i \nabla_j \left( 3 \, A + 5 \, C - 2 \, R^2 \, \Box \, D \right) = 0 \, .
\end{eqaed}
Although these equations have an unfamiliar form, the terms involving gradients must vanish separately, as discussed in detail in~\cite{Basile:2018irz}. For $\ell = 0$ nothing depends on internal coordinates, the terms involving $\nabla_\mu \nabla_i$ and $\nabla_i \nabla_j$ become empty and $D$ also disappears. In this case one is thus led to the simplified system
\begin{eqaed}\label{eq:l0_scalar_system}
	\left(L^2 \, \Box - 4 - 3 \, \sigma_3 \right) A + \frac{7\alpha}{2} \, \sigma_3 \, \varphi & = 0 \, , \\
	\left(L^2 \, \Box - \tau_3 - 2 \, \alpha^2 \, \sigma_3 \right) \varphi + 2 \, \alpha \, \sigma_3 \, A & = 0 \, , \\
	L^2 \, \Box \, B - 8 \, \sigma_3 \, A + 4 \, \alpha \, \sigma_3 \, \varphi & = 0 \, ,
\end{eqaed}
to be supplemented by the linear relation
\begin{eqaed}\label{eq:linear_constraint_a-c}
	A = - 7 \, C \, ,
\end{eqaed}
and the last column of the resulting mass matrix vanishes, so that there is a vanishing eigenvalue whose eigenvector is proportional to $B$. This perturbation is however pure gauge, since eq.~\eqref{eq:b_scalar_contribution} implies that the corresponding field strength vanishes identically. Leaving it aside, one can work with the reduced mass matrix determined by the other two equations, whose eigenvalues are
\begin{eqaed}\label{eq:l0_masses_orientifold}
	\left(\alpha^2 + \frac{3}{2} \right) \sigma_3 + \frac{\tau_3}{2} + 2
	\pm \frac{1}{2} \, \sqrt{\Delta} \, ,
\end{eqaed}
where the discriminant
\begin{eqaed}\label{eq:l0_delta_orientifold}
	\Delta \equiv 4 \alpha^4 \, \sigma_3^2 + 16 \, \alpha^2 \left(\sigma_3 + \frac{\tau_3}{4} - 1\right) \sigma_3 + \left(3 \, \sigma_3 - \tau_3 + 4 \right)^2 \, .
\end{eqaed}
There are regions of instability as one varies the parameters $\sigma_3$, $\tau_3$ of eq.~\eqref{eq:sigma_tau_orientifold}, but for the actual orientifold models, where $\left(\beta \, , \sigma_3 \, , \tau_3\right) = \left(1 \, , \frac{3}{2} \, , \frac{9}{2}\right)$, the two eigenvalues evaluate to $12$ and $4$, and thus lie well above the BF bound. To reiterate, there are no unstable scalar modes for the orientifold models in the $\ell = 0$ sector for the internal $S^7$. In view of the ensuing discussion, let us add that the stability persists for convex potentials, with $\tau_3 > 0$, independently of $\sigma_3$.

\begin{figure}[!ht]
	\centering
	\includegraphics[width=0.8\textwidth]{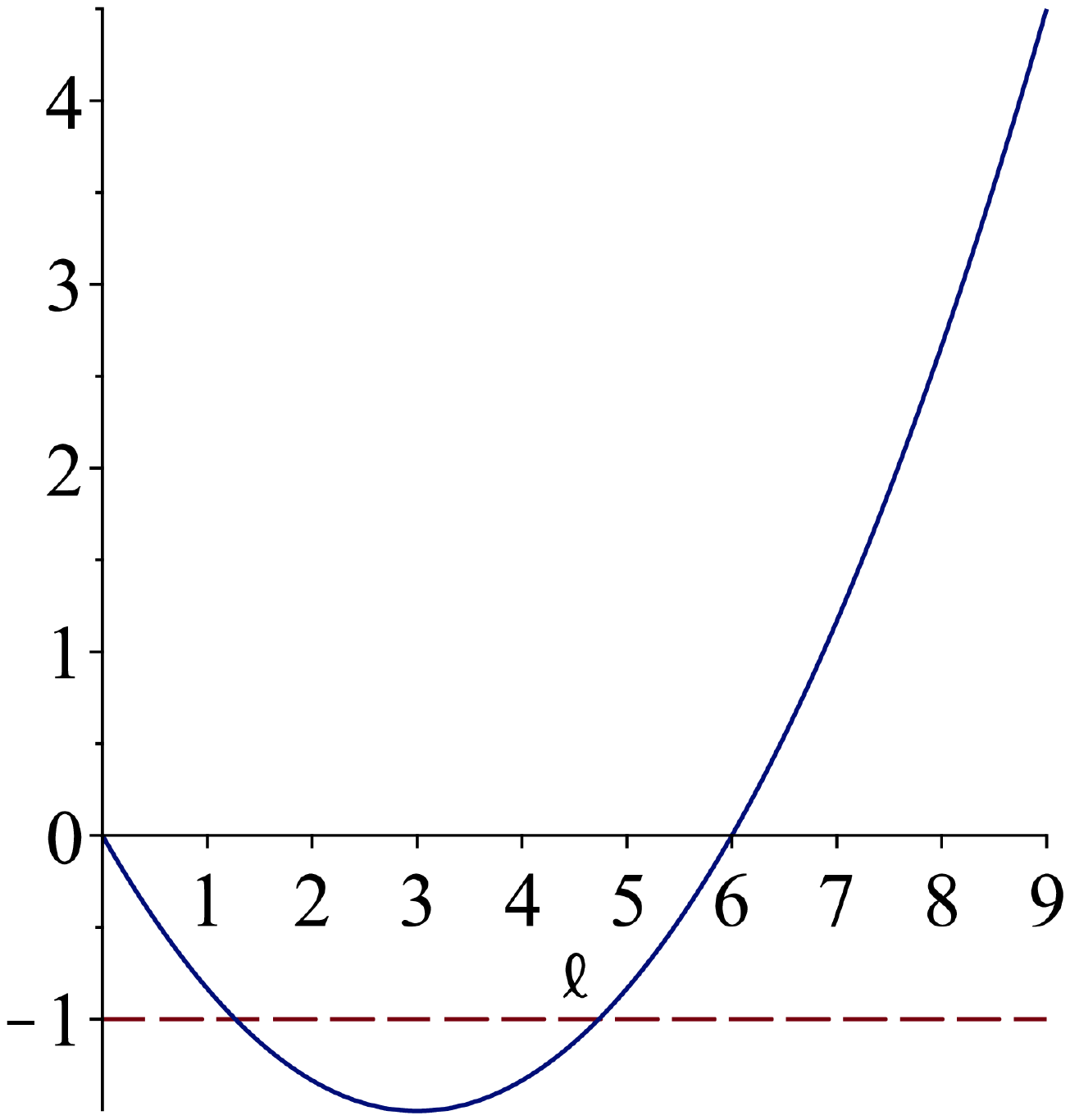}
	\vspace{-140pt}
	\caption{violations of the scalar BF bound in the orientifold models. The dangerous eigenvalue is displayed in units of $\frac{1}{L^2}$, and the BF bound is $-1$ in this case. Notice the peculiar behavior, already spotted in~\cite{Gubser:2001zr}, whereby the squared masses decrease initially, rather than increasing, as $\ell$ increases between $1$ and $3$.}
	\label{fig:bad_eigenvalue}
\end{figure}

For $\ell \neq 0$ the system becomes more complicated, since it now includes the two algebraic constraints
\begin{eqaed}\label{eq:algebraic_constraints_orientifold}
	A + 7 \, C + 2 \, \Lambda_7 \, D & = 0 \, , \\
	2 \, A - B + 6 \, C + 12 \, D & = 0 \, ,
\end{eqaed}
and the five dynamical equations
\begin{eqaed}\label{eq:scalar_dynamical_eqs_orientifold}
	\left( \Box - \frac{\Lambda_7}{R^2} - \frac{4}{L^2} \right) A + \mathcal{R}^+_\text{O} \left(\frac{7 \alpha}{2} \, \varphi - 3 \, A \right) - \, \frac{3 \, \Lambda_7}{2 \, R^2} \, B & = 0 \, , \\
	\left( \Box - \frac{\Lambda_7}{R^2} \right) B + 4 \, \mathcal{R}^+_\text{O} \left(\alpha \, \varphi - 2 \, A \right)& = 0 \, , \\
	\left( \Box - \frac{\Lambda_7 + 9}{R^2} - \frac{7}{L^2} \right) \, C - \, \frac{\Lambda_7}{2} \left( 4 \, \mathcal{R}^+_\text{O} \, D - \frac{1}{R^2} \, B \right) - \, \frac{\alpha}{2} \, \mathcal{R}^+_\text{O} \, \varphi & = 0 \, , \\
	\Box \, D - \frac{3}{2 R^2} \, A - \frac{5}{2 R^2} \, C & = 0 \, , \\
	\left( \Box - \frac{\Lambda_7}{R^2} - V''_0 \right) \varphi + \frac{\alpha \, \Lambda_7}{R^2} \, B - 2 \, \mathcal{R}^+_\text{O} \left(\alpha^2 \, \varphi - \alpha \, A\right) & = 0 \, . 
\end{eqaed}
Let us first observe that this set of seven equations for the five unknowns $(A \, , \, B, \, C, \, D, \, \varphi)$ is consistent: one can indeed verify that the algebraic constraints of eq.~\eqref{eq:algebraic_constraints_orientifold} are identically satisfied by the system in eq.~\eqref{eq:scalar_dynamical_eqs_orientifold}. One can thus concentrate on the equations relating $A$, $\varphi$ and $B$, which do not involve the other fields and read
\begin{importantbox}
	\begin{eqaed}\label{eq:final_scalar_system_orientifold}
		\left( L^2 \, \Box - \frac{\Lambda_7}{3} \, \left(\sigma_3 - 1\right) - 4 - 3 \, \sigma_3 \right) A + \frac{7 \alpha}{2} \, \sigma_3 \, \varphi - \frac{\Lambda_7}{2} \left(\sigma_3 - 1\right) B & = 0 \, , \\
		\left( L^2 \, \Box - \frac{\Lambda_7}{3} \, \left(\sigma_3 - 1\right) - \tau_3 - 2 \, \alpha^2 \, \sigma_3 \right) \varphi + 2 \, \alpha \, \sigma_3 \, A - \frac{\alpha \, \Lambda_7}{3} \left(\sigma_3 - 1\right) B & = 0 \, , \\
		\left(L^2 \, \Box - \frac{\Lambda_7}{3} \left(\sigma_3 - 1 \right) \right) B - 8 \, \sigma_3 \, A + 4 \, \alpha \, \sigma_3 \, \varphi & = 0 \, ,
	\end{eqaed}
\end{importantbox}
to then determine $C$ and $D$ via the algebraic constraints. The mass matrix of interest is now $3 \times 3$, and in all cases one is to compare its eigenvalues with the Breitenlohner-Freedman (BF) bound for scalar perturbations, which in this $\ads_3 \times \ess^7$ case reads
\begin{eqaed}\label{eq:scalar_bf_bound_orientifold}
	m^2 \, L^2 \geq - 1 \, .
\end{eqaed}
One is thus led, in agreement with~\cite{Gubser:2001zr}\footnote{For an earlier analysis in general dimensions, see~\cite{DeWolfe:2001nz}. A subsequent analysis for two internal sphere factors was performed in~\cite{Hong:2005fi}. In supersymmetric cases~\cite{Kim:1985ez}, recently techniques based on Exceptional Field Theory have proven fruitful~\cite{Malek:2019eaz, Cesaro:2021haf}.}, to the simple results
\begin{eqaed}\label{eq:scalar_masses_orientifold}
	\left(\frac{\ell \left(\ell + 6\right)}{6} + 4 \ , \, \frac{\left(\ell+6\right) \left(\ell + 12\right)}{6} \ , \, \frac{\ell \left(\ell - 6\right)}{6} \right)
\end{eqaed}
for the seven-sphere, and thus the BF bound is violated by the third eigenvalue for $\ell = 2 \, , \, 3 \, , \, 4$, as displayed in fig.~\ref{fig:bad_eigenvalue}. Decreasing the value of $\alpha$ could remove the problem for $\ell = 4$, but the instability would still be present for $\ell = 2 \, , \, 3$. On the other hand, increasing the value of $\alpha$ instabilities would appear also for higher values\footnote{For recent results on unstable modes of non-vanishing angular momentum in $\ads$ compactifications, see~\cite{Malek:2020mlk}.} of $\ell$.

\begin{figure}[!ht]
	\centering
	\includegraphics[width=0.8\textwidth]{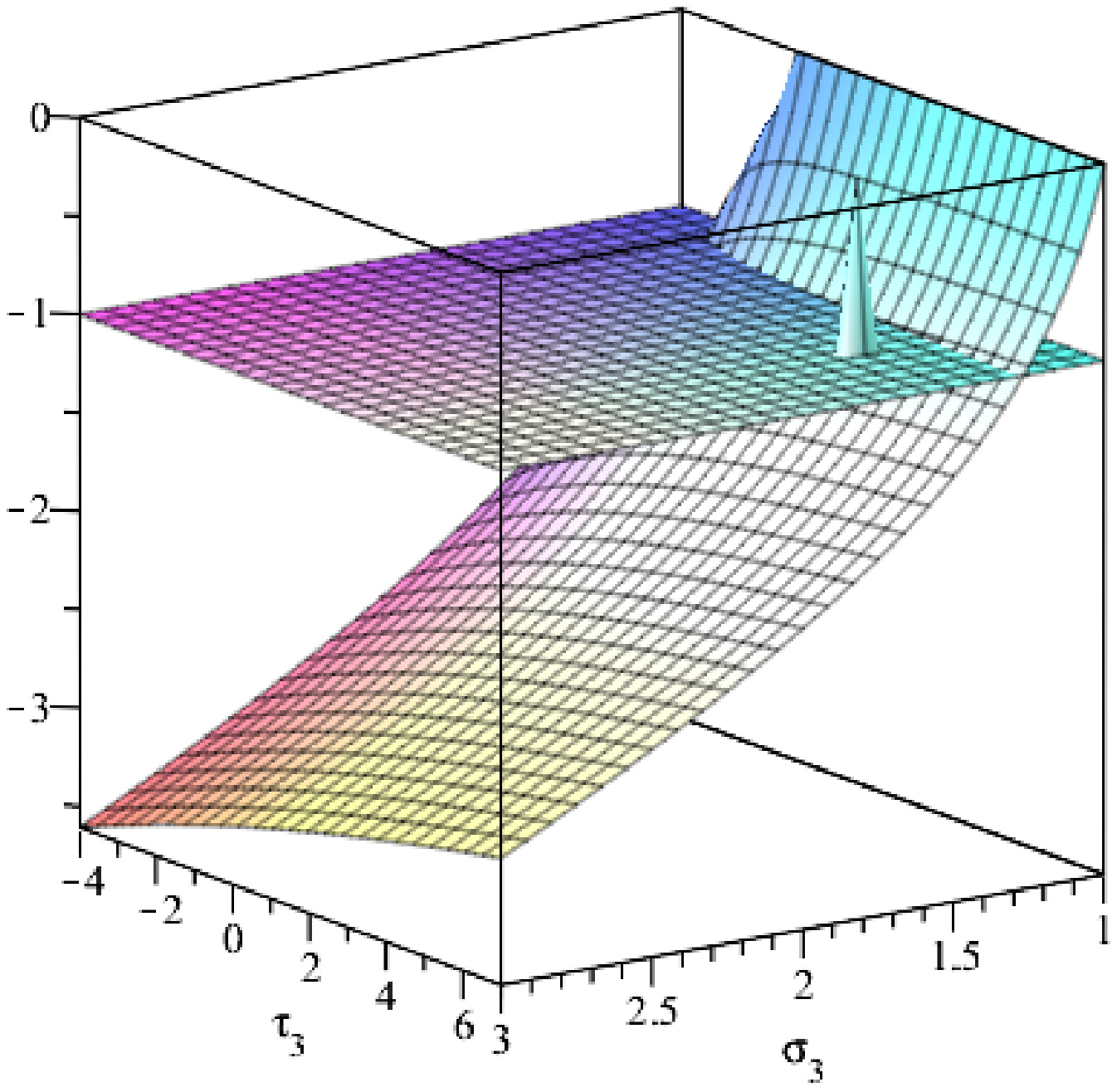}
	\vspace{-130pt}
	\caption{comparison between the lowest eigenvalue $m^2 \, L^2$ and the BF bound, which is $-1$ in this case. There are regions of stability for values of $\sigma_3$ close to $1$, which correspond to $\frac{R^2}{L^2} > 9$ and negative values of $V_0$. The example displayed here refers to $\ell = 3$, which corresponds to the minimum in fig.~\ref{fig:bad_eigenvalue}, and the peak identifies the tree-level values $\sigma_3 = \frac{3}{2}$, $\tau_3 = \frac{9}{2}$.}
	\label{fig:bad_eigenvalue_gen}
\end{figure}

\begin{figure}[!ht]
	\centering
	\includegraphics[width=0.8\textwidth]{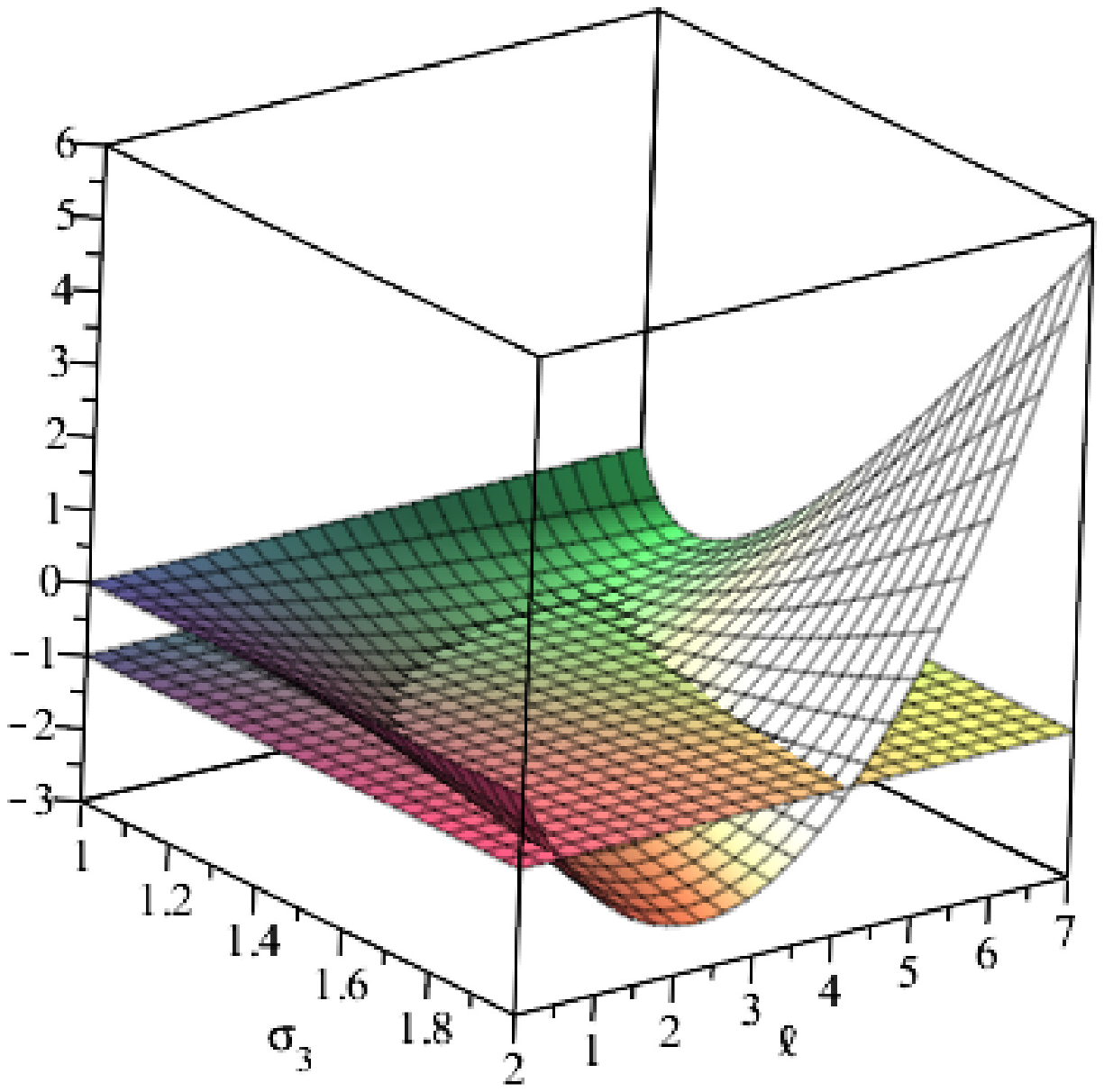}
	\vspace{-140pt}
	\caption{a different view. Comparison between the lowest eigenvalue of $m^2 \, L^2$ and the BF bound, which is $-1$ in this case, as functions of $\ell$ and $\sigma_3$, for $\tau_3 = \frac{9}{2}$. There are regions of stability for values of $\sigma_3$ close to 1, which correspond to large values for the ratio $\frac{R^2}{L^2}$ and to negative values of $V_0$.}
	\label{fig:bad_eigenvalue_different}
\end{figure}

One could now wonder whether there exist regions within the parameter space spanned by $\sigma_3$ and $\tau_3$ where the violation does not occur. We did find them, for all dangerous values of $\ell$, for values of $\sigma_3$ that are close to one, and therefore for negative $V_0$, and for positive $\tau_3$, \textit{i.e.} for potentials that are convex close to the background configuration. These results are displayed in figs.~\ref{fig:bad_eigenvalue_gen} and~\ref{fig:bad_eigenvalue_different}.

\subsubsection{Scalar perturbations in the heterotic model}\label{sec:scalar_ads_heterotic}

\begin{figure}[!ht]
	\centering
	\includegraphics[width=0.8\textwidth]{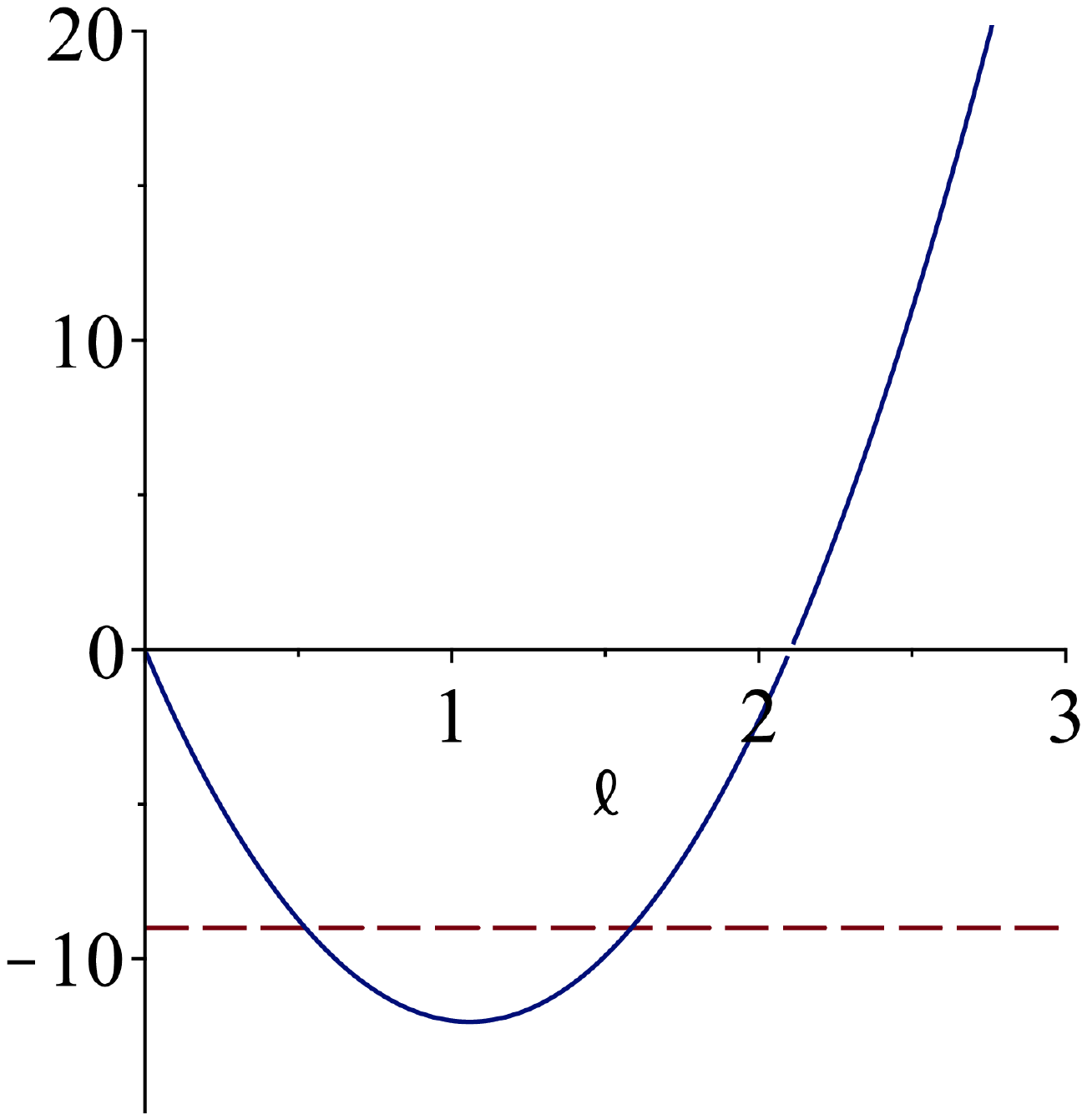}
	\vspace{-140pt}
	\caption{violations of the BF bound in the heterotic model. The dangerous eigenvalue is displayed in units of $\frac{1}{L^2}$, and the BF bound is $-9$ in this case.}
	\label{fig:bad_eigenvalue_het}
\end{figure}

Let us now move on to the stability analysis of scalar perturbations the heterotic model. Proceeding as in the preceding section, we let
\begin{eqaed}\label{eq:b_scalar_het}
	b_{ij} = \epsilon_{ijk} \, \nabla^k B \, ,
\end{eqaed}
a choice that also identically satisfies
\begin{eqaed}\label{eq:divless_b_het}
	\nabla^i b_{ij} = 0 \, .
\end{eqaed}
In addition, let us parametrize scalar metric perturbations as
\begin{eqaed}\label{eq:scalar_metric_het}
	h_{\mu \nu} & = \lambda_{\mu \nu} \, A \, , \\
	h_{\mu i} & = L^2 \, \nabla_\mu \nabla_i D \, , \\
	h_{ij} & = \gamma_{ij} \, C \, ,
\end{eqaed}
along the lines of the preceding section. For scalar perturbations one arrives again at seven equations for five unknowns, and one can verify that the system is consistent. All in all, one can thus work with $C$, $\varphi$ and $B$, restricting the attention to
\begin{importantbox}
	\begin{eqaed}\label{eq:final_scalar_system_het}
		\left( L^2 \, \Box - \Lambda_3 \left(\sigma_7 - 3 \right) - 5 \, \sigma_7 - 12 \right) C + \frac{5 \alpha}{2} \, \sigma_7 \, \varphi - \frac{3 \, \Lambda_3}{2} \left(\sigma_7 - 3\right) B & = 0 \, , \\
		\left( L^2 \, \Box - \Lambda_3 \left(\sigma_7 - 3\right) - \tau_7 - 2 \, \alpha^2 \, \sigma_7 \right) \varphi + 6 \, \alpha \, \sigma_7 \, C + \alpha \, \Lambda_3 \left(\sigma_7 - 3\right) B & = 0 \, , \\
		\left(L^2 \, \Box - \Lambda_3 \left(\sigma_7 - 3 \right) \right) B - 8 \, \sigma_7 \, C + 4 \, \alpha \, \sigma_7 \, \varphi& = 0 \, ,
	\end{eqaed}
\end{importantbox}
here expressed in terms of the two variables $\sigma_7$ and $\tau_7$ of eq.~\eqref{eq:sigma_tau_heterotic}, to then determine $A$ and $D$ algebraically. For $\ell = 0$ $B$ again decouples, and the eigenvalues of the corresponding reduced mass matrix are
\begin{eqaed}\label{eq:l0_scalar_masses_het}
	\left(\alpha^2 + \frac{5}{2} \right) \sigma_7 + \frac{\tau_7}{2} + 6 \pm \frac{1}{2} \sqrt{\Delta} \, ,
\end{eqaed}
with
\begin{eqaed}\label{eq:delta_het}
	\Delta \equiv \left(4 \alpha^4 + 40 \alpha^2 + 25 \right) \sigma_7^2 + 4 \left( \alpha^2 - \frac{5}{2}\right) \left(\tau_7 - 12 \right) \sigma_7 + \left(\tau_7 - 12\right)^2 \, .
\end{eqaed}
In particular, in the heterotic model they read $24 \left(4 \pm \sqrt{6} \right) > 0$. We can now move on to the $\ell \neq 0$ case, where the three scalars $(C \, , \, \phi, \, B)$ all contribute, so that one is led to a $3 \times 3$ mass matrix. In most of the parameter space, two eigenvalues are not problematic, but there is one dangerous eigenvalue, depicted in fig.~\ref{fig:bad_eigenvalue_het}, which corresponds to $\ell=1$ and $k=0$ in the expression
\begin{eqaed}\label{eq:actual_eigvals_het}
	64 + 12 \, \Lambda_3 - 16 \, \sqrt{34 + 15 \, \Lambda_3} \, \cos \left( \frac{\delta - 2 \pi \, k}{3} \right) \, ,
\end{eqaed}
where
\begin{eqaed}\label{eq:het_small_delta}
	\delta \equiv \arg \left(152 - 45 \, \Lambda_3 + 3 \, i \, \sqrt{3 \left(5 \, \Lambda_3 + 3\right) \left( \left( 5 \, \Lambda_3 + 14 \right)^2 + 4 \right)} \right) \, .
\end{eqaed}

\begin{figure}[!ht]
	\centering
	\includegraphics[width=0.8\textwidth]{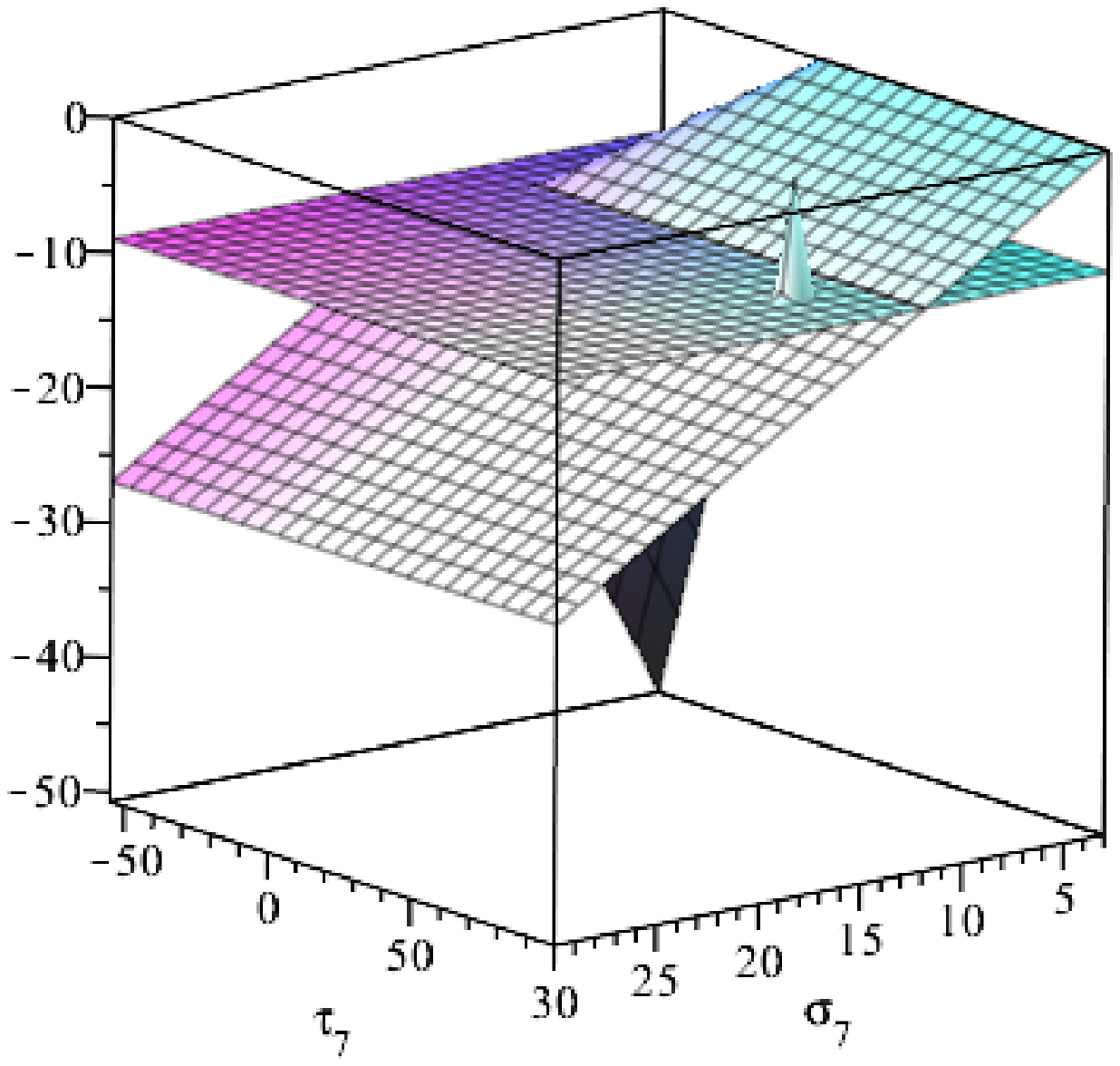}
	\vspace{-140pt}
	\caption{comparison between the lowest eigenvalue $m^2 \, L^2$ and the BF bound, which is $-9$ in this case. There are regions of stability for values of $\sigma_7$ close to $3$, which correspond to $\frac{R^2}{L^2} > 9$, and to negative values of $V_0$. The example displayed here refers to $\ell = 1$, which corresponds to the minimum in fig.~\ref{fig:bad_eigenvalue_het}, and the peak identifies the tree-level values $\sigma_7 = 15$, $\tau_7 = 75$.}
	\label{fig:bad_eigenvalue_gen_het}
\end{figure}

\begin{figure}[!ht]
	\centering
	\includegraphics[width=0.8\textwidth]{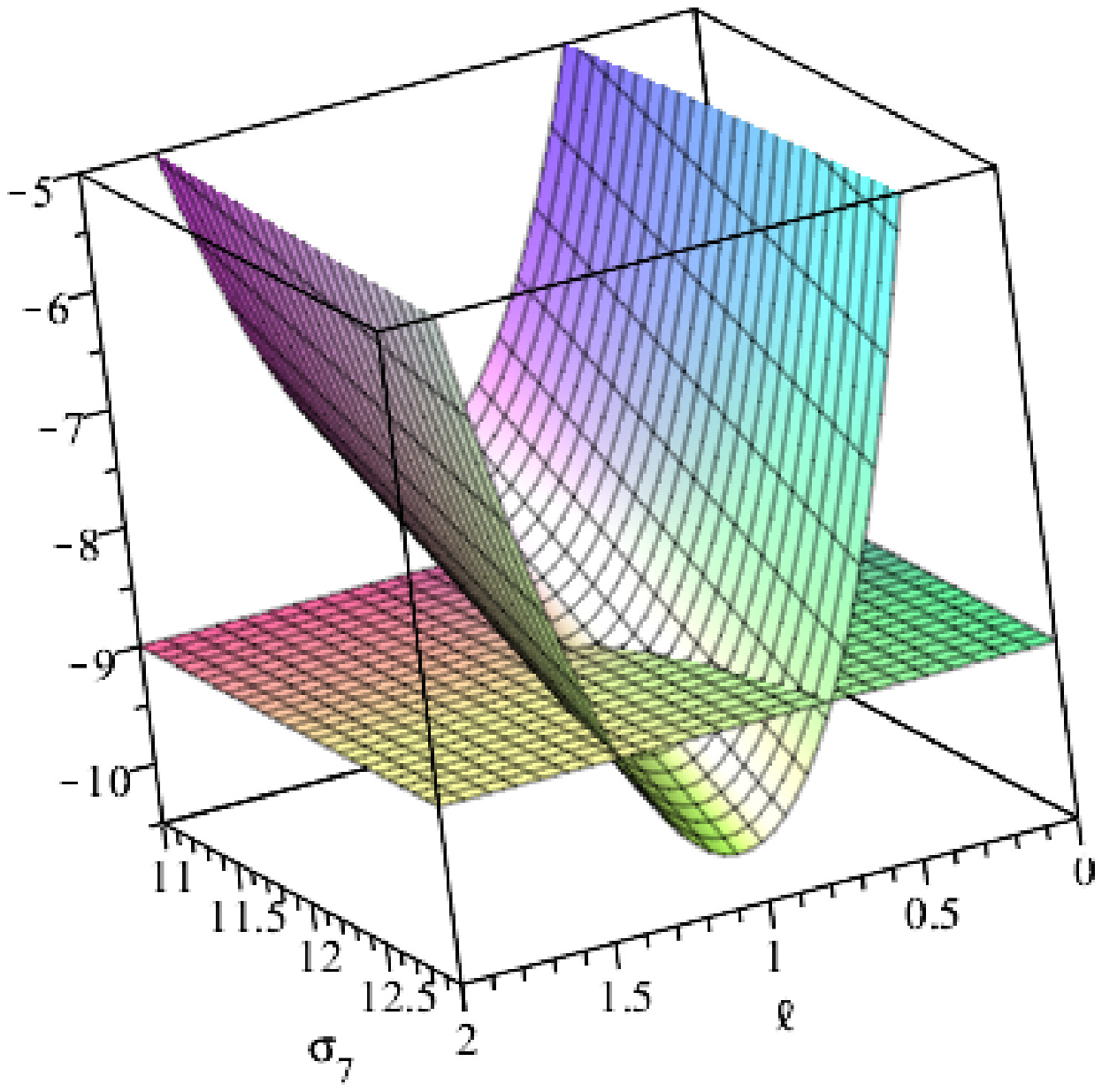}
	\vspace{-140pt}
	\caption{a different view. Comparison between the lowest eigenvalue $m^2 \, L^2$ and the BF bound, which is $-9$ in this case, as functions of $\ell$ and $\sigma_7$, for $\tau_7 = 75$. There are regions of stability for values of $\sigma_3$ below $12$, which correspond to relatively large values for the ratio $\frac{R^2}{L^2}$ and to negative values of $V_0$.}
	\label{fig:bad_eigenvalue_het_different}
\end{figure}

Still, there is again a stability region for values of $\sigma_7$ that are close to $12$, for negative $V_0$, and typically for positive $\tau_7$, \textit{i.e.} for potentials that are convex close to the background configuration. These results are displayed in figs.~\ref{fig:bad_eigenvalue_gen_het} and~\ref{fig:bad_eigenvalue_het_different}.

\subsubsection{Removing the unstable modes}\label{sec:removing_unstable_modes}

Since the number of unstable modes is finite, one can try to eliminate the unstable modes present, in the orientifold models, for $\ell=2,3,4$ by projections in the internal $\ess^7$. According to the results in Appendix~\ref{sec:tensor_harmonics}, scalar spherical harmonics of order $\ell$ correspond to harmonic polynomials of degree $\ell$, and the issue is how to project out the dangerous ones. While in the orientifold models it is not clear how to identify a suitable projection that leaves no fixed sub-varieties\footnote{Projections that leave a sub-variety fixed could entail subtleties related to twisted states that become massless.}~\cite{Basile:2018irz}, in the heterotic case one could eliminate the bad eigenvalue by a $\mathbb{Z}_2$ antipodal projection in the internal sphere $S^3$, which can be identified with the $SU(2)$ group manifold. This operation has no fixed points, and reduces the internal space to the $SO(3)$ group manifold, without affecting the massless vectors with $\ell=1$ that we have identified. However, non-perturbative instabilities would be in principle relevant to the story in this case, and we shall analyze them in detail in Section~\ref{Section4}. Curvature corrections and string loop corrections would also deserve a closer look, since they could drive the potential to a nearby stability domain, providing an interesting alternative for these $\adsts$ solutions.

Let us conclude with a few remarks. To begin with, a suitable choice of internal manifold could rid the $\ads$ flux compactifications of perturbative instabilities altogether, but in general the study of tensor and vector perturbations would become more involved. Moreover, in case the instabilities that we have discusses were not present, one would need to take into account the fluctuations of the remaining degrees of freedom of the low-energy effective theory, which include non-Abelian gauge fields that couple to the gravitational sector. At any rate, one would eventually also have to exclude non-perturbative instabilities, the analysis of which is the subject of the following section.

\section{Quantum stability: bubbles and flux tunneling}\label{Section4}

In this section we carry on the analysis of instabilities of the $\adsts$ flux compactifications that we have introduced in Section~\ref{Section2}, presenting the results of~\cite{Antonelli:2019nar}. Specifically, we address in detail their non-perturbative instabilities, which manifest themselves as (charged) vacuum bubbles at the semi-classical level, and we compute the corresponding decay rates. We find that this tunneling process reduces the flux number $n$, thus driving the vacua toward stronger couplings and higher curvatures, albeit at a rate that is exponentially suppressed in $n$. We also recast these effects in terms of branes\footnote{For a recent investigation along these lines in the context of the (massive) type IIA superstring, see~\cite{Apruzzi:2019ecr}.}, drawing upon the analogy with the supersymmetric case where BPS brane stacks generate supersymmetric near-horizon $\ads$ throats. While $\text{NS}5$-branes in the heterotic model appear more difficult to deal with in this respect, in the orientifold models $\text{D}1$-brane stacks provide a natural canditate for a microscopic description of these flux vacua and of their instabilities. Indeed, non-supersymmetric analogues of $\ads_5 \times \ess^5$ vacua in type 0 strings, where tachyon condensation breaks conformal invariance of the dual gauge theory, were described in terms of $\text{D}3$-branes in~\cite{Klebanov:1998yya}. In the non-tachyonic type $0'\text{B}$ orientifold model this rôle is played by the dilaton potential, which generates a running of the gauge coupling~\cite{Angelantonj:2000kh,Angelantonj:1999qg,Dudas:2000sn}. As a result, the near-horizon geometry is modified, and one recovers $\ads_5 \times \ess^5$ only in the limit\footnote{It is worth noting that this large-$N$ limit is not uniform, since factors of $\frac{1}{N}$ are accompanied by factors that diverge in the near-horizon limit. In principle, a resummation of $\frac{1}{N}$ corrections could cure this problem.} of infinitely many $\text{D}3$-branes, when the supersymmetry-breaking dilaton potential ought to become negligible. In contrast, $\text{D}1$-branes and $\text{NS}5$-branes should underlie the $\ads_3 \times \ess_7$ and $\ads_7\times \ess_3$ solutions found in~\cite{Mourad:2016xbk}\footnote{One could expect that solutions with different internal spaces, discussed in Section~\ref{Section2}, arise from near-horizon throats of brane stacks placed on conical singularities~\cite{Klebanov:1998hh}.}. This might appear somewhat surprising, since $\text{D}p$-brane stacks in type II superstrings do not exhibit near-horizon geometries of this type for $p \neq 3$, instead dressing them with singular warp factors. Correspondingly, the dual gauge theory is non-conformal~\cite{Itzhaki:1998dd}. While the emergence of a conformal dual involving $\text{D}1$-branes and $\text{NS}5$-branes in non-supersymmetric cases would be an enticing scenario, it is first necessary to establish whether brane descriptions of the $\adsts$ solutions hold ground in these models. In this section we provide some evidence to this effect, and in Section~\ref{Section5} we address this issue in more detail. In particular, matching the gravitational decay rates that we shall compute in Section~\ref{sec:gravity_decay_rate} to the results of the respective brane instanton computations, in Section~\ref{sec:consistency} we find consistency conditions that single out fundamental branes as the localized sources that mediate flux tunneling in the settings at stake.

We begin in Section~\ref{sec:flux_overview} with brief overview of flux tunneling. Then, in Section~\ref{sec:flux_tunneling} we study it in the context of the $\adsts$ solutions that we have described in Section~\ref{Section2}, and we present the computation of the resulting semi-classical decay rate within their low-energy description. In Section~\ref{sec:brane_pic} we introduce the microscopic picture, studying probe $\text{D}1$-branes and $\text{NS}5$-branes in the $\ads$ throat, which we develop in Section~\ref{sec:consistency} deriving consistency conditions from decay rates. We conclude in Section~\ref{sec:brane_instantons} presenting explicit expressions for the decay rates in the orientifold models and in the heterotic model.

\subsection{Flux tunneling}\label{sec:flux_overview}

Introducing charged localized sources of codimension one (``membranes'') in gravitational systems with Abelian gauge (form) fields, a novel decay mechanism arises for meta-stable flux vacua~\cite{Brown:1987dd, Brown:1988kg}, whereby charged membranes nucleate in space-time, sourcing vacuum bubbles that expand carrying away flux. In the semi-classical limit, the resulting process can be analyzed via instanton computations~\cite{Coleman:1977py, Callan:1977pt, Coleman:1980aw}, albeit the resulting (Euclidean) equations of motion are modified by the contribution due to membranes\footnote{We shall not discuss the Gibbons-Hawking-York boundary term, which is to be included at any rate to consistently formulate the variational problem.}~\cite{BlancoPillado:2009di}, which arises from actions of the form
\begin{eqaed}\label{eq:brane_action_gen}
	S_{\text{membrane}} = - \int_\mathcal{W} d^{p+1}x \, \sqrt{- j^*g} \, \tau_p + \mu_p \int_\mathcal{W} B_{p+1}
\end{eqaed}
for $(p+2)$-dimensional space-times supported by flux configurations of a $(p+1)$-form field $B_{p+1}$, where $j$ describes the embedding of the world-volume $\mathcal{W}$ in space-time and, in general, the tension $\tau_p$ can depend on the bulk scalar fields, if any. Typically one expects that maximally symmetric instanton configurations dominate the decay rate associated to processes of this type, and in practical terms one is thus faced with a shooting problem where, in addition to the initial conditions of the (Euclidean) fields, one is to determine the nucleation radius of the bubble\footnote{For a detailed exposition of the resulting (distributional) differential equations, see~\cite{Brown:2010mf}.}.

\subsubsection{Small steps and giant leaps: the thin-wall approximation}\label{sec:thin-wall}

Since flux numbers are typically quantized, even simple toy models result in rather rich landscapes of geometries supported by fluxes~\cite{Bousso:2000xa}, and it has been argued~\cite{Brown:2010bc, Brown:2010mf, Brown:2010mg} that flux tunneling in multi-flux landscapes is dominated by ``giant leaps'', where a sizable fraction of the initial flux is discharged, while in single-flux landscape ``small steps'' dominate, and thus the thin-wall approximation is expected to capture the correct leading-order physics. Therefore, we shall focus on the latter case, since the $\adsts$ solutions discussed in Section~\ref{Section2} are supported by a single flux parameter $n$, and we shall consider thin-wall bubbles with charge $\delta n \ll n$. Within this approximation, one can neglect the back-reaction of the membrane and the resulting space-time geometry is obtained gluing the initial and final states along the bubble wall, which expands at the speed of light.

\subsubsection{Bubbles of nothing}\label{sec:thick-wall}

In addition to flux tunneling, bubbles of nothing~\cite{Witten:1981gj} provide often controlled decay channels in which semi-classical computations are expected to be reliable, and whose existence in the absence of supersymmetry appears quite generic~\cite{Dibitetto:2020csn, GarciaEtxebarria:2020xsr, Draper:2021gmq, Draper:2021qtc}. Although one expects that extreme ``giant leaps'', which discharge almost all of the initial flux, lie outside of the semi-classical regime, it is conceivable that the limit in which all of the initial flux is discharged corresponds to a bubble of nothing. Indeed, some evidence to this effect was presented in~\cite{Brown:2011gt}, and, at least in the case of $\ads$ landscapes, holographic arguments also provide some hints in this direction~\cite{Antonelli:2018qwz}.

\subsection{Bubbles and branes in \texorpdfstring{$\ads$}{AdS} compactifications}\label{sec:flux_tunneling}

Let us now move on to study flux tunneling in the $\adsts$ solutions that we have described in Section~\ref{Section2}. These solutions feature perturbative instabilities carrying internal angular momenta~\cite{Gubser:2001zr,Basile:2018irz}, but we shall not concern ourselves with their effects, imposing unbroken spherical symmetry at the outset. Alternatively, as we have mentioned, one could replace the internal sphere with an Einstein manifold, if any, whose Laplacian spectrum does not contain unstable modes, or with an orbifold that projects them out. This can be simply achieved with an antipodal $\mathbb{Z}_2$ projection in the heterotic model, albeit a microscopic interpretation in terms of fundamental branes appears more subtle in this case, while an analogous operation in the orientifold models appears more elusive~\cite{Basile:2018irz}. However, as we shall see in the following, even in the absence of classical instabilities the $\adsts$ solutions would be at best meta-stable, since they undergo flux tunneling.

\subsubsection{Vacuum energy within dimensional reduction}\label{sec:dim_reduction}

In order to appreciate this, it is instructive to perform a dimensional reduction over the sphere following~\cite{BlancoPillado:2009di}, retaining the dependence on a dynamical radion field $\psi$ in a similar vein to our analysis of $\ds$ instabilities in Section~\ref{Section2}. The ansatz
\begin{eqaed}\label{eq:reduction_ansatz_einstein}
	ds^2 = e^{- \frac{2q}{p} \psi(x)} \, \widetilde{ds}^2_{p+2}(x) + e^{2\psi(x)} \, R_0^2 \, d\Omega^2_q \, ,
\end{eqaed}
where $R_0$ is an arbitrary reference radius, is warped in order to select the $(p+2)$-dimensional Einstein frame, described by $\widetilde{ds}^2_{p+2}$. Indeed, placing the dilaton and the form field on shell results in the dimensionally reduced action
\begin{eqaed}\label{eq:reduced_action}
	S_{p+2} = \frac{1}{2\kappa_{p+2}^2} \int d^{p+2} x \, \sqrt{-\widetilde{g}} \, \left(\widetilde{R} - 2 \widetilde{\Lambda}\right) \, ,
\end{eqaed}
where the $(p+2)$-dimensional Newton's constant is
\begin{eqaed}\label{eq:red_newton_constant}
	\frac{1}{\kappa_{p+2}^2} = \frac{\Omega_q R_0^q}{\kappa_D^2} \, ,
\end{eqaed}
while the ``physical'' cosmological constant $\Lambda = -\frac{p(p+1)}{2L^2}$, associated to the frame used in the preceding section, is related to $\widetilde{\Lambda}$ according to
\begin{eqaed}\label{eq:tilde_lambda}
	\widetilde{\Lambda} = \Lambda \, e^{- \frac{2q}{p} \psi} \, ,
\end{eqaed}
which is a constant when the radion is on-shell, and
\begin{eqaed}\label{eq:on-shell_radion}
	e^{\psi} = \frac{R}{R_0} \propto n^{\frac{\gamma}{\left(q-1\right)\gamma-\alpha}} \, .
\end{eqaed}
Let us remark that the dimensionally reduced action of eq.~\eqref{eq:reduced_action} does not necessarily capture a sensible low-energy regime, since in the present settings there is no scale separation between space-time and the internal sphere. Moreover, as we have discussed in Section~\ref{Section3}, in general one cannot neglect the instabilities arising from fluctuations with non-vanishing angular momentum. On the other hand, the resulting vacuum energy (density)
\begin{eqaed}\label{eq:tilde_vacuum_energy}
	\widetilde{E}_0 & = \frac{2 \widetilde{\Lambda}}{2\kappa_{p+2}^2} = - \, \frac{p(p+1) \, \Omega_q R_0^q}{2\kappa_D^2 \, L^2} \left(\frac{R}{R_0}\right)^{- \frac{2q}{p}} \\
	& \propto - \, n^{- \frac{2(D-2)}{p (q-1-\frac{\alpha}{\gamma})}}
\end{eqaed}
is actually sufficient to dictate whether $n$ increases or decreases upon flux tunneling. In particular, the two signs present in eq.~\eqref{eq:tilde_vacuum_energy} and the requirement that flux tunneling decreases the vacuum energy imply that this process drives the (false) vacua to lower values of $n$, eventually reaching outside of the perturbative regime where the semi-classical analysis is expected to be reliable.

\subsubsection{Decay rates: gravitational computation}\label{sec:gravity_decay_rate}

Let us now compute the decay rate associated to flux tunneling in the semi-classical regime. To this end, standard instanton methods~\cite{Coleman:1977py,Callan:1977pt,Coleman:1980aw} provide most needed tools, but in the present case one is confined to the thin-wall approximation, which entails a flux variation\footnote{On the other hand, as we have mentioned, the extreme case $\delta n = n$ would correspond to the production of a bubble of nothing~\cite{Witten:1981gj}.} $\delta n \ll n$, and the tension $\tau$ of the resulting bubble, which cannot be computed within the formalism, rests on the tension of the corresponding membrane and on its back-reaction\footnote{It is common to identify the tension of the bubble with the ADM tension of a brane soliton solution~\cite{BlancoPillado:2009di}. In our case this presents some challenges, as we shall discuss in Section~\ref{Section5}.}. However, the probe limit, in which the membrane does not affect the radion potential due to changing $n$, identifies the tension of the bubble with that of the membrane, and it can be systematically improved upon~\cite{Brown:2010bc} adding corrections to this equality.

In order to proceed, we work within the dimensionally reduced theory in $\ads_{p+2}$, using coordinates such that the relevant instanton is described by the Euclidean metric
\begin{eqaed}\label{eq:cdl_metric}
	ds^2_E = d\xi^2 + \rho^2(\xi) \, d\Omega^2_{p+1} \, ,
\end{eqaed}
so that the Euclidean on-shell (bulk) action takes the form
\begin{eqaed}\label{eq:on-shell_S_E}
	S_E = 2 \, \Omega_{p+1} \int d\xi \, \rho(\xi)^{p+1} \left( \widetilde{E}_0 - \frac{p(p+1)}{2\kappa_{p+2}^2 \, \rho(\xi)^2} \right) \, ,
\end{eqaed}
with the vacuum energy $\widetilde{E}_0$, along with the associated curvature radius $\widetilde{L}$, defined piece-wise by its values inside and outside of the bubble. Then, the energy constraint
\begin{eqaed}\label{eq:energy_constraint_cdl}
	(\rho')^2 = 1 - \frac{2\kappa_{p+2}^2}{p(p+1)} \, \widetilde{E}_0 \, \rho^2 = 1 + \frac{\rho^2}{\widetilde{L}^2} \, ,
\end{eqaed}
which stems from the Euclidean equations of motion, allows one to change variables in eq.~\eqref{eq:on-shell_S_E}, obtaining
\begin{eqaed}\label{eq:on-shell_S_E_final}
	S_E = - \, \frac{2p(p+1) \, \Omega_{p+1}}{2\kappa_{p+2}^2} \int d\rho \, \rho^{p-1} \sqrt{1+  \frac{\rho^2}{\widetilde{L}^2}} \, .
\end{eqaed}
This expression defines the (volume term of the) exponent $B \equiv S_{\text{inst}} - S_{\text{vac}}$ in the semi-classical formula for the decay rate (per unit volume),
\begin{eqaed}\label{eq:decay_rate_total}
	\frac{\Gamma}{\text{Vol}} \sim \left(\text{det} \right) \times e^{-B} \, , \qquad B = B_{\text{area}} + B_{\text{vol}} \, ,
\end{eqaed}
in the standard fashion~\cite{Coleman:1977py, Callan:1977pt}. The thin-wall bubble is a $(p+1)$-sphere of radius $\widetilde{\rho}$, over which the action has to be extremized, and therefore the area term of the exponent $B$ reads
\begin{eqaed}\label{eq:surface_B}
	B_{\text{area}} \sim \widetilde{\tau} \, \Omega_{p+1} \, \widetilde{\rho}^{\,p+1} \, ,
\end{eqaed}
where the tension $\widetilde{\tau} = \tau \, e^{- (p+1) \frac{q}{p} \psi}$ is measured in the $(p+2)$-dimensional Einstein frame. On the other hand, in the thin-wall approximation the volume term becomes
\begin{eqaed}\label{eq:volume_B}
	B_{\text{vol}} & = \frac{2p(p+1) \, \Omega_{p+1}}{2\kappa_{p+2}^2} \int_0^{\widetilde{\rho}} d\rho \, \rho^{p-1} \left[\sqrt{1+  \frac{\rho^2}{\widetilde{L}_{\text{vac}}^2}} - \sqrt{1+  \frac{\rho^2}{\widetilde{L}_{\text{inst}}^2}} \, \right] \\
	& \sim - \, \epsilon \, \widetilde{\text{Vol}}(\widetilde{\rho}) \, ,
\end{eqaed}
where the energy spacing
\begin{eqaed}\label{eq:spacing_energy}
	\epsilon & \sim \frac{d\widetilde{E}_0}{dn} \, \delta n \propto n^{- \frac{2(D-2)}{p(q-1-\frac{\alpha}{\gamma})}-1} \, \delta n
\end{eqaed}
and the volume $\widetilde{\text{Vol}}(\widetilde{\rho})$ enclosed by the bubble is computed in the $(p+2)$-dimensional Einstein frame,
\begin{eqaed}\label{eq:volumes_relation}
	\widetilde{\text{Vol}}(\widetilde{\rho}) & = \widetilde{L}^{p+2} \, \Omega_{p+1} \, \mathcal{V}\left( \frac{\widetilde{\rho}}{\widetilde{L}} \right) \, , \\
	\mathcal{V}(x) & \equiv \frac{x^{p+2}}{p+2} \, _2F_1\left(\frac{1}{2}, \frac{p+2}{2} ; \frac{p+4}{2};-x^2 \right) \, , \\
	x & \equiv \frac{\widetilde{\rho}}{\widetilde{L}} \, .
\end{eqaed}
All in all, the thin-wall exponent\footnote{Notice that eq.~\eqref{eq:total_B-factor} takes the form of an effective action for a $(p+1)$-brane in $\ads$ electrically coupled to $H_{p+2}$. This observation is the basis for the microscopic picture that we shall present shortly.}
\begin{eqaed}\label{eq:total_B-factor}
	B \sim \tau \, \Omega_{p+1} \, L^{p+1} \left( x^{p+1} - (p+1)\beta \, \mathcal{V}(x) \right) \, , \qquad \beta \equiv \frac{\epsilon \, \widetilde{L}}{(p+1) \widetilde{\tau}}
\end{eqaed}
attains a local maximum at $x = \frac{1}{\sqrt{\beta^2 - 1}}$ for $\beta > 1$. On the other hand, for $\beta \leq 1$ the exponent is unbounded, since $B\rightarrow \infty$ as $x\rightarrow \infty$, and thus the decay rate is completely suppressed. Hence, it is crucial to study the large-flux scaling of $\beta$, which plays a rôle akin to an extremality parameter for the bubble. In particular, if $\beta$ scales with a negative power of $n$ nucleation is suppressed, whereas if it scales with a positive power of $n$ the extremized exponent $B$ approaches zero, thus invalidating the semi-classical computation. Therefore, the only scenario in which nucleation is both allowed and semi-classical at large $n$ is when $\beta > 1$ and is flux-independent. Physically, the bubble is super-extremal and has an $n$-independent charge-to-tension ratio. Since
\begin{eqaed}\label{eq:beta_cdl_value}
	\beta = v_0 \, \frac{\Omega_q \, \delta n}{2\kappa^2_{D} \tau} \, g_s^{- \frac{\alpha}{2}}  \, ,
\end{eqaed}
where the flux-independent constant
\begin{eqaed}\label{eq:v0_parameter}
	v_0 \equiv \sqrt{\frac{2(D-2)\gamma}{(p+1) ((q-1)\gamma-\alpha)}} \, ,
\end{eqaed}
this implies the scaling\footnote{Notice that in the gravitational picture the charge of the membrane does not appear. Indeed, its contribution arises from the volume term of eq.~\eqref{eq:volume_B} in the thin-wall approximation.}
\begin{eqaed}\label{eq:tau_scaling}
	\tau = T \, g_s^{- \frac{\alpha}{2}} \, ,
\end{eqaed}
where $T$ is flux-independent and $\alpha$ denotes the coupling between the dilaton and the form field in the notation introduced in Section~\ref{Section2}. In Section~\ref{sec:consistency} we shall verify that this is precisely the scaling expected from $\text{D}p$-branes and $\text{NS}5$-branes.

\subsubsection{Bubbles as branes}\label{sec:brane_pic}

Let us now proceed to describe a microscopic picture, studying probe branes in the $\adsts$ geometry and matching the semi-classical decay rate of eq.~\ref{eq:total_B-factor} to a (Euclidean) world-volume action. While a more complete description to this effect would involve non-Abelian world-volume actions coupled to the complicated dynamics driven by the dilaton potential, one can start from the simpler setting of brane instantons and probe branes moving in the $\adsts$ geometry. This allows one to retain computational control in the large-$n$ limit, while partially capturing the unstable dynamics at play. When framed in this fashion, instabilities suggest that the non-supersymmetric models at stake are typically driven to time-dependent configurations\footnote{Indeed, as we have discussed in Section~\ref{Section1}, cosmological solutions of non-supersymmetric models display interesting features~\cite{Sagnotti:2015asa,Gruppuso:2015xqa,Gruppuso:2017nap,Mourad:2017rrl,Basile:2018irz}. Similar considerations on flux compactifications can be found in~\cite{Sethi:2017phn}.}, in the spirit of the considerations of~\cite{Basile:2018irz}.

We begin our analysis considering the dynamics of a $p$-brane moving in the $\ads_{p+2} \times \ess^q$ geometry of eq.~\eqref{eq:ads_s_solution}. In order to make contact with $\text{D}$-branes in the orientifold models and $\text{NS}5$-branes in the heterotic model, let us consider a generic string-frame world-volume action of the form
\begin{eqaed}\label{eq:world-volume_action}
	S_p = - \, T_p \int d^{p+1} \zeta \, \sqrt{-j^* g_S} \, e^{- \sigma \phi} + \mu_p \int B_{p+1} \, ,
\end{eqaed}
specified by an embedding $j$ of the brane in space-time, which translates into the $D$-dimensional and $(p+2)$-dimensional Einstein-frame expressions
\begin{eqaed}\label{eq:einstein-frame_actions_brane}
	S_p & = - \, T_p \int d^{p+1} \zeta \, \sqrt{-j^* g} \, e^{\left(\frac{2(p+1)}{D-2} - \sigma \right) \phi} + \mu_p \int B_{p+1} \\
	& = - \, T_p \int d^{p+1} \zeta \, \sqrt{-j^* \widetilde{g}} \, e^{\left(\frac{2(p+1)}{D-2} - \sigma \right) \phi - (p+1)\frac{q}{p}\psi} + \mu_p \int B_{p+1} \, .
\end{eqaed}
Since the dilaton is constant in the $\adsts$ backgrounds that we consider, from eq.~\eqref{eq:einstein-frame_actions_brane} one can read off the effective tension
\begin{eqaed}\label{eq:effective_tension}
	\tau_p = T_p \, g_s^{\frac{2(p+1)}{D-2} - \sigma} \, .
\end{eqaed}
While in this action $T_p$ and $\mu_p$ are independent of the background, for the sake of generality we shall not assume that in non-supersymmetric models $T_p = \mu_p$, albeit this equality is supported by the results of~\cite{Dudas:2001wd}.

\subsubsection{Microscopic branes from semi-classical consistency}\label{sec:consistency}

In this section we reproduce the decay rate that we have obtained in Section~\ref{sec:gravity_decay_rate} with a brane instanton computation\footnote{For more details, we refer the reader to~\cite{Brown:1987dd,Brown:1988kg,Maldacena:1998uz,Seiberg:1999xz}.}. Since flux tunneling preserves the symmetry of the internal manifold, the Euclidean branes are uniformly distributed over it, and are spherical in the Wick-rotated $\ads$ geometry. The Euclidean $p$-brane action of eq.~\eqref{eq:einstein-frame_actions_brane}, written in the $D$-dimensional Einstein frame, then reads
\begin{eqaed}\label{eq:brane_instanton_action}
	S_p^E & = \tau_p \, \text{Area} - \mu_p \, c \, \text{Vol} \\
	& = \tau_p \, \Omega_{p+1} \, L^{p+1} \left( x^{p+1} - (p+1) \, \beta_p \, \mathcal{V}(x) \right) \, ,
\end{eqaed}
where $v_0$ is defined in eq.~\eqref{eq:v0_parameter}, and
\begin{eqaed}\label{eq:beta_p_def}
	\beta_p \equiv v_0 \, \frac{\mu_p}{T_p} \, g_s^{\sigma - \frac{2(p+1)}{D-2} - \frac{\alpha}{2}} \,.
\end{eqaed}
This result matches in form the thin-wall expression in eq.~\eqref{eq:total_B-factor}, up to the identifications of the tensions $\tau$, $\tau_p$ and the parameters $\beta$, $\beta_p$. As we have argued in Section~\ref{sec:gravity_decay_rate}, the former is expected to be justified in the thin-wall approximation. Furthermore, according to the considerations that have led us to eq.~\eqref{eq:tau_scaling}, it is again reasonable to assume that $\beta_p$ does not scale with the flux, which fixes the exponent $\sigma$ to
\begin{eqaed}\label{eq:sigma_value}
	\sigma = \frac{2(p+1)}{D-2} + \frac{\alpha}{2} \, .
\end{eqaed}
This is the value that we shall use in the following. Notice that for $\text{D}p$-branes in ten dimensions, where $\alpha=\frac{3-p}{2}$, this choice gives the correct result $\sigma = 1$, in particular for $\text{D}1$-branes in the orientifold models. Similarly, for $\text{NS}5$-branes in ten dimensions, eq.~\eqref{eq:sigma_value} also gives the correct result $\sigma = 2$. This pattern persists even for the more ``exotic'' branes of~\cite{Bergshoeff:2005ac,Bergshoeff:2006gs,Bergshoeff:2011zk,Bergshoeff:2012jb,Bergshoeff:2015cba}, and it would be interesting to explore this direction further. Notice that in terms of the string-frame value $\alpha_S$, eq.~\eqref{eq:sigma_value} takes the simple form
\begin{importantbox}
	\begin{eqaed}\label{eq:sigma_string-frame}
		\sigma = 1 + \frac{\alpha_S}{2} \, .
	\end{eqaed}
\end{importantbox}
Moreover, from eqs.~\eqref{eq:effective_tension} and~\eqref{eq:sigma_value} one finds that
\begin{eqaed}\label{eq:tau_p_scaling_sigma}
	\tau_p = T_p \, g_s^{- \frac{\alpha}{2}} 
\end{eqaed}
scales with the flux with the same power as $\tau$, as can be seen from eq.~\eqref{eq:tau_scaling}. Since the flux dependence of the decay rates computed extremizing eqs.~\eqref{eq:total_B-factor} and~\eqref{eq:brane_instanton_action} is determined by the respective tensions $\tau$ and $\tau_p$, they also scale with the same power of $n$. Together with eq.~\eqref{eq:sigma_string-frame}, this provides evidence to the effect that, in the present setting, vacuum bubbles can be identified with fundamental branes, namely $\text{D}p$-branes in the orientifold models and $\text{NS}5$-branes in the heterotic model.

Requiring furthermore that the decay rates computed extremizing eqs.~\eqref{eq:total_B-factor} and~\eqref{eq:brane_instanton_action} coincide, one is led to $\beta = \beta_p$, which implies
\begin{eqaed}\label{eq:beta_match}
	\mu_p = \frac{\Omega_q \, \delta n}{2\kappa^2_{D}} = \delta \left( \frac{1}{2\kappa_D^2} \int_{\ess^q} f \star H_{p+2} \right) \, ,
\end{eqaed}
where $\delta$ denotes the variation across the bubble wall, as expected for electrically coupled objects.

\subsubsection{Decay rates: extremization}\label{sec:brane_instantons}

Extremizing the Euclidean action of eq.~\eqref{eq:brane_instanton_action} over the nucleation radius, one obtains the final result for the semi-classical tunneling exponent
\begin{importantbox}
	\begin{eqaed}\label{eq:decay_rate_p}
		S_p^E & = T_p \, L^{p+1} \, g_s^{- \frac{\alpha}{2}} \, \Omega_{p+1} \, \mathcal{B}_p\left(v_0 \, \frac{\mu_p}{T_p}\right) \\
		& \propto n^{\frac{(p+1)\gamma+\alpha}{(q-1)\gamma-\alpha}} \, ,
	\end{eqaed}
\end{importantbox}
where we have introduced
\begin{eqaed}\label{eq:Bp_function}
	\mathcal{B}_p(\beta) \equiv \frac{1}{(\beta^2 - 1)^{\frac{p+1}{2}}} - \frac{p+1}{2} \, \beta \, \int_0^{ \frac{1}{\beta^2 - 1}} \frac{u^{\frac{p}{2}}}{\sqrt{1+u}} \, du \, .
\end{eqaed}
This expression includes a complicated flux-independent pre-factor, but it always scales with a positive power of $n$, consistently with the semi-classical limit. For the sake of completeness, let us provide the explicit result for non-supersymmetric string models, where the microscopic picture goes beyond the world-volume actions of eq.~\eqref{eq:world-volume_action}. Notice that we do not assume that $\mu_p = T_p$ in the non-supersymmetric setting, for the sake of generality. However, as we have already remarked in eq.~\eqref{eq:total_B-factor}, the tunneling process is allowed also in this case. This occurs because $v_0 > 1$, and thus also $\beta > 1$, in the supersymmetry-breaking backgrounds that we consider, since using eq.~\eqref{eq:v0_parameter} one finds
\begin{eqaed}\label{eq:v0_orientifold}
	(v_0)_{\text{orientifold}} = \sqrt{\frac{3}{2}}
\end{eqaed}
for the orientifold models, while
\begin{eqaed}\label{eq:v0_heterotic}
	(v_0)_{\text{heterotic}} = \sqrt{\frac{5}{3}}
\end{eqaed}
for the heterotic model. For $\text{D}1$-branes in the orientifold models, eq.~\eqref{eq:decay_rate_p} yields
\begin{eqaed}\label{eq:decay_rate_D1}
	S_1^E & = \frac{T_1 \, L^2}{\sqrt{g_s}} \, \Omega_2 \, \mathcal{B}_1\left(\sqrt{\frac{3}{2}} \, \frac{\mu_1}{T_1}\right) \\
	& = \frac{\pi}{9\sqrt{2}} \, \mathcal{B}_1\left(\sqrt{\frac{3}{2}} \, \frac{\mu_1}{T_1}\right) T_1 \, \sqrt{T} \, \sqrt{n} \, , 
\end{eqaed}
and $S_1^E \approx 0.1 \, T_1 \, \sqrt{T} \, \sqrt{n}$ if $\mu_1=T_1$. For the heterotic model, using eq.~\eqref{eq:decay_rate_p} the Euclidean action of $\text{NS}5$-branes evaluates to
\begin{eqaed}\label{eq:decay_rate_NS5}
	S_5^E & = \frac{T_5 \, L^6}{\sqrt{g_s}} \, \Omega_6 \, \mathcal{B}_5\left(\sqrt{\frac{5}{3}} \, \frac{\mu_5}{T_5}\right) \\
	& = \frac{9216 \, \pi^3}{125} \, \mathcal{B}_5\left(\sqrt{\frac{5}{3}} \, \frac{\mu_5}{T_5}\right) T_5 \, T \, n^4 \, , 
\end{eqaed}
and $S^E_5 \approx 565.5 \, T_5 \, T \, n^4$ if $\mu_5 = T_5$. In the presence of large fluxes the tunneling instability is thus far milder in the heterotic model.

To conclude, the results in this section provide evidence to the effect that the non-supersymmetric $\ads$ flux compactifications that we have described in Section~\ref{Section2} are non-perturbatively unstable, and the flux tunneling process that they undergo can be described in terms of (stacks of) fundamental branes, namely $\text{D}1$-branes for the $\ads_3 \times \ess^7$ solutions of the orientifold models, and $\text{NS}5$-branes for the $\ads_7 \times \ess^3$ solutions of the heterotic model. In the following section we shall expand upon this picture, studying the Lorentzian evolution of the branes after a tunneling event occurs and relating the resulting dynamics to interactions between branes and to the weak gravity conjecture~\cite{ArkaniHamed:2006dz}. In addition, we shall recover the relevant $\adsts$ solutions as near-horizon geometries of the full the gravitational back-reaction of the branes, thus further supporting the idea that these solutions are built up from stacks of parallel fundamental branes.

\section{Brane dynamics: probes and back-reaction}\label{Section5}

In this section we elaborate in detail on the microscopic picture of non-perturbative instabilities of the $\adsts$ solutions that we have introduced in the preceding section. The results that we have described hitherto suggest that the $\adsts$ geometries at stake can be built up from stacks of parallel fundamental branes, an enticing picture that could, at least in principle, shed light on the high-energy regime of the settings at hand. In particular, our proposal can potentially open a computational window beyond the semi-classical regime, perhaps providing also a simpler realization of $\ads_3$/CFT$_2$ duality\footnote{The alternative case of $\ads_7$ could be studied, in principle, via $\text{M}5$-brane stacks.}. Moreover, in principle one could investigate these non-perturbative instabilities recasting them as holographic RG flows in a putative dual gauge theory~\cite{Antonelli:2018qwz}. In order to further ground this proposal, in Section~\ref{sec:probe_branes} we study the Lorentzian evolution of the expanding branes after a nucleation event takes place, identifying the relevant dynamics and comparing it to the supersymmetric case, and the resulting interaction potentials imply a version of the weak gravity conjecture for extended objects~\cite{Ooguri:2016pdq}. Furthermore, we briefly comment on a recent proposal~\cite{Banerjee:2018qey, Banerjee:2019fzz, Banerjee:2020wix} which rests on the observation that branes nucleating amidst $\ads \; \to \; \ads$ transitions host $\ds$ geometries on their world-volume, referring the reader to~\cite{Basile:2020mpt, Basile:2020xwi} for more details. Then, in Section~\ref{sec:backreaction} we investigate the gravitational back-reaction of stacks of parallel branes within the low-energy effective theory described in Section~\ref{Section2}, deriving a reduced dynamical system that captures the relevant dynamics and recovering an attractive near-horizon $\adsts$ throat. In order to provide a more intuitive understanding of this result, we compare the asymptotic behavior of the fields to the corresponding ones for $\text{D}3$-branes in the type IIB superstring and for the four-dimensional Reissner-Nordstr\"om black hole. The latter represents a particularly instructive model, where one can identify the physical origin of singular perturbations. However, away from the stack the resulting space-time exhibits a space-like singularity at a finite transverse geodesic distance~\cite{Antonelli:2019nar}, as in\footnote{Indeed, our results suggest that the solutions of~\cite{Dudas:2000ff}, which are not fluxed, correspond to $8$-branes.}~\cite{Dudas:2000ff}, which hints at the idea that, in the presence of dilaton tadpoles, any breaking of ten-dimensional Poincar\'e invariance is accompanied by a finite-distance ``pinch-off'' singularity determined by the residual symmetry. Physically, this corresponds to the fact that branes are not isolated objects in these settings, since in the case of the orientifold models non-supersymmetric projections bring along additional (anti-)$\text{D}$-branes that interact with them. In the heterotic model, this rôle is played at leading order by the one-loop vacuum energy. Finally, in Section~\ref{sec:black_branes} and Section~\ref{sec:non_extremal_probes} we extend our considerations to the case of non-extremal branes, focusing on the uncharged of $\text{D}8$-branes in the orientifold models in order to compare probe-brane computations with the corresponding string amplitudes.

\subsection{The aftermath of tunneling}\label{sec:probe_branes}

\begin{figure}[!ht]
\centering
\includegraphics[width=.7\textwidth]{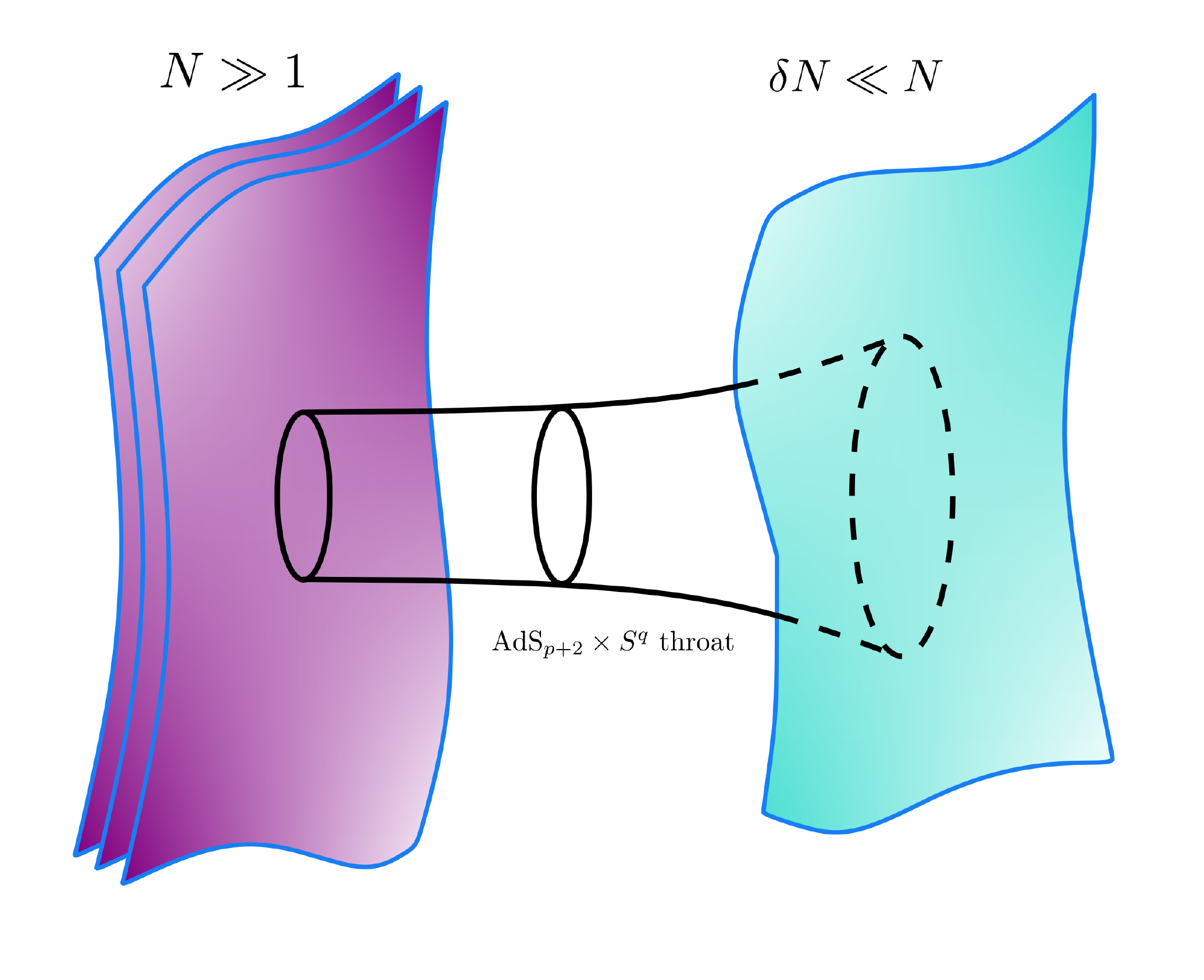}
\caption{a schematic depiction of the interaction between a heavy stack of $N \gg 1$ branes and $\delta N<< N$ light branes in the probe regime. The heavy stack sources the background geometry in which the probe branes move.}
\label{fig:throat_branes}
\end{figure}

After a nucleation event takes place, the dynamics is encoded in the Lorentzian evolution of the bubble. Its counterpart in the microscopic brane picture is the separation of pairs of branes and anti-branes, which should then lead to brane-flux annihilation\footnote{For a discussion of this type of phenomenon in Calabi-Yau compactifications, see~\cite{Kachru:2002gs}.}, with negatively charged branes absorbed by the stack and positively charged ones expelled out of the $\adsts$ near-horizon throat. The resulting picture is shown in fig.~\ref{fig:throat_branes}. In order to explore this perspective, we now study probe (anti-)branes moving in the $\adsts$ geometry. To this end, it is convenient to work in Poincar\'e coordinates, where the $D$-dimensional Einstein-frame metric reads
\begin{eqaed}\label{eq:poincare_metric_adsxs}
	ds^2 = \frac{L^2}{z^2} \left(dz^2 + dx^2_{1,p} \right) + R^2 \, d\Omega^2_q \, , \qquad dx^2_{1,p} \equiv \eta_{\mu \nu} \, dx^\mu dx^\nu \, ,
\end{eqaed}
embedding the world-volume of the brane according to the parametrization
\begin{eqaed}\label{eq:brane_embedding}
	j \, : \quad x^\mu = \zeta^\mu \, , \qquad z = Z(\zeta) \, , \qquad \theta^i = \Theta^i(\zeta) \, .
\end{eqaed}
Furthermore, when the brane is placed at a specific point in the internal sphere\footnote{One can verify that this ansatz is consistent with the equations of motion for linearized perturbations.}, $\Theta^i(\zeta) = \theta_0^i$, the Wess-Zumino term gives the volume enclosed by the brane in $\ads$. As a result, the action that we have introduced in the preceding section evaluates to
\begin{eqaed}\label{eq:brane_action_adsxs}
	S_p & = - \tau_p \int d^{p+1}\! \zeta \, \left(\frac{L}{Z}\right)^{p+1} \left[ \sqrt{ 1 + \eta^{\mu\nu} \, \partial_\mu Z \, \partial_\nu Z } - \frac{c \, L}{p+1} \, \frac{\mu_p}{\tau_p}\right]
\end{eqaed}
in the notation of Section~\ref{Section2}, so that rigid, static branes are subject to the potential
\begin{importantbox}
	\begin{eqaed}\label{eq:brane_potential}
		V_{\text{probe}}(Z) & = \tau_p \, \left(\frac{L}{Z}\right)^{p+1} \left[1 - \frac{c \, L \, g_s^{\frac{\alpha}{2}}}{p+1} \, \frac{\mu_p}{T_p} \right] \\
		& = \tau_p \, \left(\frac{L}{Z}\right)^{p+1} \left[1 - v_0 \, \frac{\mu_p}{T_p} \right] \, .
	\end{eqaed}
\end{importantbox}
The potential in eq.~\eqref{eq:brane_potential} indicates how rigid probe branes are affected by the $\adsts$ geometry, depending on the value of $v_0$. In particular, if $v_0 \, \frac{\abs{\mu_p}}{T_p} > 1$ positively charged branes are driven toward small $Z$ and thus exit the throat, while negatively charged ones are driven in the opposite direction.

Small deformations $\delta Z$ of the brane around the rigid configuration at constant $Z$ satisfy the linearized equations of motion
\begin{eqaed}\label{eq:linearized_eom_deformation}
	- \, \partial_\mu \partial^\mu \delta Z \sim \, \frac{p+1}{Z} \left(1 - v_0 \, \frac{\mu_p}{T_p} \right) - \, \frac{(p+1)(p+2)}{Z^2} \left(1 - v_0 \, \frac{\mu_p}{T_p} \right) \delta Z \, ,
\end{eqaed}
where the constant first term on the right-hand side originates from the potential of eq.~\eqref{eq:brane_potential} and affects rigid displacements, which behave as
\begin{eqaed}\label{eq:zero-mode_evolution}
	\frac{\delta Z}{Z} \sim \frac{p+1}{2} \left(1 - v_0 \, \frac{\mu_p}{T_p} \right) \left(\frac{t}{Z}\right)^2
\end{eqaed}
for small times $\frac{t}{Z} \ll 1$. On the other hand, for non-zero modes $\delta Z \propto e^{i \mathbf{k}_0 \cdot \mathbf{x} - i \omega_0 t}$ one finds the approximate dispersion relation
\begin{eqaed}\label{eq:deformations_improrer_dispersion}
	\omega^2_0 = \mathbf{k}^2_0 + \frac{(p+1)(p+2)}{Z^2} \left(1 - v_0 \, \frac{\mu_p}{T_p} \right) \, ,
\end{eqaed}
which holds in the same limit, so that $Z$ remains approximately constant. In terms of the proper, red-shifted frequency $\omega_z = \sqrt{g^{tt}} \, \omega_0$ and wave-vector $\mathbf{k}_z = \sqrt{g^{tt}} \, \mathbf{k}_0$ for deformations of $Z$ in $\ads$, eq.~\eqref{eq:deformations_improrer_dispersion} becomes
\begin{eqaed}\label{eq:deformations_dispersion}
	\omega^2_z = \mathbf{k}^2_z + \frac{(p+1)(p+2)}{L^2} \left(1 - v_0 \, \frac{\mu_p}{T_p} \right) \, .
\end{eqaed}
The dispersion relation of eq.~\eqref{eq:deformations_dispersion} displays a potential long-wavelength instability toward deformations of positively charged branes, which can drive them to grow in time, depending on the values of $v_0$ and the charge-to-tension ratio $\frac{\mu_p}{T_p}$. Comparing with eqs.~\eqref{eq:brane_potential} and~\eqref{eq:zero-mode_evolution}, one can see that this instability toward ``corrugation'' is present if and only if the branes are also repelled by the stack.

To conclude our analysis of probe-brane dynamics in the $\adsts$ throat, let us also consider small deformations $\delta \Theta$ in the internal sphere. They evolve according to the linearized equations of motion
\begin{eqaed}\label{eq:sphere_deformations_eom}
	- \, \partial_\mu \partial^\mu \delta \Theta = 0 \, ,
\end{eqaed}
so that these modes are stable at the linearized level.

\subsubsection{Weak gravity from supersymmetry breaking}\label{sec:wgc}

In the ten-dimensional orientifold models, in which the corresponding branes are $\text{D}1$-branes, $v_0 = \sqrt{\frac{3}{2}}$, so that even extremal $\text{D}1$-branes with\footnote{As we have anticipated, verifying the charge-tension equality in the non-supersymmetric case presents some challenges. We shall elaborate upon this issue in Section~\ref{sec:infinity}.} $\mu_p = T_p$ are crucially repelled by the stack, and are driven to exit the throat toward $Z \to 0$. On the other hand $\overline{\text{D}1}$-branes, which have negative $\mu_p$, are always driven toward $Z \to +\infty$, leading to annihilation with the stack. This dynamics is the counterpart of flux tunneling in the probe-brane framework, and eq.~\eqref{eq:v0_parameter} suggests that while the supersymmetry-breaking dilaton potential allows for $\ads$ vacua of this type, it is also the ingredient that allows BPS branes to be repelled. Physically, $\text{D}1$-branes are mutually BPS, but they interact with the $\overline{\text{D}9}$-branes that fill space-time. This resonates with the fact that, as we have argued in Section~\ref{sec:gravity_decay_rate}, the large-$n$ limit ought to suppress instabilities, since in this regime the interaction with $\overline{\text{D}9}$-branes is expected to be negligible~\cite{Angelantonj:1999qg,Angelantonj:2000kh}. Furthermore, the dispersion relation of eq.~\eqref{eq:deformations_dispersion} highlights an additional instability toward long-wavelength deformations of the branes, of the order of the $\ads$ curvature radius. Similarly, in the heterotic model $v_0 = \sqrt{\frac{5}{3}}$, so that negatively charged $\text{NS}5$-branes are also attracted by the stack, while positively charged ones are repelled and unstable toward sufficiently long-wavelength deformations, and the corresponding physical interpretation would involve interactions mediated by the quantum-corrected vacuum energy.

\begin{figure}[!ht]
	\centering
	\includegraphics[width=.75\textwidth]{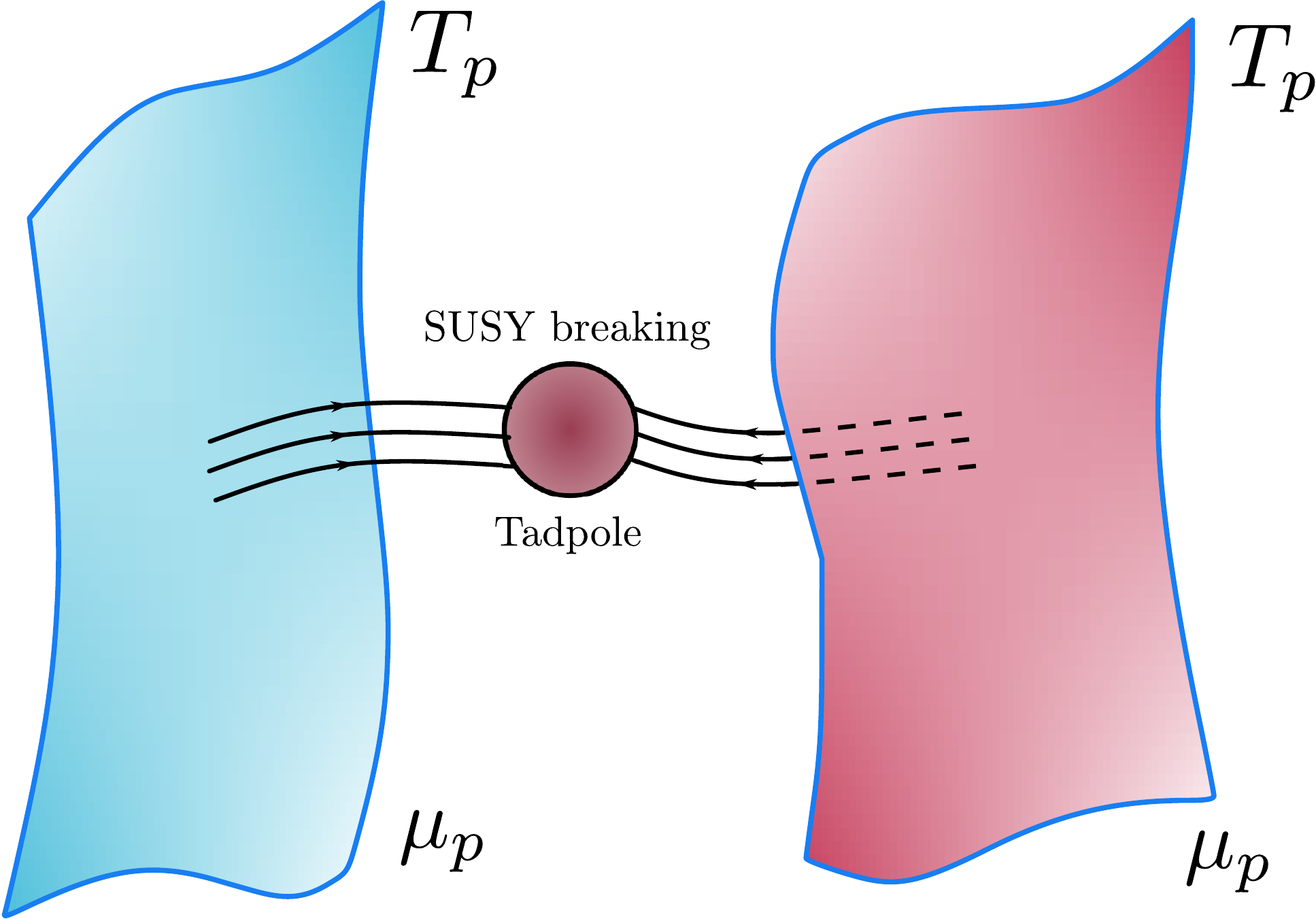}
	\caption{a schematic depiction of the interaction between extremal branes mediated by supersymmetry breaking. Its low-energy manifestation reflects the renormalization of the effective charge-to-tension ratio of eqs.~\eqref{eq:v0_parameter},~\eqref{eq:v0_orientifold} and~\eqref{eq:v0_heterotic}.}
	\label{fig:parallel_branes}
\end{figure}

Moreover, the appearance of $v_0 > 1$ in front of the charge-to-tension ratio $\frac{\mu_p}{T_p}$ is suggestive of a dressed extremality parameter, which can be thought of, \textit{e.g.}, as an effective enhancement of the charge-to-tension ratio due to both dimensional reduction and supersymmetry breaking. This behavior, depicted in fig.~\ref{fig:parallel_branes}, resonates with considerations stemming from the weak gravity conjecture~\cite{ArkaniHamed:2006dz}, since the presence of branes which are (effectively) lighter than their charge would usually imply a decay channel for extremal or near-extremal objects. While non-perturbative instabilities of non-supersymmetric $\ads$ due to brane nucleation have been thoroughly discussed in the literature~\cite{Maldacena:1998uz,Seiberg:1999xz,Ooguri:2016pdq}, we stress that in the present case this phenomenon arises from fundamental branes interacting in the absence of supersymmetry. Therefore, one may attempt to reproduce this result via a string amplitude computation, at least for $\text{D}1$-branes in the orientifold models, but since the relevant annulus amplitude vanishes~\cite{Dudas:2001wd} the leading contribution would involve ``pants'' amplitudes and is considerably more complicated\footnote{The systematics of computations of this type in the bosonic case were developed in~\cite{Bianchi:1988fr}.}. On the other hand, in the non-extremal case one has access to both a probe-brane setting, which involves the gravitational back-reaction of $\text{D}8$-branes, and to a string amplitude computation, and we shall pursue this direction\footnote{Related results in Scherk-Schwartz compactifications have been obtained in~\cite{Bonnefoy:2018tcp, Bonnefoy:2020fwt}.} in Section~\ref{sec:black_branes}.

\begin{center}
	\begin{figure}[!ht]
		\centering
		\includegraphics[width=9cm]{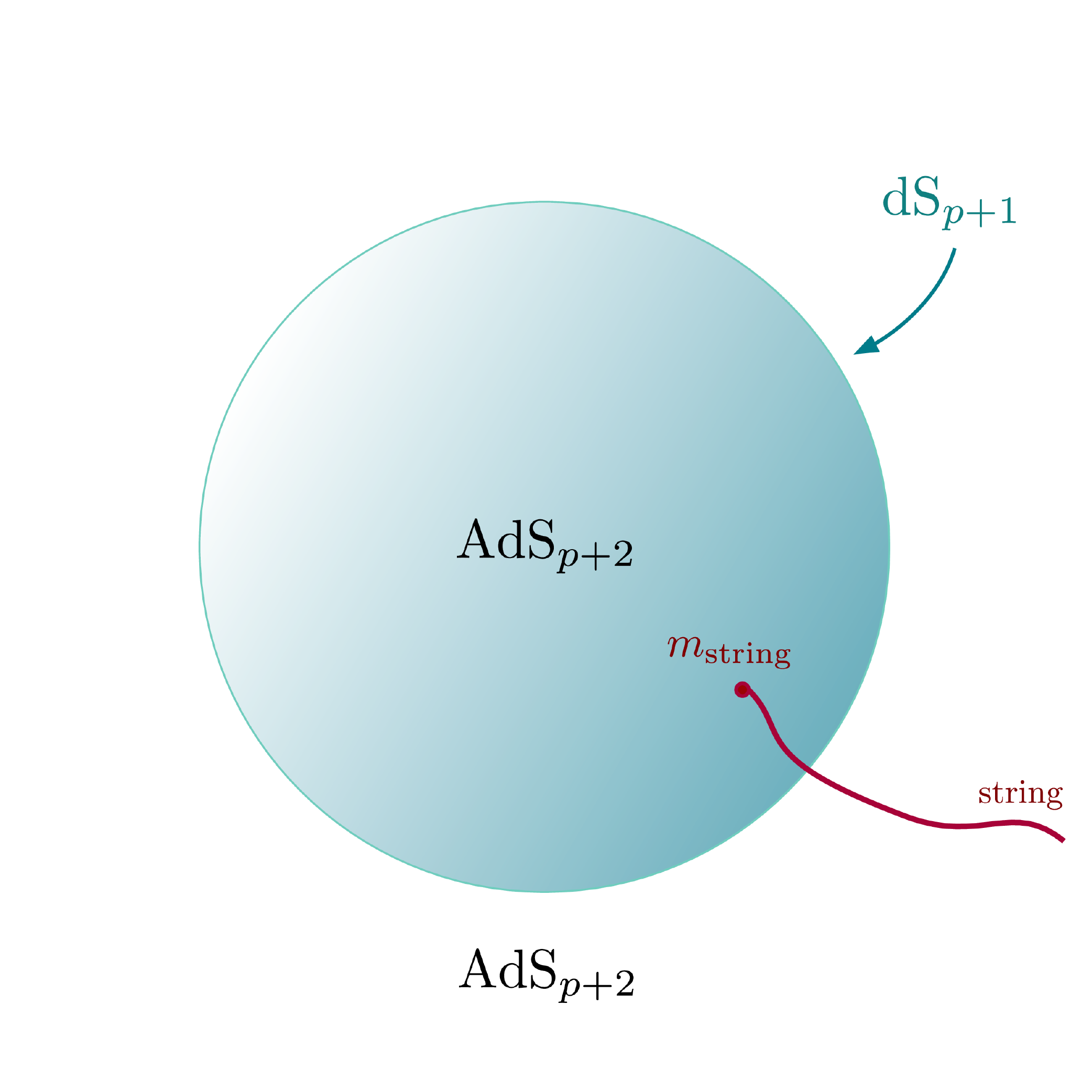}
		\caption{a bubble which interpolates between two ${\rm AdS}_{p+2}$ space-times, hosting a ${\rm dS}_{p+1}$ geometry on its world-volume. Open strings with a single endpoint attached to the bubble wall give rise to massive particles on the world-volume~\cite{Banerjee:2019fzz, Basile:2020mpt, Basile:2020xwi}.\label{Fig:BW}}
	\end{figure}
\end{center}

To conclude, let us comment on how brane nucleation in $\ads$ provides an intriguing construction of $\ds$ configurations. According to the proposal of~\cite{Banerjee:2018qey, Banerjee:2019fzz, Banerjee:2020wix}, a thin-wall bubble nucleating between two $\ads_{p+2}$ space-times hosts a $\ds_{p+1}$ geometry on its wall\footnote{For some earlier works along these lines, see~\cite{Kaloper:1999sm, Shiromizu:1999wj, Vollick:1999uz, Gubser:1999vj, Hawking:2000kj}.}, as schematically depicted in fig.~\ref{Fig:BW}.

The settings that we have described in this section provide a top-down string embedding of this scenario, embodied by D1-branes in the orientifold models and NS5-branes in the heterotic model. The resulting cosmologies appear under parametric control~\cite{Basile:2020mpt, Basile:2020xwi}, and they feature small cosmological constants relative to the brane-world Planck scale.

As a final comment, let us remark that the nucleation of bubbles of nothing~\cite{Witten:1981gj} offers another enticing possibility to construct $\ds$ configurations~\cite{Dibitetto:2020csn}. While, to our knowledge, realizations of this type of scenario in string theory have been investigated breaking supersymmetry in lower-dimensional settings~\cite{Horowitz:2007pr}\footnote{Some lower-dimensional toy models offer flux landscapes where more explicit results can be obtained~\cite{BlancoPillado:2010df, Brown:2010mf, Brown:2011gt}.}, recent results indicate that within the relevant context the nucleation of bubbles of nothing is quite generic~\cite{GarciaEtxebarria:2020xsr}.

\subsection{Gravitational back-reaction}\label{sec:backreaction}

In this section we study the background geometry sourced by a stack of branes in the class of low-energy effective theories that we have described in Section~\ref{Section2}. The dilaton potential brings along considerable challenges in this respect, both conceptual and technical. To begin with, there is no maximally symmetric vacuum that could act as a background, and thus in the presence of branes there is no asymptotic infinity of this type\footnote{Even if one were to envision a pathological Minkowski solution with ``$\phi = -\infty$'' as a degenerate background (for instance, by introducing a cut-off), no asymptotically flat solution with $\phi \; \to \; -\infty$ can be found.}. We find, instead, that the geometry away from the branes ``pinches off'' at a finite geodesic distance, and exhibits a curvature singularity where $\phi \to +\infty$. This resonates with the findings of~\cite{Dudas:2000ff}, and indeed we do reconstruct the solutions therein in the case $p = 8$. These results suggest that, due to their interactions with the dilaton potential, branes cannot be described as isolated objects in these models, reflecting the probe-brane analysis of Section~\ref{sec:probe_branes}. Consequently, identifying a sensible background string coupling or sensible asymptotic charges, such as the brane tension, appears considerably more difficult with respect to the supersymmetric case.

Despite these challenges, one can gain some insight studying the asymptotic geometry near the branes, where an $\adsts$ throat develops, and near the outer singularity, where the geometry pinches off. In Section~\ref{sec:core} we shall argue that the $\adsts$ solutions discussed in Section~\ref{Section2} can arise as near-horizon ``cores'' of the full geometry, investigating an attractor-like behavior of radial perturbations which is characteristic of extremal objects and arises after a partial fine-tuning, reminiscent of the BPS conditions on asymptotic charges in supersymmetric cases. This feature is reflected by the presence of free parameters in the asymptotic geometry away from the branes, which we construct in Section~\ref{sec:infinity}.

\subsubsection{Reduced dynamical system: extremal case}\label{sec:extremal_reduced_dynamical_system}

Let us begin imposing $SO(1,p) \times SO(q)$ symmetry, so that the metric is characterized by two dynamical functions $v(r) \, , b(r)$ of a transverse radial coordinate $r$. Specifically, without loss of generality we shall consider the ansatz
\begin{eqaed}\label{eq:brane_full_ansatz}
	ds^2 & = e^{\frac{2}{p+1}v - \frac{2q}{p}b} \, dx^2_{1,p} + e^{2v-\frac{2q}{p}b} \, dr^2 +e^{2b} \, R^2_0 \, d\Omega_q^2 \, , \\
	\phi & = \phi(r) \, , \\
	H_{p+2} & = \frac{n}{f(\phi)(R_0 \, e^b)^q} \, \Vol_{p+2} \, , \qquad \Vol_{p+2} = e^{2v - \frac{q}{p}(p+2)b} \, d^{p+1} x\,\wedge\, dr \, ,
\end{eqaed}
where $R_0$ is an arbitrary reference radius and the form field automatically solves its field equations. This gauge choice simplifies the equations of motion, which can be recast in terms of a constrained Toda-like system~\cite{Klebanov:1998yya,Dudas:2000sn, Antonelli:2020kik}. Indeed, substituting the ansatz of eq.~\eqref{eq:brane_full_ansatz} in the field equations and taking suitable linear combinations, the resulting system can be derived by the ``reduced'' action
\begin{importantbox}
	\begin{eqaed}\label{eq:toda_action}
		S_{\text{red}} = \int dr \left[ \frac{4}{D-2} \left(\phi'\right)^2 - \frac{p}{p+1} \left(v'\right)^2 + \frac{q(D-2)}{p} \left(b'\right)^2 - \, U \right] \, ,
	\end{eqaed}
\end{importantbox}
where the potential is given by
\begin{eqaed}\label{eq:toda_potential}
	U = - \,  T \, e^{\gamma \phi + 2v - \frac{2q}{p}b} - \frac{n^2}{2R_0^{2q}} \, e^{-\alpha\phi + 2v-\frac{2q(p+1)}{p}b} + \frac{q(q-1)}{R_0^2} \, e^{2v-\frac{2(D-2)}{p}b} \, ,
\end{eqaed}
and the equations of motion are supplemented by the zero-energy constraint
\begin{eqaed}\label{eq:toda_constraint}
	\frac{4}{D-2} \left(\phi'\right)^2 - \, \frac{p}{p+1} \left(v'\right)^2 + \frac{q(D-2)}{p} \left(b' \right)^2 + U = 0 \, .
\end{eqaed}

\subsubsection{\texorpdfstring{$\ads \times \ess$}{AdS x S} throat as a near-horizon geometry}\label{sec:core}

Let us now apply the results in the preceding section to recast the $\adsts$ solutions discussed in Section~\ref{Section2} as a near-horizon limit of the geometry described by eqs.~\eqref{eq:toda_action} and~\eqref{eq:toda_constraint}. To begin with, one can verify that the $\adsts$ solution now takes the form\footnote{Up to the sign of $r$ and rescalings of $R_0$, this realization of $\adsts$ with given $L$ and $R$ is unique.}
\begin{eqaed}\label{eq:ads_s_toda}
	\phi = \phi_0 \, , \qquad
	e^v = \frac{L}{p+1} \, \left(\frac{R}{R_0}\right)^{\frac{q}{p}} \, \frac{1}{-r} \, , \qquad
	e^b = \frac{R}{R_0} \, ,
\end{eqaed}
where we have chosen negative values $r < 0$. This choice places the core at $r \; \to \; -\infty$, with the horizon infinitely far away, while the outer singularity lies either at some finite $r = r_0$ or emerges as\footnote{In either case we shall find that the geodesic distance is finite.} $r \; \to \; +\infty$. The metric of eq.~\eqref{eq:brane_full_ansatz} can then be recast as $\adsts$ in Poincar\'e coordinates rescaling $x$ by a constant and substituting
\begin{eqaed}\label{eq:toda_ads_s_diffeo}
	r \; \mapsto \; - \, \frac{z^{p+1}}{p+1} \, .
\end{eqaed}
In supersymmetric cases, infinitely long $\ads$ throats behave as attractors going toward the horizon from infinity, under the condition on asymptotic parameters that specifies extremality. Therefore we proceed by analogy, studying linearized radial perturbations $\delta \phi \, , \delta v \, , \delta b$ around eq.~\eqref{eq:ads_s_toda} and comparing them to cases where the full geometry is known. To this end, notice that the potential of eq.~\eqref{eq:toda_potential} is factorized,
\begin{eqaed}\label{eq:toda_potential_factorized}
	U(\phi, v, b) \equiv e^{2v} \, \widehat{U}(\phi, b) \, ,
\end{eqaed}
so that perturbations $\delta v$ of $v$ do not mix with perturbations $\delta \phi \, , \delta b$ of $\phi$ and $b$ at the linearized level. In addition, since the background values of $\phi$ and $b$ are constant in $r$, the constraint obtained linearizing eq.~\eqref{eq:toda_constraint} involves only $v$, and reads
\begin{eqaed}\label{eq:toda_linearized_constraint}
	\frac{2p}{p+1} \, v' \, \delta v' = \partial_v U \bigg|_{\adsts} \delta v = 2 \, U \bigg|_{\adsts} \delta v \,  = \frac{2p}{(p+1) r^2} \, \delta v \, ,
\end{eqaed}
so that
\begin{eqaed}\label{eq:toda_v_mode}
	\delta v \sim \text{const.} \times (-r)^{-1} \, .
\end{eqaed}
Thus, the constraint of eq.~\eqref{eq:toda_v_mode} retains only one mode $\sim (-r)^{\lambda_0}$ with respect to the linearized equation of motion for $\delta v$, with exponent $\lambda_0 = -1$.

On the other hand, $\phi$ and $b$ perturbations can be studied using the canonically normalized fields
\begin{eqaed}\label{eq:chi_field}
	\chi \equiv \left(\sqrt{\frac{8}{D-2}} \, \delta \phi \ , \, \sqrt{\frac{2q(D-2)}{p}} \, \delta b \right) \, ,
\end{eqaed}
in terms of which one finds
\begin{eqaed}\label{eq:chi_system}
	\chi'' \sim - \, \frac{1}{r^2} \, H_0 \, \chi \, ,
\end{eqaed}
where the Hessian
\begin{eqaed}\label{eq:toda_hessian}
	H_{ab} \equiv \frac{\partial^2 U}{\partial \chi_a \partial \chi_b}\bigg|_{\adsts} \equiv \frac{1}{r^2} \left(H_0\right)_{ab} \, , \qquad \left(H_0\right)_{ab} = \mathcal{O}\!\left(r^0\right) \, .
\end{eqaed}
The substitution $t = \log(-r)$ then results in the autonomous system
\begin{eqaed}\label{eq:toda_core_autonomous}
	\left(\frac{d^2}{dt^2} - \frac{d}{dt}\right) \chi = - \, H_0 \, \chi \, ,
\end{eqaed}
so that the modes scale as $\chi \propto (-r)^{\lambda_i}$, where the $\lambda_i$ are the eigenvalues of the block matrix
\begin{eqaed}\label{eq:block_matrix}
	\mqty( 1 & -H_0\\1 & 0) \, .
\end{eqaed}
In turn, these are given by
\begin{eqaed}\label{eq:toda_ads_s_eigenvalues}
	\lambda^{(\pm)}_{1 \, , \, 2} & = \frac{1 \pm \sqrt{1 - 4 \, h_{1 \, , \, 2}}}{2} \, , \\
	h_{1 \, , \, 2} & \equiv \frac{\tr(H_0) \pm \sqrt{\tr(H_0) - 4 \det(H_0)}}{2} \, ,
\end{eqaed}
where the trace and determinant of $H_0$ are given by
\begin{eqaed}\label{eq:tr_det_H0}
	\tr(H_0) & = - \, \frac{\alpha  \left(\gamma \, (\alpha +\gamma ) (D-2)^2-16\right) + 16 \, \gamma \left(p+1\right) \left(q-1\right)}{8 \, (p+1) \, (\left(q-1\right)\gamma-\alpha)} \, , \\
	\det(H_0) & = \frac{\alpha \, \gamma \, (D-2)^2 \left(\left(p+1\right)\gamma+\alpha \right)}{4 \, (p+1)^2 \, (\left(q-1\right)\gamma-\alpha)} \, .
\end{eqaed}
In the case of the orientifold models, one obtains the eigenvalues
\begin{eqaed}\label{eq:orientifold_eigenvalues}
	\frac{1 \pm \sqrt{13}}{2} \, , \qquad \frac{1 \pm \sqrt{5}}{2} \, ,
\end{eqaed}
while in the heterotic model one obtains the eigenvalues
\begin{eqaed}\label{eq:heterotic_eigenvalues}
	\pm \, 2 \,\sqrt{\frac{2}{3}} \, , \qquad 1 \pm 2 \, \sqrt{\frac{2}{3}} \, .
\end{eqaed}
All in all, in both cases one finds three negative eigenvalues and two positive ones, signaling the presence of three attractive directions as $r \; \to \; -\infty$. The remaining unstable modes should physically correspond to deformations that break extremality, resulting in a truncation of the $\adsts$ throat and in the emergence of an event horizon at a finite distance, and it should be possible to remove them with a suitable tuning of the boundary conditions at the outer singularity. In the following section we shall argue for this interpretation of unstable modes in the throat.

\subsubsection{Comparison with known solutions}\label{sec:comparison}

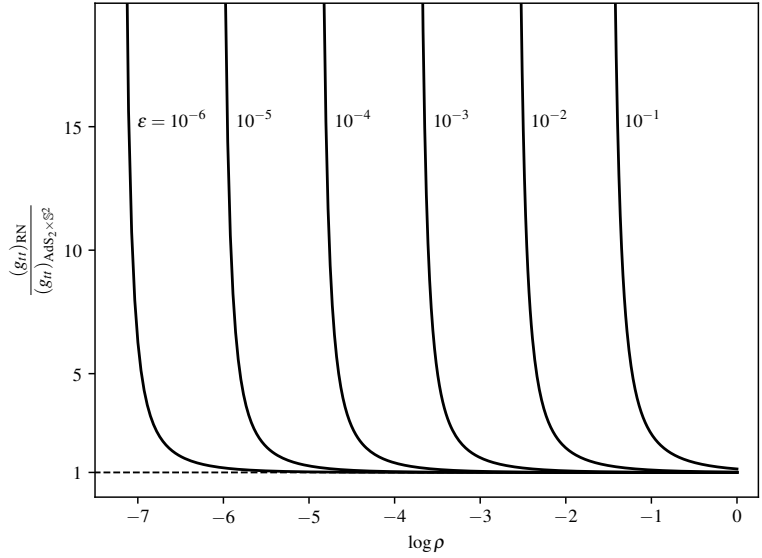
\begin{figure}[!ht]
	\centering
	\scalebox{0.7}{\input{throatplot.pgf}}
	\caption{a plot of the ratio of the Reissner-Nordstr\"om red-shift factor to the one of the corresponding $\mathrm{AdS}_2 \times \ess^2$, for various values of the extremality parameter $\epsilon \equiv \frac{m}{e}-1$. Only values outside of the event horizon are depicted. As extremality is approached, the horizon recedes to infinity and the geometry develops an approximate $\adsts$ throat, marked by $(g_{tt})_{\text{RN}} \approx (g_{tt})_{\ads_2 \times \ess^2}$, whose length in units of $\log \rho$ grows asymptotically linearly in $-\log \epsilon$.}
	\label{fig:throat_RN}
\end{figure}

In order to highlight the physical origin of the unstable modes, let us consider the Reissner-Nordstr\"om black hole in four dimensions, whose metric in isotropic coordinates takes the form
\begin{eqaed}\label{eq:RN_metric}
	ds^2_{\text{RN}} = - \, \frac{g(\rho)^2}{f(\rho)^2} \, dt^2 + f(\rho)^2 \left(d\rho^2 + \rho^2 \, d\Omega^2_2\right) \, ,
\end{eqaed}
where
\begin{eqaed}\label{eq:RN_f_g}
	f(\rho) & \equiv 1 + \frac{m}{\rho} + \frac{m^2}{4\rho^2} - \frac{e^2}{4\rho^2} \, , \\
	g(\rho) & \equiv 1 - \frac{m^2}{4\rho^2} + \frac{e^2}{4\rho^2} \, .
\end{eqaed}
The extremal solution, for which $m = e$, develops an infinitely long $\ads_2 \times \ess^2$ throat in the near-horizon limit $\rho \; \to \; 0$, and radial perturbations of the type
\begin{eqaed}\label{eq:ads2xs2_pert}
	ds^2_{\text{pert}} = - \, \frac{4\rho^2}{m^2} \, e^{2 \, \delta a(\rho)} \, dt^2 + \frac{m^2}{4\rho^2} \, e^{2 \, \delta b(\rho)} \left( d\rho^2 + \rho^2 \, d\Omega^2_2\right)
\end{eqaed}
solve the linearized equations of motion with power-law modes $\sim \rho^{\lambda_{\text{RN}}}$, with eigenvalues
\begin{eqaed}\label{eq:}
	\lambda_{\text{RN}} = -2 \, , \, 1 \, , \, 0 \, .
\end{eqaed}
The zero-mode reflects invariance under shifts of $\delta a$, while the unstable mode reflects a breaking of extremality. Indeed, writing $m \equiv e \, (1 + \epsilon)$ the $\frac{\rho}{m} \ll 1\, , \epsilon \ll 1$ asymptotics of the red-shift $g_{tt}$ take the schematic form
\begin{eqaed}\label{eq:RN_redshift}
	\frac{(g_{tt})_{\text{RN}}}{(g_{tt})_{\ads_2 \times \ess^2}} \sim \text{regular} + \epsilon \left( - \frac{1}{\rho^2} + \frac{3}{m \rho} + \text{regular} \right) + o(\epsilon) \, ,
\end{eqaed}
so that for $\epsilon = 0$ only a regular series in positive powers of $\rho$ remains. Geometrically, near extremality an approximate $\adsts$ throat exists for some finite length, after which it is truncated by a singularity corresponding to the event horizon. As $\epsilon$ decreases, this horizon recedes and the throat lengthens, with the length in $\log \rho$ growing as $-\log \epsilon$. This is highlighted numerically in the plot of fig.~\ref{fig:throat_RN}.

A similar analysis for BPS $\text{D}3$-branes in type IIB supergravity~\cite{Horowitz:1991cd} yields the eigenvalues $-8 \, , \, -4 \, , \, 4 \, , \, 0 \, , 0$, suggesting again that breaking extremality generates unstable directions, and that a fine-tuning at infinity removes them leaving only the attractive ones. Notice that the zero-modes correspond to constant rescalings of $x^\mu$, which is pure gauge, and to shifts of the asymptotic value of the dilaton.

\subsubsection{The pinch-off singularity}\label{sec:infinity}

Let us now proceed to address the asymptotic geometry away from the core. Since the dynamical system at hand is not integrable in general\footnote{In the supersymmetric case the contribution arising from the dilaton tadpole is absent, and the resulting system is integrable. Moreover, for $p = 8 \, , q = 0$ the system is also integrable, since only the dilaton tadpole contributes.}, we lack a complete solution of the equations of motion stemming from eq.~\eqref{eq:toda_action}, and therefore we shall assume that the dilaton potential overwhelms the other terms of eq.~\eqref{eq:toda_potential} for large (positive) $r$, to then verify it \textit{a posteriori}. In this fashion, one can identify the asymptotic equations of motion
\begin{eqaed}\label{eq:asymptotic_toda_system}
	\phi'' & \sim \frac{\gamma(D-2)}{8} \, T \, e^{\gamma \phi + 2v - \frac{2q}{p}b} \, , \\
	v'' & \sim - \, \frac{p+1}{p} \, T \, e^{\gamma \phi + 2v - \frac{2q}{p}b} \, , \\
	b'' & \sim - \, \frac{1}{D-2} \, T \, e^{\gamma \phi + 2v - \frac{2q}{p}b} \, ,
\end{eqaed}
whose solutions
\begin{eqaed}\label{eq:asymptotic_toda_solutions}
	\phi & \sim \frac{\gamma(D-2)}{8} \, y + \phi_1 r + \phi_0 \, , \\
	v & \sim - \, \frac{p+1}{p} \, y + v_1 r + v_0 \, , \\
	b & \sim - \, \frac{1}{D-2} \, y + b_1 r + b_0
\end{eqaed}
are parametrized by the constants $\phi_{1 , 0} \, , v_{1 , 0} \, , b_{1 , 0}$ and a function $y(r)$ which is not asymptotically linear (without loss of generality, up to shifts in $\phi_1$, $v_1$, $b_1$). Rescaling $x$ and redefining $R_0$ in eq.~\eqref{eq:brane_full_ansatz} one can set \textit{e.g.} $b_0 = v_0 = 0$. The equations of motion and the constraint then reduce to
\begin{eqaed}\label{eq:toda_y_eom}
	y'' \sim \widehat{T} \, e^{ \Omega \, y + L \, r } \, , \qquad
	\frac{1}{2} \, \Omega \left(y'\right)^2 + L \, y' \sim \widehat{T} \, e^{ \Omega \, y + L \, r } - M \, ,
\end{eqaed}
where\footnote{Notice that $\Omega = \frac{D-2}{8} \left(\gamma^2 - \gamma_c^2 \right)$, where the critical value $\gamma_c$ defined in~\cite{Basile:2018irz} marks the onset of the ``climbing'' phenomenon described in~\cite{Dudas:2010gi,Sagnotti:2013ica, Condeescu:2013gaa, Fre:2013vza} use different notations.}
\begin{eqaed}\label{eq:L_M_That}
	\widehat{T} & \equiv T \, e^{\gamma \phi_0 + 2v_0 - \frac{2q}{p} \, b_0} \, , \qquad
	\Omega \equiv \frac{D-2}{8} \, \gamma^2 - \, \frac{2(D-1)}{D-2} \, , \\
	L & \equiv \gamma \, \phi_1 + 2 \, v_1 - \frac{2q}{p} \, b_1 \, , \qquad
	M \equiv \frac{4}{D-2} \, \phi_1^2 - \frac{p}{p+1} \, v_1^2 + \frac{q(D-2)}{p} \, b_1^2 \, .
\end{eqaed}
The two additional exponentials in eq.~\eqref{eq:toda_potential}, associated to flux and internal curvature contributions, are both asymptotically $\sim \exp\left( \Omega_{n,c} \, y + L_{n,c} \, r \right)$, with corresponding constant coefficients $\Omega_{n,c}$ and $L_{n,c}$. Thus, if $y$ grows super-linearly the differences $\Omega - \Omega_{n,c}$ determine whether the dilaton potential dominates the asymptotics. On the other hand, if $y$ is sub-linear the dominant balance is controlled by the differences $L - L_{n,c}$. In the ensuing discussion we shall consider the former case\footnote{The sub-linear case is controlled by the parameters $\phi_1 \, , v_1 \, , b_1$, which can be tuned as long as the constraint is satisfied. In particular, the differences $L - L_{n,c}$ do not contain $v_1$.}, since it is consistent with earlier results~\cite{Dudas:2000ff}, and, in order to study the system in eq.~\eqref{eq:toda_y_eom}, it is convenient to distinguish the two cases $\Omega = 0$ and $\Omega \neq 0$. Moreover, we have convinced ourselves that the tadpole-dominated system of eq.~\eqref{eq:asymptotic_toda_system} is actually integrable, and its solutions behave indeed in this fashion. As a final remark, us observe that, on account of eq.~\eqref{eq:asymptotic_toda_solutions}, the warp exponents of the longitudinal sector $dx_{p+1}^2$ and the sphere sector $R_0 \, d\Omega_q^2$ are asymptotically equal,
\begin{eqaed}\label{eq:warping_asymptotic_equality}
	\frac{2}{p+1} v - \frac{2q}{p} b \sim 2b \, .
\end{eqaed}
This is to be expected, since if one takes a solution with $q=0$ and replaces
\begin{eqaed}\label{eq:longitudinal_sphere_replacement}
	dx_{p+1}^2 \; \to \; dx_{p'+1}^2 + R_0^2 \, d\Omega_{p-p'}^2
\end{eqaed}
for some $p' < p$ and large $R_0$, and then makes use of the freedom to rescale $R_0$ shifting $b$ by a constant (which does not affect the leading asymptotics), one obtains another asymptotic solution with lower $p' < p$, whose warp factors are both equal to the one of the original solution.

\subsubsection{Pinch-off in the orientifold models}\label{sec:orientifold_pinch-off}

In the orientifold models $\Omega = 0$, since the exponent $\gamma = \gamma_c$ attains its ``critical'' value~\cite{Basile:2018irz} in the sense of~\cite{Dudas:2010gi}. The system in eq.~\eqref{eq:toda_y_eom} then yields
\begin{eqaed}\label{eq:omega0_sol}
	y & \sim \frac{\widehat{T}}{L^2} \, e^{L \, r} \, , \quad  M=0 \, , \qquad L > 0 \, , \\
	y & \sim \frac{\widehat{T}}{2} \, r^2 \, ,\quad M = \widehat{T} \,, \qquad L = 0 \, .
\end{eqaed}
These conditions are compatible, since the quadratic form $M$ has signature $(+,-,+)$ and thus the equation $M = \widehat{T} > 0$ defines a one-sheeted hyperboloid that intersects any plane, including $\{L = 0 \}$. The same is also true for the cone $\{M = 0\}$.

In both solutions the singularity arises at finite geodesic distance
\begin{eqaed}\label{eq:finite_distance_singularity_orientifold}
	R_c \equiv \int^{\infty} dr \, e^{v - \frac{q}{p} b} < \infty \, ,
\end{eqaed}
since at large $r$ the warp factor
\begin{eqaed}\label{eq:geodesic_radius_warping_orientifold}
	v - \frac{q}{p} \, b \sim - \, \frac{D-1}{D-2} \, y \, .
\end{eqaed}
In the limiting case $L = 0$, where the solution is quadratic in $r$, due to the discussion in the preceding section this asymptotic behavior is consistent, up to the replacement of $dx_9^2$ with $dx_2^2 + R_0^2 \, d\Omega_7^2$, with the full solution found in~\cite{Dudas:2000ff}, whose singular structure is also reconstructed in our analysis for $p = 8$, $q = 0$, $L = 0$. The existence of a closed-form solution in this case rests on the integrability of the corresponding Toda-like system, since neither the flux nor the internal curvature are present.

\subsubsection{Pinch-off in the heterotic model}\label{sec:heterotic_pinch-off}

\begin{figure}[!ht]
	\centering
    \scalebox{0.9}{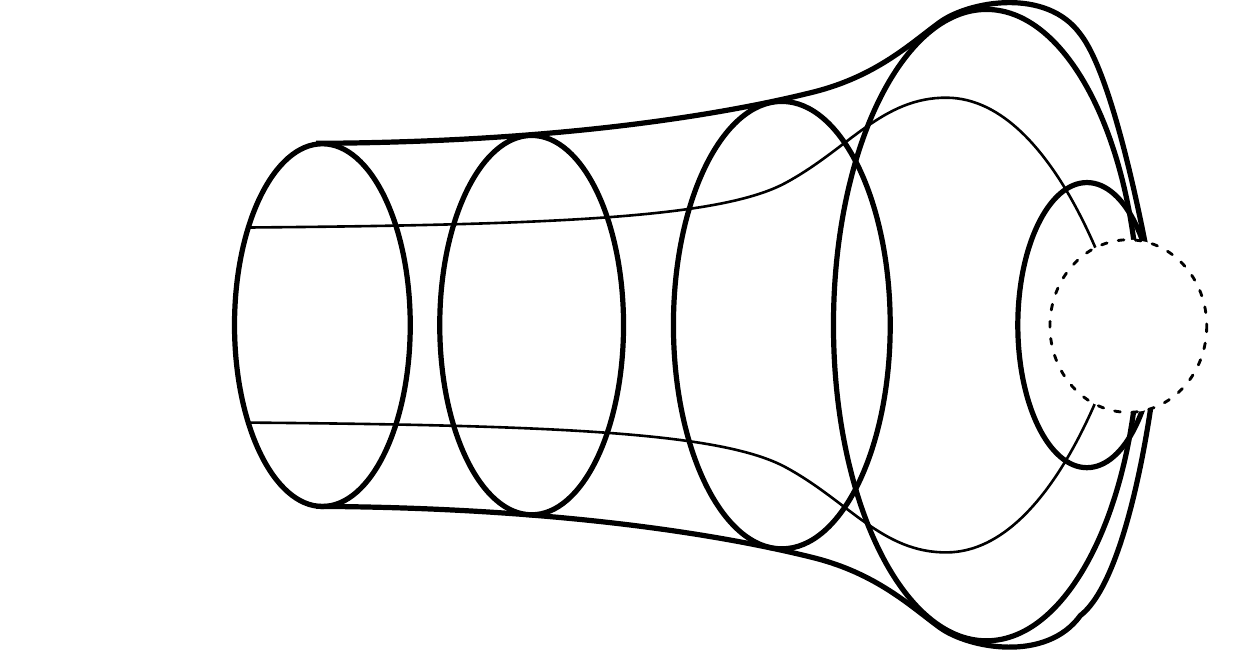}
	\caption{a schematic depiction of the expected structure of the complete geometry sourced by the branes, displaying only geodesic radial distance and the $\ess^q$ radius. The geometry interpolates between the $\adsts$ throat and the pinch-off singularity (dashed circle).}
	\label{fig:geometry}
\end{figure}

In the heterotic model $\Omega = 4$, and therefore one can define
\begin{eqaed}\label{eq:Y_def}
	Y \equiv y + \frac{L}{\Omega} \, r\,,
\end{eqaed}
removing the $L \, r$ terms from the equations. One is then left with the first-order equation
\begin{eqaed}\label{eq:Y_eq}
	\frac{1}{2} \, {Y'}^2 - \, \frac{\widehat{T}}{\Omega} \, e^{\Omega \, Y} = E \, ,
\end{eqaed}
which implies the second-order equation of motion, where the ``energy''
\begin{eqaed}\label{eq:energy_toda_Y}
	E \equiv \frac{M}{2\Omega} - \frac{L^2}{2\Omega^3} \, .
\end{eqaed}
The solutions of eq.~\eqref{eq:Y_eq} depend on the sign of $E$, and one can verify that, if $r \; \to \; +\infty$, $Y$ grows at most linearly. On the other hand, super-linear solutions develop a singularity at a finite radius $r = r_0$, and they all take the form
\begin{eqaed}\label{eq:Y_log_blowup}
	Y \sim - \, \frac{2}{\Omega} \log \left(r_0 - r \right) \, ,
\end{eqaed}
which is actually the exact solution of eq.~\eqref{eq:Y_eq} for $E = 0$. The geodesic distance to the singularity
\begin{eqaed}\label{eq:finite_distance_singularity_het}
	R_c \equiv \int^{r_0} dr \, e^{v - \frac{q}{p} b} < \infty
\end{eqaed}
is again finite, since from eqs.~\eqref{eq:geodesic_radius_warping_orientifold} and~\eqref{eq:Y_log_blowup}
\begin{eqaed}\label{eq:geodesic_radius_warping_heterotic}
	v - \frac{q}{p} \, b \sim \frac{2}{\Omega} \, \frac{D-1}{D-2} \, \log\left( r_0-r \right) = \frac{9}{16} \, \log\left( r_0-r \right) \, .
\end{eqaed}
In terms of the geodesic radial coordinate $\rho_c < R_c$, the asymptotics\footnote{More precisely, the asymptotics for the metric in eq.~\eqref{eq:het_DM_asymptotics} refer to the exponents in the warp factors, which are related to $v$ and $b$. Subleading terms could lead to additional prefactors in the metric.} are
\begin{eqaed}\label{eq:het_DM_asymptotics}
	\phi & \sim - \, \frac{4}{5} \, \log\left( R_c - \rho_c \right) \, , \qquad
	ds^2 \sim \left(R_c - \rho_c \right)^{\frac{2}{25}} \left( dx_6^2 + R_0^2 \, d\Omega_3^2 \right) + d\rho^2 \, .
\end{eqaed}
While these results are at most qualitative in this asymptotic region, since curvature corrections and string loop corrections are expected to be relevant, they again hint at a physical picture whereby space-time pinches off at finite distance in the presence of (exponential) dilaton potentals, while branes dictate the symmetries of the geometry, as depicted in fig.~\ref{fig:geometry}. In this context, the nine-dimensional Dudas-Mourad solutions correspond to (necessarily uncharged) $8$-branes\footnote{In particular, on account of the analysis that we described in the preceding section, it is reasonable to expect that in the orientifold models the Dudas-Mourad solution corresponds to $\text{D}8$-branes.}. This picture highlights the difficulties encountered in defining tension and flux as asymptotic charges, but analogous quantities might appear as parameters in the sub-leading portion of the solution, of which there are indeed two. They ought to be matched with the $\adsts$ core, and we shall elaborate on this issue in Section~\ref{sec:black_branes}. For the time being, let us recall that the results of~\cite{Dudas:2001wd}, based on string perturbation theory, suggest that at least the $\text{D}1$-branes that we consider are extremal, albeit the presence of dilaton tadpoles makes this lesson less clear.

As a final comment, let us add that cosmological counterparts, if any, of these solutions, whose behavior appears milder, can be expected to play a rôle when the dynamics of pinch-off singularities are taken into account, and they could be connected to the hints of spontaneous compactification discussed in Section~\ref{Section3}. Indeed, as already stressed in~\cite{Basile:2018irz}, the general lesson is that non-supersymmetric settings are dynamically driven toward time-dependent configurations, and this additional potential instability might be mitigated to an arbitrarily large extent studying the dynamics deep inside the $\ads$ throat, the deeper the more any effect of an asymptotic collapse is red-shifted.

\subsubsection{Black branes: back-reaction}\label{sec:black_branes}

Let us conclude our discussion on back-reactions extending the machinery that we have developed in Section~\ref{sec:extremal_reduced_dynamical_system} to the case of non-extremal branes. Including a ``blackening'' factor entails the presence of an additional dynamical function, and thus after gauge-fixing radial diffeomorphisms one is left with four dynamical functions (including the dilaton). Specifically, in order to arrive at a generalization of the Toda-like system of eqs.~\eqref{eq:toda_action} and~\eqref{eq:toda_constraint}, the correct ansatz takes the form
\begin{eqaed}\label{eq:brane_separated_ansatz}
	ds^2_D & = e^{2(A(r)+p C(r) + q B(r))} \, dr^2 - \, e^{2 A(r)} \, dt^2 + e^{2 C(r)} \, d\mathbf{x}_p^2 + e^{2B(r)} \, R^2_0 \, d\Omega_q^2 \, , \\
	\phi & = \phi(r) \, , \\
	H_{p+2} & = \frac{n}{f(\phi)(R_0 \, e^B)^q} \, \Vol_{p+2} \, , \qquad \Vol_{p+2} = e^{2A + 2p C + q B} \, dr \, \wedge \, dt \wedge d^{p} x \, .
\end{eqaed}
Then, one can verify that, after removing the mixing terms via the substitution
\begin{eqaed}\label{eq:non-extremal_toda_sub}
	A = (1-q) \, a - \, \frac{p}{q} \, c\, , \qquad
	B = a + b - \, \frac{p}{q} \, c \, , \qquad
	C = c \, ,
\end{eqaed}
the resulting reduced equations of motion stem from the Toda-like action
\begin{importantbox}
	\begin{eqaed}\label{eq:non-extremal_toda_action}
		S_{\text{red}} = \int dr & \left[ \frac{4}{D-2} \left(\phi'\right)^2 + q(q-1) \left(a'^2 - b'^2 \right) + \frac{p(D-2)}{q} \left(c'\right)^2 - \, U \right]
	\end{eqaed}
\end{importantbox}
where the effective potential now reads
\begin{eqaed}\label{eq:non-extremal_toda_potential}
	U & = -\,  T \, e^{\gamma \phi + 2a + 2q b - \frac{2p}{q}c} - \frac{n^2}{2R_0^{2q}} \, e^{-\alpha\phi - 2(q-1)a + \frac{2p(q-1)}{q}c} + \frac{q(q-1)}{R_0^2} \, e^{2(q-1)b} \, ,
\end{eqaed}
and the equations of motion are to be supplemented by the zero-energy constraint
\begin{eqaed}\label{eq:non-extremal_toda_constraint}
	\frac{4}{D-2} \left(\phi'\right)^2 + q(q-1) \left(a'\right)^2 - \, q(q-1)\left(b'\right)^2 + \frac{p(D-2)}{q} \left(c'\right)^2 + U = 0 \, .
\end{eqaed}
Changing variables in eq.~\ref{eq:brane_separated_ansatz} in order to match the ansatz of eq.~\ref{eq:brane_full_ansatz}, and substituting the resulting expressions in eqs.~\eqref{eq:non-extremal_toda_action} and~\eqref{eq:non-extremal_toda_constraint}, one recovers the Toda-like system that describes extremal branes. Hence, the generalized system that we have derived can in principle describe the back-reaction of non-extremal branes, which ought to exhibit Rindler geometries in the near-horizon limit. On the other hand, one can verify that the tadpole-dominated asymptotic system reproduces the behavior of eq.~\eqref{eq:asymptotic_toda_solutions}, thus suggesting that the pinch-off singularities described in the preceding sections are generic and do not depend on the gravitational imprint of the sources that are present in space-time, rather only on the residual symmetry left unbroken.

\subsection{Black branes: dynamics}\label{sec:non_extremal_probes}

Let us now extend the considerations of Section~\ref{sec:probe_branes} to the non-extremal case, studying potentials between non-extremal brane stacks and between stacks of different types and dimensions. While probe-brane computations are rather simple to perform using the back-reacted geometries that we described in the preceding section, they pertain to regimes in which the number of $p$-branes $N_p$ in one stack is much larger than the number of $q$-branes $N_q$ in the other stack. However, with respect to the extremal case, the leading contribution to the string amplitude for brane scattering corresponds to the annulus, which is non-vanishing and does not entail the complications due to orientifold projections, anti-branes and Riemann surfaces of higher Euler characteristic. This setting therefore offers the opportunity to compare probe computations with string amplitude computations. Specifically, we shall consider the uncharged $8$-branes in the orientifold models, since their back-reacted geometry is described by the static Dudas-Mourad solution\footnote{The generalization to non-extremal $p$-branes of different dimensions would entail solving non-integrable systems, whose correct boundary conditions are not well-understood hitherto. Moreover, a reliable probe-brane regime would exclude the pinch-off asymptotic region, thereby requiring numerical computations.}~\cite{Dudas:2000ff} that we have described in Section~\ref{Section2}. Furthermore, the other globally known back-reacted geometry in this setting pertains to extremal $\text{D}1$-branes, and $8$-branes are the only probes (of different dimension) whose potential can be reliably computed in this case, since they can wrap the internal $\ess^7$ in the near-horizon $\ads_3 \times \ess^7$ throat. On the other hand, while probe computations in the heterotic model can be performed with no further difficulties, their stringy interpretation appears more subtle, since it would involve $\text{NS}5$-branes or non-supersymmetric dualities. Nevertheless, probe-brane calculations in this setting yield attractive potentials for $8$-branes and fundamental strings, as in the orientifold models, while $\text{NS}5$-branes are repelled. In addition, in some cases the potential scales with a positive power of $g_s$. Otherwise, the instability appears to be still under control, since probes would reach the strong-coupling regions in a parametrically large time for $g_s \ll 1$.

\subsubsection{Brane probes in the Dudas-Mourad geometry}\label{sec:probe_dm}

Let us consider a stack of $N_p$ probe D$p$-branes, with $p \leq 8$, embedded in the Dudas-Mourad geometry parallel to the $8$-branes\footnote{While the number $N_8$ of $8$-branes does not appear explicitly in the solution, there is a single free parameter $g_s \equiv e^{\Phi_0}$, which one could expect to be determined by $N_8$ analogously to the extremal case, with $g_s \ll 1$ for $N_8 \gg 1$.}, at a position $y$ in the notation of Section~\ref{Section2}. We work in units where $\alpha_{\text{O}} = 1$ for clarity. This setting appears to be under control as long as the (string-frame) geodesic coordinate
\begin{eqaed}\label{eq:geodesic_r}
	r \equiv \frac{1}{\sqrt{g_s}} \, \int_0^y \frac{du}{u^{\frac{1}{3}}} \, e^{- \, \frac{3}{8} \, u^2}
\end{eqaed}
is far away from its endpoints $r=0$, $r=R_c$. Such an overlap regime exists provided that $g_s \equiv e^{\Phi_0} \ll 1$, and thus both curvature corrections and string loop corrections are expected to be under control.

\begin{figure}[!ht]
	\begin{center}
		\includegraphics[width=0.9\linewidth]{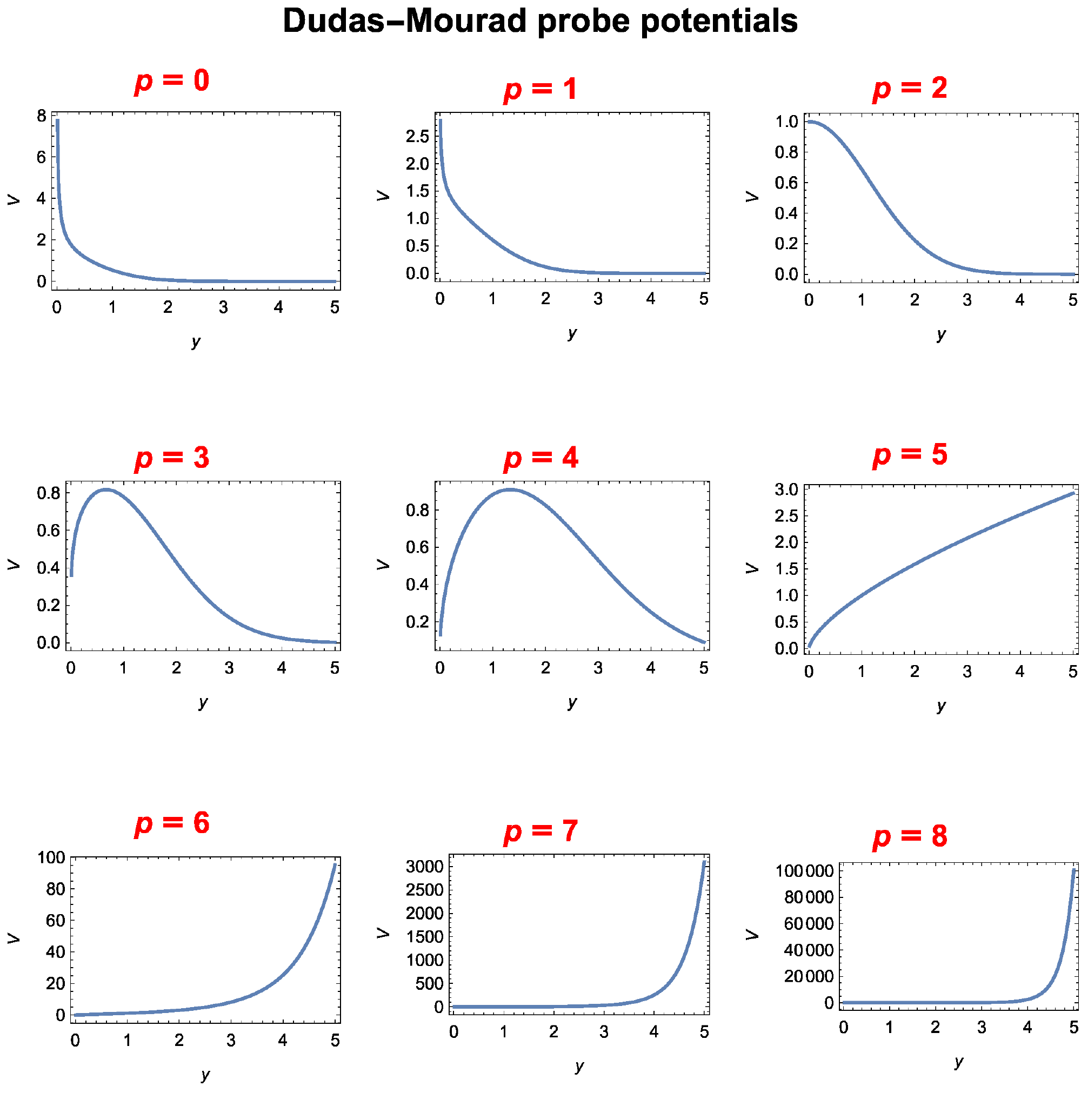}
	\end{center}
	\caption{probe potentials for $g_s = 1$ and $p \leq 8$. For $p < 3$ the probe stack is repelled by the 8-branes, while for $p > 4$ it is attracted to the $8$-branes. A string amplitude computation yields a qualitatively similar behavior, despite the string-scale breaking of supersymmetry.}
	\label{fig:probe_potential_dm_y}
\end{figure}

\begin{figure}[!ht]
	\begin{center}
		\includegraphics[width=0.9\linewidth]{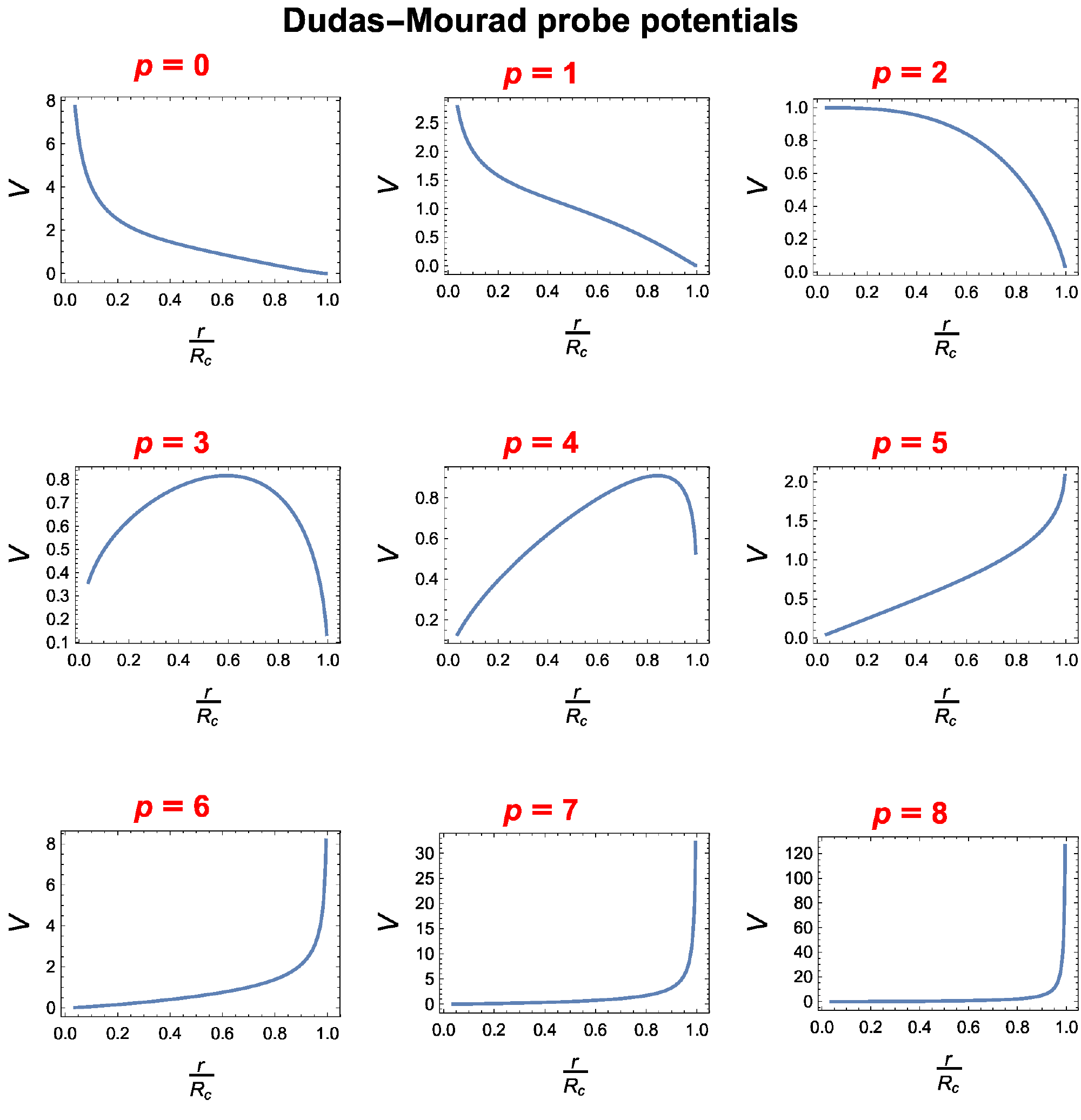}
	\end{center}
	\caption{probe potentials for $g_s = 1$ and $p \leq 8$, plotted as functions of the geodesic coordinate along the compact direction.}
	\label{fig:probe_potential_dm_r}
\end{figure}

Writing the string-frame metric as
\begin{eqaed}\label{eq:dm_metric_AB}
	ds^2_{10} = e^{2A(y)} \, dx_{1,8}^2 + e^{2B(y)} \, dy^2
\end{eqaed}
the DBI action evaluates to
\begin{eqaed}\label{eq:dbi_dm}
	S_p & = - \, N_p \, T_p \int d^{p+1}x \, e^{(p+1)A(y) - \Phi(y)} \equiv - \, N_p \, T_p \int d^{p+1}x \, V_{p8} \, ,
\end{eqaed}
where the probe potential per unit tension
\begin{importantbox}
	\begin{eqaed}\label{eq:probe_potential_dm}
		V_{p8} = {g_s}^{\frac{p-3}{4}} \, y^{\frac{2}{9} \left(p-2\right)} \, e^{\frac{p-5}{8} \, y^2}
	\end{eqaed}
\end{importantbox}
displays a non-trivial dependence on $p$, and is depicted in figs.~\ref{fig:probe_potential_dm_y} and~\ref{fig:probe_potential_dm_r}. Similarly, probe $\text{NS}5$-branes are subject to the potential $V_{58} = \frac{\sech^2 \, y}{\sqrt{g_s}}$. In particular, if the potential drives probes toward $y \; \to \; \infty$ it is repulsive, since the corresponding pinch-off singularity derived in the preceding sections agrees with the Dudas-Mourad geometry in this regime. All in all, for $p < 3$ probes are repelled by the $8$-branes, while for $p > 4$ they are attracted to the $8$-branes. The cases $p = 3 \, , \, 4$ feature unstable equilibria\footnote{Notice that, in the absence of fluxes, brane polarization~\cite{Myers:1999ps, Kachru:2002gs} would not suffice to stabilize these equilibria.} which appear to be within the controlled regime, but the large-separation behavior, to be compared to a string amplitude computation, appears repulsive.
As we have anticipated, the analogous computation for branes probing the back-reacted geometry sourced by other non-extremal branes appears considerably more challenging. This is due to the fact that even if the reduced dynamical system derived in the preceding sections were solved numerically in a reliable regime, the asymptotic boundary conditions corresponding to uncharged branes are not yet understood. While this is the case also for extremal branes, one can make progress observing that in the probe regime the scale of the dimensions transverse to the extremal stack should be large enough to ensure that the near-horizon limit is reliable. The exponential term in eq.~\eqref{eq:probe_potential_dm} is actually universal, since repeating the above probe-brane computation for the generic pinch-off singularity of eq.~\eqref{eq:omega0_sol} in the orientifold models\footnote{As we have discussed in Section~\ref{sec:black_branes}, the leading-order behavior of the pinch-off singularity is expected to be applicable to the non-extremal case, since it is dominated by the dilaton potential.} yields the same result, with the potential at large separation repulsive for $p < 5$ and attractive for $p > 5$, while the case $p = 5$ requires subleading, presumably power-like, terms in the metric. However, we do not expect these cases to provide reliable insights, since the pinch-off singularity lies beyond the controlled regime.

In order to verify that this construction is at least parametrically under control, one ought to verify that the probe-brane stack remains in the controlled region for parametrically large times. To this end, let us consider the reduced dynamical system that describes motion along $y$, with the initial conditions $y(0) = y_0 \, , \, \dot{y}(0) = 0$. The corresponding Lagrangian reads
\begin{eqaed}\label{eq:dbi_dm_reduced}
	\mathcal{L}_{\text{red}} = - T_p \, N_p \, V_{p8} \, \sqrt{1 - e^{2\left(B - A \right)} \, \dot{y}^2} \, ,
\end{eqaed}
and, since the corresponding Hamiltonian
\begin{eqaed}\label{eq:dbi_hamiltonian}
	H_{\text{red}} = \frac{T_p \, N_p \, V_{p8}}{ \sqrt{1 - e^{2\left(B - A \right)} \, \dot{y}^2}} = T_p \, N_p \, V_{p8}(y_0)
\end{eqaed}
is conserved, solving the equation of motion by quadrature gives
\begin{eqaed}\label{eq:time_separated}
	t = \int_{y_0}^y \frac{e^{B(u)-A(u)}}{\sqrt{1 - \left(\frac{V_{p8}(u)}{V_{p8}(y_0)}\right)^2}} \, du = g_s^{- \frac{3}{4}} \int_{y_0}^y \frac{e^{- \frac{u^2}{2}}}{u^{\frac{5}{9}} \, \sqrt{1 - \left(\frac{u}{y0}\right)^{\frac{4}{9}\left(p-2\right)} \, e^{\frac{p-5}{4} \left(u^2 - y_0^2\right)}}} \, ,
\end{eqaed}
which is indeed parametrically large in string units.

\subsubsection{String amplitude computation}\label{sec:string_amplitude}

Let us now compare the probe-brane result of eq.~\eqref{eq:probe_potential_dm} with a string amplitude computation. As we have anticipated, in the non-extremal case the relevant amplitude for the leading-order interaction between stacks of $N_p$ $\text{D}p$-branes and $N_q$ $\text{D}q$-branes\footnote{The ensuing string amplitude computation is expected to be reliable as long as $N_p$ and $N_q$ are $\mathcal{O}\!\left(1\right)$, complementary to the probe regimes $N_p \gg N_q$ and $N_p \ll N_q$.}, with $p < q$ for definiteness, is provided by the annulus amplitude, whose transverse-channel integrand in the present cases takes the form~\cite{Dudas:2001wd}
\begin{eqaed}\label{eq:A_pq}
	N_p \, N_q \, \widetilde{A}_{pq} \propto N_p \, N_q \left( V_{8-q+p} \, O_{q-p} - O_{8-q+p} \, V_{q-p} \right) \, ,
\end{eqaed}
where the characters are evaluated at $\mathfrak{q} = e^{-2\pi \ell}$ and we have omitted the overall unimportant positive normalization, which encodes the tensions and depends on whether both stacks consist of non-extremal branes or one stack consists of extremal branes. In suitable units for the transverse separation $r$ bewtween the two stacks, the potential $V_{pq}$ takes the form
\begin{eqaed}\label{eq:string_potential}
	V_{pq} \propto - \, N_p \, N_q \int_0^\infty \frac{d\ell}{\ell^{\frac{9-q}{2}}} \, \frac{\widetilde{A}_{pq}}{\eta^{8-q+p}} \left( \frac{2\eta}{\vartheta_2} \right)^{\frac{q-p}{2}} e^{- \, \frac{r^2}{\ell}} \, .
\end{eqaed}
For large $r$, the integral is dominated by the large $\ell$ region, where the integrand asymptotes to $\mathfrak{q}^{-\frac{1}{3}} \, \widetilde{A}_{pq}$, with
\begin{eqaed}\label{eq:amplitude_integrand_asymptotics}
	\widetilde{A}_{pq} & \propto V_{8-q+p} \, O_{q-p} - O_{8-q+p} \, V_{q-p}
	\sim 2 \left( 4 - q + p \right) \mathfrak{q}^{\frac{1}{3}} \, ,
\end{eqaed}
so that the overall sign of the potential is the sign of $q - p - 4$. Thus, for large $r$ and $q < 7$ one finds
\begin{importantbox}
	\begin{eqaed}\label{eq:large_r_string_potential}
		V_{pq} \propto \left(q - p - 4 \right) \frac{N_p \, N_q}{r^{7-q}} \, ,
	\end{eqaed}
\end{importantbox}
which is repulsive for $p < q - 4$ and attractive for $p > q - 4$. While the integral of eq.~\eqref{eq:string_potential} diverges for $q \geq 7$, a distributional computation for $q = 7 \, , \, 8$ yields a finite force stemming from potentials that behave as $\left(p - 3 \right) \log(r)$ and $\left(p - 4 \right) r$ respectively. Therefore, the only case that can be compared with a reliable probe-brane computation is $q = 8$, where the potential behaves as $(p - 4) \, r$ and is thus repulsive for $p < 4$ and attractive for $p > 4$, consistently with the results in the preceding section.

\subsubsection{Probe 8-branes in \texorpdfstring{$\ads \times \ess$}{AdS x S} throats}\label{sec:probe_8}

To conclude, let us thus consider $N_8$ $8$-branes embedded in the near-horizon $\ads_3 \times \ess^7$ geometries sourced by $N_1 \gg N_8$ extremal $\text{D}1$-branes in the orientifold models and, for the sake of completeness, by $N_5 \gg N_8$ $\text{NS}5$-branes in the heterotic model. Other than the interaction potential bewteen two extremal stacks, which we have computed in Section~\ref{sec:probe_branes}, this is the only case where a probe-brane potential can be reliably computed, since the $8$-branes can wrap the internal spheres without collapsing in a vanishing cycle, leaving only one dimension across which to separate from the stack. Moreover, this is the only case where computations can be performed in the opposite regime $N_1 \, , \, N_5 \ll N_8$, as we have described in Section~\ref{sec:probe_dm}. Since the $8$-branes are uncharged, the respective potentials $V_{81} \, , \, V_{85}$ arise from the DBI contribution only, and one finds
\begin{eqaed}\label{eq:8-brane_probe_potentials}
	V_{81} & = N_8 \, T_8 \, R^7 \left( \frac{L}{Z} \right)^2 \, , \qquad
	V_{85} = N_8 \, T_8 \, R^3 \left( \frac{L}{Z} \right)^6 \, , \\ 
\end{eqaed}
where we have omitted the \textit{a priori} unknown (and unimportant) scaling with $g_s$. These potentials are thus attractive, which may appear in contradiction with the results in the preceding sections, where both $\text{D}1$-branes and $\text{NS}5$-branes are repelled by the $8$-branes. However, let us observe that, since the $8$-branes wrap the internal spheres, in the large-separation regime they ought to behave as uncharged $1$-branes and $5$-branes respectively, consistently with an attractive interaction. Furthermore, when expressed in terms of the geodesic coordinate $r = L \, \log\left( \frac{Z}{L} \right)$, the potentials of eq.~\eqref{eq:8-brane_probe_potentials} decay exponentially in $r$.

All in all, the results in this section further support the idea that brane dynamics plays a crucial rôle in elucidating the fate of string models with broken supersymmetry. Whenever available, microscopic information such as the scaling of the tensions of fundamental branes and the string amplitude computation of eq.~\eqref{eq:string_potential} appear to be consistent with the low-energy effective theory introduced in Section~\ref{Section2}. The resulting picture builds an intuitive understanding of the high-energy behavior of the settings at stake, and points to some avenues to more quantitative results in this respect. In particular, the interpretation of the $\ads_3 \times \ess^7$ solution introduced in Section~\ref{Section2} as the near-horizon limit of the back-reacted geometry sourced by $\text{D}1$-branes, which subsequently nucleate and are repelled by each other, suggests that an holographic approach could expose some intriguing lessons~\cite{Antonelli:2018qwz, Basile:2020xwi}. Notwithstanding the important issue of corroborating our proposals quantitatively, based on our considerations one can build an intuitive physical picture, whereby charged branes are gradually expelled from the original stack until only a single brane remains. A world-sheet analysis of such an end-point to flux tunneling would presumably involve an analysis along the lines of~\cite{Giribet:2018ada}, albeit in the absence of supersymmetry its feasibility remains opaque.

\section{Conclusions}\label{Section8}

We can now summarize the main points that we have discussed in this review, collecting our considerations and results.

To begin with, in Section~\ref{Section1} we have presented a brief overview of three ten-dimensional string models with broken supersymmetry and their construction in terms of vacuum amplitudes. These comprise two orientifold models, the $USp(32)$ model of~\cite{Sugimoto:1999tx} and the $U(32)$ model of~\cite{Sagnotti:1995ga, Sagnotti:1996qj}, and the $SO(16) \times SO(16)$ heterotic model of~\cite{AlvarezGaume:1986jb, Dixon:1986iz}, and their perturbative spectra feature no tachyons. On the other hand, the perturbative expansion of these models around flat space-time involves gravitational tadpoles, whose back-reaction appears dramatic and is, at present, not completely under control. 

In Section~\ref{Section2} we have described a family of effective theories which describes their low-energy physics. In particular, their actions contain exponential potentials for the dilaton, whose presence tends to drive the dynamics toward runaway. In order to counteract this behavior, the resulting classical solutions that have been found entail warped space-time geometries~\cite{Dudas:2000ff, Antonelli:2019nar} or compactifications supported by fluxes~\cite{Mourad:2016xbk}. We have described in detail the Dudas-Mourad solutions of~\cite{Dudas:2000ff}, which comprise nine-dimensional static solutions and ten-dimensional cosmological solutions, and general Freund-Rubin flux compactifications, which include the $\adsts$ solutions found in~\cite{Mourad:2016xbk} and their generalizations studied in~\cite{Antonelli:2019nar}. Whenever $\ds$ solutions of this type are allowed, they \textbf{always contain instabilities} in the dilaton-radion sector, but in the string models that we have introduced in Section~\ref{Section1} they \textbf{do not arise}. In order to assess whether more general settings could give rise to $\ds$ solutions, we have considered warped flux compactifications, deriving an \textbf{expression for the space-time cosmological constant} which underlies an extended \textbf{no-go result} that we have connected to recent conjectures about $\ds$ solutions and the swampland~\cite{Obied:2018sgi, Ooguri:2018wrx}.

In Section~\ref{Section3} we have studied in detail the classical stability of the solutions discussed in Section~\ref{Section2}, deriving the linearized equations of motion for field perturbations. In particular, in the case of the Dudas-Mourad solutions we have recast the resulting equations as Schr\"odinger-like problems, whose Hamiltonians can be decomposed in terms of creation and annihilation operators. These solutions are stable at the classical level, with the exception of an intriguing logarithmic \textbf{growth of the homogeneous tensor mode in the cosmological case}~\cite{Basile:2018irz}, which we are tempted to interpret as a \textbf{tendency of space-time toward spontaneous compactification}. However, let us remark that from the perspective of the underlying string models these solutions entail sizeable curvature corrections or string loop corrections, thus potentially compromising some of these lessons. this issue does not appear to affect the $\adsts$ solutions, which for large fluxes are expected to be under control globally, but their Kaluza-Klein spectra contain \textbf{unstable modes in the (space-time) scalar sector}~\cite{Basile:2018irz} for a finite number of internal angular momenta. One can then attempt to remove them with suitable freely acting projections on the internal spheres, or choosing a different internal manifold altogether, and for the heterotic model one can achieve this with an antipodal $\mathbb{Z}_2$ projection on the internal $\ess^3$.

In Section~\ref{Section4} we have focused on some non-perturbative instabilities of the $\ads$ compactifications discussed in Section~\ref{Section2}, which undergo flux tunneling~\cite{Antonelli:2019nar} gradually discharging space-time. This process is exponentially unlikely for large fluxes, and it entails the nucleation of charged bubbles which then expand, reaching the (conformal) boundary in a finite time. Motivated by the qualitative properties of these bubbles, we have developed a picture involving fundamental branes, matching bulk gravitational computations to brane instanton computations of decay rates and deriving consistency conditions. In particular, we have found that the (oppositely charged pairs of) branes that mediate flux tunneling ought to be \textbf{$\text{D}$1-branes in the orientifold models and $\text{NS}$5-branes in the heterotic model}, but our results apply also to ``exotic'' branes~\cite{Bergshoeff:2005ac,Bergshoeff:2006gs,Bergshoeff:2011zk,Bergshoeff:2012jb,Bergshoeff:2015cba} whose tensions scales according to different powers of the string coupling.

In Section~\ref{Section5} we have kept developing the brane picture presented in Section~\ref{Section4}, studying the Lorentzian evolution undergone by branes after nucleation. In the non-supersymmetric models described in Section~\ref{Section1}, rigid fundamental branes are subject to a non-trivial potential which encodes an \textbf{enhanced charge-to-tension ratio} that is greater than its bare counterpart, thus \textbf{verifying the weak gravity conjecture in these settings}. In addition to their expansion, positively charged branes are driven toward \textbf{long-wavelength world-volume deformations}, while negatively charged branes are not affected by instabilities of this type. In addition, we have briefly commented on \textbf{dS brane-world constructions}, applying the proposal recently revisited in~\cite{Banerjee:2018qey, Banerjee:2019fzz, Banerjee:2020wix, Basile:2020mpt} to the non-supersymmetric string models discussed in Section~\ref{Section1}. Taking into account back-reactions ought to lead to the Einstein equations on the world-volume~\cite{Banerjee:2019fzz, Banerjee:2020wix} at low energies, and thus the complete effective field theory would involve gravity coupled to (non-)Abelian gauge fields and matter. Moreover, models of this type appear to accommodate massive particles of arbitrarily small, if unlikely, masses via open strings stretching between expanding branes~\cite{Basile:2020mpt}. We would like to further explore these enticing constructions. Moreover, in order to further develop the connection between the $\adsts$ solutions discussed in Section~\ref{Section2} and the corresponding branes, we have investigated in detail the \textbf{full back-reacted geometries} sources by the latter, which feature $\adsts$ as \textbf{attractive near-horizon throats} and strongly coupled regions where, classically, space-time \textbf{``pinches off'' at a finite transverse geodesic distance}. This result generalizes the analogous behavior of the static solutions of~\cite{Dudas:2000ff}, which is indeed reproduces for $8$-branes and appears to depend only on the residual symmetry left unbroken by the branes. Therefore, the forces exerted on nucleated brane stacks afford an interpretation as the force between two stacks in the probe-brane regime in which one contains significantly more branes than the other. Finally, we turned to the non-extremal case, deriving a system of dynamical equations for the \textbf{back-reaction of non-extremal branes} and studying their dynamics in some probe-brane regimes, namely $\text{D}p$-branes probing the static Dudas-Mourad geometry in the orientifold models and, in a complementary regime, $8$-branes probing the $\ads_3 \times \ess^7$ throat sourced by $\text{D}1$-branes. We have compared the resulting interaction potentials to a \textbf{string amplitude computation}, finding \textbf{qualitative agreement} whenever both results are reliable.

\paragraph{Outlook.---}\label{sec:outlook}

The results that we have discussed in this review suggest a tantalizing, if still elusive, picture of the rich dynamics that underpins supersymmetry breaking in string theory. Even on a fundamental level, the back-reaction of the gravitational degrees of freedom intrinsic to string theory appears dramatic to such an extent that \textit{bona fide} vacua seem either completely absent or necessarily strongly coupled. As a result, all the effective static space-times that we have investigated show a tendency to end in a singularity at a finite distance, and their existence appear to rest on the presence of localized sources that act as a symmetry-breaking compass. Hence, the oft-fruitful paradigm of studying a system via its effect on probe sources has proven all the more necessary in this context, and in particular, as we have described, it holds some potentially intriguing lessons to be unveiled: from a theoretical perspective, the rich dynamics of non-supersymmetric branes hints at a deeper connection with the microscopic interactions of open strings, and thus with holography, that could lead to further quantitative progress on the ultimate fate of non-supersymmetric string ``vacua''. On the other hand, from a phenomenological perspective, the very same dynamics appears to be able to accommodate naturally interesting cosmological models with a number of desired features. Indeed, the simplest configurations lead to higher-dimensional cosmologies, modified power spectra and point to a tendency toward spontaneous compactification, while more elaborate constructions lead to lower-dimensional $\ds$ brane-world scenarios. In addition, the extremely stringent constraints of T-duality in cosmological backgrounds~\cite{Veneziano:1991ek, Meissner:1991zj, Meissner:1996sa, Hohm:2015doa, Hohm:2019ccp, Hohm:2019jgu} point to another intriguing approach, which spurred its own novel line of research~\cite{Krishnan:2019mkv, Wang:2019dcj, Bernardo:2019bkz, 2020arXiv201215677C, Hronek:2020xxi, Wang:2020eln, Nunez:2020hxx, Bernardo:2020zlc, Basile:2020dzh, Basile:2021krk, Bonezzi:2021mub, Bernardo:2021xtr, Quintin:2021eup, Codina:2021cxh}. All in all, it seems clear that, among the long-standing issues with supersymmetry breaking, instabilities often arise from an attempt to force naturally dynamical systems into static configurations, while the most coveted phenomenology reflects the accelerated expansion of our universe. Therefore, embracing instabilities as a starting point in this respect could help to shed some light on these matters, which are of primary interest for applications to fundamental physics, but also of intrinsic value for a deeper understanding of string theory on a foundational level.


\appendix

\section{Tensor spherical harmonics: a primer}\label{sec:tensor_harmonics}

In this appendix we review some results that were needed for our stability analysis in Section~\ref{Section3}, starting from an ambient Euclidean space. In Section~\ref{sec:scalar_harmonics} we build scalar spherical harmonics, and in Section~\ref{sec:higher_harmonics} we extend our considerations to tensors of higher rank. The results agree with the constructions presented in~\cite{Rubin:1983be, Rubin:1984tc}\footnote{For a more recent analysis in the case of the five-sphere, see~\cite{vanNieuwenhuizen:2012zk}.}.

\subsection{Scalar spherical harmonics}\label{sec:scalar_harmonics}

Let $Y^1,\dots Y^{n+1}$ be Cartesian coordinates of $\mathbb{R}^{n+1}$, so that the unit sphere $\ess^{n}$ is described by the constraint
\begin{eqaed}\label{eq:sphere_constraint}
	\delta_{IJ} \, Y^I \, Y^J = r^2
\end{eqaed}
on the radial coordinate $r$, solved by spherical coordinates $y^i$ according to
\begin{eqaed}\label{eq:cartesian_to_spherical}
	Y^I = r \, \widehat{Y}^I(y) \, .
\end{eqaed}
The scalar spherical harmonics on $\ess^{n}$ can be conveniently constructed starting from harmonic polynomials of degree $\ell$ in the ambient Euclidean space $\mathbb{R}^{n+1}$. A harmonic polynomial of degree $\ell$ takes the form
\begin{eqaed}\label{eq:harmonic_poly}
	H_{(n)}^\ell(Y) = \alpha_{I_1 \dots I_\ell} \, Y^{I_1} \dots Y^{I_\ell} \, ,
\end{eqaed}
and is therefore determined by a completely symmetric and trace-less tensor $\alpha_{I_1\dots I_\ell}$ of rank $\ell$, as can be clearly seen applying to it the Euclidean Laplacian
\begin{eqaed}\label{eq:cartesian_laplacian}
	\nabla^2_{n+1} = \sum_{I=1}^{n+1} \frac{\partial^2}{\partial Y_I^2} \, .
\end{eqaed}
The scalar spherical harmonics ${\cal Y}_{(n)}^{I_1\dots I_\ell}$ are then defined restricting the $H_{(n)}^{\ell}(Y)$ to the unit sphere $S^n$, or equivalently as
\begin{eqaed}\label{eq:harmonic_sphere}
	H_{(n)}^\ell(\widehat{Y}(y)) = r^\ell \, \alpha_{I_1 \dots I_\ell} \, \mathcal{Y}_{(n)}^{I_1 \dots I_\ell}(y) \, .
\end{eqaed}
As a result, the Euclidean metric can be recast as
\begin{eqaed}\label{eq:metric_spherical_coords}
	ds^2_{n+1} = dr^2 + r^2 \, d\Omega_n^2 \, ,
\end{eqaed}
and the scalar Laplacian decomposes according to
\begin{eqaed}\label{eq:spherical_laplacian}
	0 = \nabla^2_{n+1} H_{(n)}^\ell(Y) = \frac{1}{r^n} \, \frac{\partial}{\partial r} \left( r^n \, \frac{\partial H_{(n)}^\ell (Y)}{\partial r} \right) + \frac{1}{r^2} \, \nabla^2_{\ess^n} \, H_{(n)}^\ell(Y) \, ,
\end{eqaed}
where
\begin{eqaed}\label{eq:harmonic_simple}
	\frac{\partial H_{(n)}^\ell(Y)}{\partial r} = \frac{\ell}{r} \, H_{(n)}^\ell(Y)
\end{eqaed}
for the homogeneous polynomials $H_{(n)}^{\ell}(Y)$. All in all
\begin{importantbox}
	\begin{eqaed}\label{eq:scalar_eigvals}
		\nabla^2_{\ess^n} \, \mathcal{Y}_{(n)}^{I_1 \dots I_\ell} = - \, \ell \left(\ell + n - 1\right) \mathcal{Y}_{(n)}^{I_1 \dots I_\ell} \, ,
	\end{eqaed}
\end{importantbox}
and the degeneracy of the scalar spherical harmonics for any given $\ell$ is the number of independent components of a corresponding completely symmetric and trace-less tensor, namely
\begin{eqaed}\label{eq:degeneracy}
	\frac{\left(n + 2\ell - 1 \right) \left(n + \ell - 2 \right)!}{\ell! \left(n-1 \right)!} \, .
\end{eqaed}

\subsection{Spherical harmonics of higher rank}\label{sec:higher_harmonics}

In discussing more general tensor harmonics, it is convenient to notice that, in the coordinate system of eq.~\eqref{eq:metric_spherical_coords}, the non-vanishing Christoffel symbols $\widetilde {\Gamma}_{IJ}^K$ for the ambient Euclidean space read
\begin{eqaed}\label{eq:ambient_christoffels}
	\widetilde{\Gamma}_{ij}^r = - \, r \, g_{ij} \, , \qquad \widetilde{\Gamma}_{jr}^i = \frac{1}{r} \, \delta_i^j \, , \qquad \widetilde{\Gamma}_{ij}^k = \Gamma_{ij}^k \, ,
\end{eqaed}
where the labels $i,j,k$ refer, as above, to the $n$-sphere, whose Christoffel symbols are denoted by $\Gamma_{ij}^k$.

The construction extends nicely to tensor spherical harmonics, which can be defined starting from generalized harmonic polynomials, with one proviso. The relation in eq.~\eqref{eq:cartesian_to_spherical} and its differentials imply that the actual spherical components of tensors carry additional factors of $r$, one for each covariant tensor index, with respect to those na\"{i}vely inherited from the Cartesian coordinates of the Euclidean ambient space, as we shall now see in detail. To begin with, vector spherical harmonics arise from one-forms in ambient space, built from harmonic polynomials of the type
\begin{eqaed}\label{eq:vector_harmonic_poly}
	H_{(n) \, J}^\ell(Y) = \alpha_{I_1 \dots I_\ell \, , \, J} \, Y^{I_1} \dots Y^{I_\ell} \, ,
\end{eqaed}
where the coefficients $\alpha_{I_1 \dots I_\ell\,,\,J}$ are completely symmetric and trace-less in any pair of the first $\ell$ indices. They are also subject to the condition
\begin{eqaed}\label{eq:tangency_condition}
	Y^J \, H_{(n) \, J}^\ell(Y) = 0 \, ,
\end{eqaed}
since the radial component, which does not pertain to the sphere $S^n$, ought to vanish. This implies that the complete symmetrization of the coefficients vanishes identically,
\begin{eqaed}\label{eq:tangency_alphas}
	\alpha_{(I_1 \dots I_\ell \, , \, J)} = 0 \, ,
\end{eqaed}
and on account of the symmetry in the first $\ell$ indices. As a result, $H_{n\,,\,J}^{\ell}(Y)$ is thus transverse in the ambient space,
\begin{eqaed}\label{eq:transverse_conditions}
	\partial^J H_{(n) \, J}^\ell(Y) = 0 \, .
\end{eqaed}
Moreover, any Euclidean vector $V$ such that $V_I\, Y^I = 0$ couples with differentials according to the general rule inherited from eq.~\eqref{eq:cartesian_to_spherical},
\begin{eqaed}\label{eq:cartesian_differential}
	V_I \, dY^I = V_I \, r \, d\widehat{Y}^I \, ,
\end{eqaed}
so that the actual sphere components, which are associated to $d \widehat{Y}^I$, include an additional power of $r$, and the vector spherical harmonics $\mathcal{Y}_{(n)\, i}^{I_1\dots I_\ell \, , \, J}$ are thus obtained from
\begin{eqaed}\label{eq:vector_harmonics}
	r^{\ell + 1} \, \mathcal{Y}_{(n) \, i}^{I_1 \dots I_\ell \, , \, J} \, \alpha_{I_1 \dots I_\ell \, , \, J} \, dy^i = r \, H_{(n) \, J}^\ell(Y) \, d\widehat{Y}^J \, .
\end{eqaed}
Therefore,
\begin{eqaed}\label{eq:radial_term}
	\nabla_r \nabla_r \left( r \, H_{(n) \, J}^\ell(Y) \right) = \left( \frac{\partial}{\partial r} - \frac{1}{r}\right)^2 \left( r \, H_{(n) \, J}^\ell(Y) \right) = \frac{\ell \left(\ell - 1 \right)}{r} \, H_{(n) \, J}^\ell(Y) \, ,
\end{eqaed}
while the remaining contributions to the Laplacian give
\begin{eqaed}\label{eq:sphere_term}
	\frac{1}{r^2} \, \nabla^2_{\ess^n} \left(r \, H_{(n) \, J}^\ell(Y)\right) + \frac{n \left(\ell + 1\right) - n - 1}{r} \left(r \, H_{(n) \, J}^\ell(Y)\right) \, ,
\end{eqaed}
taking into account the Christoffel symbols in eq.~\eqref{eq:ambient_christoffels}. Since the total Euclidean Laplacian vanishes by construction, adding eqs.~\eqref{eq:radial_term} and~\eqref{eq:sphere_term} finally results in
\begin{importantbox}
	\begin{eqaed}\label{eq:vector_eigvals}
		\nabla^2_{\ess^n} \, \mathcal{Y}_{(n) \, i}^{I_1 \dots I_\ell \, , \, J} = - \left( \ell \left( \ell + n - 1 \right) - 1 \right) \mathcal{Y}_{(n) \, i}^{I_1 \dots I_\ell \, , \, J} \, ,
	\end{eqaed}
\end{importantbox}
with $\ell \geq 1$.

In a similar fashion, the spherical harmonics $\mathcal{Y}_{(n)\, i_1\dots i_p}^{I_1\dots I_\ell \, , \, J_1 \dots J_p}$, corresponding to generic higher-rank transverse tensors which are also trace-less in any pair of symmetric $I$-indices, can be described starting from harmonic polynomials of the type $H_{(n) \, J_1\dots J_p}^{\ell}(Y)$, and satisfy
\begin{importantbox}
	\begin{eqaed}\label{eq:tensor_eigvals}
		\nabla^2_{\ess^n} \, \mathcal{Y}_{(n) \, i_1 \dots i_p}^{I_1 \dots I_\ell \, , \, J_1 \dots J_p} = - \left( \ell \left( \ell + n - 1 \right) - p \right) \mathcal{Y}_{(n) \, i_1 \dots i_p}^{I_1 \dots I_\ell \, , \, J_1 \dots J_p} \, ,
	\end{eqaed}
\end{importantbox}
with $\ell \geq p$.

In Young tableaux language, the scalar harmonics correspond to trace-less single-row diagrams of the type
\beq
\ytableausetup
{mathmode, boxsize=2em}
\begin{ytableau}
	I_1 & I_2 & \none[\dots]
	& I_l
\end{ytableau} \ \ \ ,
\eeq
while the independent vectors associated to vector harmonics correspond to two-row trace-less hooked diagrams of the type
\beq
\ytableausetup
{mathmode, boxsize=2em}
\begin{ytableau}
	I_1 & I_2 & \none[\dots]
	& I_l \\
	J
\end{ytableau} \ \ \ ,
\eeq
as we have explained. Similarly, the independent tensor perturbations of the metric in the internal space correspond to trace-less diagrams of the type
\beq
\ytableausetup
{mathmode, boxsize=2em}
\begin{ytableau}
	I_1 & I_2 & \none[\dots]
	& I_l \\
	J_1 & J_2
\end{ytableau} \ \ \ ,
\eeq
while the independent perturbations associated to a $(p+1)$-form gauge field in the internal space correspond, in general, to multi-row diagrams of the type
\beq
\ytableausetup
{mathmode, boxsize=2em}
\begin{ytableau}
	I_1 & I_2 & \none[\dots]
	& I_l \\
	J_1 \\
	\none[\vdots] \\
	J_{p+1}
\end{ytableau} \ \ \  .
\eeq
The degeneracies of these representations can be related to the corresponding Young tableaux, as in~\cite{ma2007group}. The structure of the various types of harmonics, which are genuinely different for large enough values of $n$, reflects nicely the generic absence of mixings between different classes of perturbations.


\section*{Acknowledgements}
This review originates from the author's Ph.D. Thesis, defended at Scuola Normale Superiore, Pisa under the supervision of A. Sagnotti. I would like to express my gratitude to him for his invaluable patience and profound insights. I would also like to thank C. Angelantonj, E. Dudas and J. Mourad for their enlightening feedback on my work, and A. Campoleoni, G. Bogna and S. Raucci for discussions and suggestions on this review.

%

The work of I.B.\ was supported by the Fonds de la Recherche Scientifique - FNRS under Grants No.\ F.4503.20 (``HighSpinSymm'') and T.0022.19 (``Fundamental issues in extended gravitational theories'').

\printbibliography
%
%

\end{document}

%% file: throatplot.pgf
\begingroup%
\makeatletter%
\begin{pgfpicture}%
\pgfpathrectangle{\pgfpointorigin}{\pgfqpoint{6.400000in}{4.800000in}}%
\pgfusepath{use as bounding box, clip}%
\begin{pgfscope}%
\pgfsetbuttcap%
\pgfsetmiterjoin%
\definecolor{currentfill}{rgb}{1.000000,1.000000,1.000000}%
\pgfsetfillcolor{currentfill}%
\pgfsetlinewidth{0.000000pt}%
\definecolor{currentstroke}{rgb}{1.000000,1.000000,1.000000}%
\pgfsetstrokecolor{currentstroke}%
\pgfsetdash{}{0pt}%
\pgfpathmoveto{\pgfqpoint{0.000000in}{0.000000in}}%
\pgfpathlineto{\pgfqpoint{6.400000in}{0.000000in}}%
\pgfpathlineto{\pgfqpoint{6.400000in}{4.800000in}}%
\pgfpathlineto{\pgfqpoint{0.000000in}{4.800000in}}%
\pgfpathclose%
\pgfusepath{fill}%
\end{pgfscope}%
\begin{pgfscope}%
\pgfsetbuttcap%
\pgfsetmiterjoin%
\definecolor{currentfill}{rgb}{1.000000,1.000000,1.000000}%
\pgfsetfillcolor{currentfill}%
\pgfsetlinewidth{0.000000pt}%
\definecolor{currentstroke}{rgb}{0.000000,0.000000,0.000000}%
\pgfsetstrokecolor{currentstroke}%
\pgfsetstrokeopacity{0.000000}%
\pgfsetdash{}{0pt}%
\pgfpathmoveto{\pgfqpoint{0.800000in}{0.528000in}}%
\pgfpathlineto{\pgfqpoint{5.760000in}{0.528000in}}%
\pgfpathlineto{\pgfqpoint{5.760000in}{4.224000in}}%
\pgfpathlineto{\pgfqpoint{0.800000in}{4.224000in}}%
\pgfpathclose%
\pgfusepath{fill}%
\end{pgfscope}%
\begin{pgfscope}%
\pgfsetbuttcap%
\pgfsetroundjoin%
\definecolor{currentfill}{rgb}{0.000000,0.000000,0.000000}%
\pgfsetfillcolor{currentfill}%
\pgfsetlinewidth{0.803000pt}%
\definecolor{currentstroke}{rgb}{0.000000,0.000000,0.000000}%
\pgfsetstrokecolor{currentstroke}%
\pgfsetdash{}{0pt}%
\pgfsys@defobject{currentmarker}{\pgfqpoint{0.000000in}{-0.048611in}}{\pgfqpoint{0.000000in}{0.000000in}}{%
\pgfpathmoveto{\pgfqpoint{0.000000in}{0.000000in}}%
\pgfpathlineto{\pgfqpoint{0.000000in}{-0.048611in}}%
\pgfusepath{stroke,fill}%
}%
\begin{pgfscope}%
\pgfsys@transformshift{1.120000in}{0.528000in}%
\pgfsys@useobject{currentmarker}{}%
\end{pgfscope}%
\end{pgfscope}%
\begin{pgfscope}%
\definecolor{textcolor}{rgb}{0.000000,0.000000,0.000000}%
\pgfsetstrokecolor{textcolor}%
\pgfsetfillcolor{textcolor}%
\pgftext[x=1.120000in,y=0.430778in,,top]{\color{textcolor}\rmfamily\fontsize{10.000000}{12.000000}\selectfont \(\displaystyle -7\)}%
\end{pgfscope}%
\begin{pgfscope}%
\pgfsetbuttcap%
\pgfsetroundjoin%
\definecolor{currentfill}{rgb}{0.000000,0.000000,0.000000}%
\pgfsetfillcolor{currentfill}%
\pgfsetlinewidth{0.803000pt}%
\definecolor{currentstroke}{rgb}{0.000000,0.000000,0.000000}%
\pgfsetstrokecolor{currentstroke}%
\pgfsetdash{}{0pt}%
\pgfsys@defobject{currentmarker}{\pgfqpoint{0.000000in}{-0.048611in}}{\pgfqpoint{0.000000in}{0.000000in}}{%
\pgfpathmoveto{\pgfqpoint{0.000000in}{0.000000in}}%
\pgfpathlineto{\pgfqpoint{0.000000in}{-0.048611in}}%
\pgfusepath{stroke,fill}%
}%
\begin{pgfscope}%
\pgfsys@transformshift{1.760000in}{0.528000in}%
\pgfsys@useobject{currentmarker}{}%
\end{pgfscope}%
\end{pgfscope}%
\begin{pgfscope}%
\definecolor{textcolor}{rgb}{0.000000,0.000000,0.000000}%
\pgfsetstrokecolor{textcolor}%
\pgfsetfillcolor{textcolor}%
\pgftext[x=1.760000in,y=0.430778in,,top]{\color{textcolor}\rmfamily\fontsize{10.000000}{12.000000}\selectfont \(\displaystyle -6\)}%
\end{pgfscope}%
\begin{pgfscope}%
\pgfsetbuttcap%
\pgfsetroundjoin%
\definecolor{currentfill}{rgb}{0.000000,0.000000,0.000000}%
\pgfsetfillcolor{currentfill}%
\pgfsetlinewidth{0.803000pt}%
\definecolor{currentstroke}{rgb}{0.000000,0.000000,0.000000}%
\pgfsetstrokecolor{currentstroke}%
\pgfsetdash{}{0pt}%
\pgfsys@defobject{currentmarker}{\pgfqpoint{0.000000in}{-0.048611in}}{\pgfqpoint{0.000000in}{0.000000in}}{%
\pgfpathmoveto{\pgfqpoint{0.000000in}{0.000000in}}%
\pgfpathlineto{\pgfqpoint{0.000000in}{-0.048611in}}%
\pgfusepath{stroke,fill}%
}%
\begin{pgfscope}%
\pgfsys@transformshift{2.400000in}{0.528000in}%
\pgfsys@useobject{currentmarker}{}%
\end{pgfscope}%
\end{pgfscope}%
\begin{pgfscope}%
\definecolor{textcolor}{rgb}{0.000000,0.000000,0.000000}%
\pgfsetstrokecolor{textcolor}%
\pgfsetfillcolor{textcolor}%
\pgftext[x=2.400000in,y=0.430778in,,top]{\color{textcolor}\rmfamily\fontsize{10.000000}{12.000000}\selectfont \(\displaystyle -5\)}%
\end{pgfscope}%
\begin{pgfscope}%
\pgfsetbuttcap%
\pgfsetroundjoin%
\definecolor{currentfill}{rgb}{0.000000,0.000000,0.000000}%
\pgfsetfillcolor{currentfill}%
\pgfsetlinewidth{0.803000pt}%
\definecolor{currentstroke}{rgb}{0.000000,0.000000,0.000000}%
\pgfsetstrokecolor{currentstroke}%
\pgfsetdash{}{0pt}%
\pgfsys@defobject{currentmarker}{\pgfqpoint{0.000000in}{-0.048611in}}{\pgfqpoint{0.000000in}{0.000000in}}{%
\pgfpathmoveto{\pgfqpoint{0.000000in}{0.000000in}}%
\pgfpathlineto{\pgfqpoint{0.000000in}{-0.048611in}}%
\pgfusepath{stroke,fill}%
}%
\begin{pgfscope}%
\pgfsys@transformshift{3.040000in}{0.528000in}%
\pgfsys@useobject{currentmarker}{}%
\end{pgfscope}%
\end{pgfscope}%
\begin{pgfscope}%
\definecolor{textcolor}{rgb}{0.000000,0.000000,0.000000}%
\pgfsetstrokecolor{textcolor}%
\pgfsetfillcolor{textcolor}%
\pgftext[x=3.040000in,y=0.430778in,,top]{\color{textcolor}\rmfamily\fontsize{10.000000}{12.000000}\selectfont \(\displaystyle -4\)}%
\end{pgfscope}%
\begin{pgfscope}%
\pgfsetbuttcap%
\pgfsetroundjoin%
\definecolor{currentfill}{rgb}{0.000000,0.000000,0.000000}%
\pgfsetfillcolor{currentfill}%
\pgfsetlinewidth{0.803000pt}%
\definecolor{currentstroke}{rgb}{0.000000,0.000000,0.000000}%
\pgfsetstrokecolor{currentstroke}%
\pgfsetdash{}{0pt}%
\pgfsys@defobject{currentmarker}{\pgfqpoint{0.000000in}{-0.048611in}}{\pgfqpoint{0.000000in}{0.000000in}}{%
\pgfpathmoveto{\pgfqpoint{0.000000in}{0.000000in}}%
\pgfpathlineto{\pgfqpoint{0.000000in}{-0.048611in}}%
\pgfusepath{stroke,fill}%
}%
\begin{pgfscope}%
\pgfsys@transformshift{3.680000in}{0.528000in}%
\pgfsys@useobject{currentmarker}{}%
\end{pgfscope}%
\end{pgfscope}%
\begin{pgfscope}%
\definecolor{textcolor}{rgb}{0.000000,0.000000,0.000000}%
\pgfsetstrokecolor{textcolor}%
\pgfsetfillcolor{textcolor}%
\pgftext[x=3.680000in,y=0.430778in,,top]{\color{textcolor}\rmfamily\fontsize{10.000000}{12.000000}\selectfont \(\displaystyle -3\)}%
\end{pgfscope}%
\begin{pgfscope}%
\pgfsetbuttcap%
\pgfsetroundjoin%
\definecolor{currentfill}{rgb}{0.000000,0.000000,0.000000}%
\pgfsetfillcolor{currentfill}%
\pgfsetlinewidth{0.803000pt}%
\definecolor{currentstroke}{rgb}{0.000000,0.000000,0.000000}%
\pgfsetstrokecolor{currentstroke}%
\pgfsetdash{}{0pt}%
\pgfsys@defobject{currentmarker}{\pgfqpoint{0.000000in}{-0.048611in}}{\pgfqpoint{0.000000in}{0.000000in}}{%
\pgfpathmoveto{\pgfqpoint{0.000000in}{0.000000in}}%
\pgfpathlineto{\pgfqpoint{0.000000in}{-0.048611in}}%
\pgfusepath{stroke,fill}%
}%
\begin{pgfscope}%
\pgfsys@transformshift{4.320000in}{0.528000in}%
\pgfsys@useobject{currentmarker}{}%
\end{pgfscope}%
\end{pgfscope}%
\begin{pgfscope}%
\definecolor{textcolor}{rgb}{0.000000,0.000000,0.000000}%
\pgfsetstrokecolor{textcolor}%
\pgfsetfillcolor{textcolor}%
\pgftext[x=4.320000in,y=0.430778in,,top]{\color{textcolor}\rmfamily\fontsize{10.000000}{12.000000}\selectfont \(\displaystyle -2\)}%
\end{pgfscope}%
\begin{pgfscope}%
\pgfsetbuttcap%
\pgfsetroundjoin%
\definecolor{currentfill}{rgb}{0.000000,0.000000,0.000000}%
\pgfsetfillcolor{currentfill}%
\pgfsetlinewidth{0.803000pt}%
\definecolor{currentstroke}{rgb}{0.000000,0.000000,0.000000}%
\pgfsetstrokecolor{currentstroke}%
\pgfsetdash{}{0pt}%
\pgfsys@defobject{currentmarker}{\pgfqpoint{0.000000in}{-0.048611in}}{\pgfqpoint{0.000000in}{0.000000in}}{%
\pgfpathmoveto{\pgfqpoint{0.000000in}{0.000000in}}%
\pgfpathlineto{\pgfqpoint{0.000000in}{-0.048611in}}%
\pgfusepath{stroke,fill}%
}%
\begin{pgfscope}%
\pgfsys@transformshift{4.960000in}{0.528000in}%
\pgfsys@useobject{currentmarker}{}%
\end{pgfscope}%
\end{pgfscope}%
\begin{pgfscope}%
\definecolor{textcolor}{rgb}{0.000000,0.000000,0.000000}%
\pgfsetstrokecolor{textcolor}%
\pgfsetfillcolor{textcolor}%
\pgftext[x=4.960000in,y=0.430778in,,top]{\color{textcolor}\rmfamily\fontsize{10.000000}{12.000000}\selectfont \(\displaystyle -1\)}%
\end{pgfscope}%
\begin{pgfscope}%
\pgfsetbuttcap%
\pgfsetroundjoin%
\definecolor{currentfill}{rgb}{0.000000,0.000000,0.000000}%
\pgfsetfillcolor{currentfill}%
\pgfsetlinewidth{0.803000pt}%
\definecolor{currentstroke}{rgb}{0.000000,0.000000,0.000000}%
\pgfsetstrokecolor{currentstroke}%
\pgfsetdash{}{0pt}%
\pgfsys@defobject{currentmarker}{\pgfqpoint{0.000000in}{-0.048611in}}{\pgfqpoint{0.000000in}{0.000000in}}{%
\pgfpathmoveto{\pgfqpoint{0.000000in}{0.000000in}}%
\pgfpathlineto{\pgfqpoint{0.000000in}{-0.048611in}}%
\pgfusepath{stroke,fill}%
}%
\begin{pgfscope}%
\pgfsys@transformshift{5.600000in}{0.528000in}%
\pgfsys@useobject{currentmarker}{}%
\end{pgfscope}%
\end{pgfscope}%
\begin{pgfscope}%
\definecolor{textcolor}{rgb}{0.000000,0.000000,0.000000}%
\pgfsetstrokecolor{textcolor}%
\pgfsetfillcolor{textcolor}%
\pgftext[x=5.600000in,y=0.430778in,,top]{\color{textcolor}\rmfamily\fontsize{10.000000}{12.000000}\selectfont \(\displaystyle 0\)}%
\end{pgfscope}%
\begin{pgfscope}%
\definecolor{textcolor}{rgb}{0.000000,0.000000,0.000000}%
\pgfsetstrokecolor{textcolor}%
\pgfsetfillcolor{textcolor}%
\pgftext[x=3.280000in,y=0.240809in,,top]{\color{textcolor}\rmfamily\fontsize{10.000000}{12.000000}\selectfont \(\displaystyle \log \rho\)}%
\end{pgfscope}%
\begin{pgfscope}%
\pgfsetbuttcap%
\pgfsetroundjoin%
\definecolor{currentfill}{rgb}{0.000000,0.000000,0.000000}%
\pgfsetfillcolor{currentfill}%
\pgfsetlinewidth{0.803000pt}%
\definecolor{currentstroke}{rgb}{0.000000,0.000000,0.000000}%
\pgfsetstrokecolor{currentstroke}%
\pgfsetdash{}{0pt}%
\pgfsys@defobject{currentmarker}{\pgfqpoint{-0.048611in}{0.000000in}}{\pgfqpoint{0.000000in}{0.000000in}}{%
\pgfpathmoveto{\pgfqpoint{0.000000in}{0.000000in}}%
\pgfpathlineto{\pgfqpoint{-0.048611in}{0.000000in}}%
\pgfusepath{stroke,fill}%
}%
\begin{pgfscope}%
\pgfsys@transformshift{0.800000in}{0.712800in}%
\pgfsys@useobject{currentmarker}{}%
\end{pgfscope}%
\end{pgfscope}%
\begin{pgfscope}%
\definecolor{textcolor}{rgb}{0.000000,0.000000,0.000000}%
\pgfsetstrokecolor{textcolor}%
\pgfsetfillcolor{textcolor}%
\pgftext[x=0.633333in,y=0.660038in,left,base]{\color{textcolor}\rmfamily\fontsize{10.000000}{12.000000}\selectfont \(\displaystyle 1\)}%
\end{pgfscope}%
\begin{pgfscope}%
\pgfsetbuttcap%
\pgfsetroundjoin%
\definecolor{currentfill}{rgb}{0.000000,0.000000,0.000000}%
\pgfsetfillcolor{currentfill}%
\pgfsetlinewidth{0.803000pt}%
\definecolor{currentstroke}{rgb}{0.000000,0.000000,0.000000}%
\pgfsetstrokecolor{currentstroke}%
\pgfsetdash{}{0pt}%
\pgfsys@defobject{currentmarker}{\pgfqpoint{-0.048611in}{0.000000in}}{\pgfqpoint{0.000000in}{0.000000in}}{%
\pgfpathmoveto{\pgfqpoint{0.000000in}{0.000000in}}%
\pgfpathlineto{\pgfqpoint{-0.048611in}{0.000000in}}%
\pgfusepath{stroke,fill}%
}%
\begin{pgfscope}%
\pgfsys@transformshift{0.800000in}{1.452000in}%
\pgfsys@useobject{currentmarker}{}%
\end{pgfscope}%
\end{pgfscope}%
\begin{pgfscope}%
\definecolor{textcolor}{rgb}{0.000000,0.000000,0.000000}%
\pgfsetstrokecolor{textcolor}%
\pgfsetfillcolor{textcolor}%
\pgftext[x=0.633333in,y=1.399238in,left,base]{\color{textcolor}\rmfamily\fontsize{10.000000}{12.000000}\selectfont \(\displaystyle 5\)}%
\end{pgfscope}%
\begin{pgfscope}%
\pgfsetbuttcap%
\pgfsetroundjoin%
\definecolor{currentfill}{rgb}{0.000000,0.000000,0.000000}%
\pgfsetfillcolor{currentfill}%
\pgfsetlinewidth{0.803000pt}%
\definecolor{currentstroke}{rgb}{0.000000,0.000000,0.000000}%
\pgfsetstrokecolor{currentstroke}%
\pgfsetdash{}{0pt}%
\pgfsys@defobject{currentmarker}{\pgfqpoint{-0.048611in}{0.000000in}}{\pgfqpoint{0.000000in}{0.000000in}}{%
\pgfpathmoveto{\pgfqpoint{0.000000in}{0.000000in}}%
\pgfpathlineto{\pgfqpoint{-0.048611in}{0.000000in}}%
\pgfusepath{stroke,fill}%
}%
\begin{pgfscope}%
\pgfsys@transformshift{0.800000in}{2.376000in}%
\pgfsys@useobject{currentmarker}{}%
\end{pgfscope}%
\end{pgfscope}%
\begin{pgfscope}%
\definecolor{textcolor}{rgb}{0.000000,0.000000,0.000000}%
\pgfsetstrokecolor{textcolor}%
\pgfsetfillcolor{textcolor}%
\pgftext[x=0.563888in,y=2.323238in,left,base]{\color{textcolor}\rmfamily\fontsize{10.000000}{12.000000}\selectfont \(\displaystyle 10\)}%
\end{pgfscope}%
\begin{pgfscope}%
\pgfsetbuttcap%
\pgfsetroundjoin%
\definecolor{currentfill}{rgb}{0.000000,0.000000,0.000000}%
\pgfsetfillcolor{currentfill}%
\pgfsetlinewidth{0.803000pt}%
\definecolor{currentstroke}{rgb}{0.000000,0.000000,0.000000}%
\pgfsetstrokecolor{currentstroke}%
\pgfsetdash{}{0pt}%
\pgfsys@defobject{currentmarker}{\pgfqpoint{-0.048611in}{0.000000in}}{\pgfqpoint{0.000000in}{0.000000in}}{%
\pgfpathmoveto{\pgfqpoint{0.000000in}{0.000000in}}%
\pgfpathlineto{\pgfqpoint{-0.048611in}{0.000000in}}%
\pgfusepath{stroke,fill}%
}%
\begin{pgfscope}%
\pgfsys@transformshift{0.800000in}{3.300000in}%
\pgfsys@useobject{currentmarker}{}%
\end{pgfscope}%
\end{pgfscope}%
\begin{pgfscope}%
\definecolor{textcolor}{rgb}{0.000000,0.000000,0.000000}%
\pgfsetstrokecolor{textcolor}%
\pgfsetfillcolor{textcolor}%
\pgftext[x=0.563888in,y=3.247238in,left,base]{\color{textcolor}\rmfamily\fontsize{10.000000}{12.000000}\selectfont \(\displaystyle 15\)}%
\end{pgfscope}%
\begin{pgfscope}%
\definecolor{textcolor}{rgb}{0.000000,0.000000,0.000000}%
\pgfsetstrokecolor{textcolor}%
\pgfsetfillcolor{textcolor}%
\pgftext[x=0.508333in,y=2.376000in,,bottom,rotate=90.000000]{\color{textcolor}\rmfamily\fontsize{10.000000}{12.000000}\selectfont \(\displaystyle \frac{(g_{tt})_{\mathrm{RN}}}{(g_{tt})_{\mathrm{AdS}_2 \times \mathbb{S}^2}}\)}%
\end{pgfscope}%
\begin{pgfscope}%
\pgfpathrectangle{\pgfqpoint{0.800000in}{0.528000in}}{\pgfqpoint{4.960000in}{3.696000in}}%
\pgfusepath{clip}%
\pgfsetrectcap%
\pgfsetroundjoin%
\pgfsetlinewidth{1.505625pt}%
\definecolor{currentstroke}{rgb}{0.000000,0.000000,0.000000}%
\pgfsetstrokecolor{currentstroke}%
\pgfsetdash{}{0pt}%
\pgfpathmoveto{\pgfqpoint{4.690528in}{4.234000in}}%
\pgfpathlineto{\pgfqpoint{4.696887in}{3.828905in}}%
\pgfpathlineto{\pgfqpoint{4.706977in}{3.325390in}}%
\pgfpathlineto{\pgfqpoint{4.717068in}{2.938122in}}%
\pgfpathlineto{\pgfqpoint{4.727159in}{2.633404in}}%
\pgfpathlineto{\pgfqpoint{4.737249in}{2.388990in}}%
\pgfpathlineto{\pgfqpoint{4.747340in}{2.189700in}}%
\pgfpathlineto{\pgfqpoint{4.757430in}{2.024880in}}%
\pgfpathlineto{\pgfqpoint{4.767521in}{1.886869in}}%
\pgfpathlineto{\pgfqpoint{4.777612in}{1.770039in}}%
\pgfpathlineto{\pgfqpoint{4.792748in}{1.625581in}}%
\pgfpathlineto{\pgfqpoint{4.807884in}{1.509282in}}%
\pgfpathlineto{\pgfqpoint{4.823020in}{1.414124in}}%
\pgfpathlineto{\pgfqpoint{4.838156in}{1.335169in}}%
\pgfpathlineto{\pgfqpoint{4.853292in}{1.268856in}}%
\pgfpathlineto{\pgfqpoint{4.868428in}{1.212562in}}%
\pgfpathlineto{\pgfqpoint{4.883564in}{1.164321in}}%
\pgfpathlineto{\pgfqpoint{4.898700in}{1.122631in}}%
\pgfpathlineto{\pgfqpoint{4.913836in}{1.086330in}}%
\pgfpathlineto{\pgfqpoint{4.928972in}{1.054508in}}%
\pgfpathlineto{\pgfqpoint{4.949153in}{1.017814in}}%
\pgfpathlineto{\pgfqpoint{4.969334in}{0.986477in}}%
\pgfpathlineto{\pgfqpoint{4.989515in}{0.959482in}}%
\pgfpathlineto{\pgfqpoint{5.009697in}{0.936050in}}%
\pgfpathlineto{\pgfqpoint{5.029878in}{0.915570in}}%
\pgfpathlineto{\pgfqpoint{5.055105in}{0.893398in}}%
\pgfpathlineto{\pgfqpoint{5.080331in}{0.874357in}}%
\pgfpathlineto{\pgfqpoint{5.105558in}{0.857881in}}%
\pgfpathlineto{\pgfqpoint{5.135830in}{0.840881in}}%
\pgfpathlineto{\pgfqpoint{5.166102in}{0.826353in}}%
\pgfpathlineto{\pgfqpoint{5.201419in}{0.811931in}}%
\pgfpathlineto{\pgfqpoint{5.241782in}{0.798122in}}%
\pgfpathlineto{\pgfqpoint{5.287190in}{0.785278in}}%
\pgfpathlineto{\pgfqpoint{5.337643in}{0.773611in}}%
\pgfpathlineto{\pgfqpoint{5.393142in}{0.763223in}}%
\pgfpathlineto{\pgfqpoint{5.453686in}{0.754131in}}%
\pgfpathlineto{\pgfqpoint{5.524320in}{0.745762in}}%
\pgfpathlineto{\pgfqpoint{5.600000in}{0.738801in}}%
\pgfpathlineto{\pgfqpoint{5.600000in}{0.738801in}}%
\pgfusepath{stroke}%
\end{pgfscope}%
\begin{pgfscope}%
\pgfpathrectangle{\pgfqpoint{0.800000in}{0.528000in}}{\pgfqpoint{4.960000in}{3.696000in}}%
\pgfusepath{clip}%
\pgfsetrectcap%
\pgfsetroundjoin%
\pgfsetlinewidth{1.505625pt}%
\definecolor{currentstroke}{rgb}{0.000000,0.000000,0.000000}%
\pgfsetstrokecolor{currentstroke}%
\pgfsetdash{}{0pt}%
\pgfpathmoveto{\pgfqpoint{3.985993in}{4.234000in}}%
\pgfpathlineto{\pgfqpoint{3.993749in}{3.710004in}}%
\pgfpathlineto{\pgfqpoint{4.002293in}{3.266225in}}%
\pgfpathlineto{\pgfqpoint{4.010837in}{2.917173in}}%
\pgfpathlineto{\pgfqpoint{4.019381in}{2.637297in}}%
\pgfpathlineto{\pgfqpoint{4.027925in}{2.409169in}}%
\pgfpathlineto{\pgfqpoint{4.036468in}{2.220558in}}%
\pgfpathlineto{\pgfqpoint{4.045012in}{2.062673in}}%
\pgfpathlineto{\pgfqpoint{4.053556in}{1.929056in}}%
\pgfpathlineto{\pgfqpoint{4.062100in}{1.814876in}}%
\pgfpathlineto{\pgfqpoint{4.070644in}{1.716460in}}%
\pgfpathlineto{\pgfqpoint{4.087732in}{1.556181in}}%
\pgfpathlineto{\pgfqpoint{4.104820in}{1.432045in}}%
\pgfpathlineto{\pgfqpoint{4.121907in}{1.333731in}}%
\pgfpathlineto{\pgfqpoint{4.138995in}{1.254393in}}%
\pgfpathlineto{\pgfqpoint{4.156083in}{1.189337in}}%
\pgfpathlineto{\pgfqpoint{4.173171in}{1.135253in}}%
\pgfpathlineto{\pgfqpoint{4.190258in}{1.089751in}}%
\pgfpathlineto{\pgfqpoint{4.207346in}{1.051065in}}%
\pgfpathlineto{\pgfqpoint{4.224434in}{1.017870in}}%
\pgfpathlineto{\pgfqpoint{4.241522in}{0.989151in}}%
\pgfpathlineto{\pgfqpoint{4.258610in}{0.964123in}}%
\pgfpathlineto{\pgfqpoint{4.275697in}{0.942167in}}%
\pgfpathlineto{\pgfqpoint{4.292785in}{0.922792in}}%
\pgfpathlineto{\pgfqpoint{4.318417in}{0.897725in}}%
\pgfpathlineto{\pgfqpoint{4.344048in}{0.876555in}}%
\pgfpathlineto{\pgfqpoint{4.369680in}{0.858510in}}%
\pgfpathlineto{\pgfqpoint{4.395312in}{0.843001in}}%
\pgfpathlineto{\pgfqpoint{4.429487in}{0.825499in}}%
\pgfpathlineto{\pgfqpoint{4.463663in}{0.810899in}}%
\pgfpathlineto{\pgfqpoint{4.497838in}{0.798602in}}%
\pgfpathlineto{\pgfqpoint{4.540558in}{0.785796in}}%
\pgfpathlineto{\pgfqpoint{4.583277in}{0.775241in}}%
\pgfpathlineto{\pgfqpoint{4.634541in}{0.764882in}}%
\pgfpathlineto{\pgfqpoint{4.694348in}{0.755239in}}%
\pgfpathlineto{\pgfqpoint{4.762699in}{0.746631in}}%
\pgfpathlineto{\pgfqpoint{4.839594in}{0.739212in}}%
\pgfpathlineto{\pgfqpoint{4.933577in}{0.732479in}}%
\pgfpathlineto{\pgfqpoint{5.044647in}{0.726821in}}%
\pgfpathlineto{\pgfqpoint{5.181349in}{0.722123in}}%
\pgfpathlineto{\pgfqpoint{5.360771in}{0.718310in}}%
\pgfpathlineto{\pgfqpoint{5.600000in}{0.715555in}}%
\pgfpathlineto{\pgfqpoint{5.600000in}{0.715555in}}%
\pgfusepath{stroke}%
\end{pgfscope}%
\begin{pgfscope}%
\pgfpathrectangle{\pgfqpoint{0.800000in}{0.528000in}}{\pgfqpoint{4.960000in}{3.696000in}}%
\pgfusepath{clip}%
\pgfsetrectcap%
\pgfsetroundjoin%
\pgfsetlinewidth{1.505625pt}%
\definecolor{currentstroke}{rgb}{0.000000,0.000000,0.000000}%
\pgfsetstrokecolor{currentstroke}%
\pgfsetdash{}{0pt}%
\pgfpathmoveto{\pgfqpoint{3.250336in}{4.234000in}}%
\pgfpathlineto{\pgfqpoint{3.252806in}{4.013486in}}%
\pgfpathlineto{\pgfqpoint{3.265031in}{3.293442in}}%
\pgfpathlineto{\pgfqpoint{3.277256in}{2.791837in}}%
\pgfpathlineto{\pgfqpoint{3.289481in}{2.427401in}}%
\pgfpathlineto{\pgfqpoint{3.301706in}{2.153619in}}%
\pgfpathlineto{\pgfqpoint{3.313931in}{1.942261in}}%
\pgfpathlineto{\pgfqpoint{3.326156in}{1.775368in}}%
\pgfpathlineto{\pgfqpoint{3.338381in}{1.641051in}}%
\pgfpathlineto{\pgfqpoint{3.350606in}{1.531182in}}%
\pgfpathlineto{\pgfqpoint{3.362831in}{1.440042in}}%
\pgfpathlineto{\pgfqpoint{3.375056in}{1.363508in}}%
\pgfpathlineto{\pgfqpoint{3.387281in}{1.298546in}}%
\pgfpathlineto{\pgfqpoint{3.399506in}{1.242880in}}%
\pgfpathlineto{\pgfqpoint{3.411731in}{1.194775in}}%
\pgfpathlineto{\pgfqpoint{3.423956in}{1.152887in}}%
\pgfpathlineto{\pgfqpoint{3.436181in}{1.116163in}}%
\pgfpathlineto{\pgfqpoint{3.448406in}{1.083767in}}%
\pgfpathlineto{\pgfqpoint{3.460631in}{1.055029in}}%
\pgfpathlineto{\pgfqpoint{3.472856in}{1.029405in}}%
\pgfpathlineto{\pgfqpoint{3.497306in}{0.985798in}}%
\pgfpathlineto{\pgfqpoint{3.521756in}{0.950235in}}%
\pgfpathlineto{\pgfqpoint{3.546205in}{0.920820in}}%
\pgfpathlineto{\pgfqpoint{3.570655in}{0.896194in}}%
\pgfpathlineto{\pgfqpoint{3.595105in}{0.875359in}}%
\pgfpathlineto{\pgfqpoint{3.619555in}{0.857572in}}%
\pgfpathlineto{\pgfqpoint{3.644005in}{0.842262in}}%
\pgfpathlineto{\pgfqpoint{3.668455in}{0.828993in}}%
\pgfpathlineto{\pgfqpoint{3.705130in}{0.812180in}}%
\pgfpathlineto{\pgfqpoint{3.741805in}{0.798317in}}%
\pgfpathlineto{\pgfqpoint{3.778480in}{0.786768in}}%
\pgfpathlineto{\pgfqpoint{3.827380in}{0.774171in}}%
\pgfpathlineto{\pgfqpoint{3.876280in}{0.764046in}}%
\pgfpathlineto{\pgfqpoint{3.937404in}{0.754008in}}%
\pgfpathlineto{\pgfqpoint{3.998529in}{0.746158in}}%
\pgfpathlineto{\pgfqpoint{4.071879in}{0.738868in}}%
\pgfpathlineto{\pgfqpoint{4.157454in}{0.732502in}}%
\pgfpathlineto{\pgfqpoint{4.267479in}{0.726676in}}%
\pgfpathlineto{\pgfqpoint{4.401953in}{0.721939in}}%
\pgfpathlineto{\pgfqpoint{4.573103in}{0.718239in}}%
\pgfpathlineto{\pgfqpoint{4.817602in}{0.715436in}}%
\pgfpathlineto{\pgfqpoint{5.184351in}{0.713711in}}%
\pgfpathlineto{\pgfqpoint{5.600000in}{0.713077in}}%
\pgfpathlineto{\pgfqpoint{5.600000in}{0.713077in}}%
\pgfusepath{stroke}%
\end{pgfscope}%
\begin{pgfscope}%
\pgfpathrectangle{\pgfqpoint{0.800000in}{0.528000in}}{\pgfqpoint{4.960000in}{3.696000in}}%
\pgfusepath{clip}%
\pgfsetrectcap%
\pgfsetroundjoin%
\pgfsetlinewidth{1.505625pt}%
\definecolor{currentstroke}{rgb}{0.000000,0.000000,0.000000}%
\pgfsetstrokecolor{currentstroke}%
\pgfsetdash{}{0pt}%
\pgfpathmoveto{\pgfqpoint{2.512732in}{4.234000in}}%
\pgfpathlineto{\pgfqpoint{2.526389in}{3.334750in}}%
\pgfpathlineto{\pgfqpoint{2.542314in}{2.688883in}}%
\pgfpathlineto{\pgfqpoint{2.558240in}{2.260979in}}%
\pgfpathlineto{\pgfqpoint{2.574165in}{1.961704in}}%
\pgfpathlineto{\pgfqpoint{2.590090in}{1.743462in}}%
\pgfpathlineto{\pgfqpoint{2.606016in}{1.578942in}}%
\pgfpathlineto{\pgfqpoint{2.621941in}{1.451529in}}%
\pgfpathlineto{\pgfqpoint{2.637867in}{1.350625in}}%
\pgfpathlineto{\pgfqpoint{2.653792in}{1.269200in}}%
\pgfpathlineto{\pgfqpoint{2.669718in}{1.202433in}}%
\pgfpathlineto{\pgfqpoint{2.685643in}{1.146928in}}%
\pgfpathlineto{\pgfqpoint{2.701569in}{1.100229in}}%
\pgfpathlineto{\pgfqpoint{2.717494in}{1.060524in}}%
\pgfpathlineto{\pgfqpoint{2.733419in}{1.026453in}}%
\pgfpathlineto{\pgfqpoint{2.749345in}{0.996974in}}%
\pgfpathlineto{\pgfqpoint{2.765270in}{0.971282in}}%
\pgfpathlineto{\pgfqpoint{2.781196in}{0.948741in}}%
\pgfpathlineto{\pgfqpoint{2.797121in}{0.928848in}}%
\pgfpathlineto{\pgfqpoint{2.813047in}{0.911196in}}%
\pgfpathlineto{\pgfqpoint{2.828972in}{0.895458in}}%
\pgfpathlineto{\pgfqpoint{2.860823in}{0.868687in}}%
\pgfpathlineto{\pgfqpoint{2.892674in}{0.846885in}}%
\pgfpathlineto{\pgfqpoint{2.924525in}{0.828894in}}%
\pgfpathlineto{\pgfqpoint{2.956376in}{0.813879in}}%
\pgfpathlineto{\pgfqpoint{2.988227in}{0.801228in}}%
\pgfpathlineto{\pgfqpoint{3.020078in}{0.790479in}}%
\pgfpathlineto{\pgfqpoint{3.067854in}{0.777175in}}%
\pgfpathlineto{\pgfqpoint{3.115630in}{0.766500in}}%
\pgfpathlineto{\pgfqpoint{3.163407in}{0.757837in}}%
\pgfpathlineto{\pgfqpoint{3.227108in}{0.748665in}}%
\pgfpathlineto{\pgfqpoint{3.306736in}{0.740016in}}%
\pgfpathlineto{\pgfqpoint{3.386363in}{0.733607in}}%
\pgfpathlineto{\pgfqpoint{3.481916in}{0.727989in}}%
\pgfpathlineto{\pgfqpoint{3.609319in}{0.722877in}}%
\pgfpathlineto{\pgfqpoint{3.768574in}{0.718900in}}%
\pgfpathlineto{\pgfqpoint{3.991530in}{0.715863in}}%
\pgfpathlineto{\pgfqpoint{4.325964in}{0.713914in}}%
\pgfpathlineto{\pgfqpoint{4.947057in}{0.712980in}}%
\pgfpathlineto{\pgfqpoint{5.600000in}{0.712828in}}%
\pgfpathlineto{\pgfqpoint{5.600000in}{0.712828in}}%
\pgfusepath{stroke}%
\end{pgfscope}%
\begin{pgfscope}%
\pgfpathrectangle{\pgfqpoint{0.800000in}{0.528000in}}{\pgfqpoint{4.960000in}{3.696000in}}%
\pgfusepath{clip}%
\pgfsetrectcap%
\pgfsetroundjoin%
\pgfsetlinewidth{1.505625pt}%
\definecolor{currentstroke}{rgb}{0.000000,0.000000,0.000000}%
\pgfsetstrokecolor{currentstroke}%
\pgfsetdash{}{0pt}%
\pgfpathmoveto{\pgfqpoint{1.775888in}{4.234000in}}%
\pgfpathlineto{\pgfqpoint{1.792191in}{3.187291in}}%
\pgfpathlineto{\pgfqpoint{1.811819in}{2.484575in}}%
\pgfpathlineto{\pgfqpoint{1.831447in}{2.050044in}}%
\pgfpathlineto{\pgfqpoint{1.851075in}{1.761103in}}%
\pgfpathlineto{\pgfqpoint{1.870703in}{1.558345in}}%
\pgfpathlineto{\pgfqpoint{1.890331in}{1.410042in}}%
\pgfpathlineto{\pgfqpoint{1.909959in}{1.297944in}}%
\pgfpathlineto{\pgfqpoint{1.929587in}{1.210925in}}%
\pgfpathlineto{\pgfqpoint{1.949214in}{1.141868in}}%
\pgfpathlineto{\pgfqpoint{1.968842in}{1.086046in}}%
\pgfpathlineto{\pgfqpoint{1.988470in}{1.040208in}}%
\pgfpathlineto{\pgfqpoint{2.008098in}{1.002058in}}%
\pgfpathlineto{\pgfqpoint{2.027726in}{0.969935in}}%
\pgfpathlineto{\pgfqpoint{2.047354in}{0.942608in}}%
\pgfpathlineto{\pgfqpoint{2.066982in}{0.919154in}}%
\pgfpathlineto{\pgfqpoint{2.086610in}{0.898862in}}%
\pgfpathlineto{\pgfqpoint{2.106237in}{0.881183in}}%
\pgfpathlineto{\pgfqpoint{2.125865in}{0.865683in}}%
\pgfpathlineto{\pgfqpoint{2.145493in}{0.852016in}}%
\pgfpathlineto{\pgfqpoint{2.165121in}{0.839904in}}%
\pgfpathlineto{\pgfqpoint{2.204377in}{0.819479in}}%
\pgfpathlineto{\pgfqpoint{2.243633in}{0.803036in}}%
\pgfpathlineto{\pgfqpoint{2.282888in}{0.789621in}}%
\pgfpathlineto{\pgfqpoint{2.322144in}{0.778553in}}%
\pgfpathlineto{\pgfqpoint{2.361400in}{0.769336in}}%
\pgfpathlineto{\pgfqpoint{2.420283in}{0.758193in}}%
\pgfpathlineto{\pgfqpoint{2.479167in}{0.749497in}}%
\pgfpathlineto{\pgfqpoint{2.557679in}{0.740668in}}%
\pgfpathlineto{\pgfqpoint{2.636190in}{0.734119in}}%
\pgfpathlineto{\pgfqpoint{2.734330in}{0.728170in}}%
\pgfpathlineto{\pgfqpoint{2.852097in}{0.723264in}}%
\pgfpathlineto{\pgfqpoint{3.009120in}{0.719127in}}%
\pgfpathlineto{\pgfqpoint{3.225027in}{0.716004in}}%
\pgfpathlineto{\pgfqpoint{3.539073in}{0.714008in}}%
\pgfpathlineto{\pgfqpoint{4.108281in}{0.713013in}}%
\pgfpathlineto{\pgfqpoint{5.600000in}{0.712803in}}%
\pgfpathlineto{\pgfqpoint{5.600000in}{0.712803in}}%
\pgfusepath{stroke}%
\end{pgfscope}%
\begin{pgfscope}%
\pgfpathrectangle{\pgfqpoint{0.800000in}{0.528000in}}{\pgfqpoint{4.960000in}{3.696000in}}%
\pgfusepath{clip}%
\pgfsetrectcap%
\pgfsetroundjoin%
\pgfsetlinewidth{1.505625pt}%
\definecolor{currentstroke}{rgb}{0.000000,0.000000,0.000000}%
\pgfsetstrokecolor{currentstroke}%
\pgfsetdash{}{0pt}%
\pgfpathmoveto{\pgfqpoint{1.040565in}{4.234000in}}%
\pgfpathlineto{\pgfqpoint{1.050551in}{3.419624in}}%
\pgfpathlineto{\pgfqpoint{1.073882in}{2.510614in}}%
\pgfpathlineto{\pgfqpoint{1.097212in}{2.001704in}}%
\pgfpathlineto{\pgfqpoint{1.120543in}{1.685900in}}%
\pgfpathlineto{\pgfqpoint{1.143873in}{1.475245in}}%
\pgfpathlineto{\pgfqpoint{1.167204in}{1.327015in}}%
\pgfpathlineto{\pgfqpoint{1.190534in}{1.218337in}}%
\pgfpathlineto{\pgfqpoint{1.213865in}{1.136027in}}%
\pgfpathlineto{\pgfqpoint{1.237195in}{1.072028in}}%
\pgfpathlineto{\pgfqpoint{1.260526in}{1.021176in}}%
\pgfpathlineto{\pgfqpoint{1.283856in}{0.980033in}}%
\pgfpathlineto{\pgfqpoint{1.307187in}{0.946232in}}%
\pgfpathlineto{\pgfqpoint{1.330517in}{0.918094in}}%
\pgfpathlineto{\pgfqpoint{1.353848in}{0.894406in}}%
\pgfpathlineto{\pgfqpoint{1.377178in}{0.874266in}}%
\pgfpathlineto{\pgfqpoint{1.400509in}{0.856995in}}%
\pgfpathlineto{\pgfqpoint{1.423839in}{0.842071in}}%
\pgfpathlineto{\pgfqpoint{1.447170in}{0.829088in}}%
\pgfpathlineto{\pgfqpoint{1.470500in}{0.817726in}}%
\pgfpathlineto{\pgfqpoint{1.517161in}{0.798893in}}%
\pgfpathlineto{\pgfqpoint{1.563822in}{0.784056in}}%
\pgfpathlineto{\pgfqpoint{1.610483in}{0.772191in}}%
\pgfpathlineto{\pgfqpoint{1.657144in}{0.762587in}}%
\pgfpathlineto{\pgfqpoint{1.727136in}{0.751344in}}%
\pgfpathlineto{\pgfqpoint{1.797127in}{0.742884in}}%
\pgfpathlineto{\pgfqpoint{1.867119in}{0.736429in}}%
\pgfpathlineto{\pgfqpoint{1.960441in}{0.730050in}}%
\pgfpathlineto{\pgfqpoint{2.077093in}{0.724541in}}%
\pgfpathlineto{\pgfqpoint{2.217076in}{0.720264in}}%
\pgfpathlineto{\pgfqpoint{2.403720in}{0.716918in}}%
\pgfpathlineto{\pgfqpoint{2.683687in}{0.714506in}}%
\pgfpathlineto{\pgfqpoint{3.150297in}{0.713199in}}%
\pgfpathlineto{\pgfqpoint{4.386814in}{0.712809in}}%
\pgfpathlineto{\pgfqpoint{5.600000in}{0.712800in}}%
\pgfpathlineto{\pgfqpoint{5.600000in}{0.712800in}}%
\pgfusepath{stroke}%
\end{pgfscope}%
\begin{pgfscope}%
\pgfpathrectangle{\pgfqpoint{0.800000in}{0.528000in}}{\pgfqpoint{4.960000in}{3.696000in}}%
\pgfusepath{clip}%
\pgfsetbuttcap%
\pgfsetroundjoin%
\pgfsetlinewidth{1.003750pt}%
\definecolor{currentstroke}{rgb}{0.000000,0.000000,0.000000}%
\pgfsetstrokecolor{currentstroke}%
\pgfsetdash{{3.700000pt}{1.600000pt}}{0.000000pt}%
\pgfpathmoveto{\pgfqpoint{0.790000in}{0.712800in}}%
\pgfpathlineto{\pgfqpoint{1.760000in}{0.712800in}}%
\pgfpathlineto{\pgfqpoint{3.040000in}{0.712800in}}%
\pgfpathlineto{\pgfqpoint{4.320000in}{0.712800in}}%
\pgfpathlineto{\pgfqpoint{5.600000in}{0.712800in}}%
\pgfusepath{stroke}%
\end{pgfscope}%
\begin{pgfscope}%
\pgfsetrectcap%
\pgfsetmiterjoin%
\pgfsetlinewidth{0.803000pt}%
\definecolor{currentstroke}{rgb}{0.000000,0.000000,0.000000}%
\pgfsetstrokecolor{currentstroke}%
\pgfsetdash{}{0pt}%
\pgfpathmoveto{\pgfqpoint{0.800000in}{0.528000in}}%
\pgfpathlineto{\pgfqpoint{0.800000in}{4.224000in}}%
\pgfusepath{stroke}%
\end{pgfscope}%
\begin{pgfscope}%
\pgfsetrectcap%
\pgfsetmiterjoin%
\pgfsetlinewidth{0.803000pt}%
\definecolor{currentstroke}{rgb}{0.000000,0.000000,0.000000}%
\pgfsetstrokecolor{currentstroke}%
\pgfsetdash{}{0pt}%
\pgfpathmoveto{\pgfqpoint{5.760000in}{0.528000in}}%
\pgfpathlineto{\pgfqpoint{5.760000in}{4.224000in}}%
\pgfusepath{stroke}%
\end{pgfscope}%
\begin{pgfscope}%
\pgfsetrectcap%
\pgfsetmiterjoin%
\pgfsetlinewidth{0.803000pt}%
\definecolor{currentstroke}{rgb}{0.000000,0.000000,0.000000}%
\pgfsetstrokecolor{currentstroke}%
\pgfsetdash{}{0pt}%
\pgfpathmoveto{\pgfqpoint{0.800000in}{0.528000in}}%
\pgfpathlineto{\pgfqpoint{5.760000in}{0.528000in}}%
\pgfusepath{stroke}%
\end{pgfscope}%
\begin{pgfscope}%
\pgfsetrectcap%
\pgfsetmiterjoin%
\pgfsetlinewidth{0.803000pt}%
\definecolor{currentstroke}{rgb}{0.000000,0.000000,0.000000}%
\pgfsetstrokecolor{currentstroke}%
\pgfsetdash{}{0pt}%
\pgfpathmoveto{\pgfqpoint{0.800000in}{4.224000in}}%
\pgfpathlineto{\pgfqpoint{5.760000in}{4.224000in}}%
\pgfusepath{stroke}%
\end{pgfscope}%
\begin{pgfscope}%
\definecolor{textcolor}{rgb}{0.000000,0.000000,0.000000}%
\pgfsetstrokecolor{textcolor}%
\pgfsetfillcolor{textcolor}%
\pgftext[x=4.755980in,y=3.300000in,left,base]{\color{textcolor}\rmfamily\fontsize{10.000000}{12.000000}\selectfont \(\displaystyle 10^{-1}\)}%
\end{pgfscope}%
\begin{pgfscope}%
\definecolor{textcolor}{rgb}{0.000000,0.000000,0.000000}%
\pgfsetstrokecolor{textcolor}%
\pgfsetfillcolor{textcolor}%
\pgftext[x=4.059766in,y=3.300000in,left,base]{\color{textcolor}\rmfamily\fontsize{10.000000}{12.000000}\selectfont \(\displaystyle 10^{-2}\)}%
\end{pgfscope}%
\begin{pgfscope}%
\definecolor{textcolor}{rgb}{0.000000,0.000000,0.000000}%
\pgfsetstrokecolor{textcolor}%
\pgfsetfillcolor{textcolor}%
\pgftext[x=3.327231in,y=3.300000in,left,base]{\color{textcolor}\rmfamily\fontsize{10.000000}{12.000000}\selectfont \(\displaystyle 10^{-3}\)}%
\end{pgfscope}%
\begin{pgfscope}%
\definecolor{textcolor}{rgb}{0.000000,0.000000,0.000000}%
\pgfsetstrokecolor{textcolor}%
\pgfsetfillcolor{textcolor}%
\pgftext[x=2.590836in,y=3.300000in,left,base]{\color{textcolor}\rmfamily\fontsize{10.000000}{12.000000}\selectfont \(\displaystyle 10^{-4}\)}%
\end{pgfscope}%
\begin{pgfscope}%
\definecolor{textcolor}{rgb}{0.000000,0.000000,0.000000}%
\pgfsetstrokecolor{textcolor}%
\pgfsetfillcolor{textcolor}%
\pgftext[x=1.854052in,y=3.300000in,left,base]{\color{textcolor}\rmfamily\fontsize{10.000000}{12.000000}\selectfont \(\displaystyle 10^{-5}\)}%
\end{pgfscope}%
\begin{pgfscope}%
\definecolor{textcolor}{rgb}{0.000000,0.000000,0.000000}%
\pgfsetstrokecolor{textcolor}%
\pgfsetfillcolor{textcolor}%
\pgftext[x=1.117229in,y=3.300000in,left,base]{\color{textcolor}\rmfamily\fontsize{10.000000}{12.000000}\selectfont \(\displaystyle \epsilon = 10^{-6}\)}%
\end{pgfscope}%
\end{pgfpicture}%
\makeatother%
\endgroup%

%% file: geometry2.pdf_tex
\begingroup%
  \makeatletter%
  \providecommand\color[2][]{%
    \errmessage{(Inkscape) Color is used for the text in Inkscape, but the package 'color.sty' is not loaded}%
    \renewcommand\color[2][]{}%
  }%
  \providecommand\transparent[1]{%
    \errmessage{(Inkscape) Transparency is used (non-zero) for the text in Inkscape, but the package 'transparent.sty' is not loaded}%
    \renewcommand\transparent[1]{}%
  }%
  \providecommand\rotatebox[2]{#2}%
  \newcommand*\fsize{\dimexpr\f@size pt\relax}%
  \newcommand*\lineheight[1]{\fontsize{\fsize}{#1\fsize}\selectfont}%
  \ifx\svgwidth\undefined%
    \setlength{\unitlength}{361.85638668bp}%
    \ifx\svgscale\undefined%
      \relax%
    \else%
      \setlength{\unitlength}{\unitlength * \real{\svgscale}}%
    \fi%
  \else%
    \setlength{\unitlength}{\svgwidth}%
  \fi%
  \global\let\svgwidth\undefined%
  \global\let\svgscale\undefined%
  \makeatother%
  \begin{picture}(1,0.51711375)%
    \lineheight{1}%
    \setlength\tabcolsep{0pt}%
    \put(0,0){\includegraphics[width=\unitlength,page=1]{geometry2.pdf}}%
    \put(0.77294008,0.26901672){\color[rgb]{0,0,0}\makebox(0,0)[lt]{\begin{minipage}{0.25538106\unitlength}\centering $\phi\rightarrow \infty$\end{minipage}}}%
    \put(0.37304658,0.48571341){\color[rgb]{0,0,0}\makebox(0,0)[lt]{\begin{minipage}{0.1967578\unitlength}\centering $\int \sqrt{g_{rr}} dr$\end{minipage}}}%
    \put(-0.02475508,0.27922251){\color[rgb]{0,0,0}\makebox(0,0)[lt]{\begin{minipage}{0.25538106\unitlength}\centering $\ads_{p+2}\times\ess^q$\end{minipage}}}%
    \put(-0.00658487,0.24092555){\color[rgb]{0,0,0}\makebox(0,0)[lt]{\begin{minipage}{0.25538106\unitlength}\centering $\phi=\phi_0$\end{minipage}}}%
    \put(0.37029824,0.05353275){\color[rgb]{0,0,0}\makebox(0,0)[lt]{\begin{minipage}{0.11829335\unitlength}\centering $\ess^q$\end{minipage}}}%
    \put(0,0){\includegraphics[width=\unitlength,page=2]{geometry2.pdf}}%
  \end{picture}%
\endgroup%